\documentclass[twoside,12pt]{article}

\usepackage{graphicx}
\usepackage{amssymb}
\usepackage{amsmath}
\usepackage{dcolumn}
\usepackage{bm}
\usepackage{sidecap}

\usepackage[labelfont={normalsize},subrefformat=parens]{subfig}
\usepackage{cite} 

\newcommand{\be}{\begin{equation}}
\newcommand{\ee}{\end{equation}}
\newcommand{\beqn}{\begin{equation}}
\newcommand{\eeqn}{\end{equation}}
\newcommand{\bea}{\begin{eqnarray}}
\newcommand{\eea}{\end{eqnarray}}

\newcommand{\vlowk}{V_{{\rm low}\,k}}
\newcommand{\tlowk}{T_{{\rm low}\,k}}
\newcommand{\vnn}{V_{\rm NN}}
\newcommand{\tnn}{T_{\rm NN}}

\newcommand{\fm}{\, \text{fm}}
\newcommand{\fmi}{\, \text{fm}^{-1}}
\newcommand{\mev}{\, \text{MeV}}
\newcommand{\la}{\langle}
\newcommand{\ra}{\rangle}

\newcommand{\ddlamfsq}{\frac{d}{d\Lambda}[f^2(p)]}
\newcommand{\Hzero}{T}
\newcommand{\flow}{s}

\newcommand{\kf}{k_{\text{F}}}
\newcommand{\Vtwo}[2]{V_{#1#2}}
\newcommand{\Vthree}{V_{123}}
\newcommand{\Ttwo}[2]{T_{#1#2}}
\newcommand{\Trel}{T_{\rm rel}}

\newcommand{\energy}[1]{E_{#1}}

\newcommand{\adaggera}{a^\dagger_{\qvec} a^{{\phantom{\dagger}}}_{\qvec}}
\newcommand{\qvec}{{\bf q}}
\newcommand{\kmax}{k_{\rm max}}
\newcommand{\vsrg}{V_s}
\newcommand{\adag}{a^\dagger}
\newcommand{\nmax}{$N_{\rm max}$}

\newcommand{\bi}{\begin{itemize}}
\newcommand{\ei}{\end{itemize}}
\newcommand{\I}{\item}
\newcommand{\lm}{\Lambda}
\newcommand{\vtn}{V_{\text{3N}}}

\newcommand{\Ref}{Ref.}

\newcommand{\NthreeLO}{N$^3$LO}

\newcommand{\infm}{\, {\rm fm}^{-1}}
\newcommand{\ad}{a^{\dagger}}
\newcommand{\lb}{\Lambda_{\rm b}}
\newcommand{\ef}{\varepsilon_{\rm F}}
\newcommand{\vbar}{\overline{V}_{\rm 3N}}

\begin{document}

\title{From low-momentum interactions to nuclear structure}

\author{S.K.\ Bogner$^{1}$, R.J.\ Furnstahl$^{2}$, A. Schwenk$^{3,4,5}$\\
$^1$
National Superconducting Cyclotron Laboratory and\\
Department of Physics and Astronomy,\\
Michigan State University, East Lansing, MI 48844, USA\\
$^2$
Department of Physics, The Ohio State University, Columbus, OH 43210, USA\\
$^3$
TRIUMF, 4004 Wesbrook Mall, Vancouver, BC, V6T 2A3, Canada\\
$^4$
ExtreMe Matter Institute EMMI,\\
GSI Helmholtzzentrum f\"ur Schwerionenforschung GmbH,\\
64291 Darmstadt, Germany\\
$^5$
Institut f\"ur Kernphysik, Technische Universit\"at Darmstadt,\\
64289 Darmstadt, Germany}

\maketitle

\begin{abstract}
We present an overview of low-momentum two-nucleon and many-body                
interactions and their use in calculations of nuclei and infinite               
matter. The softening of phenomenological and effective field theory            
(EFT) potentials by renormalization group (RG) transformations that             
decouple low and high momenta leads to greatly enhanced convergence             
in few- and many-body systems while maintaining a decreasing                    
hierarchy of many-body forces. This review surveys the RG-based                 
technology and results, discusses the connections to chiral EFT,                
and clarifies various misconceptions.                                           
\end{abstract}


\tableofcontents

 
\section{Introduction}
\label{sec:intro}

A new era is dawning for the theory of
nuclear structure and reactions.  Renewed
interest in the physics of nuclei is fueled by experiments at rare
isotope beam facilities, which open the door to new regions of exotic
nuclei; by astrophysical observations and simulations of neutron stars
and supernovae, which require controlled extrapolations of the
equation of state of nucleonic matter in density, temperature, and
proton fraction; and by studies of universal physics, which unite cold
atom and dilute neutron physics~\cite{NAP:1999,NuPECC:2004,LRP:2007,NAP:2007}.
The interplay and coalescence of different
threads: rapidly increasing computational power, effective field
theory (EFT), and renormalization group (RG) transformations are enabling the
development of new many-body methods and the revival of old ones to
successfully attack these problems.

A key to optimizing computational power for describing nuclei is a
proper choice of degrees of freedom.  While there is little doubt that
quantum chromodynamics (QCD) is the correct underlying theory of
strong interactions, the \emph{efficient} low-energy degrees of
freedom for nuclear structure are not quarks and gluons, but the
colorless hadrons of traditional nuclear phenomenology.  But this
realization is not enough.  For low-energy calculations to be
computationally efficient (or even feasible in some cases) we need to
exclude or, more generally, to \emph{decouple} the high-energy degrees
of freedom.

Progress on the nuclear many-body problem has been slowed for decades
because nucleon-nucleon (NN) potentials that reproduce elastic
scattering phase shifts typically have strong short-range repulsion
and strong short-range tensor forces. The consequence is substantial
coupling to high-energy modes, which is manifested as highly
correlated many-body wave functions and highly nonperturbative few-
and many-body systems. In recent years, new approaches to nuclear
forces grounded in RG ideas and techniques have been developed. The RG
allows continuous changes in ``resolution'' that decouple the
troublesome high-momentum modes and can be used to evolve interactions
to nuclear structure energy and momentum scales while preserving
low-energy
observables~\cite{Bogner:2003wn,Bogner:2006vp,Bogner:2006pc}. Such
potentials, known generically as ``low-momentum interactions,'' are
more perturbative and generate much less correlated wave functions%
~\cite{Nogga:2004ab,Bogner:2005sn,Bogner:2005fn,Bogner:2006ai,%
Bogner:2006tw,Hagen:2007hi,Bogner:2007rx,Bacca:2009yk,Bogner:2009un,%
Hebeler:2009iv}. This
greatly simplifies the nuclear many-body problem, making structure and
reaction calculations more convergent, while variations of the
resolution provide new tools to assess theoretical errors.

In this review, we survey the technical and phenomenological aspects
of the low-momentum methods. Although there are multiple paths to
low-momentum interactions, we focus on the RG-based techniques (known
as ``$\vlowk$'' or ``SRG'' potentials), which provide new perspectives
that mesh constructively with the developments of EFT for nuclear
forces. When combined with advances in many-body methods and the
increases in computer power, EFT and RG make feasible a
controlled description, grounded in QCD symmetries, of nuclei across
the nuclear many-body landscape. At the same time, the RG approach
leads to reinterpretations of the physics or the role of different
parts of the physics, such as what causes nuclear saturation. An
unintended consequence is that many misconceptions or
misinterpretations have arisen. A principal goal of this review is to
address these.

\subsection{Nuclear forces}
\label{subsec:forces}

\begin{figure}[t]
\centering
\includegraphics*[width=6.0in,clip=]{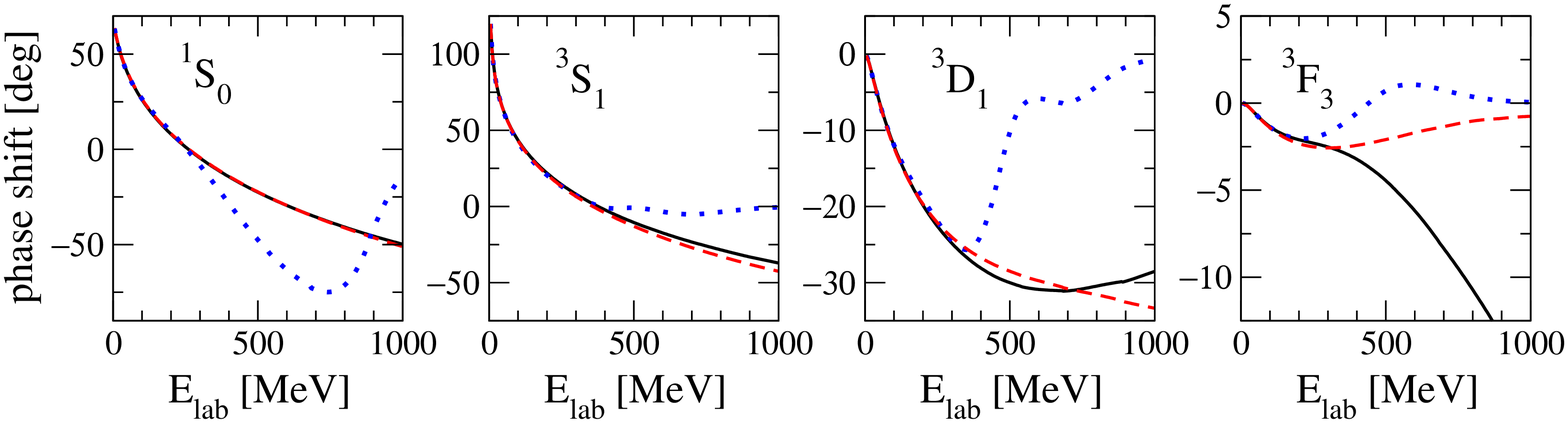}
\caption{NN phase shifts for the Argonne
$v_{18}$~\cite{Wiringa:1994wb} (solid),
CD-Bonn~\cite{Machleidt:2000ge} (dashed), and one of the chiral
N$^3$LO~\cite{Entem:2003ft} (dotted) potentials in selected channels
(using nonrelativistic kinematics).  All agree with experiment up to
about $300 \mev$. }
\label{fig:phases}
\end{figure}

Establishing an interparticle Hamiltonian, which is the most basic
precursor to many-body calculations, is a difficult and on-going
challenge for low-energy nuclear physics.  The two-body sector has
been ``solved'' in the sense that various interactions are available
that reproduce phase shifts with $\chi^2/\mbox{dof} \approx 1$ in the
elastic regime (up to roughly $300\mbox{--}350 \mev$ energy in the
laboratory frame, see Fig.~\ref{fig:phases}).  The unsettled frontier
is three- and higher-body forces, although there remain important open
questions about the systematic construction of NN potentials using EFT.

\begin{figure}[t]
 \centering
 \subfloat[][]{%
  \label{fig:nnpot-a}%
  \raisebox{.05in}{\includegraphics*[width=3.0in,clip=]{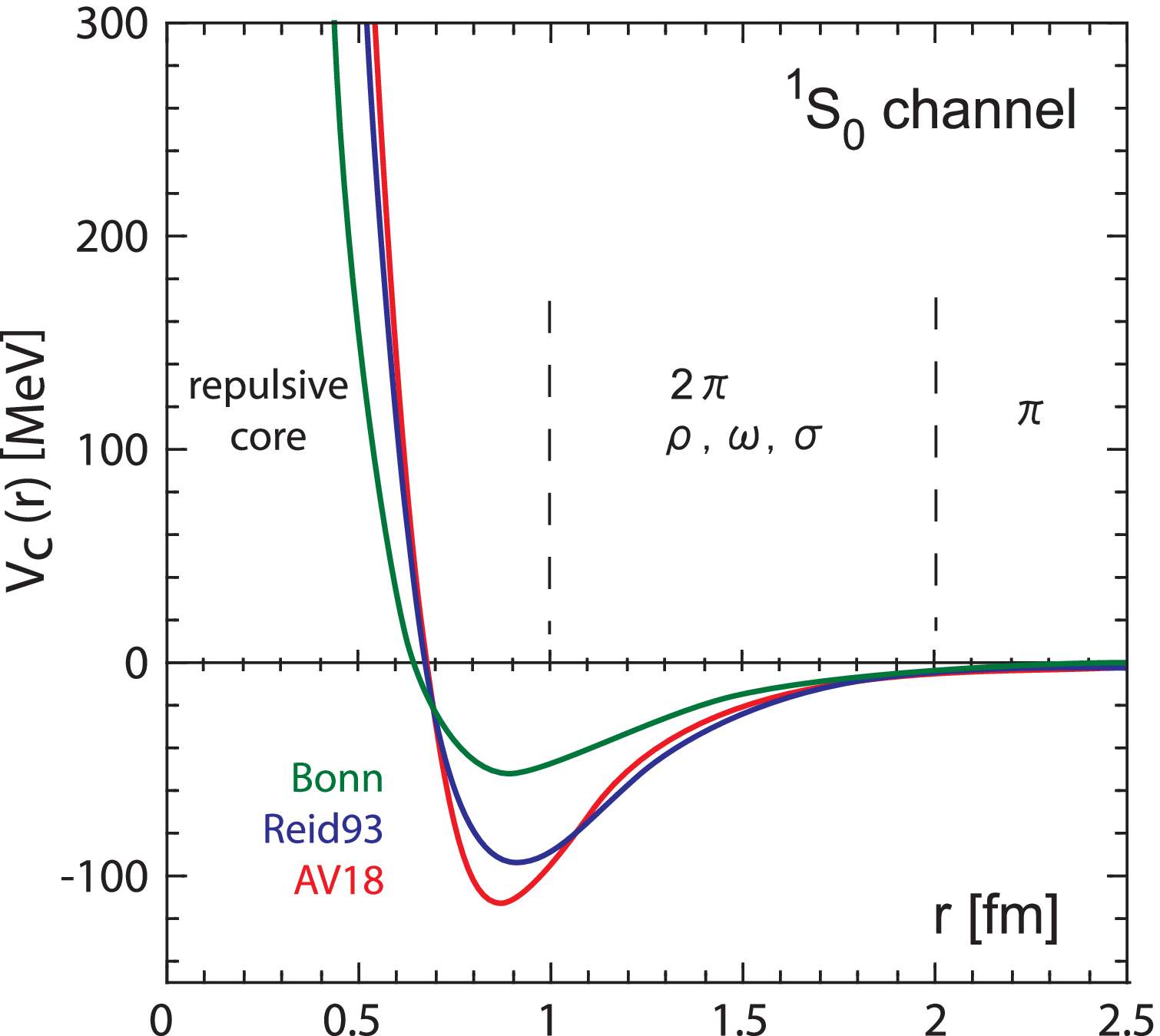}}%
 }%
 \hspace*{.4in}%
 \subfloat[][]{%
  \label{fig:nnpot-b}%
  \includegraphics[width=3.2in,clip=]{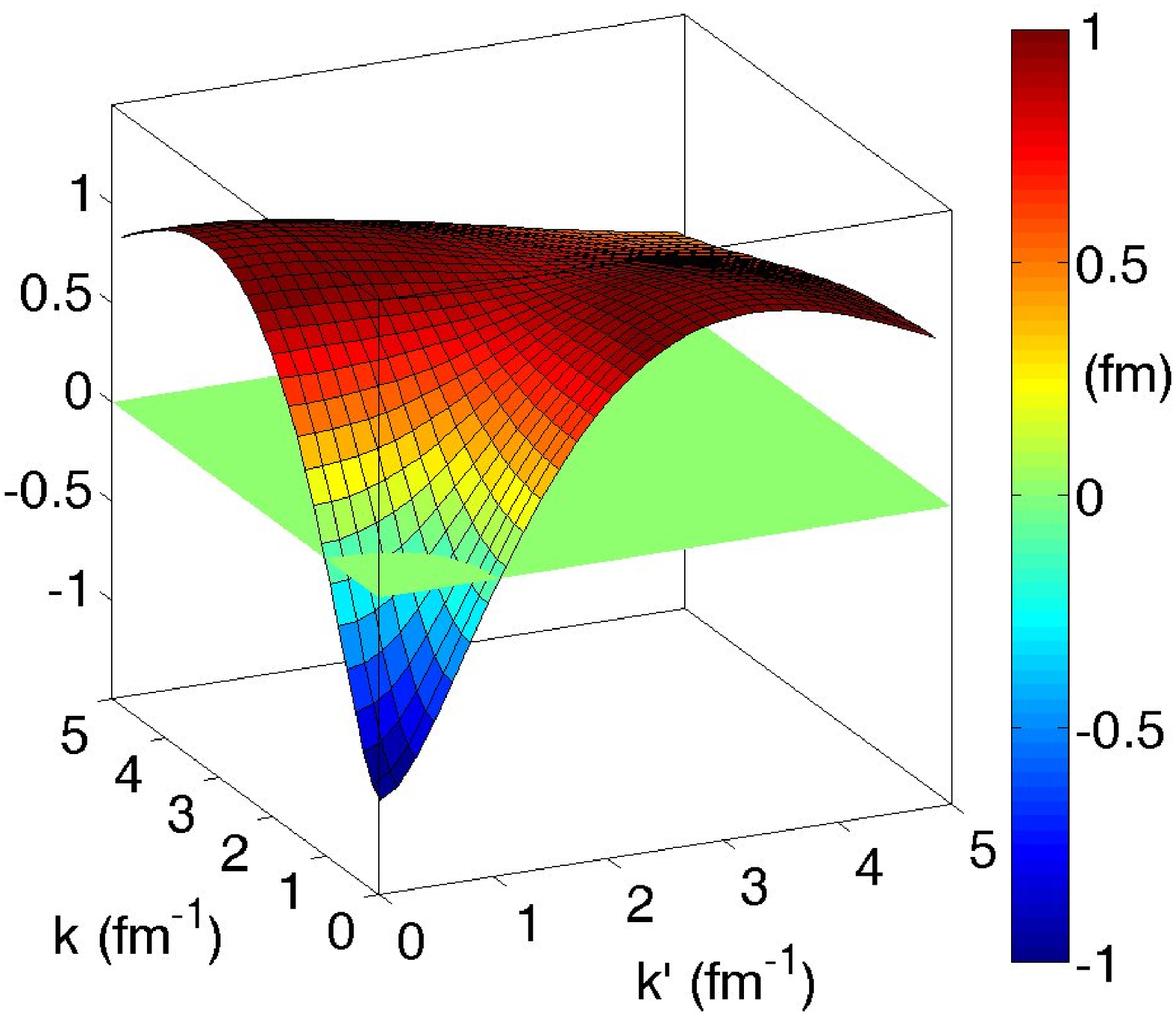}%
 }%
\caption{(a) Several phenomenological NN potentials in the
$^1$S$_0$ channel from Ref.~\cite{Aoki:2008hh}. (b)~Momentum-space
matrix elements of the Argonne $v_{18}$ (AV18) $^1$S$_0$ 
potential after Fourier (Bessel) transformation.$^1$}%
\end{figure}

Figure~\subref*{fig:nnpot-a} shows nuclear interactions in the
$^1$S$_0$ channel for several phenomenological NN potentials. The
longest range feature is one-pion exchange, which is justified by
quantum chromodynamics (via the spontaneous breaking of chiral
symmetry) and is a common feature of most potentials.  The midrange
part, which has a net attraction, has usually been associated with
two-pion and/or heavy meson exchange ($\rho$, $\omega$,
``$\sigma$''). The short-range part of the potentials in
Fig.~\subref*{fig:nnpot-a} is a
repulsive core (often called a ``hard core'').

\begin{figure}[t]
 \centering
 \subfloat[][]{%
  \label{fig:nnpotmom-a}%
  \raisebox{.2in}{\includegraphics[width=3.0in,clip=]{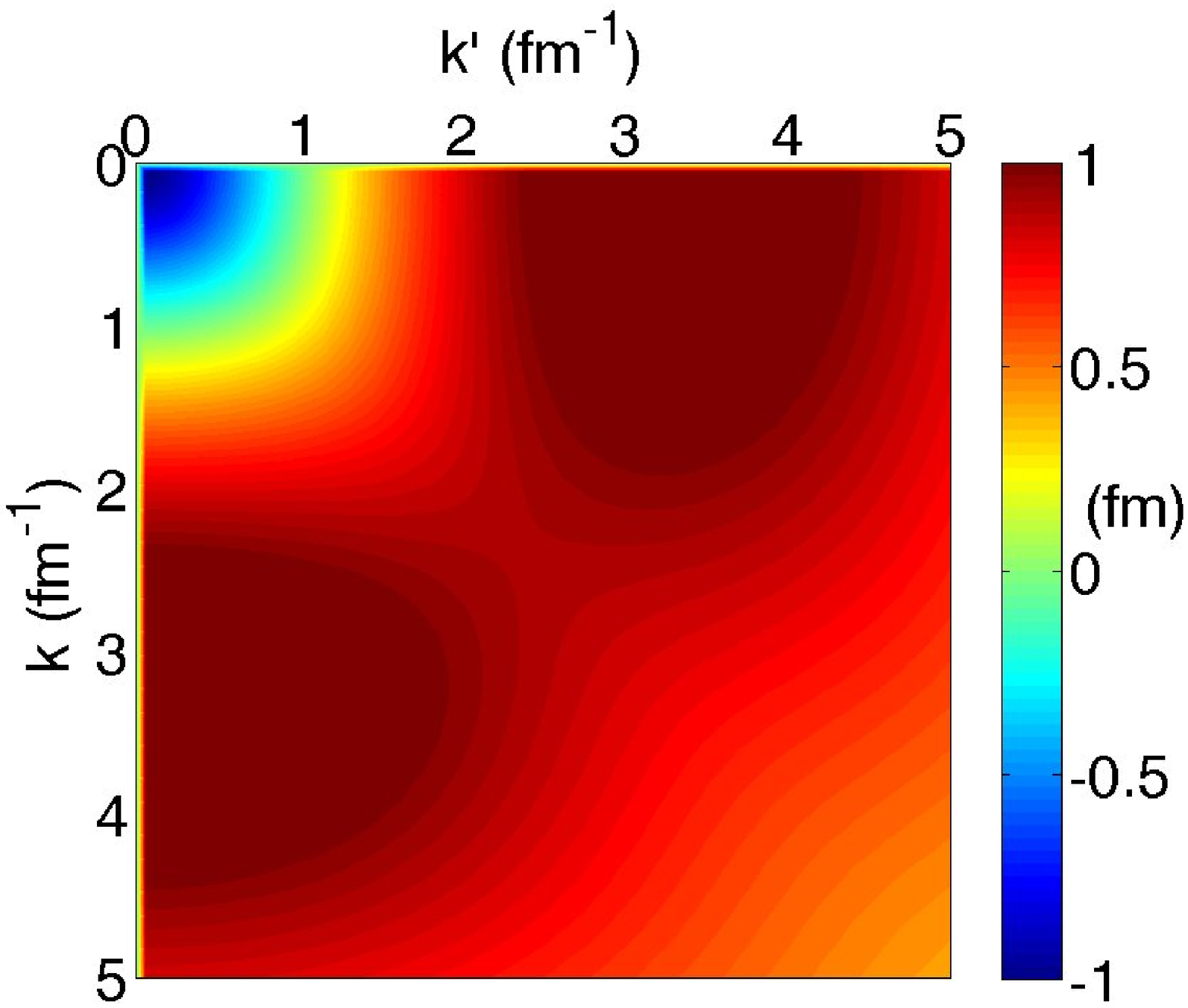}}%
 }
 \hspace*{.4in}%
 \subfloat[][]{%
  \label{fig:srcorr-a}%
  \includegraphics[width=2.8in,clip=]{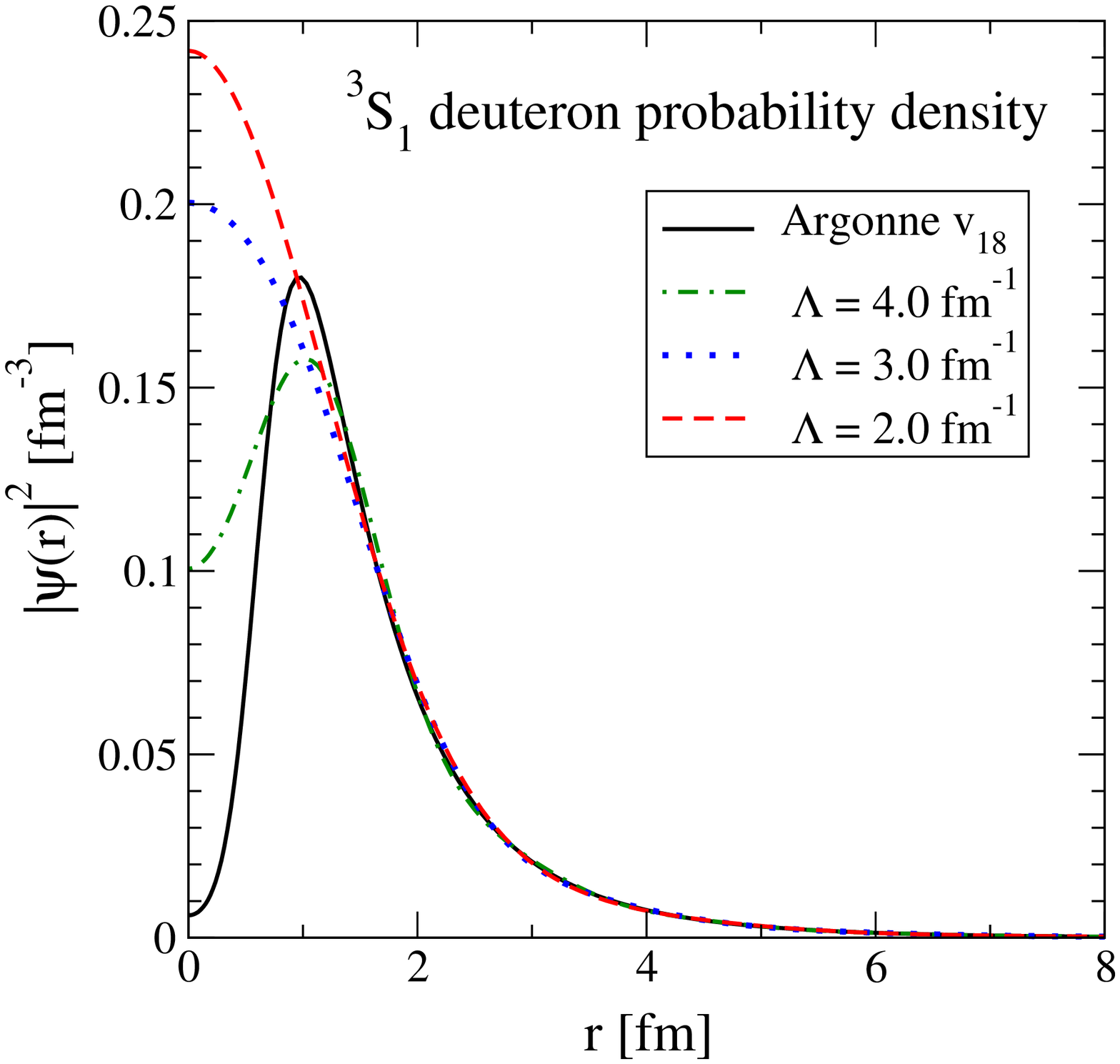}%
 }%
\caption{(a) Momentum-space matrix elements of the Argonne $v_{18}$
$^1$S$_0$ potential. (b)~Probability density for
the S-wave part of the deuteron wave function for Argonne $v_{18}$ and
smooth $\vlowk$ potentials with several cutoffs $\Lambda$. The 
probability suppression at short distances is called ``short-range
correlation''.}
\end{figure}

Nuclear structure calculations are complicated 
due to the \emph{coupling} of
low to high momenta by these potentials. This is made clear by the
Fourier transform (that is, the Bessel transform in a given partial
wave), as shown in Fig.~\subref*{fig:nnpot-b}. We feature the Argonne
$v_{18}$ potential~\cite{Wiringa:1994wb} because it is used in the
most successful high precision ($\lesssim 1\%$ accuracy) nuclear
structure calculations of nuclei with mass number
$A \leqslant 12$~\cite{Pieper:2001mp,Pieper:2004qw,Pieper:2004qh}. For our
purposes, the equivalent contour plot in Fig.~\subref*{fig:nnpotmom-a}
is a clearer representation and we use such plots throughout this
review.\footnote{In units where $\hbar = c = m = 1$ (with nucleon mass
$m$), the momentum-space potential is given in fm. In addition, we
typically express momenta in  fm$^{-1}$ (the conversion to MeV is
using $\hbar c \approx 197 \mev \fm$).} The elastic regime
for NN scattering corresponds to relative momenta $k \lesssim 2\infm$.
The strong low- to high-momentum coupling driven by the short-range
repulsion is manifested in Fig.~\subref*{fig:nnpotmom-a} by the large
regions of non-zero off-diagonal matrix elements. A consequence is a
suppression of probability in the relative wave function (``short-range
correlations''), as seen for the deuteron in Fig.~\subref*{fig:srcorr-a}.

The potentials in Fig.~\subref*{fig:nnpot-a} are partial-wave local;
that is, in each partial wave they are functions of the separation $r$
alone. This condition, which simplifies certain types of numerical
calculations,%
\footnote{For example, in current implementations of Green's Function Monte Carlo
(GFMC) calculations~\cite{Pieper:2001mp}, the potential must be
(almost) diagonal in coordinate space, such as the Argonne $v_{18}$
potential.} constrains the radial dependence to be similar to
Fig.~\subref*{fig:nnpot-a} if the potential is to reproduce elastic
phase shifts, and in particular necessitates a strong short-range
repulsion in the S-waves. The similarity of all such potentials,
perhaps combined with experience from the Coulomb potential, has led
to the (often implicit) misconception that the nuclear potential must
have this form. This prejudice has been reinforced recently by QCD
lattice calculations that apparently validate a repulsive
core~\cite{Ishii:2006ec,Hatsuda:2008zz,Ishii:2009zr,Aoki:2009ji}.

For finite-mass composite particles, locality is a feature we expect
at long distances, but non-local interactions would be more natural at
short distances. In fact, the potential  at short range
is far removed from an
observable, and locality is imposed on potentials for
convenience, not because of physical necessity.  Recall that we are
free to apply a 
short-range unitary transformation $U$ to the Hamiltonian (and to
other operators at the same time),
\be
E_n = \langle \Psi_n | H | \Psi_n \rangle
= \bigl( \langle \Psi_n | U^\dagger \bigr) \,
U H U^\dagger \, \bigl( U | \Psi_n \rangle \bigr)
= \langle \widetilde\Psi_n | \widetilde H | \widetilde \Psi_n
\rangle \,,
\ee
and the physics described by $H$ and $\widetilde H$ is
indistinguishable by experiment. Thus there are an \emph{infinite}
number of equally valid potentials, and once we allow non-locality, a
repulsive core and the strong low- to high-momentum coupling is no
longer inevitable.

The EFT approach uses this freedom to construct a systematic expansion
of the Hamiltonian. A particular EFT is associated with a momentum
scale $\lb$ that is the dividing point between resolved, long-range
physics, which is treated explicitly, and unresolved, short-range
physics, which is expanded in contact interactions. Results are given
order-by-order in $Q/\lb$, where $Q$ is a generic momentum (or light
mass) scale of the process being calculated. There is also a cutoff
$\lm$ needed to regulate the theory, which suppresses high
momenta. Thus $\lm$ acts as a resolution scale for the theory. If
$\lm$ is chosen to be less than $\lb$, then the truncation error for
the EFT will be dominated by powers of $Q/\lm$ rather than $Q/\lb$.
In principle one could take $\lm$ as large as desired but in practice
this only works if the renormalization and the numerics involved in
matching to data are sufficiently under control~\cite{Lepage:1989hf}.

In general, the forces between nucleons depend on the
resolution scale $\lm$ and are given by an effective theory for
scale-dependent two-nucleon $\vnn(\lm)$ and corresponding many-nucleon
interactions $V_{\rm 3N}(\lm), V_{\rm 4N}(\lm)$ and so
on~\cite{Bedaque:2002mn,Bogner:2003wn,Epelbaum:2008ga}.  This scale
dependence is analogous to the scale dependence of parton distribution
functions.  At very low momenta $Q \ll m_\pi$, the details of pion
exchanges are not resolved and nuclear forces can be systematically
expanded in contact interactions and their
derivatives~\cite{Bedaque:2002mn}. The corresponding pionless EFT (for
which $\lb \sim m_\pi$) is very successful in capturing universal
large scattering-length physics (with improvements by including
effective range and higher-order terms) in dilute neutron matter and
reactions at astrophysical
energies~\cite{Bedaque:2002mn,Braaten:2004rn,%
Furnstahl:2008df,Schwenk:2005ka,Platter:2009gz}.

\begin{figure}[t]
 \centering
 \subfloat[][]{%
   \label{chiralEFT-a}
   \raisebox{.25in}{\includegraphics[width=3.2in,clip=]{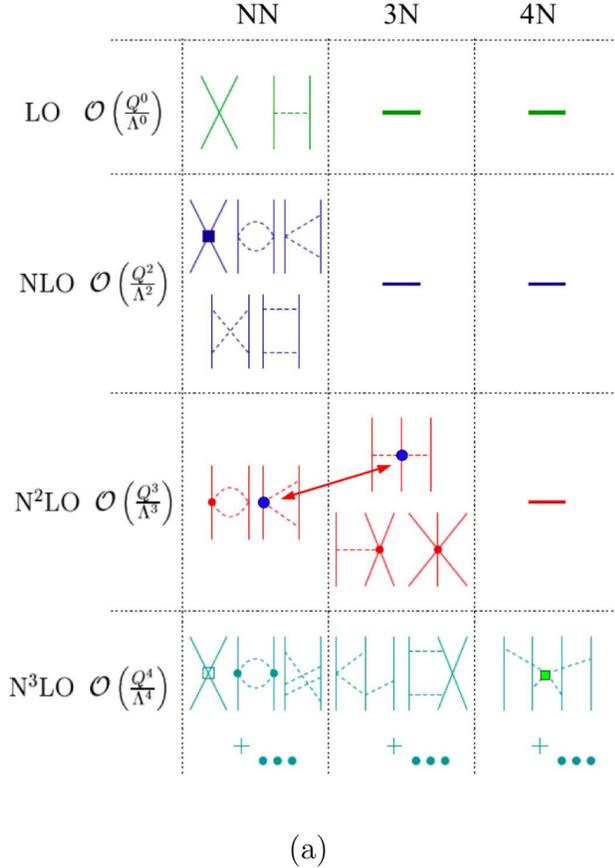}}%
 }%
 \hspace*{.4in}%
 \subfloat[][]{%
   \label{chiralEFT-b}
   \includegraphics[width=3.2in,clip=]{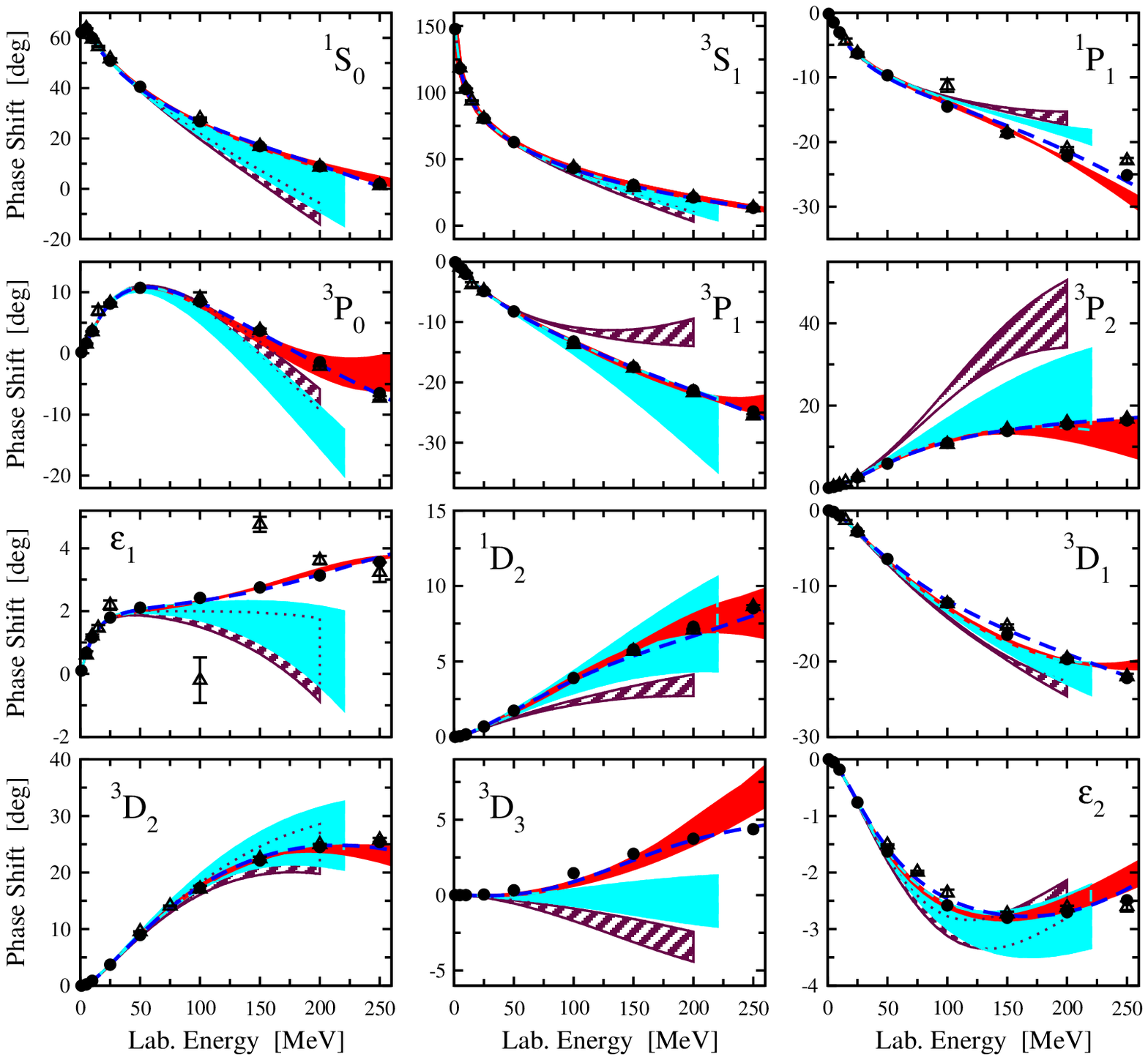}%
 }  
\caption{(a) Chiral EFT for nuclear forces. (b)~Improvement in
neutron-proton phase shifts shown by shaded bands
from cutoff variation at NLO (dashed), N$^2$LO (light), and N$^3$LO
(dark) compared to extractions from experiment (points)~\cite{Epelbaum:2008ga}.
The dashed line is from the N$^3$LO potential
of Ref.~\cite{Entem:2003ft}.}
\end{figure}

\begin{figure}[t]
 \centering
 \subfloat[][]{%
  \label{fig:3nfobs}%
  \includegraphics[width=3.8in,clip=]{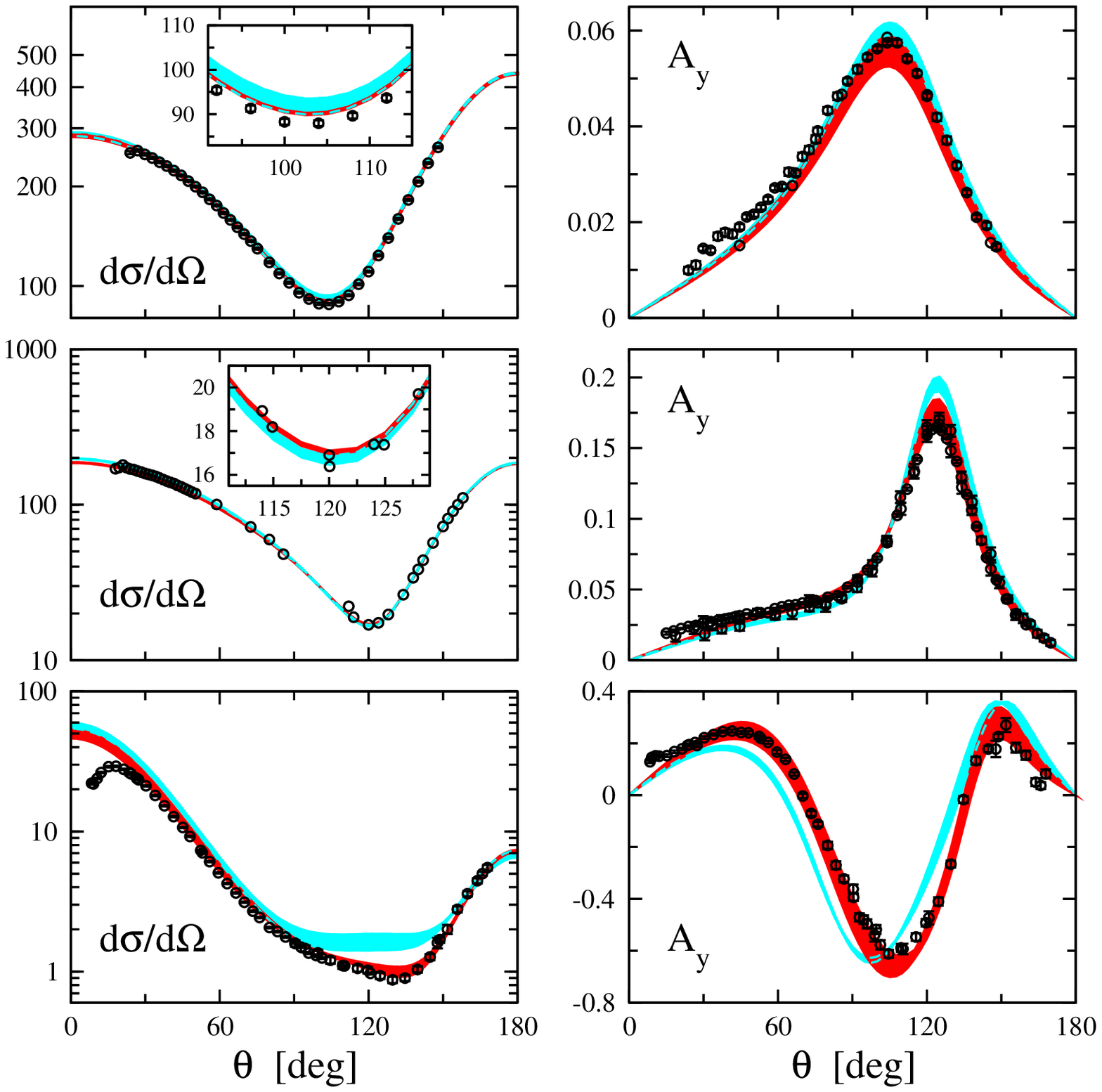}%
 }%
 \hspace*{.15in}%
 \subfloat[][]{%
  \label{fig:lithium6}%
  \raisebox{.05in}{\includegraphics[width=2.8in,clip=]{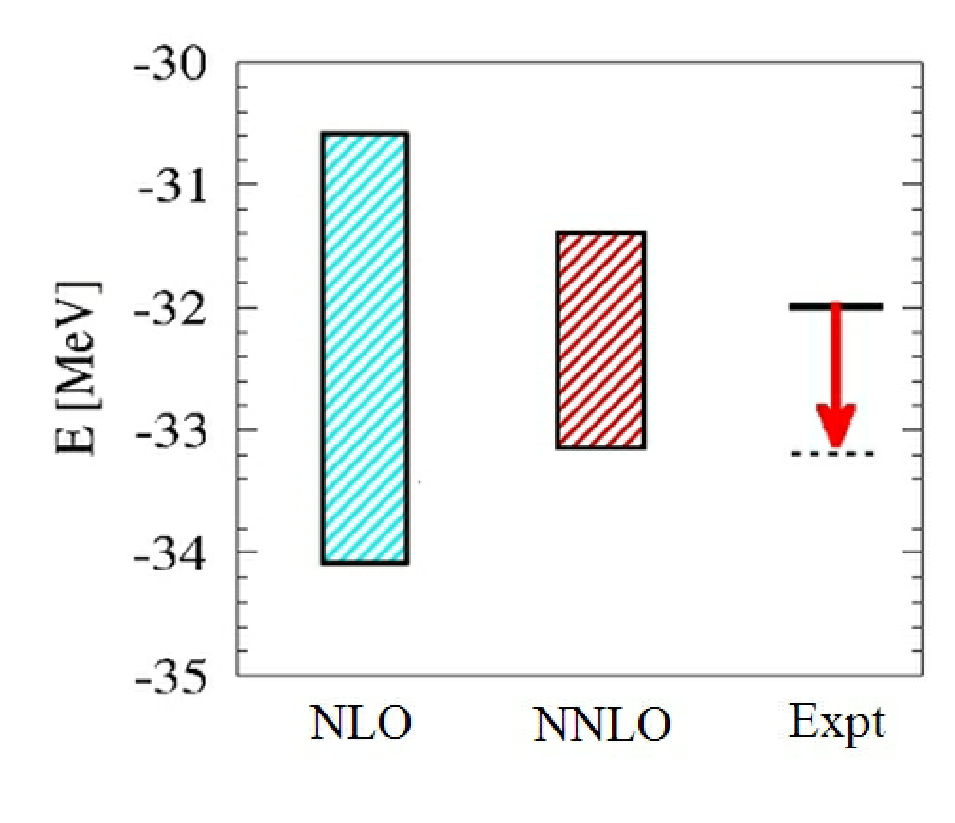}}%
 }%
\caption{(a) Differential cross section (in mb/sr) and vector
analyzing power for elastic neutron-deuteron scattering at $10 \mev$
(top) and $65 \mev$ (bottom) at  NLO (light) and N$^2$LO (dark) from
Ref.~\cite{Epelbaum:2005pn}. (b)~Ground-state energy of $^6$Li at NLO
and N$^2$LO with bands corresponding to the $\Lambda$ variation
over $500$--$600 \mev$ compared to
experiment (solid line, see Ref.~\cite{Epelbaum:2005pn} for details).}
\end{figure}

\begin{figure}[t!]
 \centering
 \subfloat[][]{%
  \label{fig:n3lopotmom-a}%
  \includegraphics[width=2.8in,clip=]{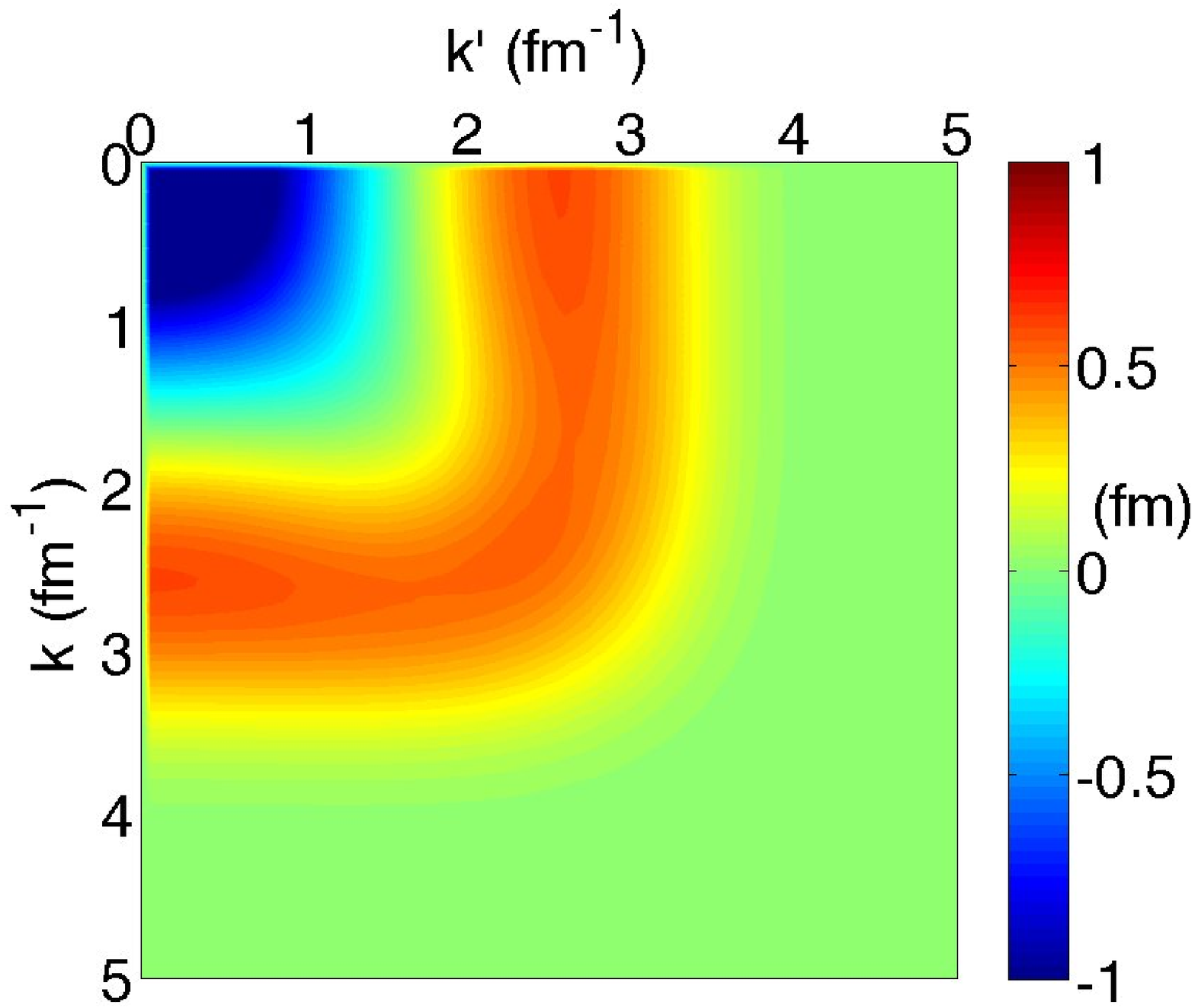}%
 }
 \hspace*{.4in}%
 \subfloat[][]{%
  \label{fig:n3lopotmom-b}%
  \includegraphics[width=2.8in,clip=]{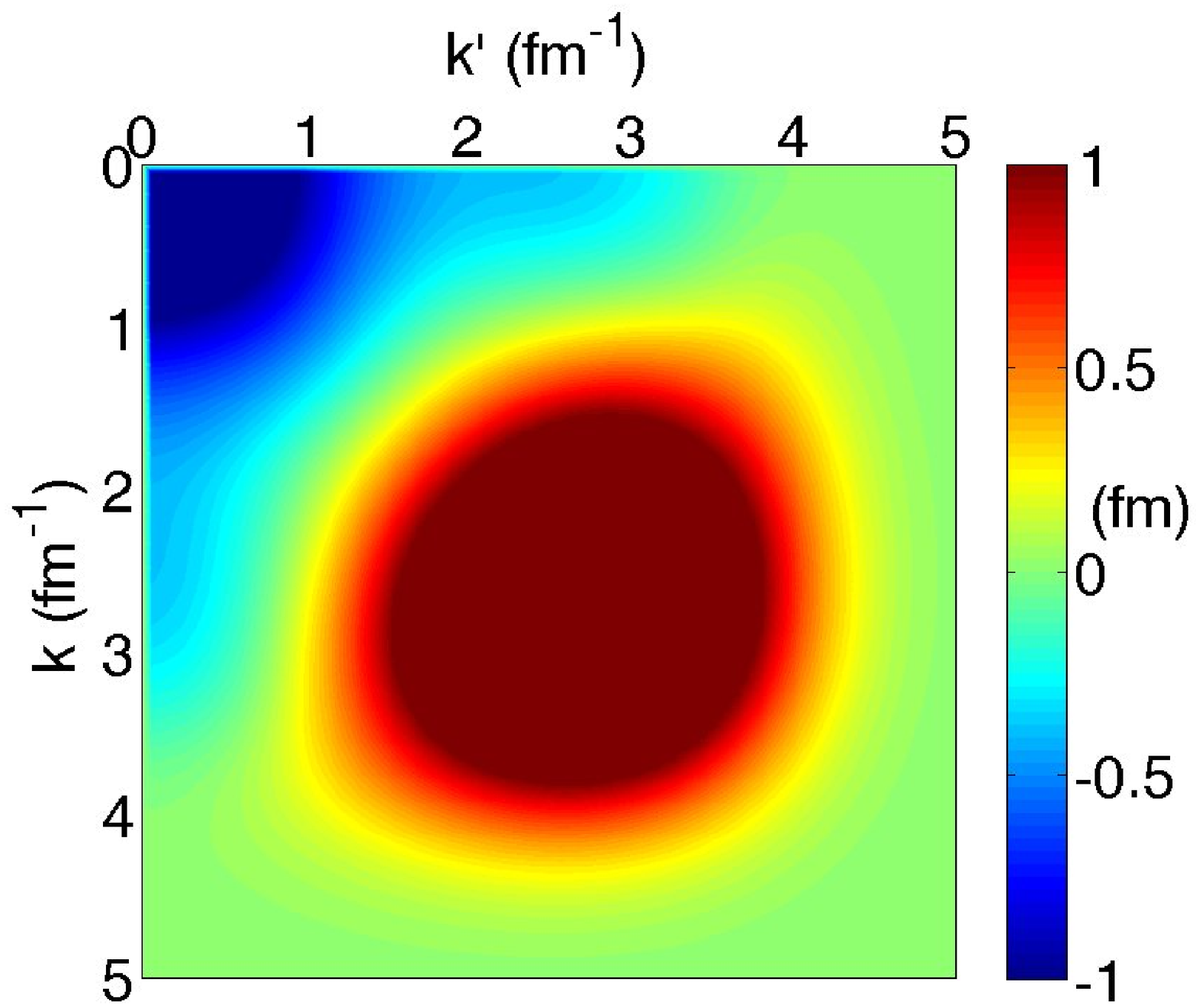}%
 }%
\caption{Two N$^3$LO $^1$S$_0$ potentials with (a) $500 \mev$ and (b)
$600 \mev$ cutoffs~\cite{Entem:2003ft}.}
\label{fig:n3lopotmom}
\end{figure}
 
For most nuclei, the typical momenta are $Q \sim m_\pi$ and therefore
pion exchanges are included explicitly in nuclear forces. 
The corresponding chiral EFT has been developed for over fifteen years as a
systematic approach to nuclear
interactions~\cite{Bedaque:2002mn,Epelbaum:2005pn,Epelbaum:2008ga,%
Machleidt:2007ms}. This
provides a unified approach to NN and many-body forces, and a pathway to
direct connections with QCD through lattice calculations (see, for
example, Ref.~\cite{Beane:2008dv}).  Examples of order-by-order
improved calculations of observables are shown in
Figs.~\subref*{chiralEFT-b}, \subref*{fig:3nfobs}, and
\subref*{fig:lithium6}. However, some open questions
remain~\cite{Epelbaum:2008ga}: understanding the power counting with
singular pion exchanges~\cite{Nogga:2005hy,PavonValderrama:2005wv,%
Valderrama:2009ei}, including $\Delta$ degrees
of freedom, the counting of relativistic $1/m$ corrections.  Resolving
these questions is important for improving the starting Hamiltonian
for low-momentum interactions, but does not affect our discussion of
RG technology.

In chiral EFT~\cite{Bedaque:2002mn,Epelbaum:2005pn,Epelbaum:2008ga,%
Machleidt:2007ms},
the expansion in powers of $Q/\lb$ has roughly $\lb \lesssim
m_\rho$. As shown in Fig.~\subref*{chiralEFT-a}, at a given order this
includes contributions from one- or multi-pion exchanges and from
contact interactions, with scale-dependent short-range couplings that are fit 
to low-energy data for each $\Lambda$ (experiment captures all short-range effects). 
There are natural sizes
to many-body force contributions that are made manifest in the EFT
power counting and which explain the phenomenological hierarchy of
many-body forces.  In addition, the EFT (extended to include chiral
perturbation theory) provides a consistent theory for multi-pion and
pion-nucleon systems and electroweak
operators, as well as for
hyperon-nucleon 
interactions~\cite{Beane:2000fx,Haidenbauer:2007ra,Epelbaum:2008ga}.

The highest-order NN interactions available to date are at
next-to-next-to-next-to-leading order, N$^3$LO or $(Q/\lb)^3$, for
several different cutoffs ($\lm = 450\mbox{--}600 \mev$) and two different
regulator schemes~\cite{Entem:2003ft,Epelbaum:2004fk}.  Representative
results for NN phase shifts at NLO, N$^2$LO, and N$^3$LO are shown in
Fig.~\subref*{chiralEFT-b}, where error bands are determined by the
spread in predictions for different $\lm \sim \lb$. Contour plots
of momentum-space matrix elements
for the softest, most commonly used N$^3$LO potential
and one with a higher cutoff are shown in
Fig.~\ref{fig:n3lopotmom}. While they are much softer than Argonne
$v_{18}$ in the $^1$S$_0$ channel, there is still considerable
off-diagonal strength above $k = 2 \infm$, which remains problematic
for nuclear structure calculations (and the coupled $^3$S$_1 -
^3$D$_1$ channel is generally worse).%
\footnote{Note that the cutoff associated with the potential in
Fig.~\subref*{fig:n3lopotmom-a} is $\lm = 500 \mev$, which might
lead one to expect no strength above $k \approx 2.5 \infm$. However,
the regulator does not sharply cut off relative momenta.}  One might
think the solution is to simply fit with a smaller $\lm$, but then the
fit worsens significantly as the truncation error grows with $Q/\lm$.

Instead we can use RG transformations to evolve to lower $\lm$ while
preserving the truncation error of the original Hamiltonian.  We will
roughly define ``low-momentum interactions'' as potentials that do not
couple $k \lesssim 2\infm$ to larger momenta.  This is a relatively
small (but significant) evolution for chiral potentials and a large
one for phenomenological potentials.  Low-momentum interactions are
sometimes categorized as phenomenological interactions or
regarded as an alternative to EFT interactions. Instead they are an
entire class of potentials associated with an initial Hamiltonian.
There is no prejudice in the RG methods as to the starting
Hamiltonian, but we favor the chiral EFT framework because of the
consistent organization of many-body forces and operators and the
capacity for systematic improvements.  We emphasize that precision
tests of all currently available Hamiltonians have only been made in
few-body systems.  Details of three-body forces and the four-body
strength, for example, are quite uncertain.%
\footnote{An exploratory calculation for $^4$He suggested a
contribution of order $100 \, {\rm keV}$ to the binding energy
from the long-range four-body force at N$^3$LO, but with substantial
uncertainty~\cite{Rozpedzik:2006yi,Epelbaum:2009zs}.}

In the past, unitary transformations were used to soften NN
potentials, which were then applied to calculations of few-body
systems and nuclear matter.  Results for observables depended on the
transformations and this was often the context for discussing
``off-shell'' effects and which was the ``true potential''.  From the
modern perspective, this approach is misleading at best. These
transformations \emph{always} lead to many-body interactions, even if
absent in the initial Hamiltonian.%
\footnote{It is possible to find NN potentials that come close to
the binding energy of $^3$H and $^4$He and other few-body nuclei
without three-body forces.  Beyond the problem of constructing
consistent electroweak currents for these interactions, this entails a fine
tuning to avoid the natural size of many-body forces, which may not
persist when these interactions are applied elsewhere.}  While this fact was
clearly recognized in past investigations \cite{Coester:1970ai},
many-body forces were usually neglected, which is what led to the
different results.  In contrast, we account for three-body forces,
but we also stress that maintaining \emph{exact} equivalence is not
required, because the original Hamiltonian already
contains truncation errors.

\subsection{Nuclear structure challenges} 
\label{subsec:challenges}

\begin{figure}[t]
 \centering
 \subfloat[][]{%
  \label{fig:scales-a}%
  \raisebox{3.1in}{\includegraphics*[width=8cm,clip=,angle=-90]{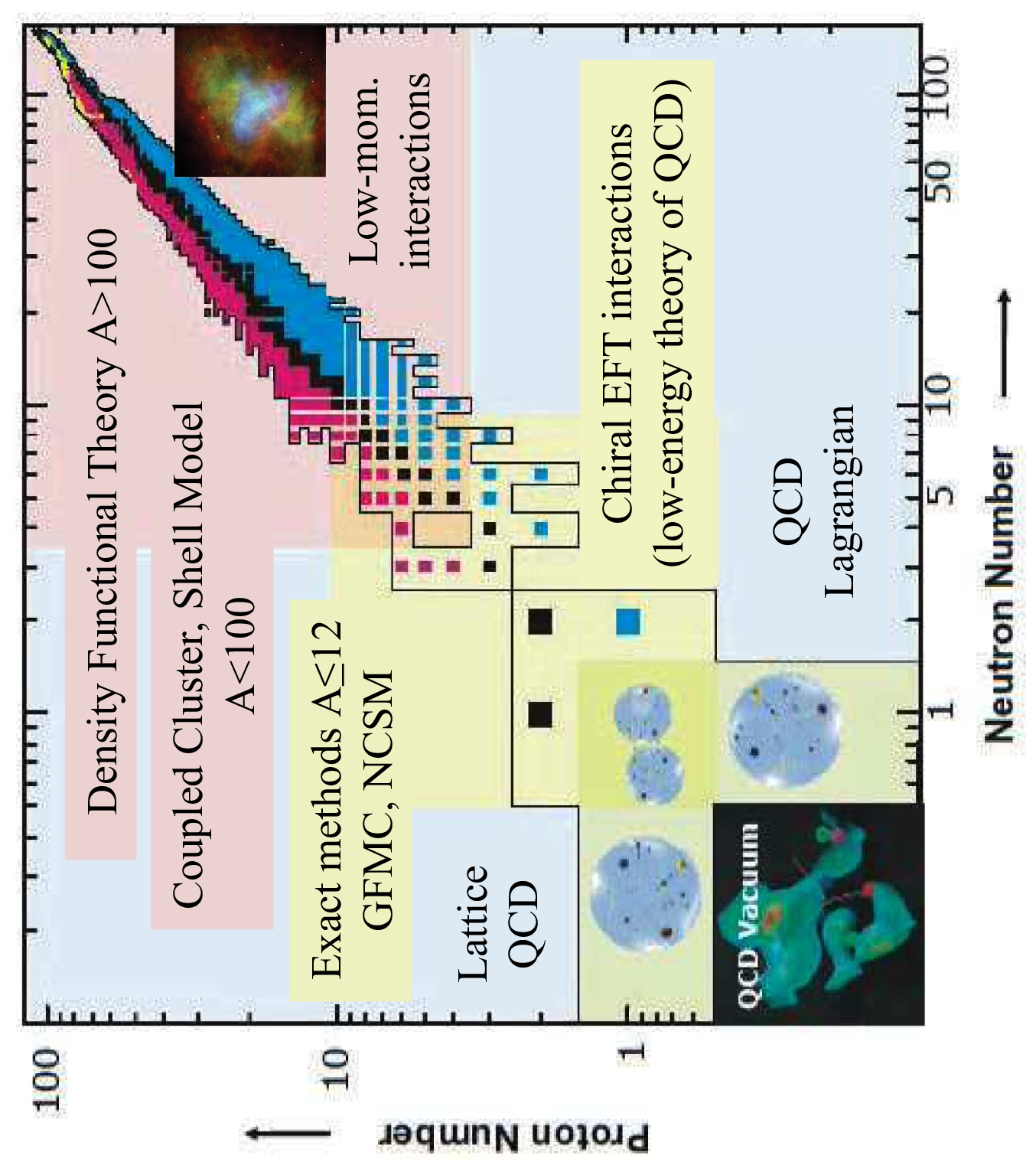}}%
 }%
 \hspace*{.4in}%
 \subfloat[][]{%
  \label{fig:scales-b}%
  \includegraphics*[width=6.8cm,clip=]{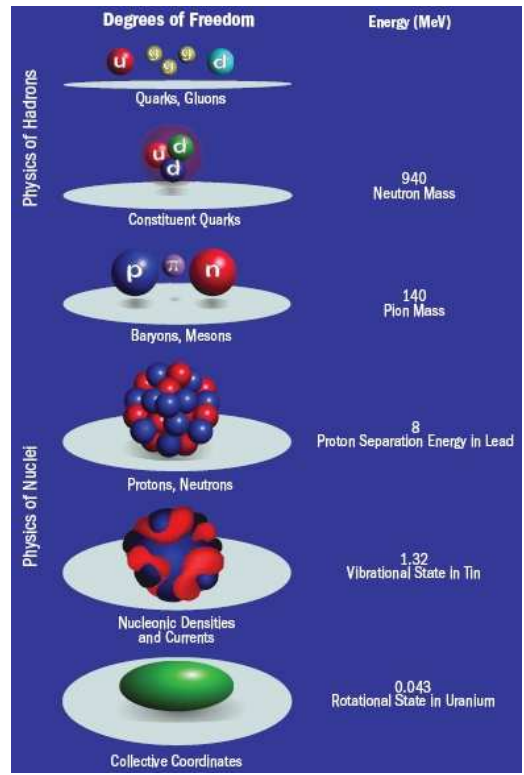}%
 }%
\caption{(a) Nuclear many-body landscape and (b)~degrees of freedom
and corresponding scales in nuclei~\cite{unedf:2007}.}
\end{figure}

The nuclear many-body landscape is illustrated in
Fig.~\subref*{fig:scales-a}, where each square represents a nucleus.
The challenge of contemporary nuclear structure physics is to describe
this entire range of nuclei and beyond to neutron stars and supernovae
in a controlled and unified way.  The logarithmic scale in proton and
neutron number emphasizes the different regions in $A$ where different
many-body methods are applicable.  Also shown is the theoretical
hierarchy that we propose to exploit: QCD $\rightarrow$ chiral EFT
$\rightarrow$ low-momentum interactions, which then feed into (most
of) the many-body methods, including some not listed.  Within the
broader challenge there are many more specific challenges, for
example, mapping out the drip-lines (the boundaries of nuclear
existence), the description of halo nuclei, the evolution of shell
structure, the characteristics of pairing, and the description of
nuclear decays and spontaneous fission.  An intertwined challenge to
controlled calculations of binding energies and spectra are nuclear
reactions.  This includes the interaction with external electroweak probes,
which require the consistent calculation of electroweak currents.

These challenges are increased by the wide range of scales for
strongly interacting particles, illustrated schematically in
Fig.~\subref*{fig:scales-b}.  In particular, we highlight the weak
binding of nuclei compared to typical QCD energy scales.  For nuclear
structure physics, we need methods appropriate to the energy and
momentum scales of nuclei, which means isolating the relevant degrees
of freedom.  As we emphasize throughout this review, the decoupling of
scales is at the heart of EFT and RG approaches, and leads us to the
low-momentum interaction technology to be documented.

Each of the many-body methods faces computational challenges,
principally to the rapid growth of resources needed with increasing
$A$. For methods such as
the No-Core Shell Model and coupled-cluster methods, the key is to
improve convergence within the spaces accessible. For heavier nuclei,
density functional theory methods offer favorable scaling, but are
challenged to reach reliable accuracy such as the sub-MeV level for
binding energies. Improved accuracy is needed in the shell model and
other methods to realize the use of nuclei as laboratories for
fundamental symmetries, such as for isospin-symmetry-breaking
corrections to superallowed decays, for neutrinoless double-beta
decay, or for octupole enhancement factors of electric dipole moments.
In all cases, we seek theoretical error estimates, particularly for
extrapolations to systems where measurements will be limited or
non-existent.

Progress toward such controlled nuclear calculations has long been
hindered by the difficulty of the nuclear many-body problem when
conventional nuclear potentials are used.  This has historically been
accepted as an unavoidable reality.  Indeed, conventional wisdom among
nuclear physicists, as summarized by Bethe in his review of over 30
years ago~\cite{Bethe:1971xm}, holds that successful nuclear matter
calculations must be highly nonperturbative in the potential.  This is
in contrast to the Coulomb many-body problem, for which Hartree-Fock
is a useful starting point and (possibly resummed) many-body
perturbation theory is an effective tool.  The possibility of a soft
potential providing a more perturbative solution to the nuclear matter
problem was discarded at that time, and saturation firmly identified
with the density dependence due to the tensor
force~\cite{Bethe:1971xm}.  Until recent RG-based
calculations~\cite{Bogner:2005sn,Bogner:2009un,Hebeler:2009iv},
subsequent work on the nuclear matter
problem~\cite{Jackson:1984ha,Machleidt:1989tm,Muther:2000qx,Akmal:1998cf}
had not significantly altered the general perspective or conclusions
of Bethe's review (although the role of three-nucleon (3N) forces has
been increasingly emphasized).

As already noted, nonperturbative behavior in the particle-particle
channel for nuclear forces arises from several sources.  First is the
strong short-range repulsion, which requires at least a summation of
particle-particle ladder diagrams~\cite{Bethe:1971xm}.  Second is the
tensor force, for example, from pion exchanges, which is highly
singular at short distances, and requires iteration in the triplet
channels~\cite{Beane:2001bc,Fleming:1999ee}.  Third is the presence of
low-energy bound states or nearly-bound states in the S-waves.  These
states imply poles in the scattering $T$ matrix that render the
perturbative Born series divergent.  All of these nonperturbative
features are present in conventional high-precision NN potentials.

The philosophy behind the standard approach to nuclear matter is to
attack these features head-on. This attitude was succinctly stated by
Bethe~\cite{Bethe:1971xm}:
\begin{quote}
``The theory must be such that it can deal with any
NN force, including hard or `soft' core, tensor
forces, and other complications.  It ought not to be necessary to tailor
the NN force for the sake of making the computation of nuclear matter
(or finite nuclei) easier, but the force should be chosen on the basis
of NN experiments (and possibly subsidiary experimental evidence, like
the binding energy of $^3$H).''
\end{quote}
In contrast, the EFT and RG perspective has a completely different
underlying philosophy, which stresses that the potential is not an
observable to be fixed from experiment (there is no ``true
potential''), but that an infinite number of potentials are capable of
accurately describing low-energy physics~\cite{Lepage:1997cs}.  In
order to be predictive and systematic, an organization (``power
counting'') must be present to permit a truncation of possible terms
in the potential.  If a complete Hamiltonian is used (including
many-body forces), then all observables should be equivalent up
to truncation errors.  The EFT philosophy implies using this freedom
to choose a convenient and efficient potential for the problems
of interest.

\begin{figure}[t]
\centering
\includegraphics*[width=6.2in,clip=]{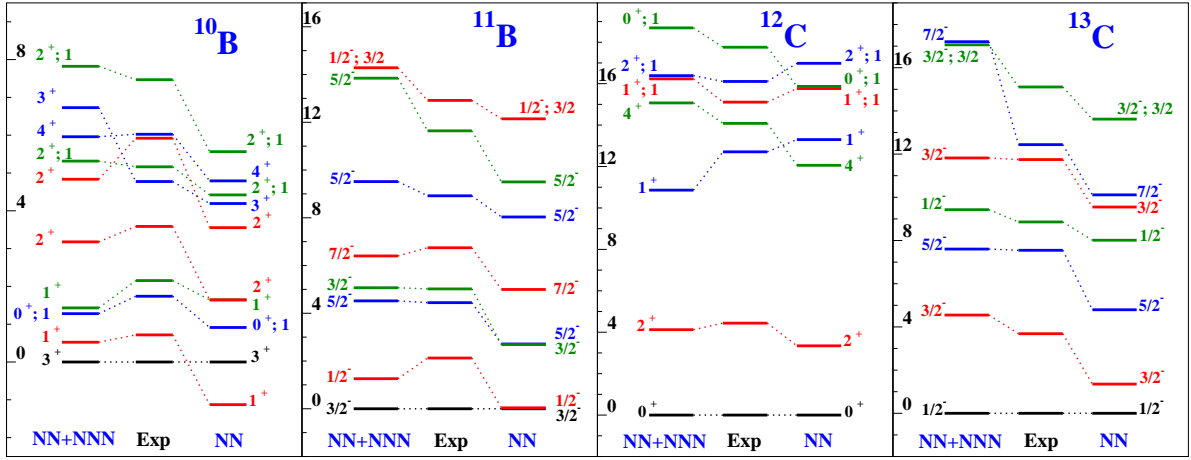}
\vspace*{.1in}
\caption{Excitation energies (in MeV) in light nuclei calculated using
the No-Core Shell Model (NCSM) with chiral EFT interactions (NN to
N$^3$LO and 3N to N$^2$LO) compared to
experiment~\cite{Navratil:2007we}.
Reprinted with permission from P.~Navratil et
al.~\cite{Navratil:2007we}, copyright (2007) by the American Physical
Society.\label{fig:n3lonuclei}}
\end{figure}

The use of energy-independent low-momentum interactions is a direct
implementation of these ideas.  Varying the cutoff can be used as a
powerful tool to study the underlying physics scales, to evaluate the
completeness of approximate calculations, and to estimate truncation
errors from omitted higher-order contributions.  These variable-cutoff
interactions reveal the resolution or scale dependence of the first
two sources of nonperturbative behavior, which are tamed as high
momenta are decoupled.  In free space, the third source of
nonperturbative behavior remains independent of the cutoff because the
pole positions of weakly and nearly bound states that necessitate fine
tuning are physical observables.  However, this fine tuning is
eliminated in the medium at sufficiently high density.  In short, a
repulsive core is not constrained by phase shifts and is essentially
removed by even a moderately low-momentum cutoff (note the $\Lambda$
dependence in Fig.~\subref*{fig:srcorr-a}), the short-range 
tensor force is tamed
by a sufficiently low cutoff, and the weakly and nearly bound states
become perturbative as a result of Pauli blocking.  For cutoffs around
$2 \infm$, which preserve phase shifts up to $330 \mev$ laboratory
energy, the Born series in nuclear matter is well converged at second
order in the potential, bringing the nuclear and Coulomb many-body
problems closer together~\cite{Bogner:2005sn}.

While \emph{evolving} a soft potential from higher momentum is a new
development in nuclear physics~\cite{Bogner:2001gq,Bogner:2003wn},
attempts to use soft potentials for nuclear matter were made in the
mid sixties and early seventies~\cite{Coester:1970ai,Haftel:1971er}.
It had long been observed that a strongly repulsive core is not
resolved until eight times nuclear saturation
density~\cite{Bethe:1971xm}.  Thus, saturation is \emph{not} driven by
a hard core (unlike liquid $^3$He).  However, these soft potentials
were abandoned because they seemed incapable of quantitatively
reproducing nuclear matter properties.  Their requiem was given by
Bethe~\cite{Bethe:1971xm}:
\begin{quote}
``Very soft potentials must be excluded because they do not give
saturation; they give too much binding and too high density. In
particular, a substantial tensor force is required.''
\end{quote}
From the EFT perspective, a failure to reproduce nuclear matter
observables should not be interpreted as showing that the low-energy
potential is wrong, but that it is incomplete. This misconception
still persists and has led to the conclusion that low-momentum NN
interactions are ``wrong'' because they do not give saturation in
nuclear matter and finite nuclei are overbound for lower cutoffs.  The
missing physics that invalidates this conclusion is many-body forces.

In a low-energy effective theory, many-body forces are inevitable; the
relevant question is how large they are.  It is established beyond
doubt that 3N forces are required to describe light
nuclei~\cite{Nogga:2000uu,Pieper:2001mp,Pieper:2004qh,Pieper:2004qw,%
Navratil:2003ef,Navratil:2007we},
as shown, for example, in Fig.~\ref{fig:n3lonuclei}.  For
variable-cutoff potentials, three-body (and higher-body)
interactions evolve naturally with the resolution scale.

\subsection{Renormalization group approaches}  
\label{subsec:RGoverview}

A fundamental tenet of renormalization theory is that the
\emph{relevant} details of high-energy physics for calculating
low-energy observables can be captured in the scale-dependent
coefficients of operators in a low-energy Hamiltonian
\cite{Lepage:1989hf}.  This principle does not mean that high-energy
and low-energy physics is automatically decoupled in every effective
theory.  In fact, it implies that we can include as much irrelevant
coupling to \emph{incorrect} high-energy physics as we want by using a
large cutoff, with no consequence to low-energy predictions (assuming
we can calculate accurately).  But this freedom also offers the
possibility of decoupling, which makes practical calculations more
tractable by restricting the necessary degrees of freedom.  This
decoupling can be efficiently achieved by evolving nuclear
interactions using RG transformations designed to handle similar
problems in relativistic field theories and critical phenomena in
condensed matter systems.%
\footnote{For an early discussion of decoupling based on Okubo unitary
transformations, see Ref.~\cite{Epelbaum:1998na}.}

\begin{figure}[t]
 \centering
 \subfloat[][]{%
  \label{fig:schematic-a}%
  \includegraphics*[width=2.2in,clip=]{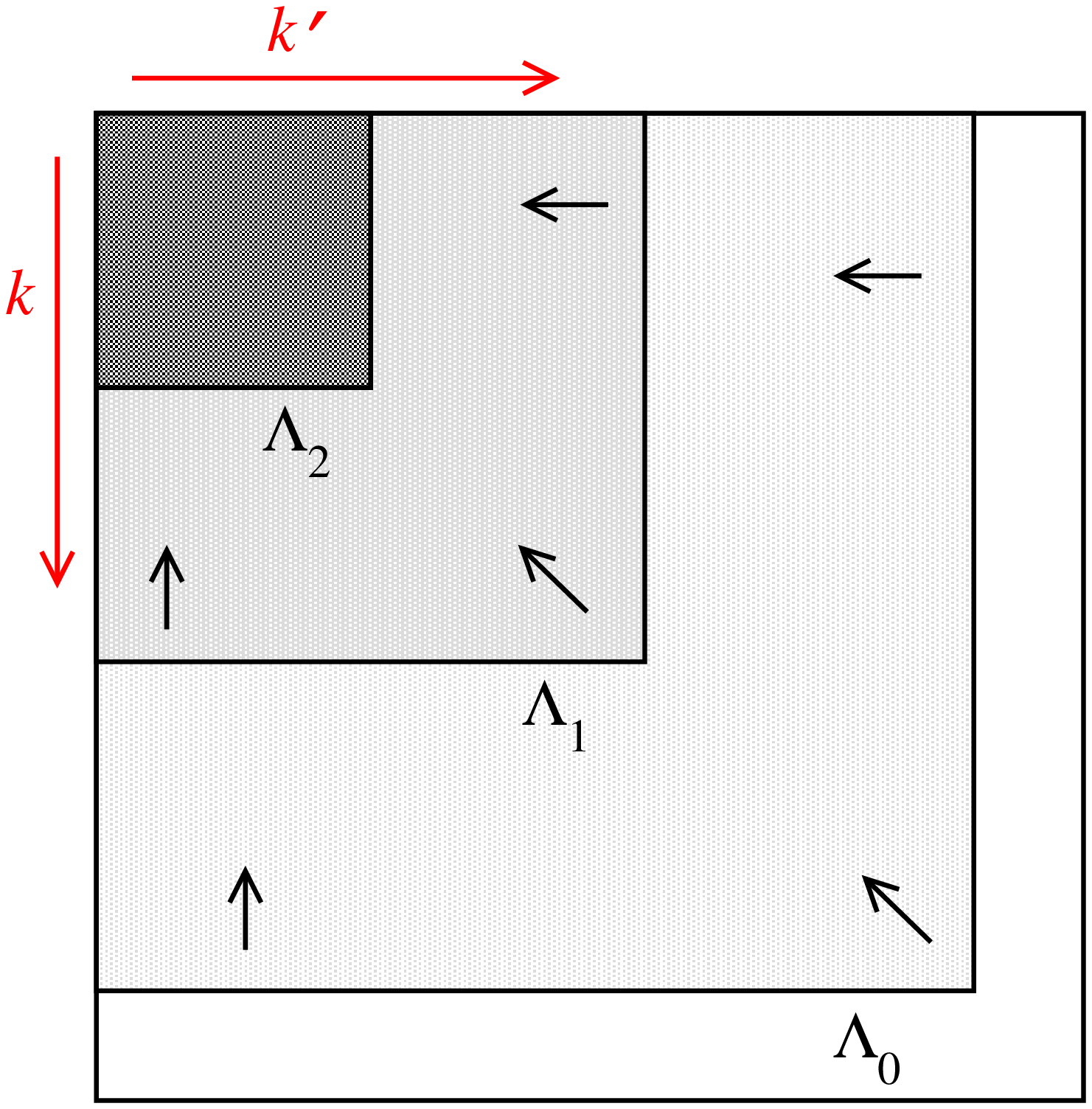}%
 }%
 \hspace*{.4in} %
 \subfloat[][]{%
  \label{fig:schematic-b}%
  \includegraphics*[width=2.2in,clip=]{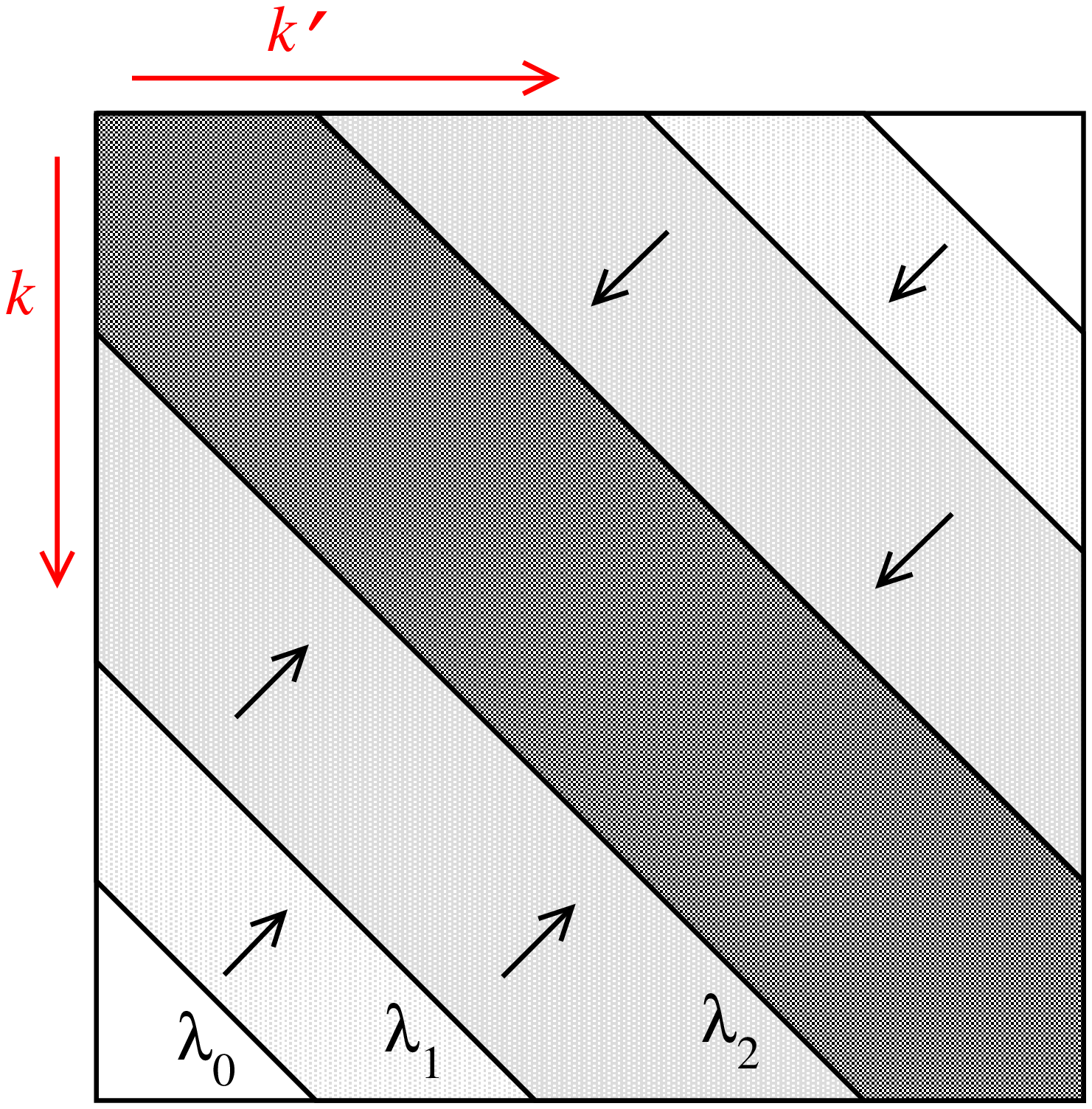}%
 }%
\caption{Schematic illustration of two types of RG evolution for NN
potentials in momentum space: (a)~$\vlowk$ running in $\Lambda$,
and (b)~SRG running in $\lambda$. At each $\Lambda_i$ or $\lambda_i$,
the matrix elements outside of the corresponding lines are zero, so
that high- and low-momentum states are decoupled.}
\label{fig:schematic}
\end{figure}

The general purpose of the RG when dealing with the
large range of scales in physical systems was eloquently
explained by David Gross~\cite{Deligne:2000}:
\begin{quote}
``At each scale, we have different degrees of freedom and different
dynamics.  Physics at a larger scale (largely) decouples from the
physics at a smaller scale. \ldots Thus, a theory at a larger scale
remembers only finitely many parameters from the theories at smaller
scales, and throws the rest of the details away.  More precisely,
when we pass from a smaller scale to a larger scale, we average over
irrelevant degrees of freedom. \ldots The general aim of the RG
method is to explain how this decoupling takes place and why exactly
information is transmitted from scale to scale through finitely many
parameters.''
\end{quote}  
The common features of RG for critical phenomena and high-energy
scattering are discussed by Steven Weinberg in an essay in
Ref.~\cite{Guth:1984rq}.  He summarizes:
\begin{quote}
``The method in its most general form can I think be understood
as a way to arrange in various theories that the degrees of freedom
that you're talking about are the relevant degrees of freedom for the
problem at hand.''
\end{quote}
This is the heart of what is done with low-momentum interaction
approaches: arrange for the degrees of freedom for nuclear structure
to be the relevant ones.  This does not mean that other degrees of
freedom cannot be used, but to again quote
Weinberg~\cite{Guth:1984rq}: ``You can use any degrees of freedom you
want, but if you use the wrong ones, you'll be sorry.''

\begin{figure}[t]
 \centering
 \subfloat[][]{%
  \label{fig:vlowksrg-a}%
  \includegraphics*[width=5.8in,clip=]{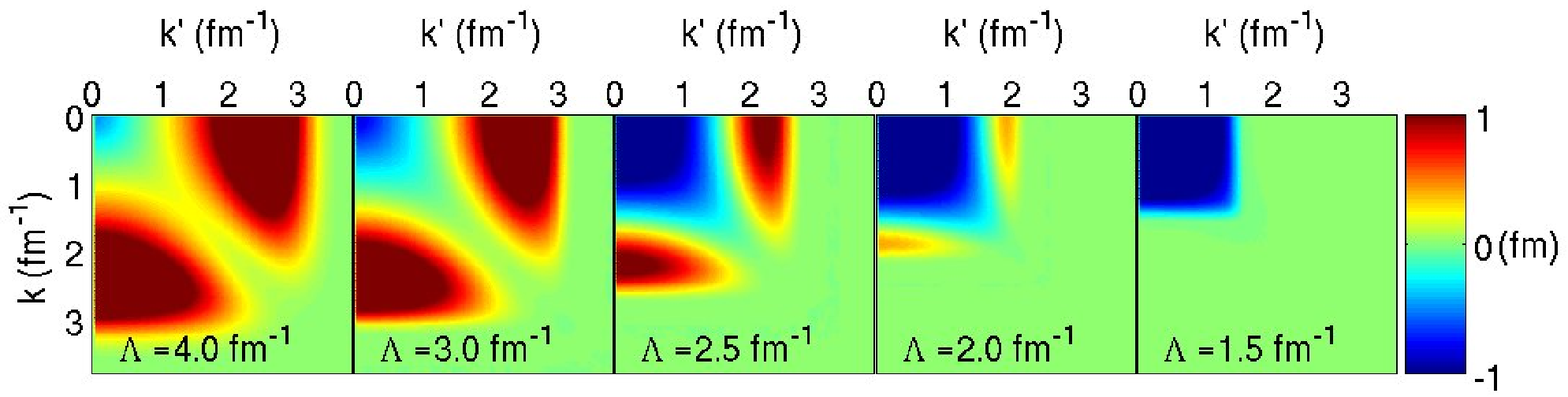}%
 }%
 \\%
 \subfloat[][]{%
  \label{fig:vlowksrg-b}%
  \includegraphics*[width=5.8in,clip=]{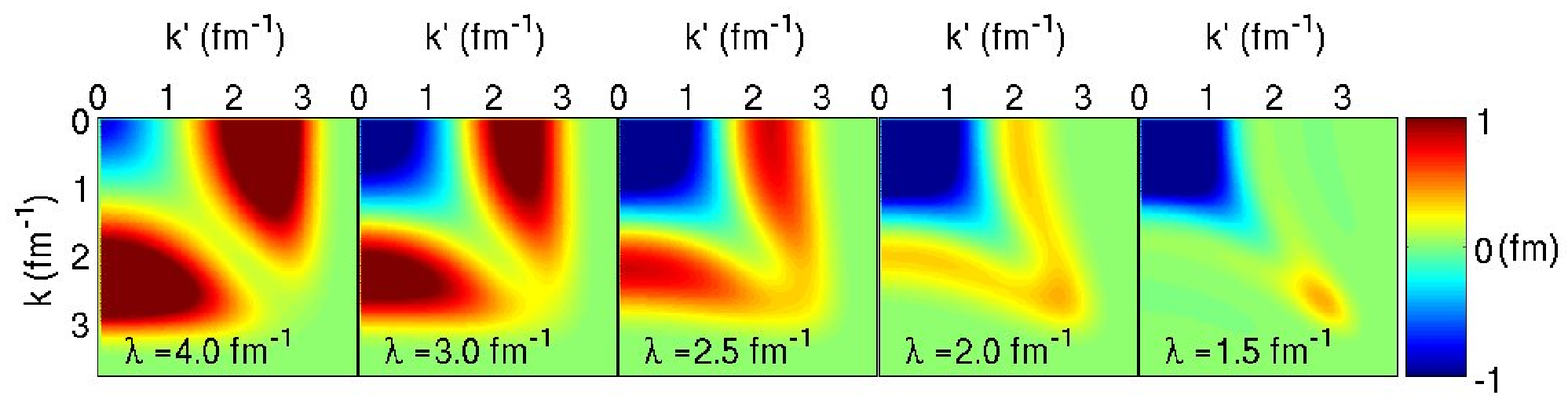}%
 }%
\caption{Two types of RG evolution applied to one of the chiral
N$^3$LO NN potentials ($550/600 \mev$) of
Ref.~\cite{Epelbaum:2004fk} in the $^3$S$_1$ channel: (a)~$\vlowk$
running in $\Lambda$, and (b)~SRG running in $\lambda$ (see
Fig.~\ref{fig:vsrg} for plots in $k^2$, which show the diagonal
width of order $\lambda^2$).}
\end{figure}

There are two major classes of RG transformations
used to construct low-momentum interactions,
which are illustrated schematically in Fig.~\ref{fig:schematic}.  In
the $\vlowk$ approach, decoupling is achieved by lowering a momentum
cutoff $\Lambda$ above which matrix elements go to zero.  In the SRG
approach, decoupling is achieved by lowering a cutoff $\lambda$ (in
energy differences $\lambda^2$) using flow equations, which means
evolving toward the diagonal in momentum space.  The technology for
carrying these out is outlined in Section~\ref{sec:technology}, but
the effects can be readily seen in the series of contour plots in
Figs.~\subref*{fig:vlowksrg-a} and \subref*{fig:vlowksrg-b}.
 
With either approach, lowering the cutoff leaves low-energy
observables unchanged by construction, but shifts contributions
between the interaction strengths and the sums over intermediate
states in loop integrals. The evolution of phenomenological or chiral
EFT interactions to lower resolution is beneficial because these
shifts can weaken or largely eliminate sources of nonperturbative
behavior, and because lower cutoffs require smaller bases in many-body
calculations, leading to improved convergence for nuclei.  The RG
cutoff variation estimates theoretical uncertainties due to
higher-order contributions, to neglected many-body interactions or to
an incomplete many-body treatment.  When initialized with different
orders of chiral EFT interactions, we have a powerful tool for
extrapolations to the extremes and for assessing the uncertainties of
key matrix elements needed in fundamental symmetry tests.

The idea of effective interactions in a limited model space is an old
and well-exploited one in nuclear physics.  However, we will emphasize
the flexibility of the RG compared to effective interaction methods.  The
continuous ``cutoff'' variation (in quotes because it may not be an
explicit cutoff) is a valuable new tool for nuclear physics (see
Section~\ref{subsec:cutoff}).  The RG methods are versatile and
suggest new ways to make progress (for example, using the in-medium
SRG, discussed in Section~\ref{subsec:normal}).  In addition, RG
combined with EFT is a natural framework for uncovering universal
behavior.
   
We note that the RG is an integral part of any EFT.  Matching of the
EFT at a given truncation level (to data or to an underlying theory)
but at different regulator cutoffs establishes the RG evolution (or
``running'') of the EFT couplings.  This includes the shift of
strength between loop integrals and couplings and between two and
many-body interactions.  However, because the EFT basis is truncated,
the error at the initial cutoff is not preserved with the running, in
contrast to the momentum-space RG evolution using $\vlowk$ or SRG
techniques, which keep all orders. 

\subsection{Scope of the review}
\label{subsec:scope}

This review focuses on the use of RG methods to derive low-momentum
interactions.  There are other approaches to softened interactions
that share many of the same advantages and issues, which we do not
address in detail.  Fortunately there are recent reviews
to cover these omissions, for example, Ref.~\cite{Epelbaum:2008ga} on
nuclear forces, Refs.~\cite{Navratil:2009ut} and
\cite{Coraggio:2008in} on ab initio and shell-model applications, and
Ref.~\cite{Drut:2009ce} on density functional theory.

In Section~\ref{sec:RG}, we discuss general principles of the RG as
applied to nuclear forces. The various technologies for evolving NN
potentials are outlined in Section~\ref{sec:technology}. 
Many-body
interactions and operators are addressed in
Section~\ref{sec:manybody}.
Sections~\ref{sec:infinite}
and~\ref{sec:finite} review many-body advances with applications to
infinite matter and finite nuclei. We conclude in
Section~\ref{sec:summary} with a summary of the main points, through a
consideration of misconceptions and clarifications, on-going
developments, and important open questions.


\section{Renormalization group: motivation and principles} 
\label{sec:RG}

\subsection{Decoupling}
\label{subsec:decoupling}

High-momentum degrees of freedom do not automatically decouple from
low-energy observables, especially for NN potentials that have
significant high-momentum off-diagonal strength (typically
beyond $k \gtrsim 2 \fmi$). The coupling between low- and
high-momentum states is a consequence of quantum fluctuations, as
manifested in intermediate state summations in perturbation theory,
e.g.,
\be
T(k',k;E) = \la k'| V^\lm |k \ra
+ \sum_{q=0}^{\Lambda} \: \frac{\la k'| V^\Lambda |q \ra \la q|
V^\Lambda |k \ra}{E - \varepsilon_{\bf q}} \, + \ldots \,.
\label{eq:2ndorderPT}
\ee
This coupling is regulated
by the cutoff or resolution scale $\Lambda$ that is
present in all NN interactions.

\begin{figure}[t]
\centering
\includegraphics*[width=6.0in,clip=]{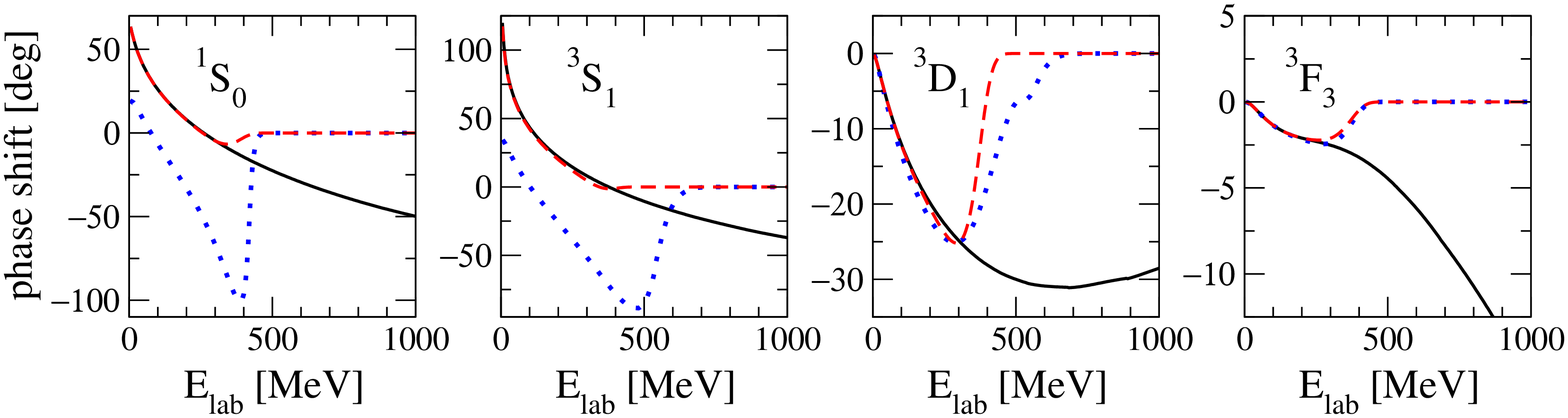}
\caption{NN phase shifts for the Argonne
$v_{18}$ potential~\cite{Wiringa:1994wb} and the SRG-evolved potential
for $\lambda = 2 \fmi$ with all momenta included (two
indistinguishable solid lines) and with the exclusion of momenta $k >
\kmax = 2.2 \fmi$ (Argonne $v_{18}$ dotted, SRG-evolved
dashed)~\cite{Bogner:2007jb}.}
\label{fig:phases_exclude}
\end{figure}

\begin{figure}[t]
\centering
\includegraphics*[width=4.0in,clip=]{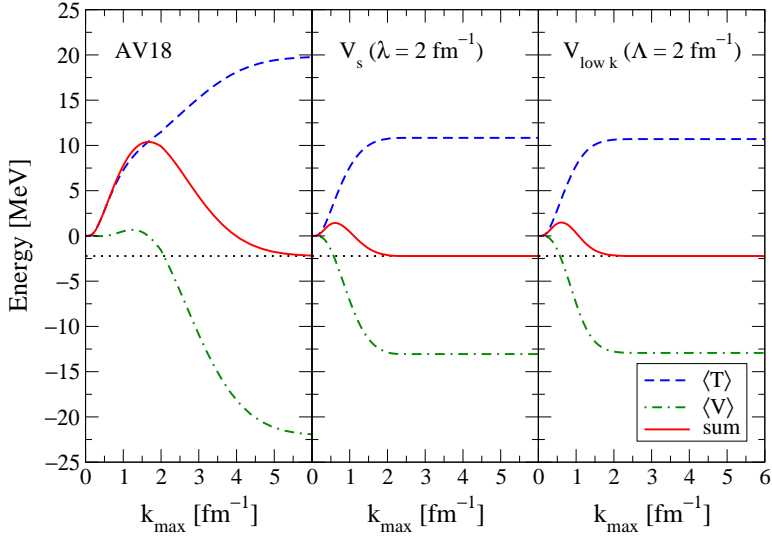}
\caption{Expectation values in the deuteron of the kinetic $\langle
T\rangle$ (dashed), 
potential $\langle V\rangle$ (dot-dashed), 
and total energy (solid) evaluated in momentum space as 
a function of the maximum momentum  $\kmax$, see
Eq.~(\ref{eq:Ed}). Results are shown
for the Argonne $v_{18}$ potential~\cite{Wiringa:1994wb} (left), the 
SRG-evolved potential $\vsrg$ for $\lambda = 2\fmi$ (middle), and the
smooth-cutoff $\vlowk$ interaction with $\Lambda=2 \fmi$
(right)~\cite{Bogner:2007jb}.}
\label{fig:deutbinding}
\end{figure}

A central theme of this review is that nonperturbative features of
nuclear forces can be radically altered by explicitly
decoupling low- and high-momentum degrees of freedom.  As we will see,
many-body calculations become more perturbative, basis expansions
converge more rapidly, and strong short-range correlations are smeared
out, rendering variational methods more effective.  However, if NN
potentials at a large scale $\lm$ are simply truncated at some $k_{\rm
max} \ll \Lambda$, the calculated low-energy observables (such as phase
shifts and the deuteron binding energy) are drastically changed, as
shown in Figs.~\ref{fig:phases_exclude}
and~\ref{fig:deutbinding}. Such a brute-force cutoff does not
disentangle high-energy (short-range) features from low-energy
(long-range) observables. To do so, it is necessary to integrate out
(and thus decouple) irrelevant high-momentum details from their
effects on low-energy observables. This is achieved by the two classes
of RG transformations illustrated in Fig.~\ref{fig:schematic} and
described in greater detail in Section~\ref{sec:technology}.

A practical test for decoupling is whether changing high-momentum
matrix elements of a given potential changes low-energy
observables. The SRG allows for particularly convincing demonstrations
of decoupling, because the unitary evolution of the Hamiltonian
preserves the deuteron binding energy and phase shifts for {\it
all} energies. The SRG-evolved potential $\vsrg$ explicitly
decouples high-energy dynamics from low-energy observables, which
means that we can set to zero the high-momentum parts (so that we have
a potential like $\vlowk$) with no significant changes in the low-energy physics.

In Ref.~\cite{Bogner:2007jb}, the decoupling of high-energy details
from low-energy phase shifts and the deuteron binding energy was
demonstrated by setting $V_s(k,k')$ to zero for all $k,k'$ above a
specified momentum $\kmax$ (using a smooth regulator function).
Phase shifts for $\kmax = 2.2 \fmi$ are shown in
Fig.~\ref{fig:phases_exclude} for the initial Argonne $v_{18}$
potential and for the SRG-evolved $\vsrg$ with $\lambda = 2
\fmi$. The phase shifts for the initial potential in the lower partial
waves bear no relation to the result without a $\kmax$ cutoff. In
contrast, the low-energy phase shifts for the SRG-evolved potential
are unchanged, even though the high-energy phase shifts above $\kmax$
are now zero.

\begin{figure}[t]
 \centering
 \subfloat[][]{%
  \label{fig:H3_srg_decoupling_N3LO_500MeV-a}%
  \includegraphics*[width=3.0in,clip=]{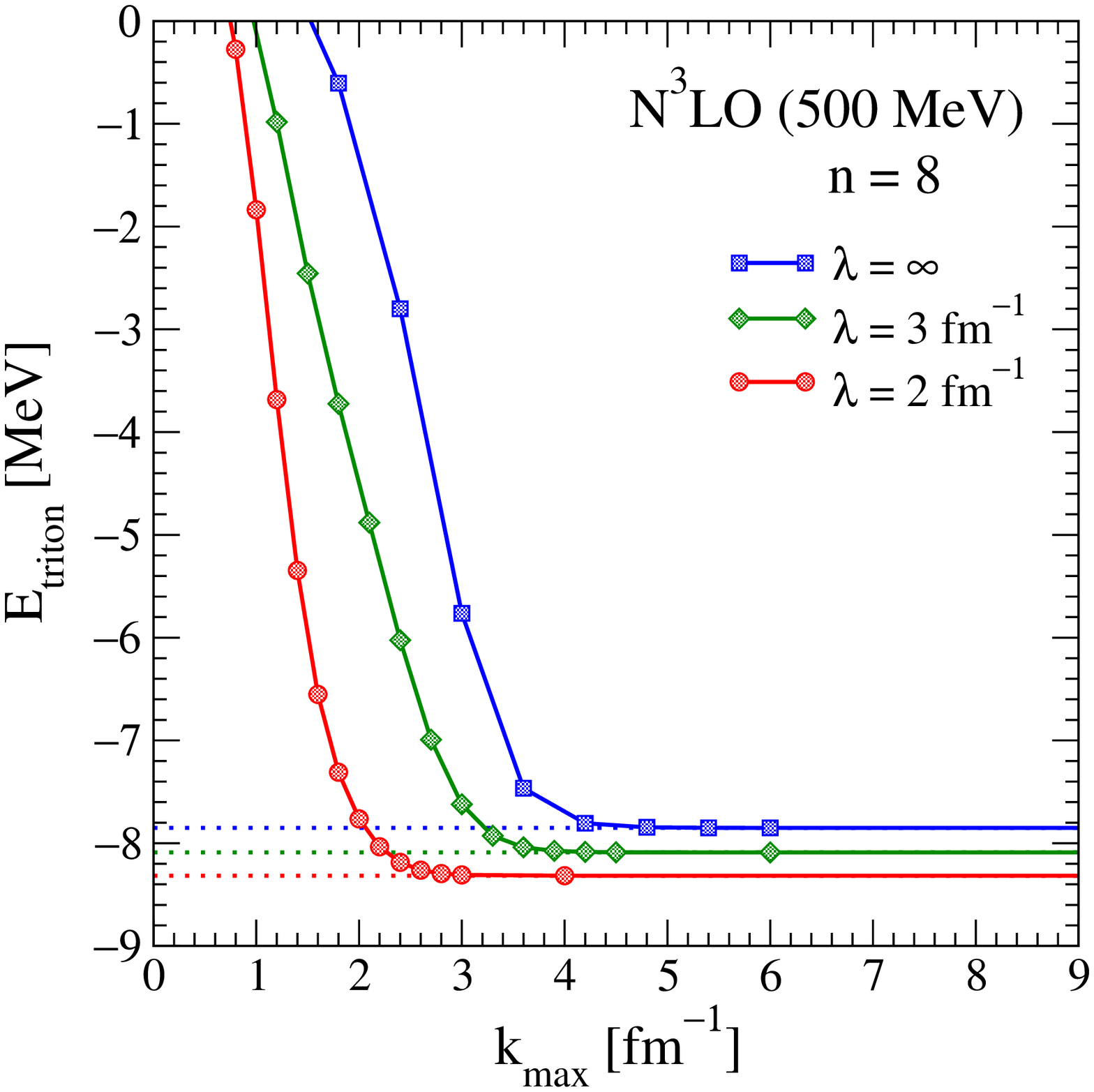}%
 }%
 \hspace*{0.4in}%
 \subfloat[][]{%
  \label{fig:H3_srg_decoupling_N3LO_500MeV-b}%
  \includegraphics*[width=3.0in,clip=]{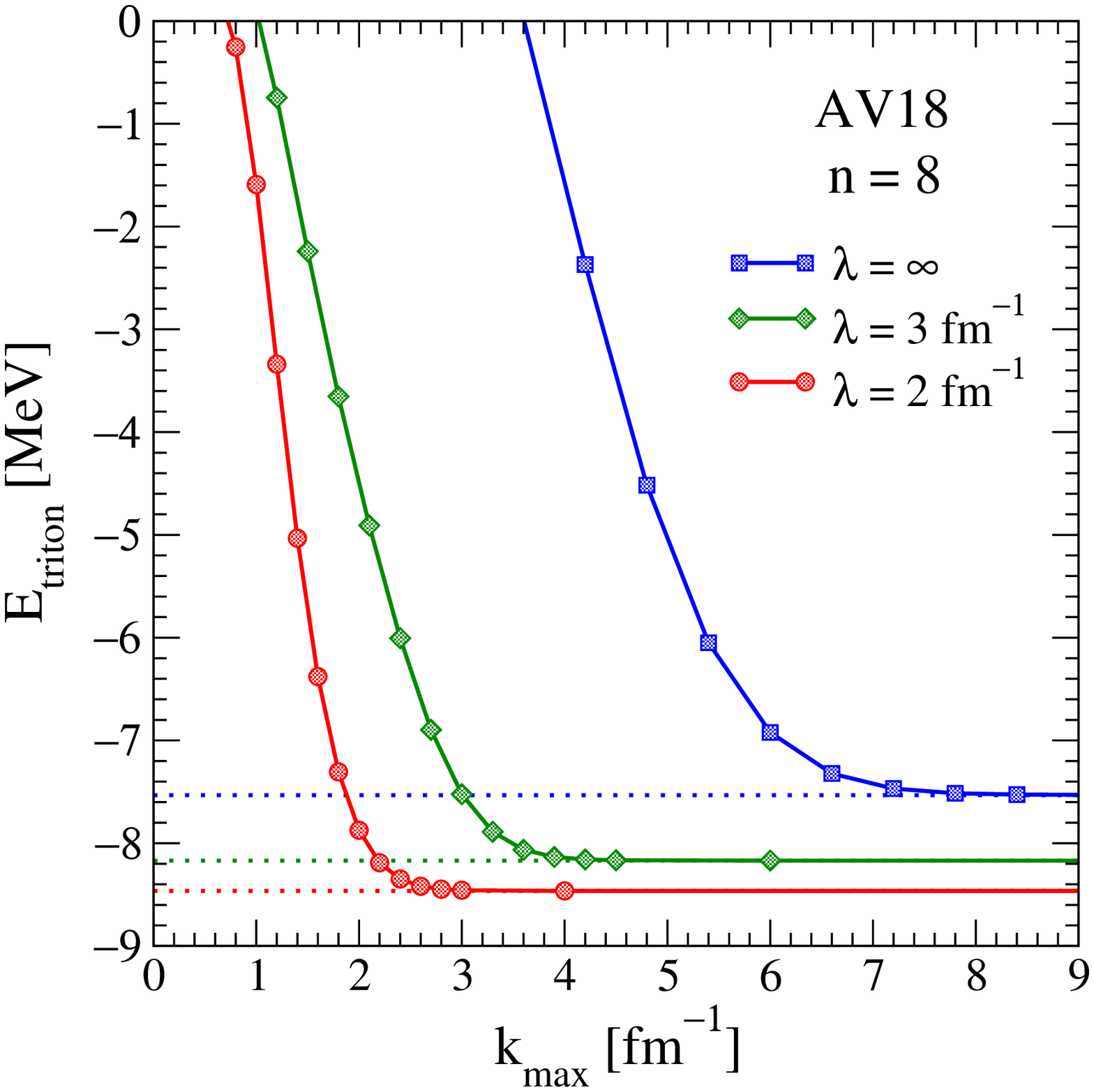}%
 }%
\caption{Ground-state energy of $^3$H as a function of the maximum
momentum $k_{\rm max}$ for three different values of
$\lambda$~\cite{Bogner:2007rx}. The cutoff function is
$\exp[-(k^2/k_{\rm max}^2)^n]$ with $n=8$. The initial potentials
are (a)~the N$^3$LO NN potential of Ref.~\cite{Entem:2003ft}
and~(b) the Argonne $v_{18}$ potential~\cite{Wiringa:1994wb}.}
\label{fig:H3_srg_decoupling_N3LO_500MeV}
\end{figure}

The deuteron binding energy provides another clear example of how the
contributions of different momentum components to a low-energy
observable depend on the resolution scale (as measured by $\lm$ or
$\lambda$, see Fig.~\ref{fig:schematic}). 
In Fig.~\ref{fig:deutbinding}, we show the kinetic,
potential, and total energy from an integration in momentum space
including momenta up to $\kmax$. That is, we plot
\be
E_d (k < \kmax) = \int_0^{\kmax} d{\bf k} \int_0^{\kmax} d{\bf k'} \,
\psi_d^\dagger({\bf k};\lambda) \left( k^2 \delta^3({\bf k} - {\bf k'})
+ V_s({\bf k},{\bf k'}) \right) \psi_d({\bf k'};\lambda) \,,
\label{eq:Ed}
\ee
where $\psi_d({\bf k};\lambda)$ is the momentum-space deuteron wave
function from the corresponding potential $V_s$ (without $\kmax$).
Figure~\ref{fig:deutbinding} shows that if one excludes momenta
greater than $2\fmi$ in the Argonne $v_{18}$ deuteron wave
function, the deuteron is $9.9
\mev$ unbound (that is, the integrated kinetic energy up to $2\fmi$ is
$11.5 \mev$ while the potential energy is $-1.6 \mev$).  In contrast,
using $\vsrg$ with $\lambda = 2\fmi$, one sees that the converged result
is dominated by contributions from much lower momenta.  Note that the
$\vsrg$ potential has no appreciable contributions above $\lambda$,
even though the near-diagonal matrix elements of the potential
$V_s(k,k')$ for $k,k' > \kmax$ are sizable.  This again validates
decoupling. It is also evident that the $\vsrg$ and $\vlowk$ results are very
similar for $\lambda \approx \Lambda$, where $\Lambda$ is the momentum
cutoff for $\vlowk$. Lastly, we mention that the decoupling carries
over to few-body systems~\cite{Bogner:2007rx,Jurgenson:2007td}, 
as shown similarly for the $^3$H ground-state energy in
Fig.~\ref{fig:H3_srg_decoupling_N3LO_500MeV}. Note that the different
converged energies for each $\lambda$ reflects the truncation of the
SRG at the NN level, see Section~\ref{subsec:intops}.

\subsection{Connections to effective field theory}
\label{subsec:eft}

\begin{figure}[t]
 \centering
 \subfloat[][]{%
  \label{fig:n3low-a}%
  \includegraphics*[width=2.4in,,clip=]{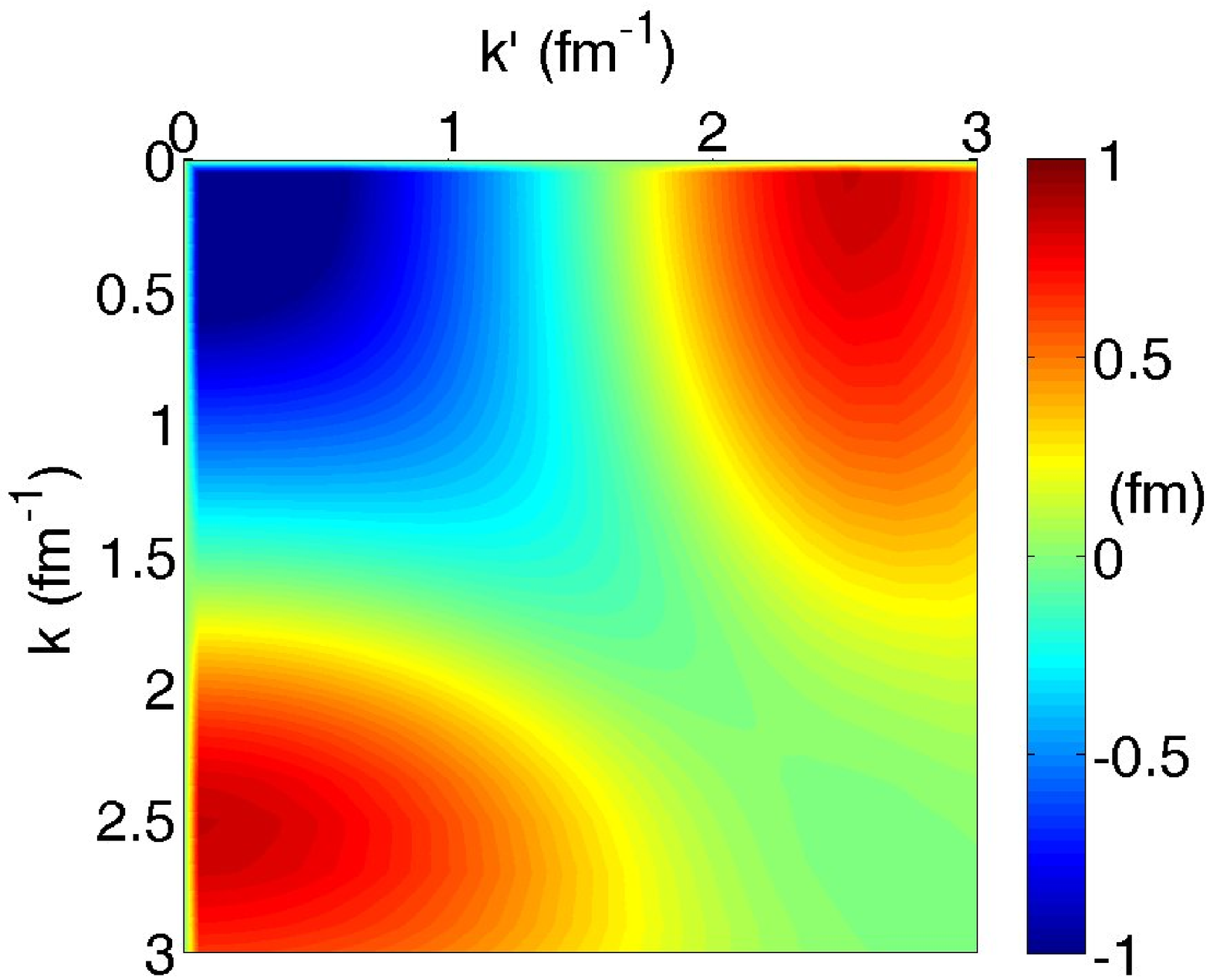}%
 }%
 \hspace*{.1in}%
 \subfloat[][]{%
  \label{fig:n3low-b}%
  \includegraphics*[width=2.4in,,clip=]{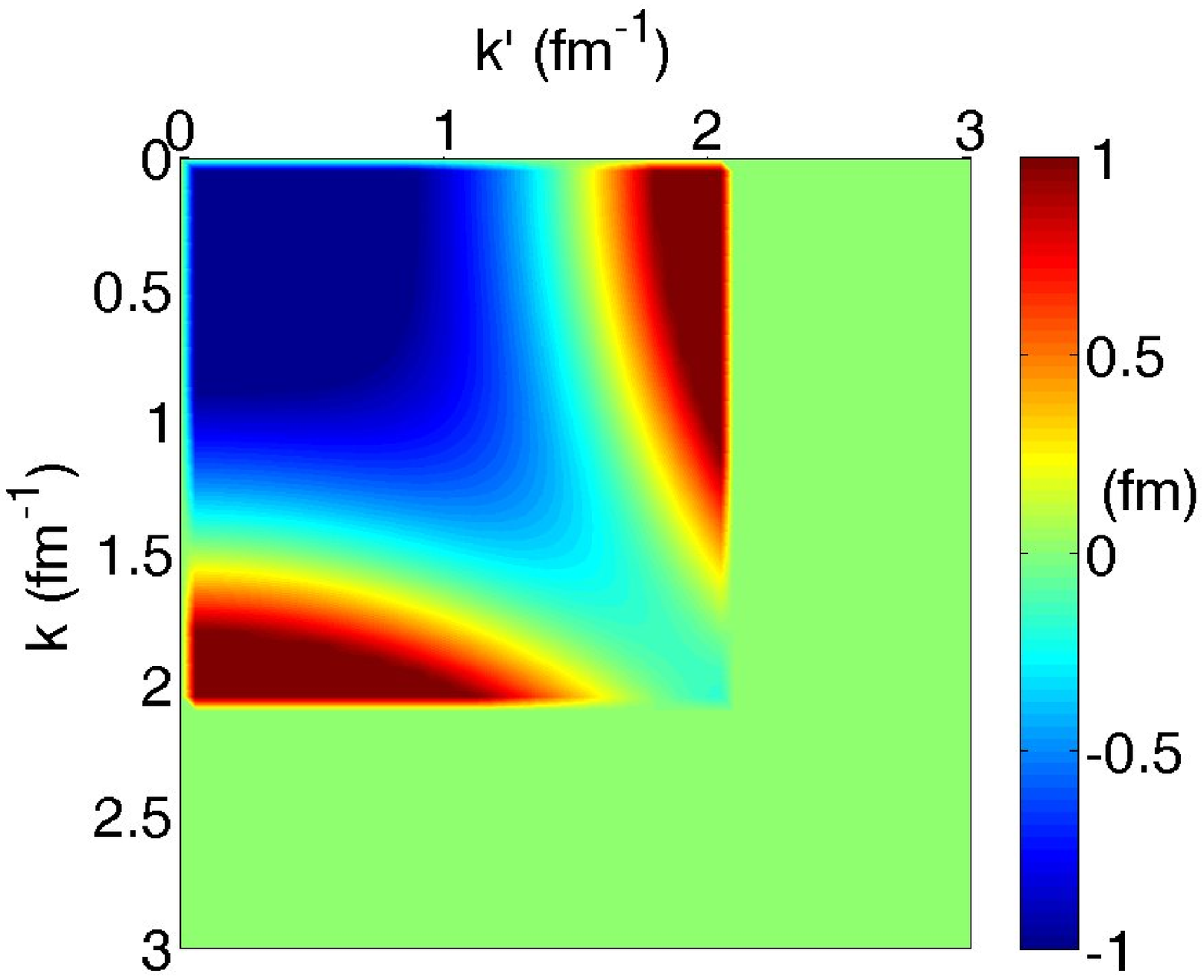}%
 }%
 \hspace*{.1in}%
 \subfloat[][]{%
  \label{fig:n3low-c}%
  \includegraphics*[width=2.4in,,clip=]{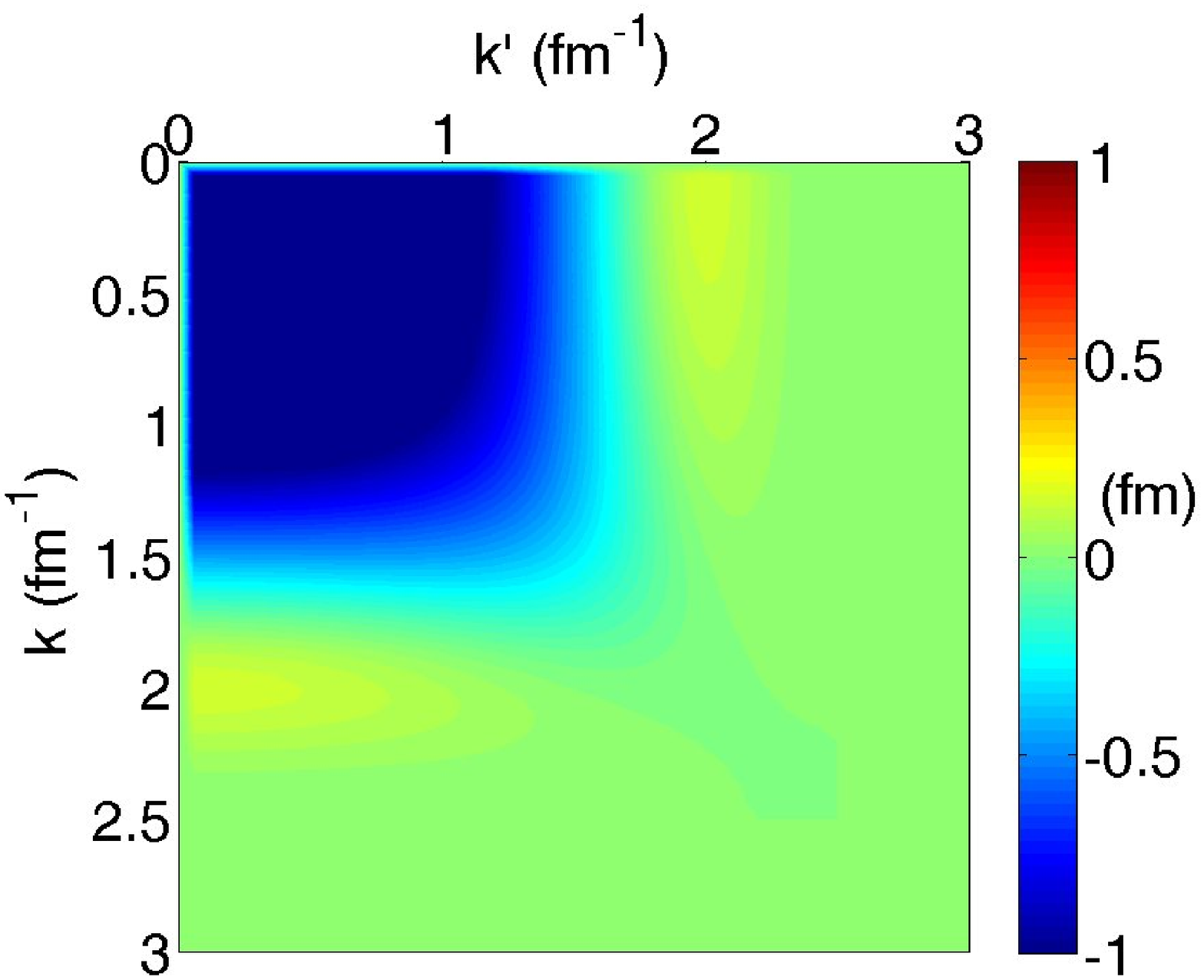}%
 }%
\caption{Momentum-space matrix elements in the $^3$S$_1$ channel for
(a) the N$^3$LO potential ($\lm = 500 \mev$) of
Ref.~\cite{Entem:2003ft}, (b) the sharp cutoff N$^3$LOW potential
($\lm = 2.0 \fmi$) of Ref.~\cite{Coraggio:2007mc}, and (c) for the
N$^3$LO potential ($\lm = 500 \mev$) evolved by a smooth $\vlowk$
to $\lm = 2.0 \fmi$.}
\label{fig:n3low}
\end{figure}

The RG plays a central role in EFT. It determines the running of
low-energy couplings (see, for example, Ref.~\cite{Bedaque:2002mn})
and the RG scaling of operators can be used to identify power counting
schemes~\cite{Birse:2005um,Birse:2009my}. In this section, we discuss
how the nonperturbative RG evolution works in the context of chiral
EFT interactions. In EFT, it is optimal to fit the low-energy
couplings at larger cutoffs, because this minimizes the truncation
errors that can scale as rapidly as $(Q/\lm)^n$, and because larger
cutoffs include maximal long-distance physics up to the given order in
the EFT expansion. This is illustrated in Fig.~\ref{fig:n3low} by
comparing an N$^3$LO potential with a larger cutoff $\lm = 500 \mev\
(2.5\infm)$ to the sharp cutoff N$^3$LOW potential of
Ref.~\cite{Coraggio:2007mc} with $\lm = 2.0 \fmi$. At the same EFT
order, the fit for a lower cutoff requires more fine-tuning and leads to a
potential that is not very smooth in momentum space.  This indicates a
breakdown of the gradient expansion.

\begin{figure}[t!]
\centering
\includegraphics[scale=0.4,clip=]{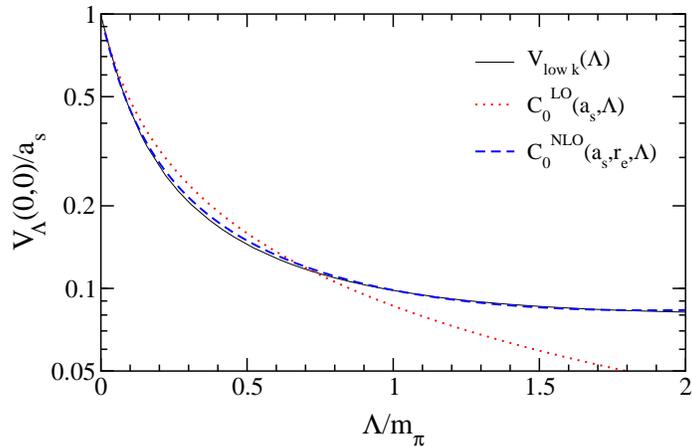}%
\caption{Flow of $\vlowk(k'=0,k=0;\lm)$ compared to the
corresponding momentum-independent contact interaction $C_0(\lm)$ at
LO and NLO, where this coupling is determined entirely from RG
invariance and fits to the scattering length $a_{\rm s}$ (at LO) plus
effective range $r_{\rm e}$ (at NLO)~\cite{Schwenk:2008su}.}
\label{connections-a}
\end{figure}

The RG evolution of EFT interactions to lower cutoffs is beneficial
because the resulting low-momentum interactions preserve the
truncation error of the initial potential and therefore include the
advantages of fitting at larger cutoffs, while they have lower
cutoffs that are advantageous for applications to nuclear structure
(see Sections~\ref{sec:infinite} and~\ref{sec:finite}).  As shown in
Fig.~\ref{fig:n3low}, the evolved interactions are smooth without the
strong momentum dependencies observed for the sharp cutoff N$^3$LOW
potential at the same cutoff scale, and the NN phase shifts are
reproduced with the same accuracy.

We can gain further insights into the interplay of the RG and EFT by
considering chiral EFT as providing a general operator basis that can be used to
expand the RG evolution. At a given order $(Q/\lb)^n$, chiral EFT
includes contributions from one- or multi-pion exchanges and from
contact interactions, with short-range couplings that depend on the
resolution or cutoff scale. As part of the RG evolution, short-range
couplings included in the initial potential evolve. This is
illustrated in Fig.~\ref{connections-a} by comparing the flow of
$\vlowk(k'=0,k=0;\lm)$ with the corresponding momentum-independent
contact interaction $C_0(\lm)$ in subsequent orders of pionless EFT.
In addition, the RG generates higher-order short-range contact
interactions so that observables are exactly reproduced and the
truncation error is unchanged. Consequently, the cutoff variation can
be used to estimate theoretical uncertainties due to higher-order
short-range many-body interactions. We will discuss using cutoff dependence as a
tool in Section~\ref{subsec:cutoff}.

\begin{figure}[t]
 \centering
 \subfloat[][]{%
  \label{fig:deuteronwfsp2-a}%
  \includegraphics*[width=2.8in,clip=]{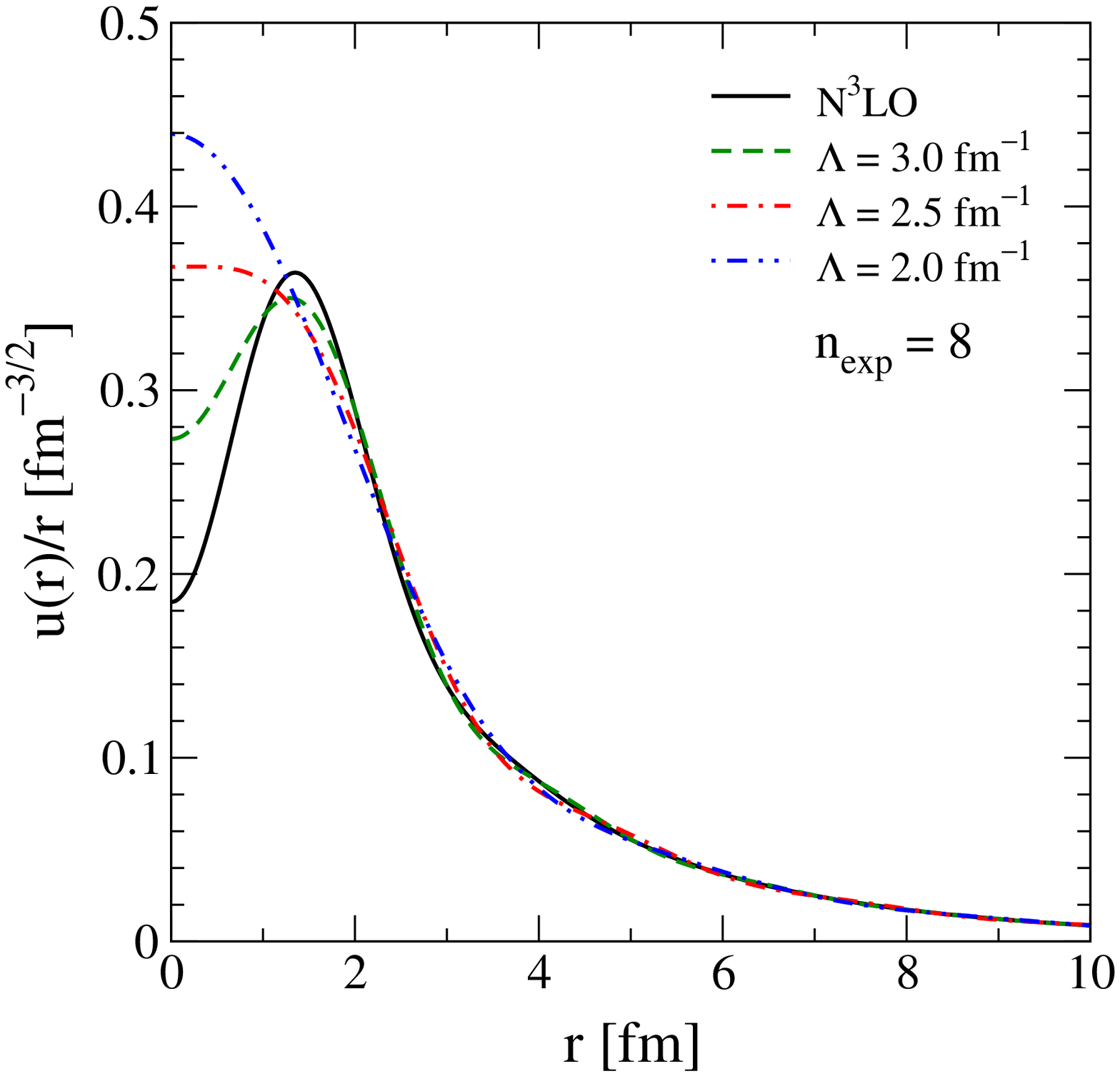}%
 }%
 \hspace*{.4in}%
 \subfloat[][]{%
  \label{fig:deuteronwfsp2-b}%
  \includegraphics*[width=2.8in,clip=]{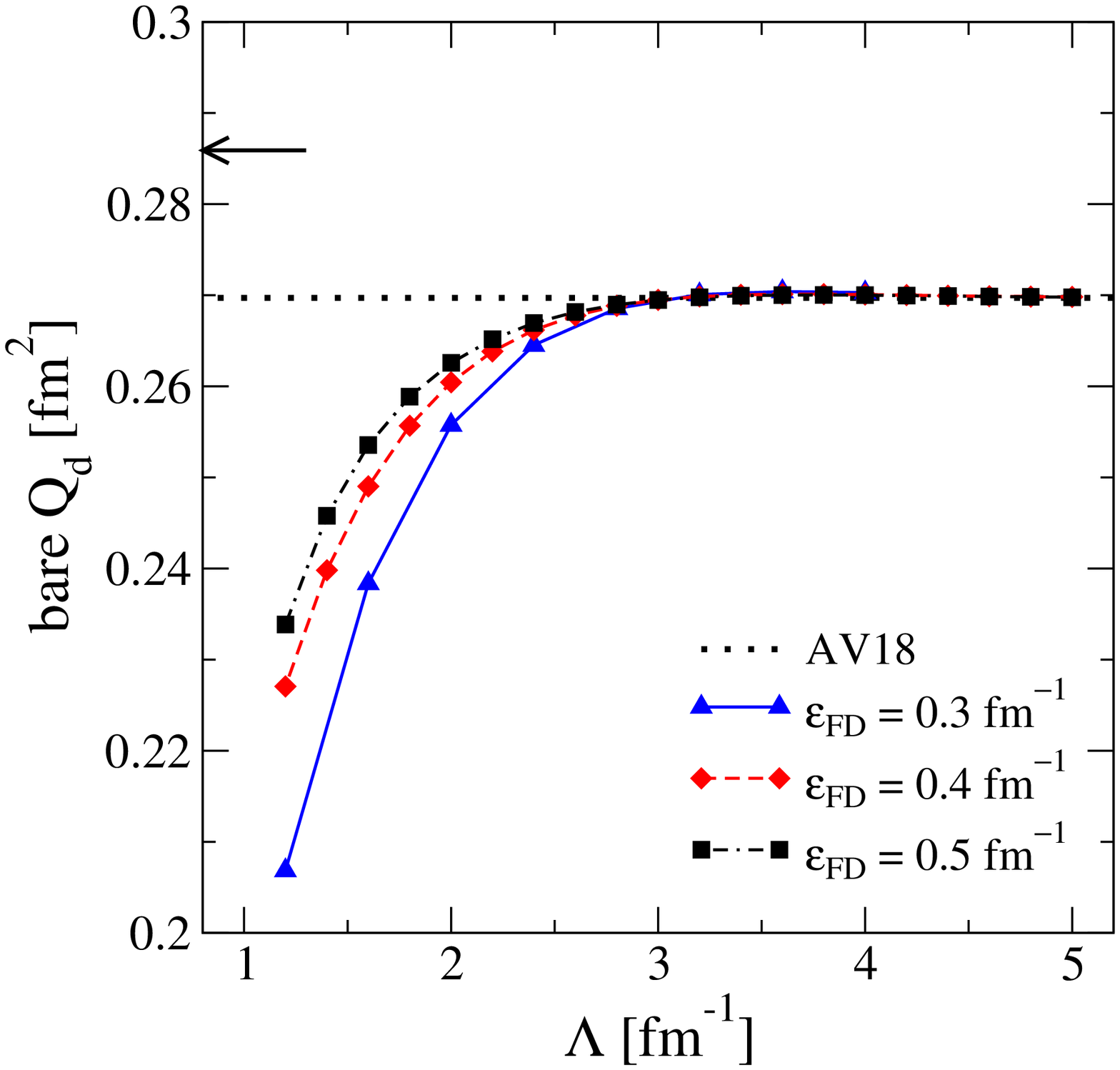}%
 }%
\caption{(a) Deuteron wave functions in coordinate space for smooth
$\vlowk$ interactions at several different cutoffs using an
exponential regulator with $n_{\rm exp} = 8$. The initial interaction
is the N$^3$LO potential of Ref.~\cite{Entem:2003ft}.
(b)~Matrix element of the bare quadrupole moment operator as a
function of the cutoff for several smooth $\vlowk$ regulators. For
details see Ref.~\cite{Bogner:2006vp}.}
\label{fig:deuteronwfsp2}
\end{figure}

For cutoffs large compared to the pion mass, the long-range parts in
the initial chiral potentials are preserved by the RG evolution. This
is demonstrated in Fig.~\subref*{fig:deuteronwfsp2-a} for deuteron
wave functions at different resolution scales.  While the short-range
behavior changes with the cutoff, the long-range parts are governed by
low-energy pion physics and remain unchanged. The results discussed in
Section~\ref{sec:manybody} for many-body interactions and operators
can be explained based on the above observations.

\subsection{Universality}
\label{subsec:universal}

\begin{figure}[t!]
\centering
\includegraphics[scale=0.45,clip=]{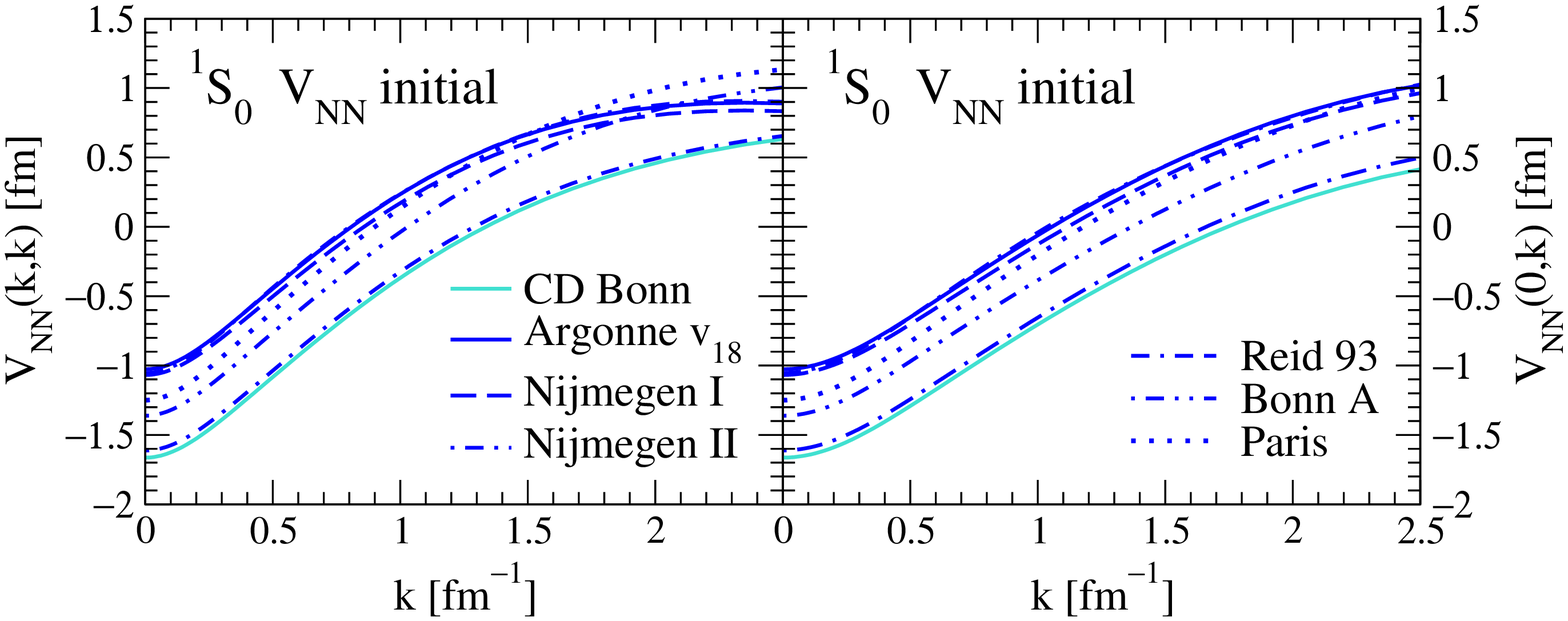} \\
\includegraphics[scale=0.45,clip=]{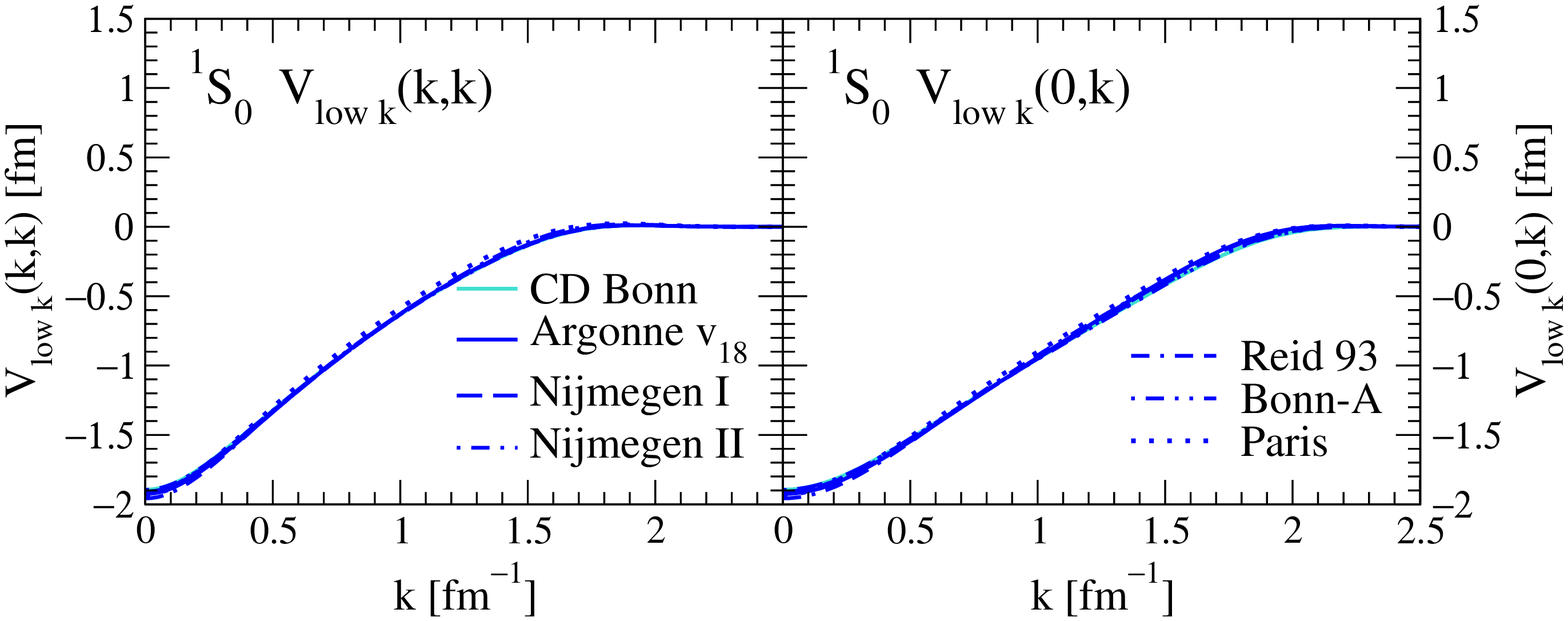}
\caption{Diagonal (left) and off-diagonal (right) momentum-space
matrix elements for various phenomenological NN potentials initially
(upper figures) and after RG evolution to low-momentum
interactions~$\vlowk$~\cite{Bogner:2003wn,Bogner:2006vp} (lower
figures) for a smooth regulator with $\Lambda = 2.0 \fmi$ and
$n_{\rm exp}=4$.\label{universality1}}
\vspace*{.2in}
\centering
\includegraphics[scale=0.45,clip=]{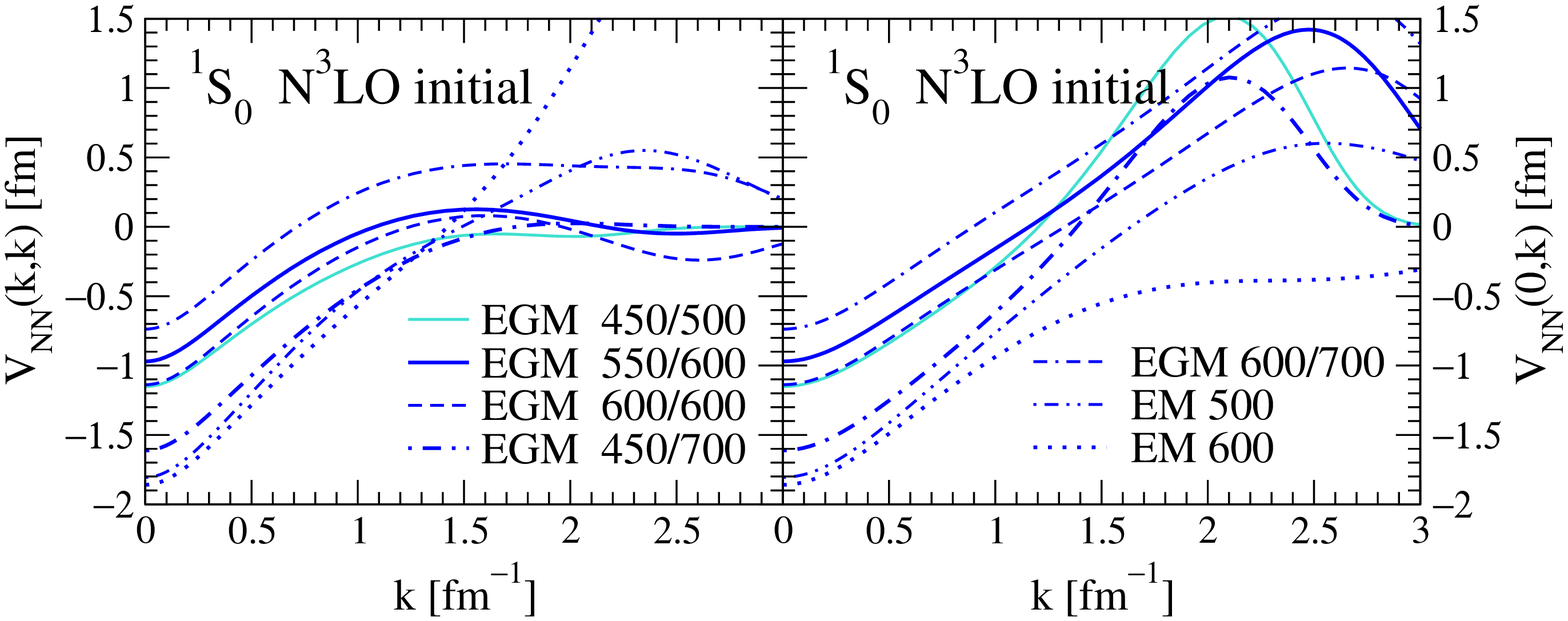} \\
\includegraphics[scale=0.45,clip=]{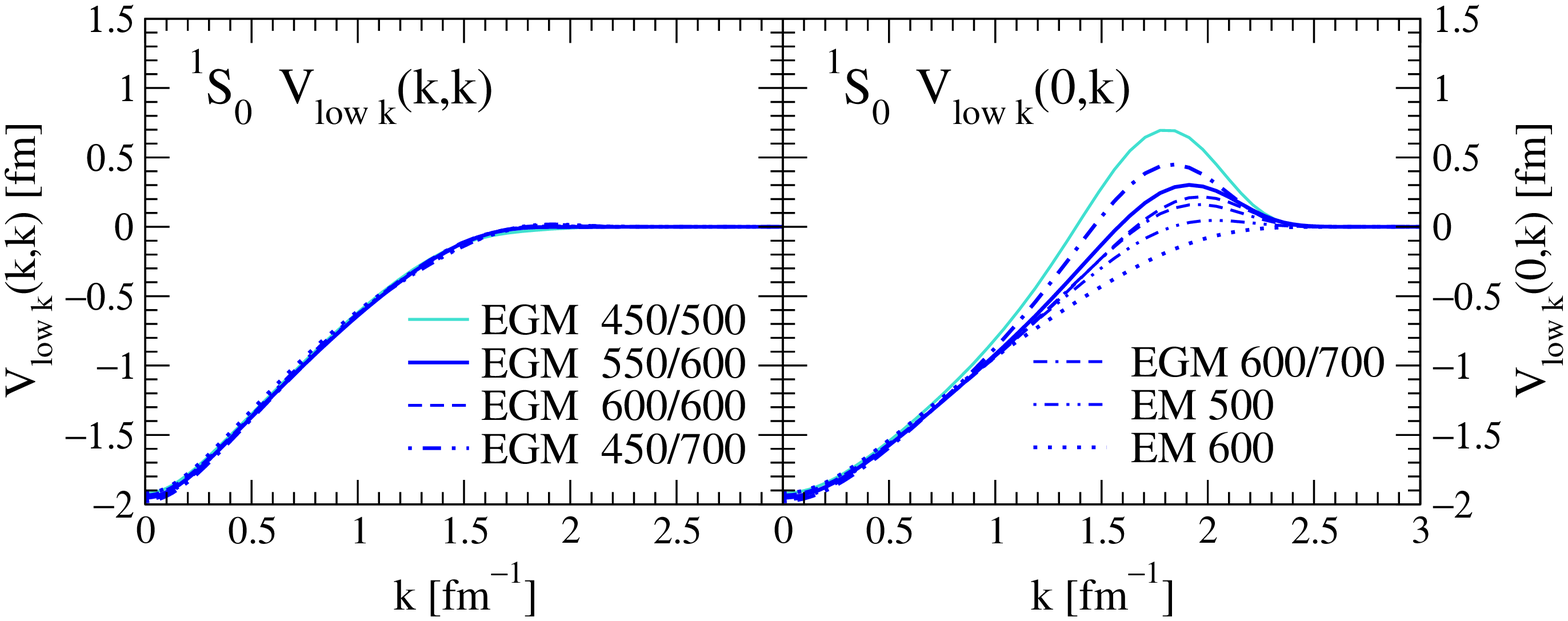}
\caption{Diagonal (left) and off-diagonal (right) momentum-space
matrix elements of different N$^3$LO NN interactions
(EM~\cite{Entem:2003ft} and EGM~\cite{Epelbaum:2004fk}) initially
(upper figures) and after RG evolution to low-momentum
interactions~$\vlowk$~\cite{Bogner:2003wn,Bogner:2006vp} (lower
figures) for a smooth regulator with $\Lambda = 2.0 \fmi$ and
$n_{\rm exp}=4$.\label{universality2}}
\end{figure}

Low-momentum interactions starting from different initial potentials
are found to be quantitatively
similar~\cite{Bogner:2003wn,Bogner:2006vp,Bogner:2006pc}. We use the
terminology universality in this context (independent of the
high-momentum details in the initial potential) and not in the sense
of critical phenomena (independent of the system
details). Figures~\ref{universality1} and \ref{universality2} show
various phenomenological NN potentials and different N$^3$LO NN
interactions of Entem and Machleidt (EM)~\cite{Entem:2003ft} and of
Epelbaum et al.~(EGM)~\cite{Epelbaum:2004fk}. Each potential
accurately reproduces low-energy NN scattering, as shown in
Fig.~\subref*{chiralEFT-b}. When the potentials are evolved to lower
cutoffs, the resulting low-momentum interactions become universal.  In
addition, the RG evolution weakens the off-diagonal coupling between
low and high momenta. Section~\ref{sec:technology} reviews different
low-momentum technologies for achieving this decoupling.

A similar universal behavior and decoupling is found for the
low-momentum parts of SRG evolved interactions, see
Fig.~\subref*{fig:compare_and_weinberg-a}.  We attribute the collapse
to universal low-momentum interactions to the common long-range pion
physics in the initial potentials, and to a similar description of
low-energy NN observables up to the resolution scale, while the
high-momentum parts are decoupled. This is similar to the inverse
scattering problem, where the theory is restricted to the energy range
set by the resolution scale.  Important open problems are to
understand the observed universality in terms of an RG scaling
analysis, to identify eigenoperators of the RG evolution, and to
extend the study of universality to 3N forces.

\subsection{Perturbativeness}
\label{subsec:perturb}

\begin{figure}[t]
 \centering
 \subfloat[][]{%
  \label{pert2nf-a}%
  \includegraphics*[width=3.0in,clip=]{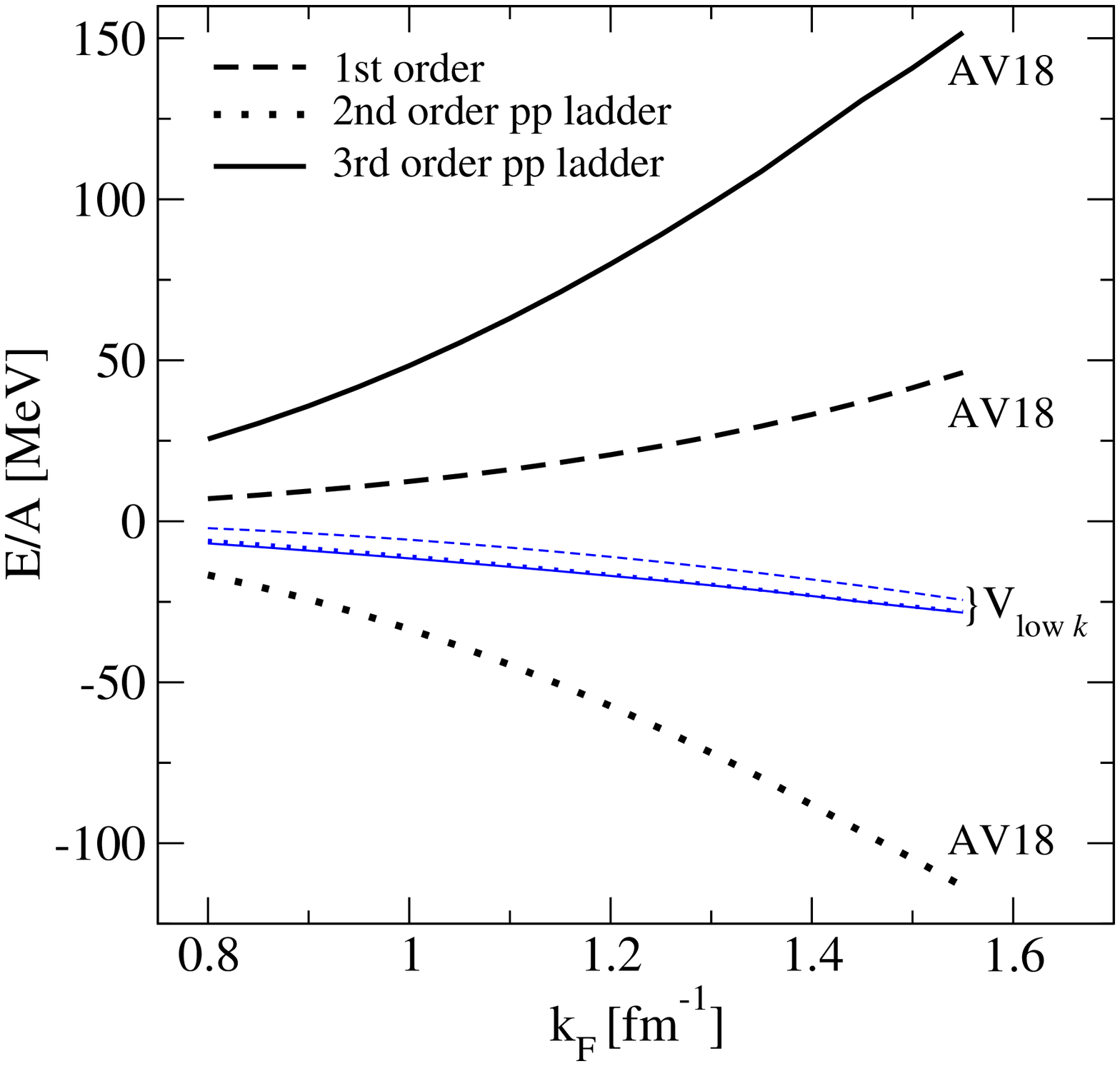}%
 }%
 \hspace*{.4in}%
 \subfloat[][]{%
  \label{pert2nf-b}%
  \includegraphics*[width=2.9in,clip=]{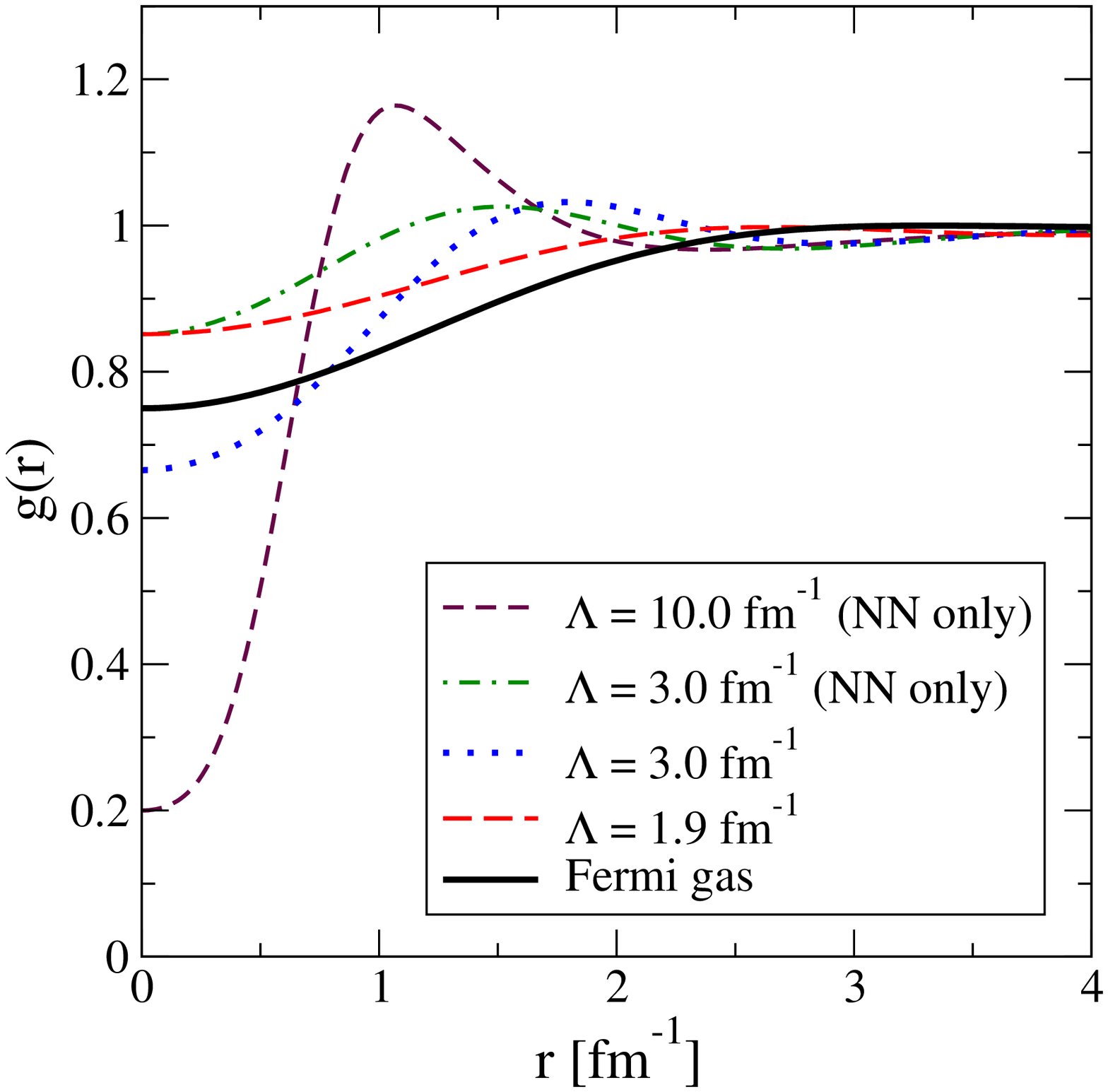}%
 }%
\caption{(a) Particle-particle contributions to the energy per nucleon
in symmetric nuclear matter as a function of the Fermi momentum $\kf$ 
for the initial Argonne $v_{18}$ potential and the RG-evolved $\vlowk$
with  $\lm=2.1 \fmi$~\cite{Bogner:2005sn}. (b)~Pair-distribution
function $g(r)$ in nuclear matter for $\kf = 1.35 \fmi$ at different
resolutions, for details see Ref.~\cite{Bogner:2005fn}.}
\label{pert2nf}
\end{figure}

Nuclear many-body calculations are complicated by strong short-range
repulsion and strong short-range tensor forces found in most
NN potential models. However, both of these
sources of non-perturbative behavior are
resolution-dependent,\footnote{This is in contrast to non-perturbative
features like low-energy bound or nearly bound states in the S-waves
and the pairing instability at finite density that are insensitive
to the short-distance details.} because they depend on the degree of
coupling between low- and high-momentum
states~\cite{Bogner:2005sn,Bogner:2006vp,Bogner:2006pc,Bogner:2006tw}.
Consequently, RG methods can be used to improve perturbative
convergence and reduce the short-range strength of the associated
correlations in the wave functions, as shown for symmetric nuclear
matter in Figs.~\subref*{pert2nf-a} and~\subref*{pert2nf-b}, respectively.

\begin{figure}[t]
 \centering
 \subfloat[][]{%
  \label{fig:etaEcmRep-a}%
  \includegraphics*[width=2.8in,clip=]{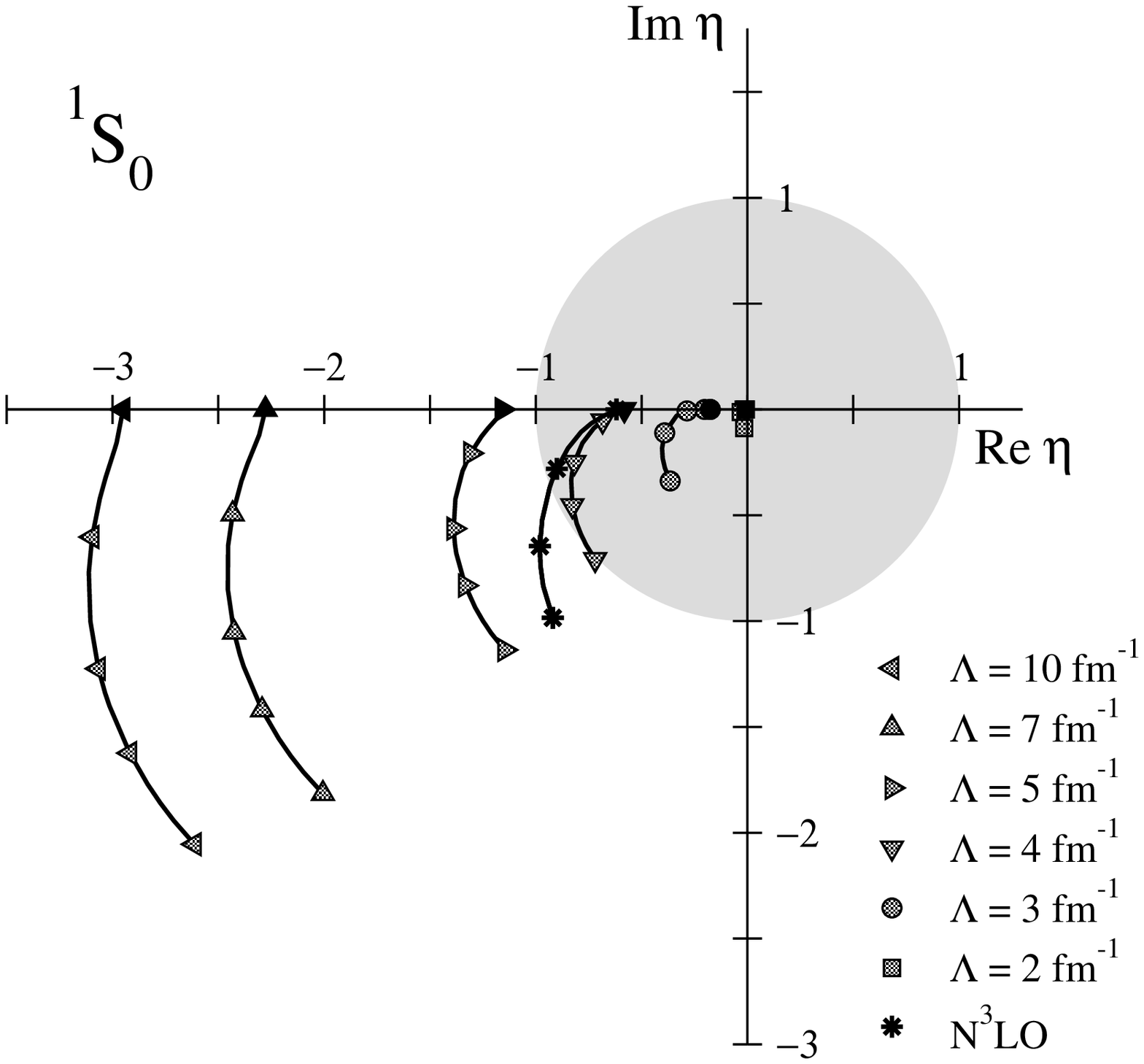}%
 }%
 \hspace*{.4in}%
 \subfloat[][]{%
  \label{fig:etaEcmRep-b}%
  \includegraphics*[width=2.8in,clip=]{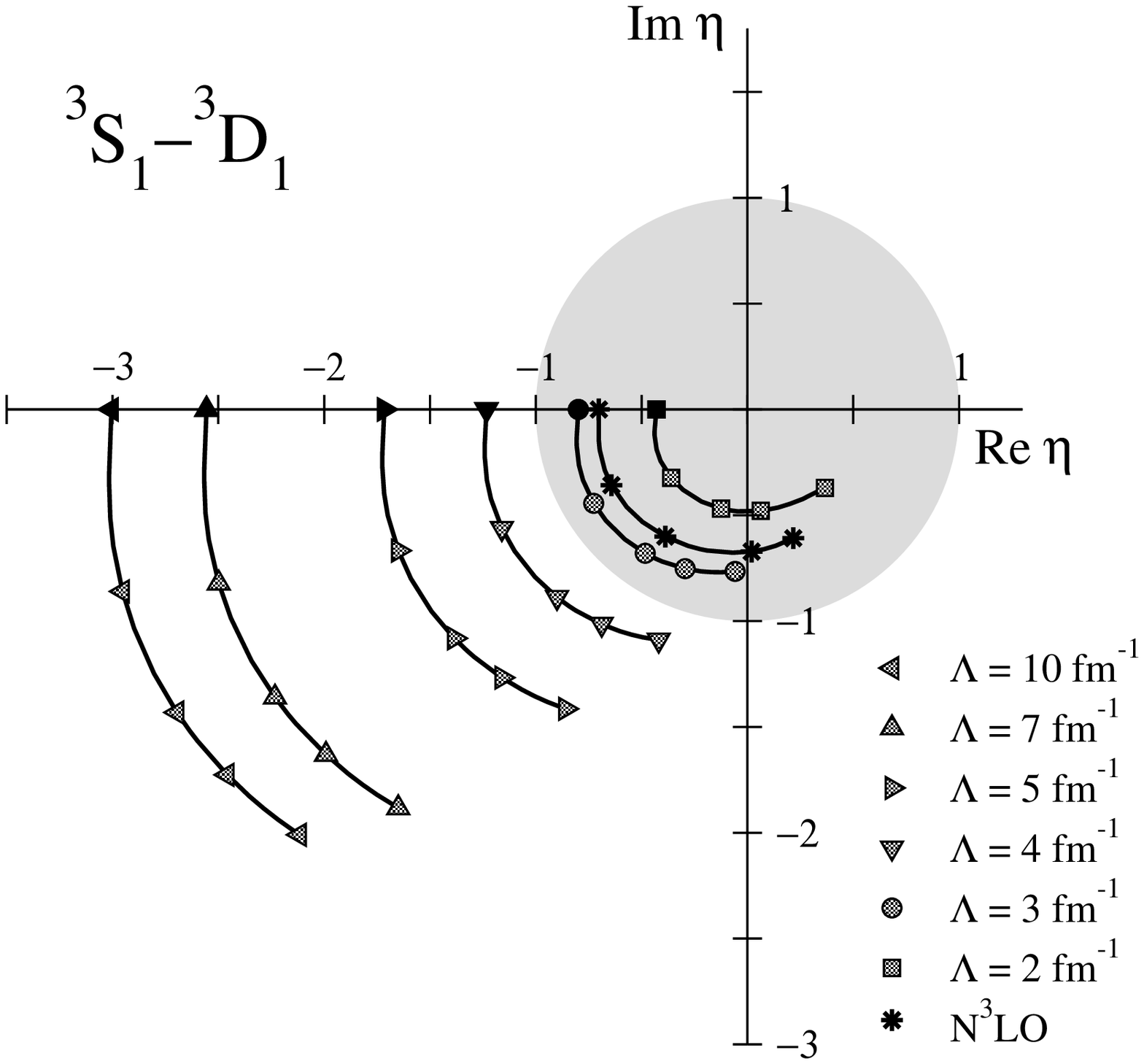}%
 }%
\caption{Trajectories of the largest repulsive Weinberg eigenvalues in
the (a) $^1S_0$ and (b) $^3$S$_1$--$^3$D$_1$ channels as a function of energy 
for $\vlowk$ evolved from the Argonne $v_{18}$ potential~\cite{Bogner:2006tw}.
The results for selected cutoffs are indicated by the different symbols. 
The positions of the symbols on each trajectory mark the eigenvalues
for center-of-mass energies $E_{\rm cm}=0, 25, 66, 100$ and $150 \mev$, starting
from the filled symbol at $0 \mev$. The trajectory with stars are 
eigenvalues for the N$^3$LO potential of Ref.~\cite{Entem:2003ft}.}
\label{fig:etaEcmRep}
\end{figure}

\begin{figure}[ht!]
 \centering
 \subfloat[][]{%
  \label{fig:chiral-a}%
  \includegraphics*[width=3.0in,clip=]{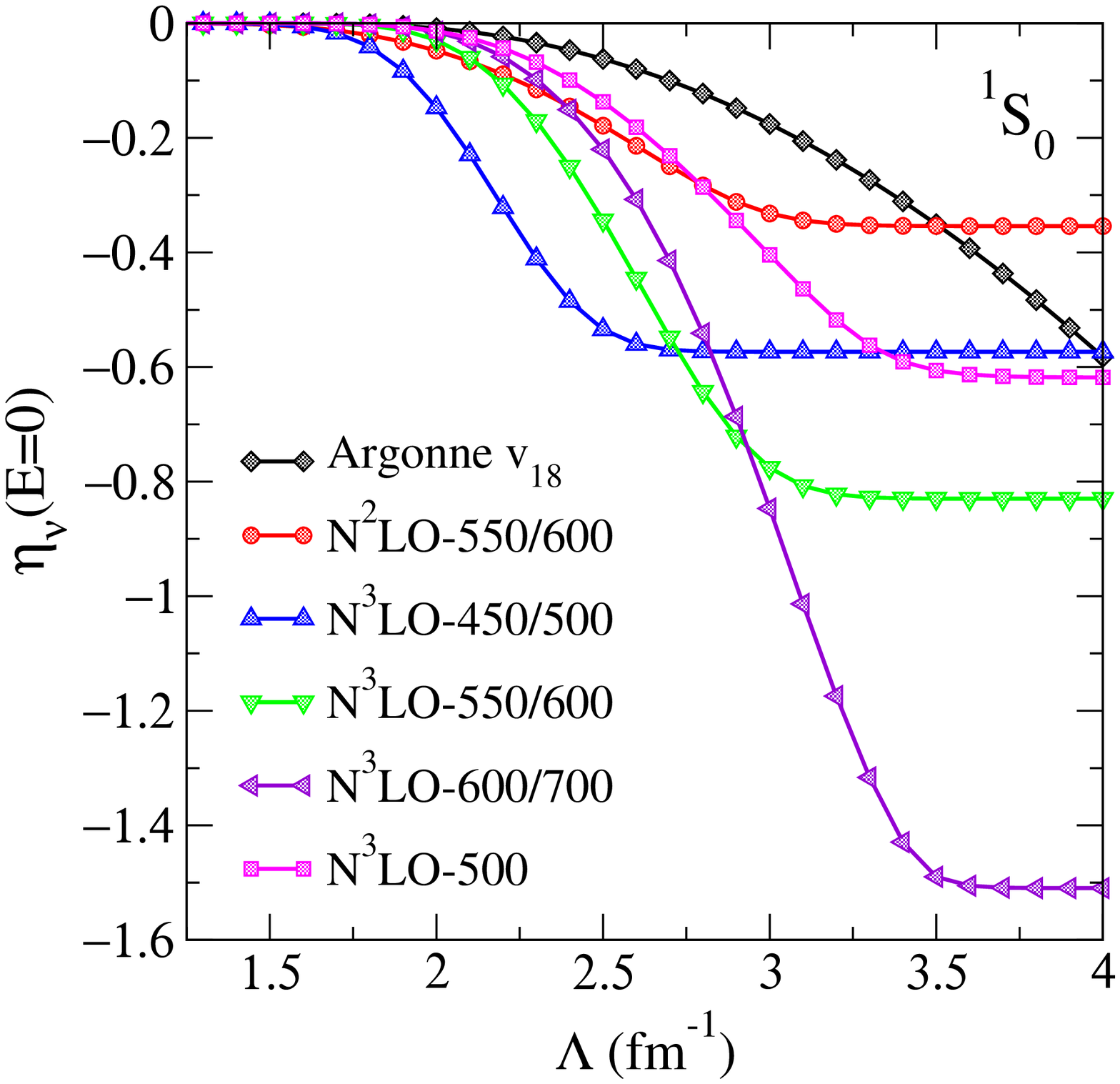}%
 }%
 \hspace*{.4in}%
 \subfloat[][]{%
  \label{fig:chiral-b}%
  \includegraphics*[width=3.0in,clip=]{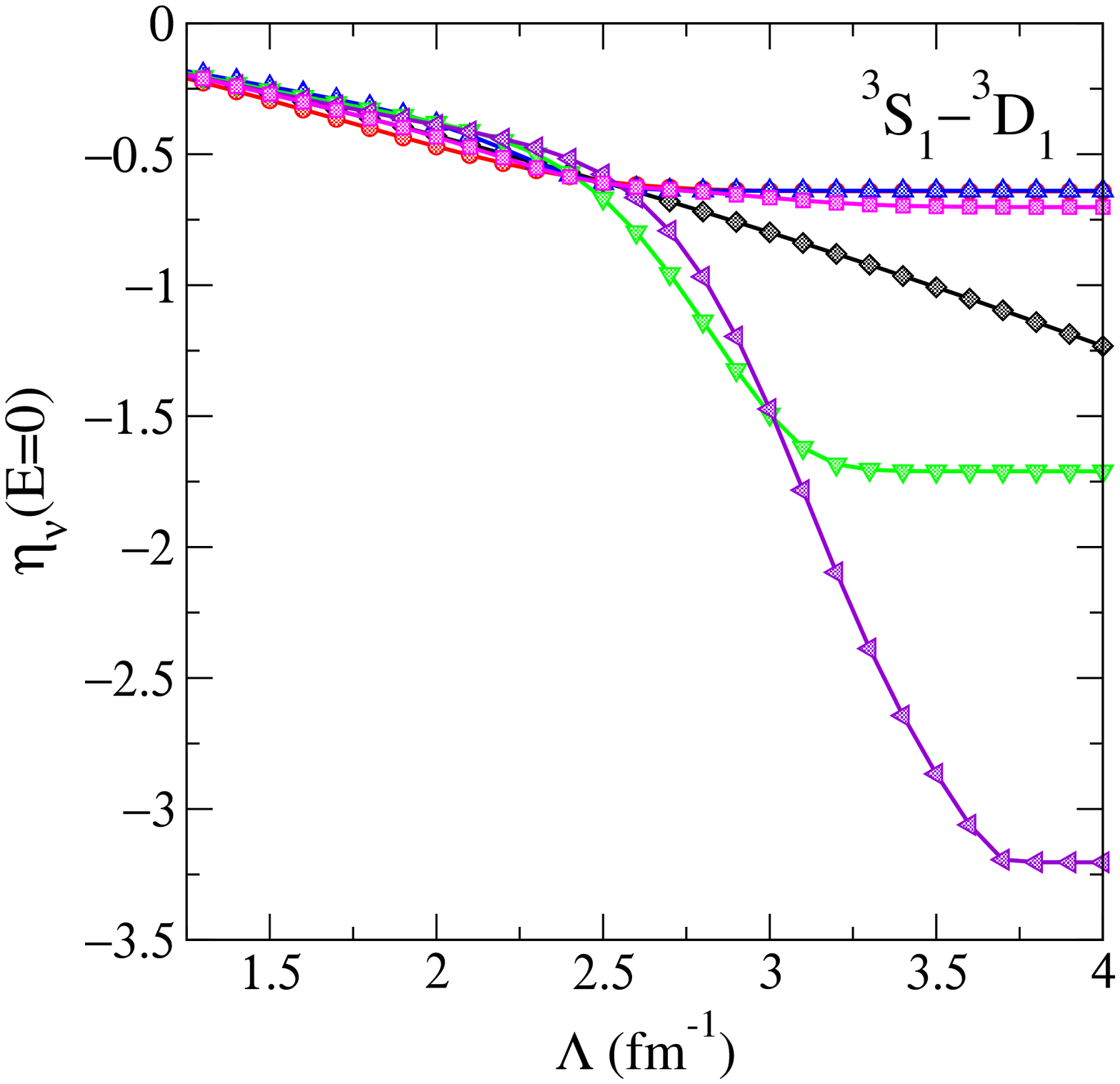}%
 }%
\caption{Largest repulsive Weinberg eigenvalues for $E=0$ in the
(a)~$^1$S$_0$ and (b)~$^3$S$_1$--$^3$D$_1$ channels as a function
of cutoff for $\vlowk$ evolved from chiral EFT
interactions~\cite{Bogner:2005sn}. Results are shown for the
N$^3$LO potential of Entem and Machleidt~\cite{Entem:2003ft}, for
N$^3$LO potentials of Epelbaum et al.~\cite{Epelbaum:2004fk}
with different cutoffs $\Lambda$/$\tilde{\Lambda}$ (as indicated in
MeV), and for an N$^2$LO potential~\cite{Epelbaum:1999dj}.  For
comparison, we have plotted the largest repulsive Weinberg
eigenvalues for $\vlowk$ evolved from the Argonne $v_{18}$
potential.}
\label{fig:chiral}
\end{figure}

We can quantify the perturbativeness of the potential as we evolve to
lower $\Lambda$ for $\vlowk$ (or $\lambda$ for SRG) interactions by
using the eigenvalue analysis introduced long ago by Weinberg
\cite{Weinberg:1963zz} and applied to $\vlowk$ and SRG potentials in
Refs.~\cite{Bogner:2005sn,Bogner:2006vp,Bogner:2006tw,Bogner:2006pc}.
Consider the
Born series for the $T$ matrix at energy $E$ with Hamiltonian $H = H_0
+ V$,
\be
 T(E) = V + V \frac{1}{E-H_0} V + \ldots \,.
\ee
By finding the eigenvalues and eigenvectors of the operator
$(E-H_0)^{-1}V$,
\be
\frac{1}{E - H_0} \, V | \Gamma_\nu \rangle
= \eta_\nu(E) | \Gamma_\nu \rangle \,,
\label{eq:Weinberg}
\ee
and then acting with $T(E)$ on the eigenvectors,
\be
T(E) | \Gamma_\nu \rangle
=  \bigl( 1 + \eta_\nu(E) + \eta_\nu^2(E) + \ldots \bigr)
V | \Gamma_\nu \rangle \,,
\ee
it follows that nonperturbative behavior at energy $E$ is signaled by
one or more eigenvalues with $|\eta_\nu(E)| \geqslant
1$~\cite{Weinberg:1963zz}. A rearrangement of Eq.~(\ref{eq:Weinberg})
gives a simple interpretation of the eigenvalue $\eta_{\nu}(E)$ as an
energy-dependent coupling that must divide $V$ to produce a solution
to the Schr\"odinger equation at energy $E$.  For negative energies, a
purely attractive $V$ gives positive real $\eta_{\nu}(E)$ values,
while a purely repulsive $V$ gives negative eigenvalues. For this
reason, we refer to negative eigenvalues as repulsive and positive
ones as attractive, although the eigenvalues become complex for
positive $E$.

Figure~\ref{fig:etaEcmRep} shows the trajectories of the largest
repulsive Weinberg eigenvalue in the $^1$S$_0$ and
$^3$S$_1$--$^3$D$_1$ channels as a function of (positive) energy for
$\vlowk$ interactions with various cutoffs evolved from the Argonne
$v_{18}$ potential. The magnitude of the largest repulsive eigenvalue
at all energies decreases rapidly as the cutoff is lowered. This
reflects the decrease of the short-range repulsion present in the
initial potential. In the $^1$S$_0$ channel, the trajectory lies
completely inside the shaded unit circle for cutoffs near $4 \fmi$ and
below, which implies that the Born series becomes perturbative with
respect to the repulsive part of the potential. In the
$^3$S$_1$--$^3$D$_1$ channel, the largest repulsive eigenvalues are
just within the unit circle for $\Lambda = 3 \fmi$. The decrease from
$\Lambda = 3 \fmi$ to $2 \fmi$ is significant for ensuring a
convergence of particle-particle ladders in nuclear matter, because
second-order short-range tensor contributions peak at
intermediate-state momenta $k \approx 2.5-3.5 \fmi$ in nuclear
matter~\cite{GerryMBbook}.

Chiral EFT interactions typically have cutoffs below $3.5 \fmi$ and
therefore are expected to be soft potentials. However, as shown in
Fig.~\ref{fig:chiral}, the N$^3$LO potentials of Epelbaum et
al.~\cite{Epelbaum:2004fk} have substantial repulsive Weinberg
eigenvalues in both S-waves. These unexpected features were traced to
the singular central and tensor interactions (at N$^3$LO) in
Ref.~\cite{Bogner:2006tw}. The results in Fig.~\ref{fig:chiral} show
there is a major decrease in the largest repulsive eigenvalues as the
cutoff is run down from $\Lambda = 3 \fmi$ to $2 \fmi$, which shows
that it is advantageous to evolve chiral EFT interactions to lower
cutoffs using the RG.

\begin{figure}[t]
 \centering
 \subfloat[][]{%
  \label{fig:weinberglambda-a}%
  \includegraphics[width=8.0cm,clip=]{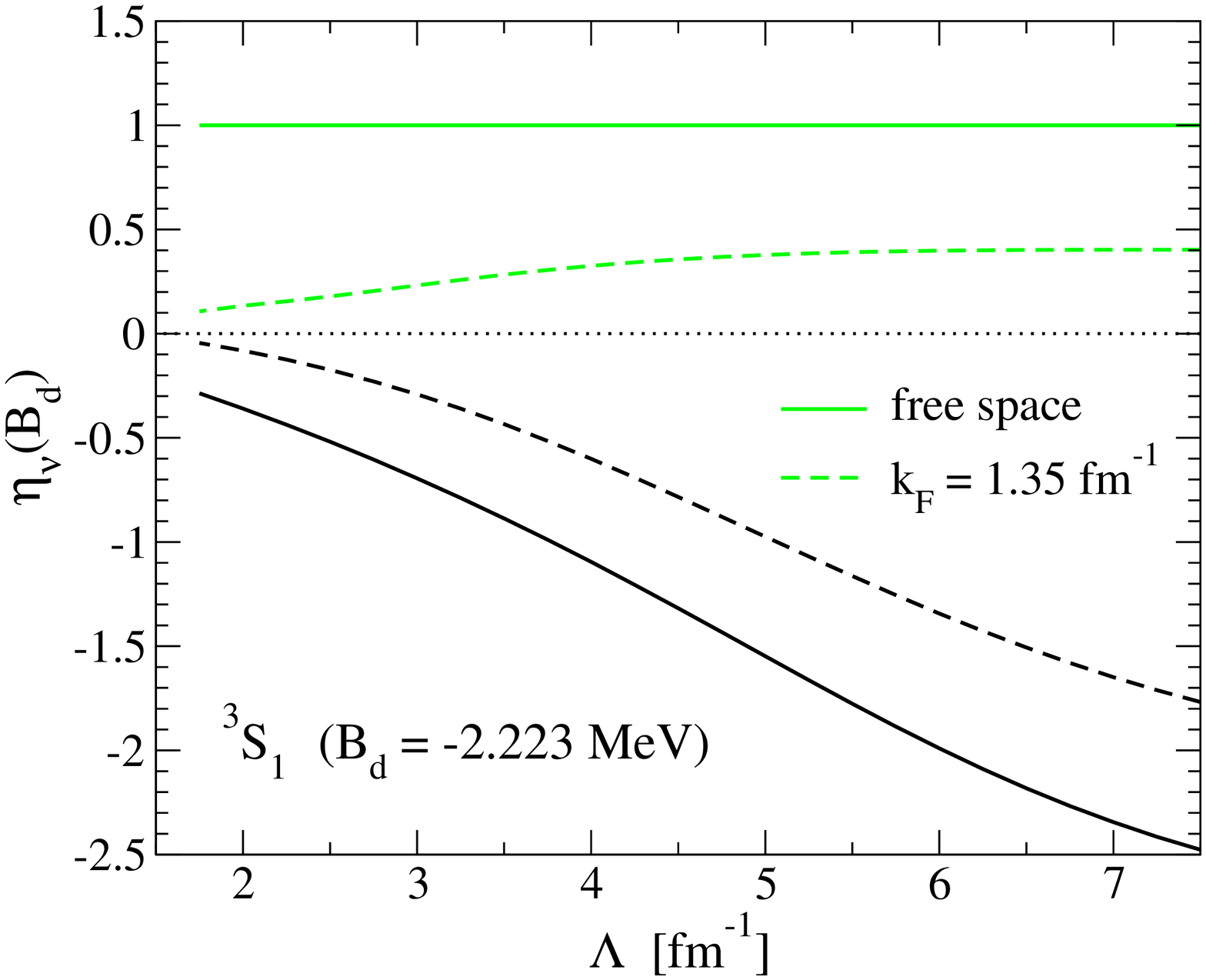}%
 }%
 \hspace*{.4in}%
 \subfloat[][]{%
  \label{fig:weinberglambda-b}%
  \includegraphics[width=8.0cm,clip=]{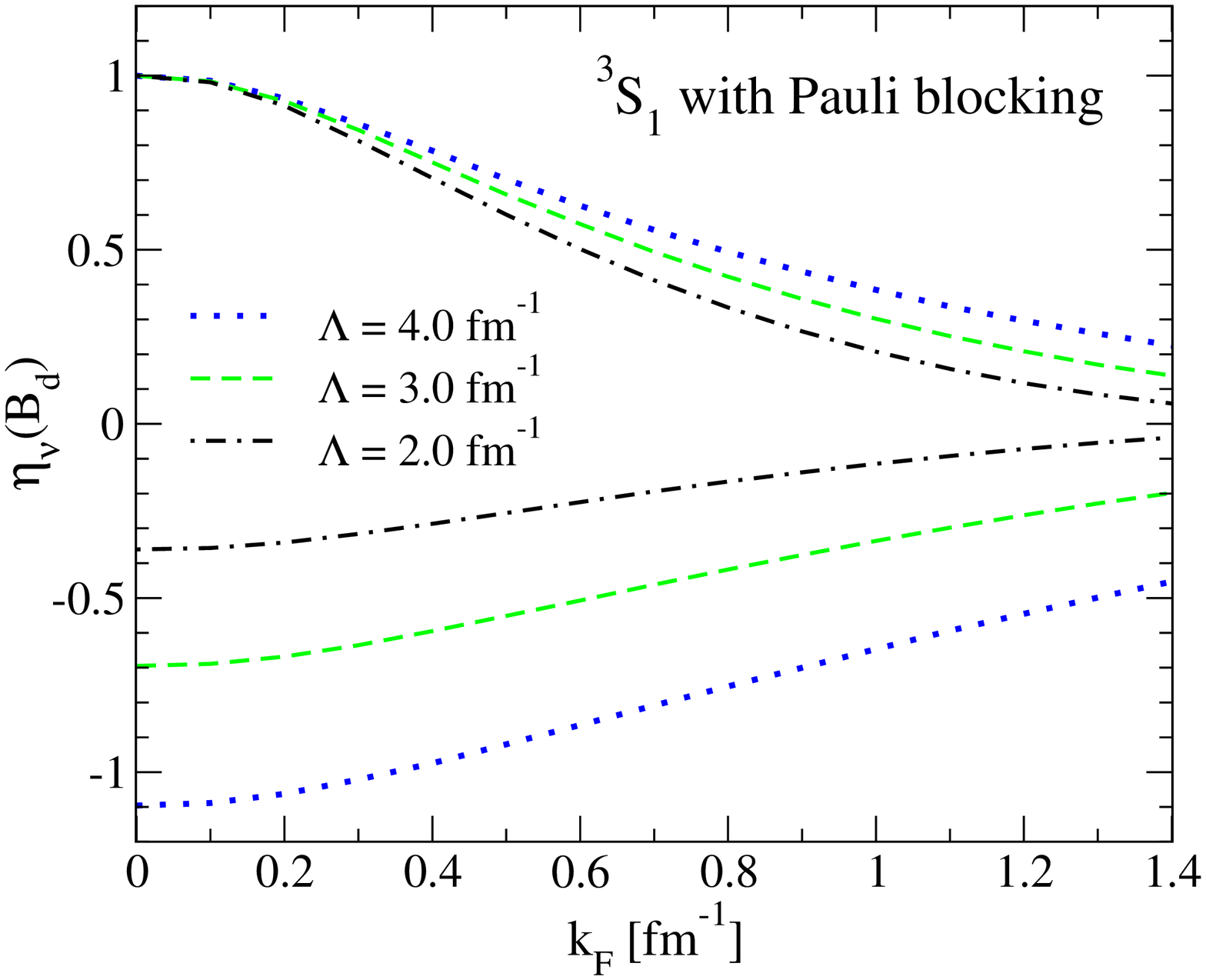}%
 }%
\caption{(a) Evolution of the two largest Weinberg eigenvalues with
$\Lambda$ in the $^3$S$_1$--$^3$D$_1$ channel in free-space (solid)
and at saturation density (dashed). The lighter curves correspond
to attractive eigenvalues and the darker curves to repulsive 
eigenvalues. (b)~Dependence on density of the two largest Weinberg
eigenvalues in the $^3$S$_1$--$^3$D$_1$ channel. Results are based
on $\vlowk$ evolved from the Argonne $v_{18}$ potential. For details
see Ref.~\cite{Bogner:2005sn}.}
\label{fig:weinberglambda}
\end{figure}

The systematics of the Weinberg eigenvalues in free space and at
finite density are compared for the $^3$S$_1$--$^3$D$_1$ channel in
Fig.~\subref*{fig:weinberglambda-a}, which shows the evolution with
$\lm$ of the two largest eigenvalues evaluated at the deuteron pole,
$E=B_d$. In free space, we observe that the repulsive eigenvalue is
initially large but decreases rapidly with cutoff, while the large
attractive eigenvalue corresponding to the deuteron pole remains
invariant, $\eta_\nu(B_d)=1$. The behavior is similar at finite
density, but now the attractive eigenvalue is tamed by Pauli-blocking
effects. This result is general at sufficient density, as shown in
Fig.~\subref*{fig:weinberglambda-b}.  We therefore reach a promising
conclusion: Large-cutoff sources of nonperturbative behavior can be
eliminated using the RG, while physical sources due to weakly and
nearly bound states are suppressed in the medium by
Pauli-blocking. These results are qualitatively unchanged with the
inclusion of low-momentum 3N
interactions~\cite{Bogner:2005sn,Bogner:2009un} (see
Section~\ref{sec:infinite}). Moreover, the Weinberg analysis has
been applied to pairing in Ref.~\cite{Ramanan:2007bb}.

In addition to improving perturbative convergence, RG-evolved
potentials result in more effective variational
calculations~\cite{Bogner:2006ai} and rapidly converging basis
expansions~\cite{Bogner:2007rx,Bacca:2009yk}, as discussed in
Section~\ref{sec:finite}. Finally, we mention that as a by-product of
the Weinberg analysis in Ref.~\cite{Bogner:2006tw}, it was found that
RG-evolved interactions can be accurately described by low-rank
separable expansions, a result that allows significant simplifications
in applications ranging from few-body scattering~\cite{Shepard:2009gi}
to the treatment of pairing in finite
nuclei~\cite{Duguet:2007be,Lesinski:2008cd}.

\subsection{Many-body interactions and operators}
\label{subsec:intops}
 
\begin{figure}[t]
\centering
\includegraphics*[width=6.2in]{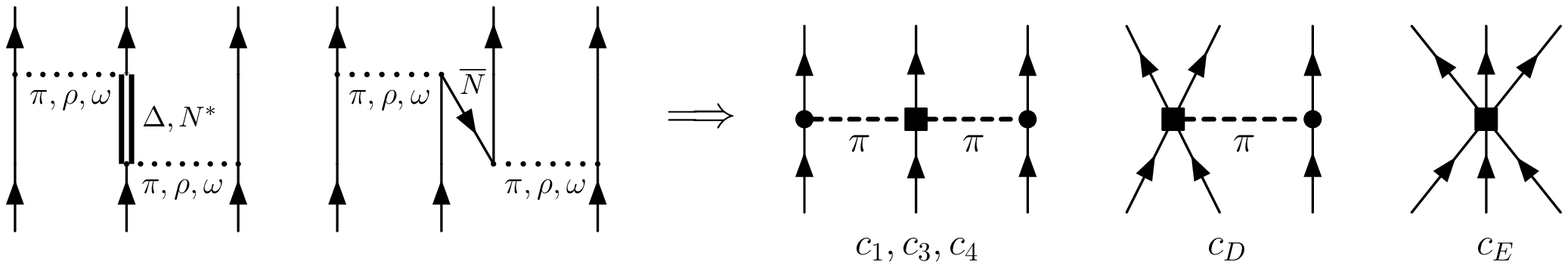}
\caption{Eliminating degrees of freedom leads to three-body forces.}
\label{fig:3NFsources}
\end{figure}

In a low-energy effective theory of finite-mass composite particles
such as atoms or nucleons, three-body (or higher-body) interactions
are defined as a contribution to the Hamiltonian that is not accounted
for by the sum of pairwise interactions. The polarization of
interacting atoms or molecules provides an intuitive example. The
Axilrod-Teller potential is a three-body version of the familiar
two-body long-range van der Waals force between atoms; its physical
origin is triple-dipole mutual
polarization~\cite{Axilrod:1943aa}. Because it contributes at third
order in perturbation theory and the fine structure constant is small,
there is a rapidly decreasing hierarchy of many-body forces. Indeed,
this contribution is usually negligible in metals and semiconductors
although not in rare gas solids~\cite{Mahan:1990aa}. For solid xenon,
its contribution is calculated to be about 10\% of the ground-state
energy~\cite{Bell:1976aa}, comparable to the typical 3N force
contribution to the $^3$H binding energy. Note that polarization with
additional atoms means that the Hamiltonian for an $A$-body system
will inevitably lead to $A$-body forces.

The EFT perspective (see Section~\ref{subsec:forces}) confirms that
operators (including the Hamiltonian) in \emph{any} low-energy
effective theory, if it is systematic rather than a model, will have
many-body components.  The generic origin of these many-body operators
is a restriction of degrees of freedom, such as 3N forces arising from
the elimination of the $\Delta$ or anti-nucleon components, as
illustrated in Fig.~\ref{fig:3NFsources}.  Integrating out or
decoupling high momentum modes with the RG is just another example.
The EFT expansion systematically includes all such contributions as a
combination of long-range (for example, due to pion exchanges) and
short-range (contact) terms as on the right side of
Fig.~\ref{fig:3NFsources}, even when the origin of the short-range
parts is unclear.  Power counting in chiral EFT establishes a
decreasing hierarchy of many-body operators (see
Fig.~\subref*{chiralEFT-a} for many-body forces) that permits
truncation at a tractable level (which in present nuclear structure
calculations means 3N forces and two-body current operators).

\begin{figure}[t]
 \centering
 \subfloat[][]{%
  \label{Tjon-a}%
  \includegraphics[width=3.2in,clip=]{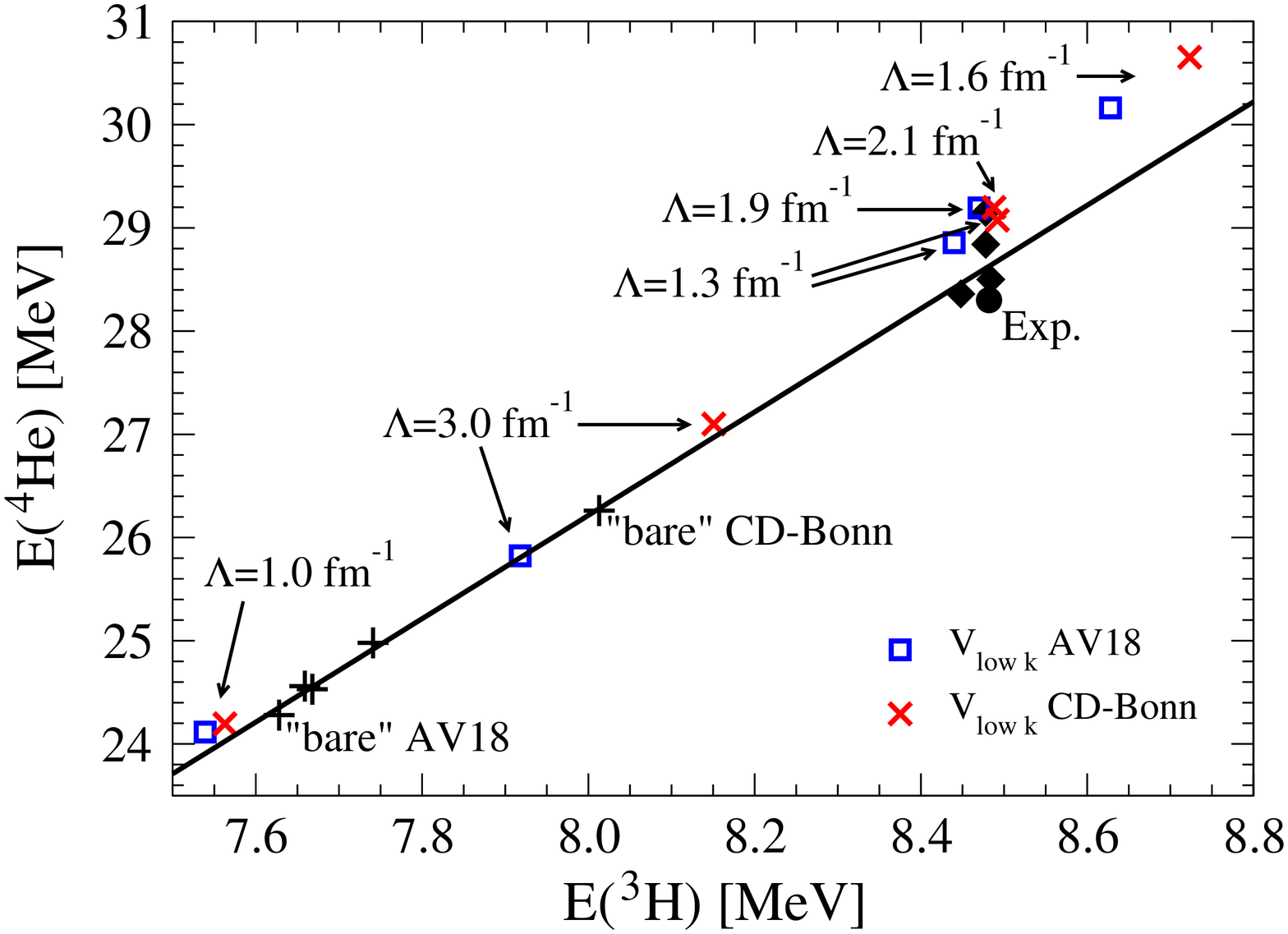}%
 }%
 \hspace*{.2in}%
 \subfloat[][]{%
  \label{Tjon-b}%
  \includegraphics[width=3.2in,clip=]{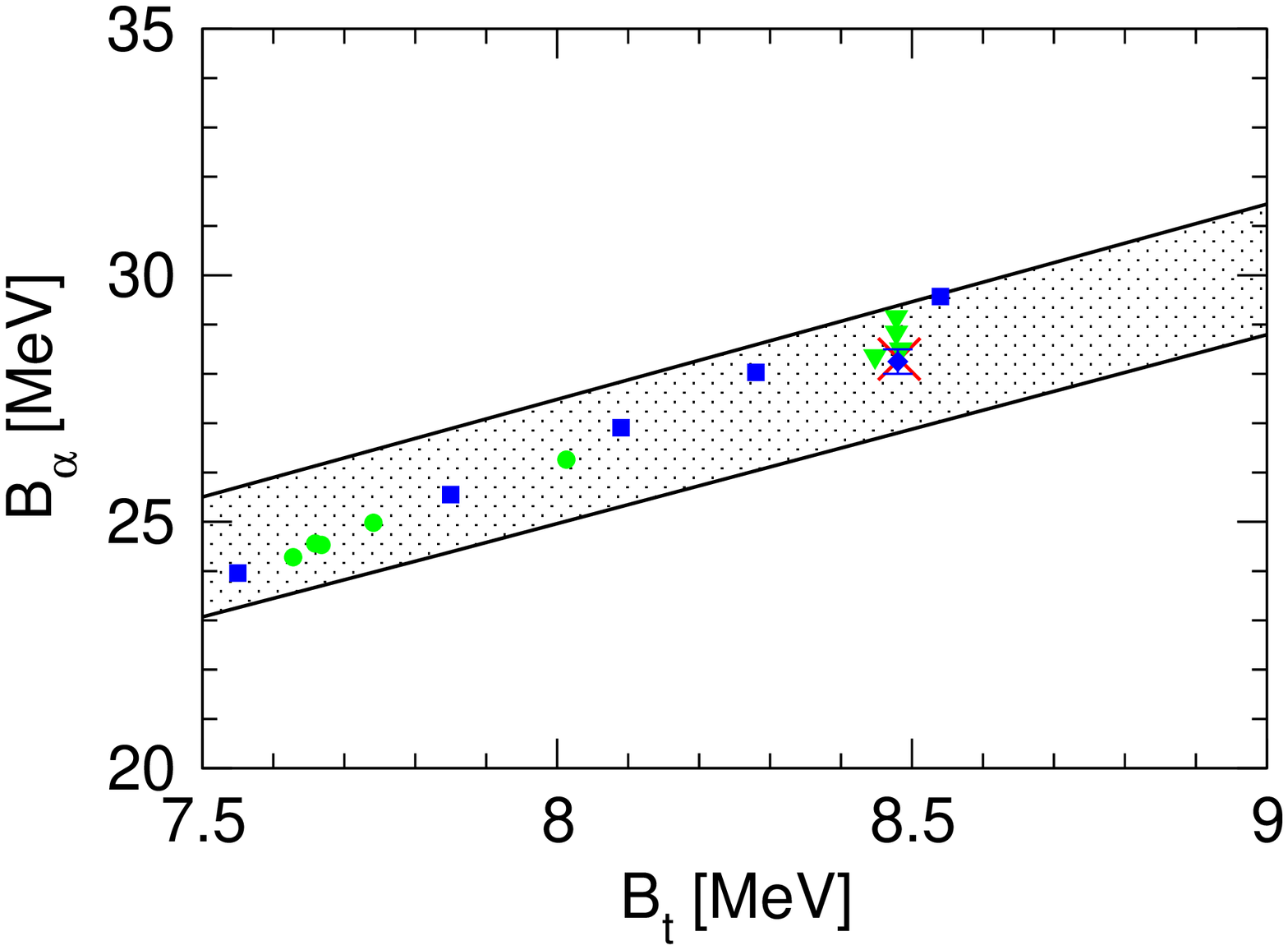}%
 }%
\caption{Correlation of $^3$H and $^4$He binding energies. (a)~The
cutoff dependence obtained from low-momentum NN interactions $\vlowk$
compared to the Tjon line~\cite{Nogga:2004ab}. (b)~This correlation is driven
by large scattering lengths, as demonstrated by the band obtained in
pionless EFT~\cite{Platter:2004zs}.\label{Tjon}}
\end{figure}

The strength of many-body components in an initial operator will shift
with any change in how high-energy degrees of freedom are coupled to
the low-energy degrees of freedom; particular examples are the running
of a $\vlowk$ cutoff $\Lambda$ or an SRG flow parameter $\lambda$.
The correlation plot in Fig.~\subref*{Tjon-a} of $^3$H and $^4$He
binding energies calculated using $\vlowk$ NN-only interactions shows
the change in the (omitted) many-body contribution with $\Lambda$.
This reproduces the empirical Tjon line from phenomenological
potentials.  The technical challenge is to carry out the evolution of
many-body forces and operators in an RG implementation.  The physics
challenge is to establish that the EFT hierarchy of many-body
components is not affected by the evolution, so that a tractable
truncation is still possible.  This typically means that the evolution
is not extended below $\Lambda$ or $\lambda$ of about $1.5 \fmi$.
(For cutoffs below the pion mass, three-body forces will increase to
leading order in pionless EFT~\cite{Bedaque:2002mn,Braaten:2004rn}.)

The consistent RG evolution of $\vlowk$ many-body interactions has not
yet been achieved. The underlying difficulty is that the technology
used to construct $\vlowk$ requires the solution of the full 3N
problem (bound state wave functions plus scattering wave functions in
all breakup channels) to consistently evolve 3N forces (without
simplifications). In $\vlowk$ calculations to date, the three-body
evolution is therefore approximated by fitting the leading chiral EFT
3N forces at each cutoff while evolving the two-body interaction
exactly~\cite{Nogga:2004ab,Bogner:2009un}.  This takes advantage of
the EFT expansion being a complete operator basis for 3N forces, so
that the leading effects of the evolution are simply a change in the
operator coefficients (see Section~\ref{subsec:3nf} for more details).

The SRG offers a new path to the consistent running of many-body
interactions and operators because they evolve through unitary
transformations that require only a representation in a convenient
basis.  This has recently been demonstrated in
practice~\cite{Jurgenson:2009qs}; we discuss the details in
Sections~\ref{subsec:evolution} and \ref{subsec:effops}.  To see how
the two-, three-, and higher-body potentials are identified and
evolved, it is useful to decompose the running SRG Hamiltonian%
\footnote{The SRG Hamiltonian is denoted equivalently by $H_\lambda$
or $H_s$, where $\lambda \equiv 1/s^{1/4}$.}  $H_\lambda$ in
second-quantized form.  Schematically (suppressing indices and sums),
\be
H_\lambda = \la T \ra \, \adag a  
+ \la V_\lambda^{(2)} \ra \, \adag \adag a a
+ \la V_\lambda^{(3)} \ra \, \adag \adag \adag a a a + \ldots \,,
\label{eq:2ndquant}
\ee
where $\adag$, $a$ are creation and annihilation operators with
respect to the vacuum in a single-particle basis.  This \emph{defines}
$\la T \ra$, $\la V_\lambda^{(2)} \ra$, $\la V_\lambda^{(3)} \ra$,
\ldots as the one-body (kinetic energy), two-body, three-body, \ldots
matrix elements at each $\lambda$.  The SRG evolution is dictated by
commutators involving $H_\lambda$ (see Eq.~\eqref{eq:commutator});
when they are evaluated using $H_\lambda$ from
Eq.~\eqref{eq:2ndquant}, we see that even if initially there are only
two-body potentials, higher-body interactions are generated with each
step in $\lambda$.  Thus, when applied in an $A$-body subspace, the SRG
will ``induce'' $A$-body forces.  But we also see that $\la T \ra$ is
fixed, $\la V_\lambda^{(2)} \ra$ is determined only in the $A=2$
subspace with no dependence on $\la V_\lambda^{(3)} \ra$, $\la
V_\lambda^{(3)} \ra$ is determined in $A=3$ by subtraction for given
$\la V_\lambda^{(2)} \ra$, and so on.  With this formulation,
$H_\lambda$ is a free-space Hamiltonian, independent of $A$.  It is
also possible to normal order in the medium, which changes the
definition of the creation and annihilation in Eq.~\eqref{eq:2ndquant}
and shifts higher-body pieces to the zero-, one-, and two-body levels
(see Section~\ref{subsec:normal}).

Although one often characterizes the changes with $\Lambda$ or
$\lambda$ by saying that the RG evolution ``induces'' many-body
forces, this should not be interpreted as meaning the initial
Hamiltonian has a ``true'' three-body force.  The initial and evolved
interactions are equally effective interactions and if the EFT
truncation error is preserved in the evolution, there is no physics
reason to prefer one over the other.  (Indeed, referring to the
initial interaction as the ``bare'' interaction is misleading in the
same way.) Each Hamiltonian has different associated many-body
components and operators, although the long-range parts are the same,
as they are not modified by the evolution.  A consequence of the
latter is that operators that probe only long distances will not
evolve, as seen by the weak cutoff dependence until small $\Lambda$
for matrix elements of the bare quadrupole moment operator in
Fig.~\subref*{fig:deuteronwfsp2-b}.

\subsection{Cutoff dependence as a tool}
\label{subsec:cutoff}

The RG equations for $\vlowk$ or SRG interactions start with a
Hamiltonian as an initial condition, which together with the
corresponding operators fixes the values of observables.  In principle
there should be no change in observables as the
Hamiltonian is evolved.  (More precisely, in the usual $\vlowk$
formulation, low-energy observables are unchanged while for the SRG
all observables are preserved.)  In practice there will be
approximations both in the implementation of the RG and then in the
subsequent calculations of nuclear structure observables.  That is,
cutoff dependence arises because of the truncation (or approximation)
of ``induced'' many-body forces (see Section~\ref{subsec:intops}) or
because of many-body approximations.  Because observables should be by
construction independent of the RG cutoff or flow parameter, we can
use changes as a diagnostic of approximations and to estimate
theoretical errors.

In more general applications of the RG, it is expected that
significant changes in the cutoff are needed to glean useful
information. In the examples considered here, however, the range of
cutoff variation is a factor of two or in some cases even
smaller.  However, even small changes in the RG parameter can make
dramatic shifts in the physics of nuclear structure observables, such
as integrating out the short-range repulsion or short-range tensor
forces.

Here are some examples of using RG cutoff dependence as a tool:
\bi
\I Cutoff dependence at different orders in an EFT expansion.  In
Figs.~\subref*{chiralEFT-b}, \subref*{fig:3nfobs}, and
\subref*{fig:lithium6}, the spread of predictions for EFT interactions
with a range of cutoffs indicates whether the results improve as
expected.  This carries over directly to low-momentum interactions
constructed from initial EFT Hamiltonians at different orders.

\I Running of ground-state energies with cutoff in few-body systems.
The variation in Fig.~\ref{fig:h3_srg} shows the scale of the (net)
omitted many-body forces, which is then an estimate of the error from
this approximation.  This allows a comparison with expectations from
EFT power counting.

\I Tjon line and the nature of correlations (see Fig.~\ref{Tjon}).
The cutoff dependence obtained from low-momentum NN interactions
$\vlowk$ in Fig.~\subref*{Tjon-a} explains the empirical (solid) Tjon
line~\cite{Nogga:2004ab}.  This correlation is driven by large
scattering lengths, as demonstrated by the band in
Fig.~\subref*{Tjon-b} obtained in pionless EFT~\cite{Platter:2004zs}.

\I Calculations of nuclear matter.  The decrease in cutoff dependence
with improved approximations (see Fig.~\ref{nm_all}) helps validate
the many-body convergence and establishes a lower bound on the error
from short-range many-body interactions.

\I Calculations of nuclei.  For example, the cutoff variation in
helium halo nuclei with NN-only interactions identifies sensitivity to
3N force effects~\cite{Hagen:2006pq,Bacca:2009yk} (see
Fig.~\ref{4-8He_v2}).  We refer to Section~\ref{subsec:abinitio} for
more examples of cutoff dependence used as a diagnostic of missing
many-body forces.

\I Identification of (non-)observables. If a quantity changes more
rapidly with the cutoff than any possible error, it is not an
observable.  An example is the D-state probability of the deuteron,
whose cutoff dependence as shown in Fig.~\ref{fig:properties}
demonstrates it is not an observable.  In contrast, the independence
of the asymptotic D/S-state ratio is evidence it is an observable.

\I Cutoff dependence can be used in EFT as a tool to determine power
counting (see Ref.~\cite{Birse:2009my} and references therein).  This
entails running a cutoff $\Lambda$ lower, rescaling in units of
$\Lambda$, and looking for fixed points of the RG flow; that is, for
flows that end in scale independence.  The EFT is then expanded around
such a fixed point.  Examples for the nuclear case are described in
Ref.~\cite{Birse:2009my}.
\ei
The common lesson here is that it is advantageous to do calculations
for a range of cutoff values.


\section{Low-momentum technology for two-nucleon interactions} 
\label{sec:technology}

In this section, we give an overview of the equations and techniques
used to derive low-momentum NN interactions (with
many-body interactions treated in Section~\ref{sec:manybody}).  We use
the equations as a guide to how renormalization group methods decouple
low- and high-energy degrees of freedom but refer to the literature
for the more technical details.

\subsection{Sharp cutoff $\vlowk$}
\label{subsec:sharp}

Imposing a sharp cutoff $\Lambda$ in relative momentum is the most
direct way to limit the resolution of an NN potential by excluding
high-momentum modes.  But to incorporate the relevant details from
such modes into low-momentum interactions, they must be integrated out
rather than simply truncated.  The $\vlowk$
approach~\cite{Bogner:2001gq, Bogner:2001jn, Bogner:2003wn} does this
by demanding that the (half-on-shell) $T$ matrix for an initial
potential $\vnn$ be unchanged in every NN partial wave as $\lm$ is
lowered, so that
\begin{align}
T(k',k;k^{2}) &= \vnn(k',k) + \frac{2}{\pi} \, \mathcal{P} 
\int_{0}^{\lm_\infty} 
\frac{\vnn(k',p) \, T(p,k;k^{2})}{k^{2}-p^{2}} \, p^{2} dp \,, \\[2mm]
&= \vlowk^\lm(k',k) +  \frac{2}{\pi} \, \mathcal{P} 
\int_{0}^{\lm} \frac{\vlowk^\lm(k',p) \, T(p,k;k^{2})}{k^{2}-p^{2}} \, 
p^{2} dp \,,
\label{eq:vlowkmatching}
\end{align}
for all $k,k' < \Lambda$.  Note that this means that $\vlowk$ only has
momentum components below the cutoff but the invariance of
$T(k',k;k^2)$ ensures that observables for these momenta (such as
phase shifts
and the deuteron binding energy) are preserved.

Imposing $dT(k',k;k^2)/d\lm=0$ on Eq.~\eqref{eq:vlowkmatching} gives
the RG equation~\cite{Bogner:2001gq, Bogner:2001jn}%
\footnote{The derivation of Eq.~(\ref{RGEsharp}) is slightly more
complicated in the presence of bound states, although the final
result remains the same~\cite{Bogner:2008xh}.}
\be
\frac{d}{d \lm} \vlowk^\lm(k',k) = \frac{2}{\pi} \frac{\vlowk^\lm(k',\lm) \,
T^\lm(\lm,k;\lm^{2})}{1-(k / \lm)^{2}} \,.
\label{RGEsharp}
\ee
Integrating Eq.~\eqref{RGEsharp} with the large cutoff initial
condition%
\footnote{The cutoff $\lm_\infty$ is chosen large enough so that the
initial potential $\vnn$ is effectively zero for $k > \lm_\infty$.
This can be very large for phenomenological potentials, for example,
$\lm_\infty \gtrsim 25\infm$ for the Argonne $v_{18}$
potential~\cite{Wiringa:1994wb}.} $V^{\lm_{\infty}}_{{\rm low}\,k}
= \vnn$ is equivalent to first re-summing high-momentum ladders in an
energy-dependent effective interaction, which is the solution to the
two-body Bloch-Horowitz equation in momentum space with the projector
$Q = \frac{2}{\pi} \int_{\lm}^{\infty} q^2 dq \, |q \ra \la q|$, and subsequently
trading energy dependence for momentum dependence by using the
equations of motion. This two-step procedure is also equivalent to the
commonly-used Lee-Suzuki transformation method used to construct
energy-independent effective
interactions~\cite{Lee1980173,Suzuki:1980yp,%
HjorthJensen:1995ap,Bogner:2003wn}. Note, however,
that the RG approach differs from Lee-Suzuki as the $Q$-space block of
the effective Hamiltonian is set to zero. As a result of this
equivalence (see Ref.~\cite{Bogner:2008xh}), the $\vlowk$ interactions
can be constructed in energy-independent form either using model-space
methods (such as Lee-Suzuki or Okubo~\cite{Okubo:1954zz} transformations) or
through the RG treatment. In practice, most calculations of $\vlowk$
have used the model-space methods\footnote{See Refs.~\cite{Bayegan:2008tf,Bayegan:2009sf} for a three-dimensional formulation that 
avoids partial-wave decompositions.} (followed by an Okubo 
Hermitization~\cite{Bogner:2003wn,Bogner:2006vp}) rather than the differential
equations of the RG because they are more robust numerically. This is
true for both sharp and smooth cutoff versions of $\vlowk$.

The evolution to lower cutoffs shifts contributions from the sum over
intermediate states to the interactions, just as RG equations in
quantum field theory shift strength from loop integrals to coupling
constants.  The change in the interactions include subleading contact
interactions to all orders.  Because only large momentum modes are
decoupled, long-range pion exchanges are preserved.  As documented in
Section~\ref{subsec:universal}, different initial $\vnn$ interactions
collapse to the same nearly universal $\vlowk$ at sufficiently small 
cutoffs ($\Lambda\sim 2.0\fmi$).

\subsection{Smooth regulators for $\vlowk$}
\label{subsec:smooth}

\begin{figure}[t]
 \centering
 \subfloat[][]{%
  \label{fig:deuteronwfs-a}%
  \includegraphics*[width=3.0in,clip=]{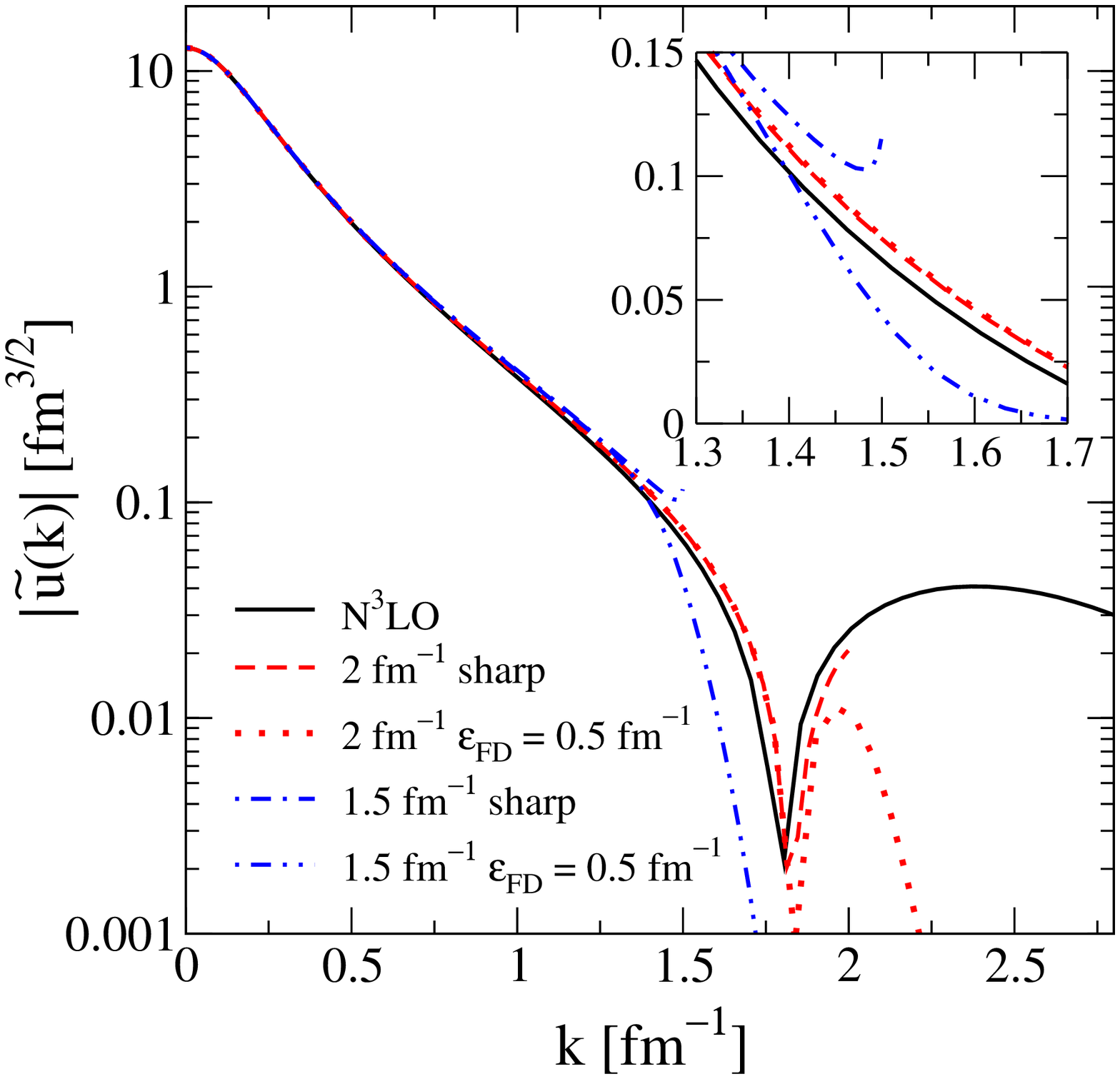}%
 }%
 \hspace*{.4in}%
 \subfloat[][]{%
  \label{fig:deuteronwfs-b}%
  \includegraphics*[width=3.0in,clip=]{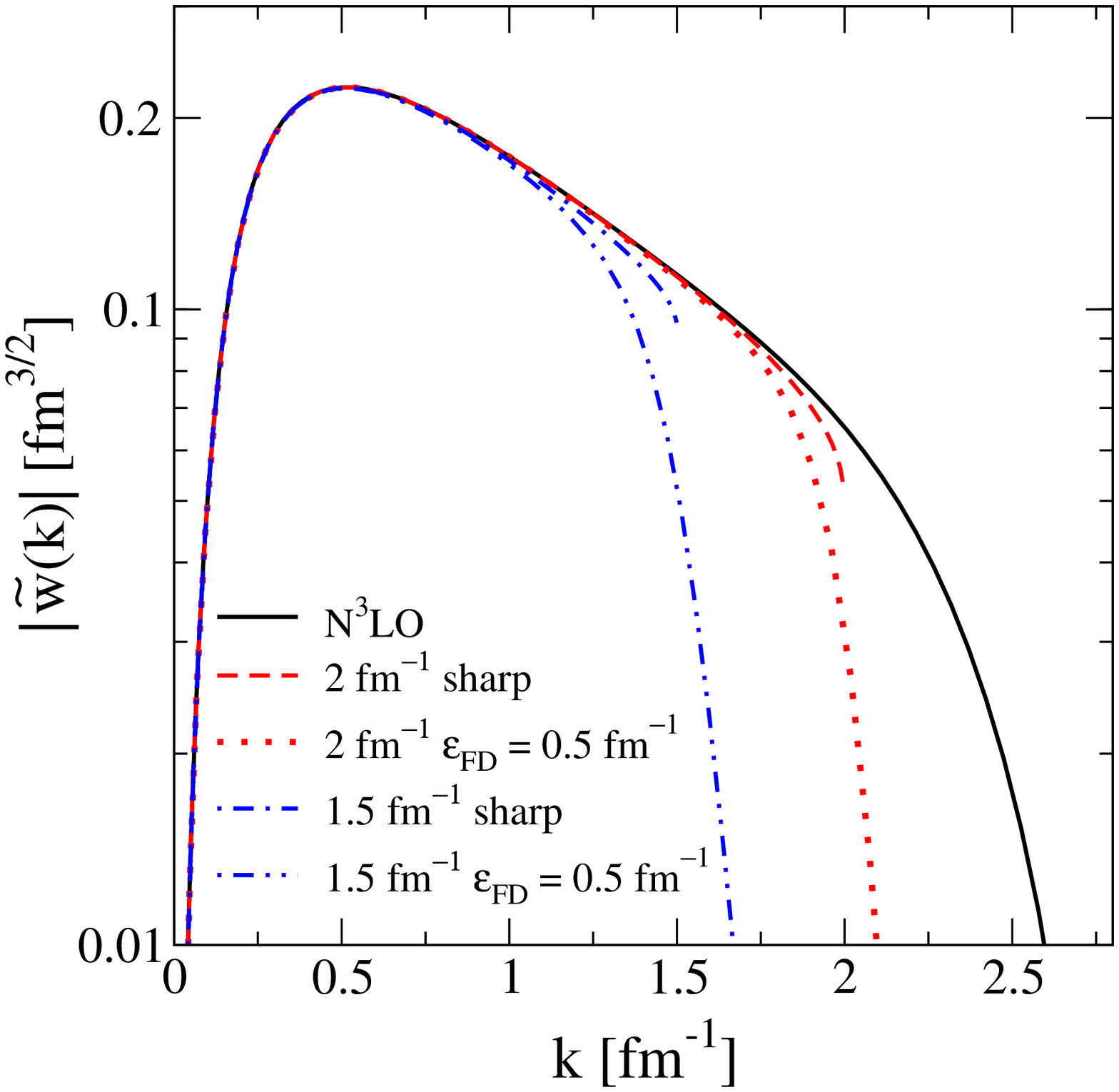}%
 }%
\caption{(a) S-state and (b)~D-state components of the deuteron wave
function in momentum space ($\widetilde{u}(k)$ and $\widetilde{w}(k)$
respectively) for the N$^3$LO potential of Ref.~\cite{Entem:2003ft}
and evolved $\vlowk$ using smooth and sharp cutoffs at $\Lambda = 2.0
\fmi$ and $1.5 \fmi$.}
\label{fig:deuteronwfs}
\end{figure}

\begin{figure}[t]
 \centering
 \subfloat[][]{%
  \label{fig:deutconvergence}%
  \includegraphics*[width=3.0in,clip=]{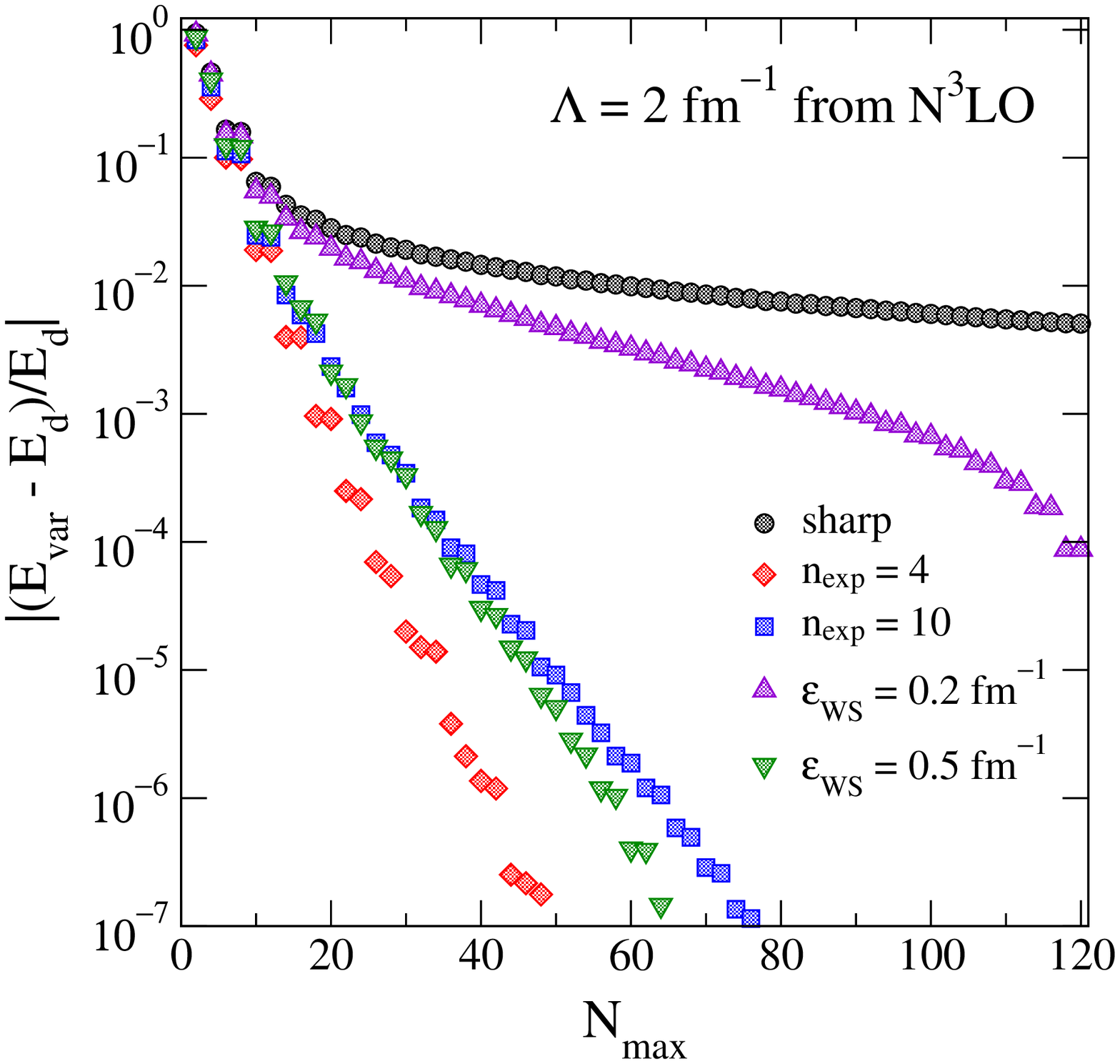}%
 }%
 \hspace*{.4in}%
 \subfloat[][]{%
  \label{fig:triton}%
  \includegraphics*[width=3.0in,clip=]{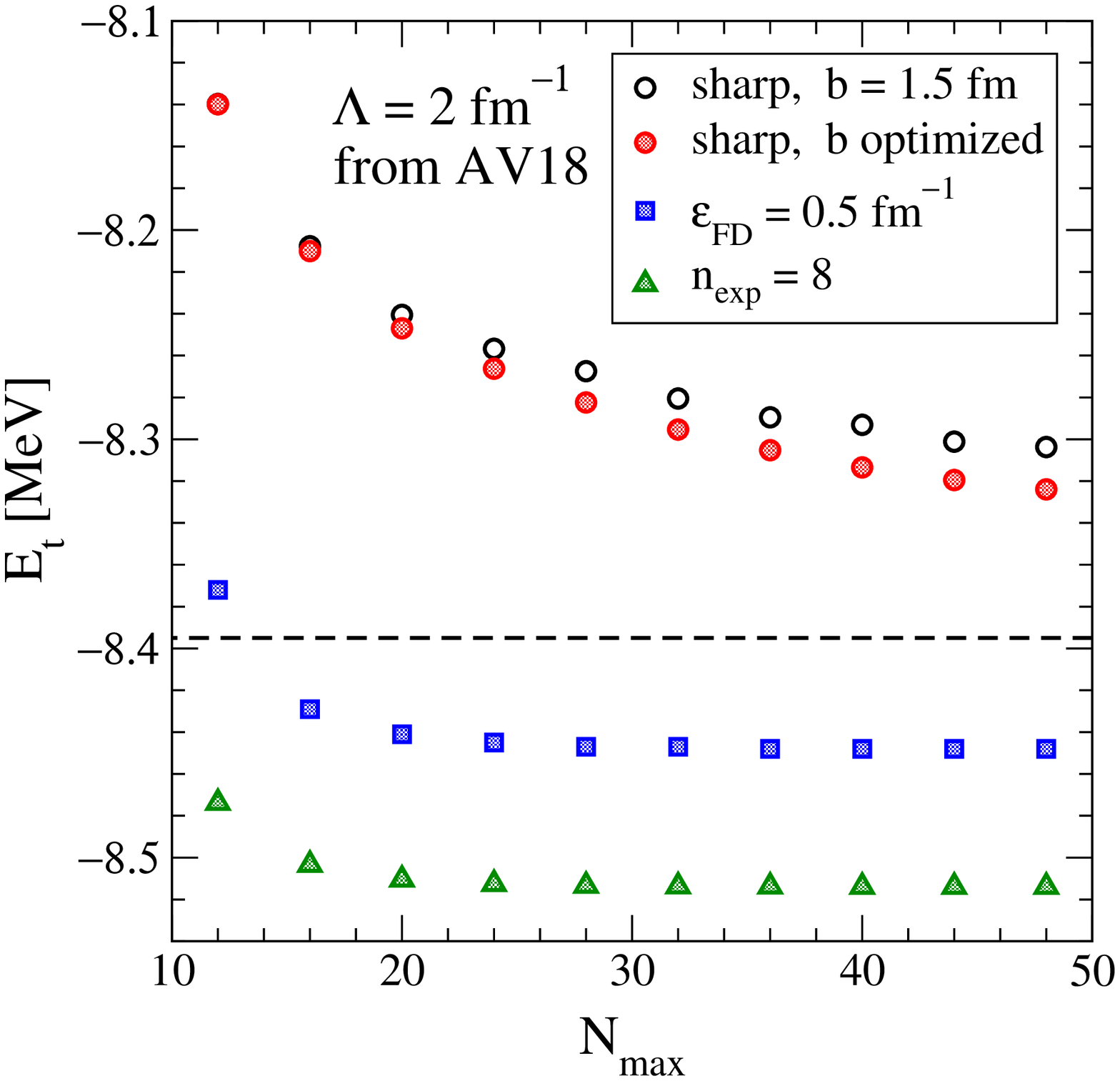}
 }%
\caption{(a) The relative error in the deuteron binding energy $E_d$
as a function of the size of the oscillator space for sharp cutoff
and various smooth regulators. (b)~The triton binding energy $E_t$
calculated from a direct
diagonalization in a harmonic-oscillator basis of the low-momentum
Hamiltonian evolved from the Argonne $v_{18}$
potential with cutoff $\Lambda = 2 \fmi$, as a
function of the size of the oscillator space ($N_{\rm max} \, \hbar
\omega$ excitations). The open circles are calculated with a sharp
cutoff for a fixed oscillator parameter $b$ while the filled ones
correspond to optimizing $b$ at each $N_{\rm max}$. The dashed line
indicates the exact Faddeev result~\cite{Nogga:2004ab}, which shows
the slow convergence at the $100 \, {\rm kev}$ level of the
diagonalization for the sharp cutoff. The squares are for a smooth
Fermi-Dirac regulator and the triangles for a  smooth exponential
regulator, which each solve the convergence problem. For details see
Ref.~\cite{Bogner:2006vp}.}
\end{figure}

The sharp cutoff RG equation for $\vlowk$ takes a particularly simple
form that preserves two-body observables for all momenta up to the
cutoff. However, a sharp momentum cutoff leads to cusp-like behavior
close to the cutoff in some channels and for the deuteron wave
function (see inset in Fig.~\subref*{fig:deuteronwfs-a}), which
becomes increasingly evident as the cutoff is lowered below $2
\fmi$. In some applications, this leads to slow convergence, for
example at the $10\mbox{--}100 \, {\rm keV}$ level in few-body calculations
using harmonic-oscillator bases, as shown in
Fig.~\subref*{fig:triton}. One would expect that such calculations for
the deuteron and the triton, which are particularly low-energy bound
states, should show improvement for cutoffs well below $2 \fmi$, but
instead a degradation was observed in Ref.~\cite{Bogner:2005fn}. This
result was attributed to the use of sharp cutoffs, and subsequent
studies~\cite{Bogner:2006ai,Bogner:2006vp} have shown that these
problems are alleviated by using a smooth-cutoff version of $\vlowk$.

The sharp cutoff $\vlowk$ interactions were originally constructed
using model-space methods (such as
Lee-Suzuki~\cite{Bogner:2001gq,Bogner:2003wn} or
Okubo~\cite{Epelbaum:1998na} transformations) that rely on the
definition of orthogonal projectors $P$ and $Q$, such that $P + Q = 1$
and $P Q = Q P = 0$. While smooth cutoffs seem incompatible with
methods requiring $PQ = 0$, it is not a conceptual problem for the RG
approach. Indeed, there is an appreciable literature on smooth-cutoff
regulators for applications of the functional or exact
RG~\cite{Liao:1999sh}.  The functional RG keeps invariant the full
generating functional, which translates into preserving all matrix
elements of the fully off-shell inter-nucleon $T$ matrix.  While this
straightforwardly leads to RG equations, it also implies an
energy-dependent interaction, which is undesirable for practical few-
and many-body calculations.  This conflict can be resolved by a
generalization of the energy-independent RG equation
derived in Section~\ref{subsec:sharp} to smooth
cutoffs~\cite{Bogner:2006vp}. It is convenient to first define the
partial-wave $\vlowk$ and the corresponding $\tlowk$ matrix in terms
of a reduced potential $v$ and a reduced $t$ matrix as
\begin{align}
\vlowk(k',k) &= f(k') \, v(k',k) \, f(k) \,, \\[2mm]
\tlowk(k',k;k^2) &= f(k') \, t(k',k;k^2) \, f(k) \,.
\end{align}
where $f(k)$ is a smooth cutoff function satisfying
\be
f(k) \stackrel{k \ll \Lambda}{\longrightarrow} 1
\qquad \text{and} \qquad
f(k) \stackrel{k \gg \Lambda}{\longrightarrow} 0 \,.
\label{eq:regulator}
\ee
Representative examples are exponentials $f(k)=\exp
[-(k^2/\lm^2)^{n_{\rm exp}}]$
and Fermi-Dirac functions $f(k) = 1/(1 + e^{(k^2-\lm^2)/\epsilon^2})$.
The reduced half-on-shell $t$ matrix obeys a Lippmann-Schwinger
equation with loop integrals smoothly cut off by $f^2(p)$,
\be
t(k',k;k^2) = v(k',k) + \frac{2}{\pi} \int_{0}^{\infty} p^2dp \,
\frac{v(k',p) \, f^2(p) \, t(p,k;k^2)}{k^2-p^2} \,.
\ee
Analogous to the RG derivation for a sharp cutoff, we impose that the
reduced half-on-shell $t$ matrix is independent of the cutoff,
$dt(k',k;k^2)/d\Lambda=0$. This choice preserves the on-shell $t$ matrix
while also maintaining energy independence.  The resulting RG equation
is
\be
\frac{d}{d\lm} \, v(k',k) = \frac{2}{\pi}\int_{0}^{\infty} p^2dp \, 
\frac{v(k',p) \, \ddlamfsq \, t(p,k;p^2)}{p^2-k^2} \,,
\label{EindepRGE}
\ee
which describes the evolution of the reduced low-momentum interaction
with the cutoff. If one takes an energy-independent NN potential as
the large-cutoff initial condition and numerically integrates
Eq.~\eqref{EindepRGE}, then the resulting $v$ preserves the
half-on-shell $\tnn$ matrix for \emph{all} external momenta,
$t(k',k;k^2)=\tnn(k',k;k^2)$. Therefore, $\vlowk(k',k)=f(k')\,
v(k',k)\,f(k)$ preserves the low-momentum half-on-shell $\tnn$ matrix
up to factors of the smooth cutoff function. In the limit $f(p)
\rightarrow \theta(\Lambda-p)$, the RG equation for a sharp cutoff,
Eq.~\eqref{RGEsharp}, is recovered.

The solution to Eq.~\eqref{EindepRGE} is non-Hermitian. However, one
can show~\cite{Bogner:2006vp} that a simple generalization of the
Okubo Hermitization transformation to smooth cutoffs is obtained by
symmetrizing the smooth-cutoff RG equation, Eq.~(\ref{EindepRGE}), to
obtain
\be
\frac{d}{d\lm} \, {v}(k',k) =
\frac{1}{\pi} \int_{0}^{\infty} p^2dp \,
\biggl[\frac{{v}(k',p) \ddlamfsq \, 
{t}(p,k;p^2)}{p^2-k^2} 
+ \frac{{t}(k',p;p^2) \, \ddlamfsq \, 
{v}(p,k)}{p^2-k'^2} \biggr] \,.
\label{EindepRGEherm}
\ee
Here, the Hermitian low-momentum interaction $\vlowk$ preserves the
low-momentum fully-on-shell $\tnn$ matrix, up to factors of the
regulator function $\tlowk(k,k;k^2)=f^2(k)\, \tnn(k,k;k^2)$, and the
deuteron binding energy. Note that demanding $dt(k,k;k^2)/d\Lambda $
does not imply a ``unique'' Hermitian RG equation. In conventional
model-space methods, this non-uniqueness corresponds to the choice in
the Hermitization procedure (or to residual unitary transformations in
the $P$-space block). In RG language, this can be viewed as demanding
$d t(k',k;k^2)/d\Lambda = (k^2-k'^2) \, \Phi(k',k)$, where
$\Phi(k',k)$ is any function satisfying $\lim_{k \to k'} (k^2-k'^2) \,
\Phi(k',k) = 0$. The above RG equation derived from the Okubo
transformation corresponds to a particular choice for $\Phi(k',k)$.

In practice, the numerical solution of Eq.~\eqref{EindepRGEherm} is
slowed by the $t$ matrix evaluations involved at each step.  In
addition, the RG equation involves two-dimensional interpolations and
principal-value integrals over narrowly peaked functions making it
easy to introduce small errors at each step that can accumulate as the
cutoff is lowered. Therefore, an alternative 3-step procedure to
construct a low-momentum, energy-independent interaction with smooth
cutoffs that is better suited for numerical calculations has been
formulated (see Ref.~\cite{Bogner:2006vp} for details):

\begin{enumerate}
\item Evolve a large-cutoff potential to a lower, smooth cutoff while
preserving the \emph{full} off-shell $T$ matrix modulo factors of the
smooth cutoff functions. This generates an energy-dependent low-momentum
interaction, and amounts to solving  the Bloch-Horowitz equation with a
smooth cutoff.
\item Convert the energy dependence to momentum dependence using
equations of motion,  which results in a non-Hermitian smooth-cutoff 
interaction that preserves the low-momentum half-on-shell $T$-matrix
modulo factors of the smooth cutoff functions.
\item Perform a similarity transformation to Hermitize the low-momentum
interaction. The resulting low-momentum interaction is
energy-independent and Hermitian, and preserves the low-momentum fully
on-shell $T$-matrix up to factors of the smooth cutoff functions.
\end{enumerate}
It has been verified numerically that Eq.~\eqref{EindepRGEherm} and
the 3-step procedure give very similar results, and both tend towards
the sharp cutoff $\vlowk$ (with Okubo Hermitization) calculated using
the Lee-Suzuki method in the limit of a sharp cutoff. However, due to
its computational advantages the 3-step method is used in most
applications, although Eq.~\eqref{EindepRGEherm} has also been used in
practical calculations~\cite{Hebeler:2006kz,Hebeler:2009dy,Hebeler:2009iv}.

The momentum space deuteron wave functions for smooth and sharp
cutoffs are contrasted in Fig.~\ref{fig:deuteronwfs}. We follow the
notation of Ref.~\cite{Epelbaum:1999dj}, with S- and D-state
components denoted in coordinate space by $u$ and $w$ respectively,
and with tildes in momentum space.  The sharp-cutoff wave functions
develop cusp-like structures in momentum space below $2 \fmi$ (see
inset in Fig.~\subref*{fig:deuteronwfs-a}), which are removed by the
Fermi-Dirac (or any other smooth) regulator. As with the sharp cutoff
$\vlowk$, the smooth-cutoff evolution removes the short-range
correlation ``wound'' in the coordinate-space deuteron wave function
at lower cutoffs, and only modifies short-distance physics as
indicated by the weak cutoff dependence of long-range operators
like the ``bare'' quadrupole moment in the deuteron, see
Fig.~\subref*{fig:deuteronwfsp2-b}. These features lead to more
perturbative behavior in nuclear matter~\cite{Bogner:2005sn} as well
as in few-body systems~\cite{Bogner:2006tw,Bogner:2006ai}.

The primary motivation for smooth cutoffs was to remedy the slow
convergence at the $10 \, {\rm keV}$ level in the deuteron and at the
$100 \, {\rm keV}$ level in the triton, when calculated in a
harmonic-oscillator basis.  In Fig.~\subref*{fig:deutconvergence}, the
relative error in the binding energy of the deuteron (with respect to
the converged result) is shown as a function of the size of the
oscillator space ($N_{\rm max} \, \hbar \omega$ excitations) for a
range of smooth regulators.  The slow convergence is evident for the
sharp cutoff, where the error is below the percent level only for the
largest space.  The Fermi-Dirac regulator eliminates this problem,
with improved convergence by increasing the smoothness parameter
$\epsilon$ from $0.2 \fmi$ to $0.5 \fmi$. Very similar results are
obtained for the exponential regulator, or any other smooth regulator
function. The convergence is also greatly improved for the triton.  In
Fig.~\subref*{fig:triton}, the triton binding energy is plotted as a
function of the size of the oscillator space.  For efficiency,
convergence for the smallest possible space is desirable, as the
computational cost grows rapidly with $N_{\rm max}$ and for larger
systems.  Convergence at the keV level is achieved by all exponential
regulators with $n_{\rm exp} \geqslant 4$ ($n_{\rm exp} = 8$ is shown) soon
after $N_{\rm max} = 20$.  The consequence in moving from sharp to
increasingly smooth regulators is also seen from the Fermi-Dirac
regulators, where $\epsilon = 0.5 \fmi$ yields very satisfactory
results.

\subsection{Similarity renormalization group (SRG)}
\label{subsec:srg}

An alternative path to decoupling high-momentum from low-momentum
physics is the similarity renormalization group (SRG), which is based
on a continuous sequence of unitary transformations that suppress
off-diagonal matrix elements, driving the Hamiltonian towards a
band-diagonal form~\cite{Glazek:1993rc,Wegner:1994,%
Szpigel:2000xj,Roth:2005pd,Kehrein:2006}.  The SRG potentials are automatically
energy independent and have the feature that high-energy phase shifts
(and other high-energy NN observables), while typically highly model
dependent (see Fig.~\ref{fig:phases}), are preserved, unlike the case
with $\vlowk$ as usually implemented.  Most important, the same
transformations renormalize all operators, including many-body
operators, and the class of transformations can be tailored for
effectiveness in particular problems.

In Ref.~\cite{Bogner:2006pc}, the first application of SRG
transformations to NN interactions using the flow equation formalism
of Wegner~\cite{Wegner:1994,Kehrein:2006} was made.  The evolution or
flow of the Hamiltonian with a parameter $\flow$ is a series of
unitary transformations,
\be
H_\flow = U_\flow H U^\dagger_\flow \equiv \Trel + V_\flow \,,
\label{eq:Hflow}
\ee
where $\Trel$ is the relative kinetic energy and $H = \Trel + V$ is
the initial Hamiltonian. Equation
(\ref{eq:Hflow}) defines the evolved potential $V_\flow$, with $\Trel$
taken to be independent of $\flow$.  Then $H_\flow$ evolves according to
\be
\frac{dH_\flow}{d\flow} = [\eta_\flow,H_\flow] \,,
\ee
with
\beqn
\eta_\flow = \frac{dU_\flow}{d\flow} \, U^\dagger_\flow 
= -\eta^\dagger_\flow \,.
\ee
Choosing $\eta_\flow$ specifies the transformation, which is taken as the 
commutator of an operator, $G_\flow$, with the Hamiltonian, 
\be
\eta_\flow = [G_\flow, H_\flow] \,,
\ee
so that
\be
\frac{dH_\flow}{d\flow} = [ [G_\flow, H_\flow], H_\flow] \,.
\label{eq:commutator}
\ee
Applications to nuclear forces 
have used $G_\flow = T_{\rm rel}$~\cite{Bogner:2006pc}, but one
could also use momentum-diagonal operators such as $\Trel{}^2$, or the
running diagonal Hamiltonian $H_D$, as advocated by
Wegner~\cite{Wegner:1994} (see also Ref.~\cite{Glazek:2008pg}). 
Taking the simplest choice $G_\flow =
T_{\rm rel}$, the flow equation in a partial wave takes the form
\be
\frac{dV_\flow(k,k')}{d\flow} = - (k^2 - k'{}^2)^2 \, V_\flow(k,k')
+ \frac{2}{\pi} \int_0^\infty q^2 dq \, (k^2 + k'{}^2 - 2q^2) \,
V_\flow(k,q) \, V_\flow(q,k') \,.
\label{eq:diffeq}
\ee
For matrix elements far from
the diagonal, the first term on the right side of
Eq.~(\ref{eq:diffeq}) evidently dominates and exponentially suppresses
these elements as $\flow$ increases,
\be
V_\flow(k,k') \approx V_{s=0}(k,k') e^{-\flow(k^2 - k'^2)^2} \,.
\label{eq:exponential}
\ee
It is convenient to switch to the flow variable $\lambda \equiv s^{-1/4}$,
which has units of fm$^{-1}$, because Eq.~(\ref{eq:exponential}) shows
that $\lambda$ is a measure of the resulting diagonal width of
$V_\flow$ in momentum space.

\begin{figure}[t]
 \centering
 \subfloat[][]{%
  \label{fig:vsrg-a}%
  \includegraphics*[width=6.2in]{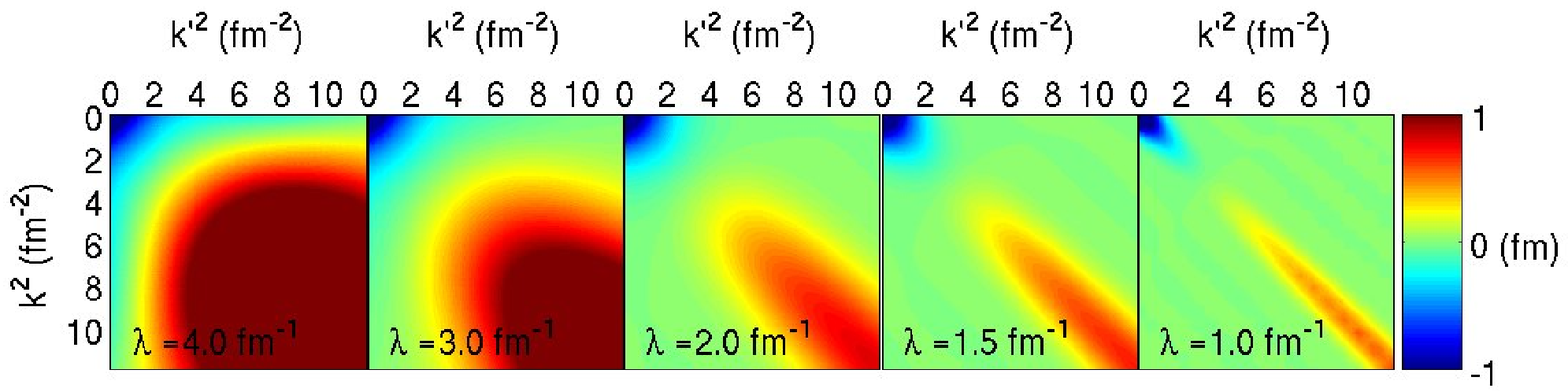}%
 }%
 \\%
 \subfloat[][]{%
  \label{fig:vsrg-b}%
  \includegraphics*[width=6.2in]{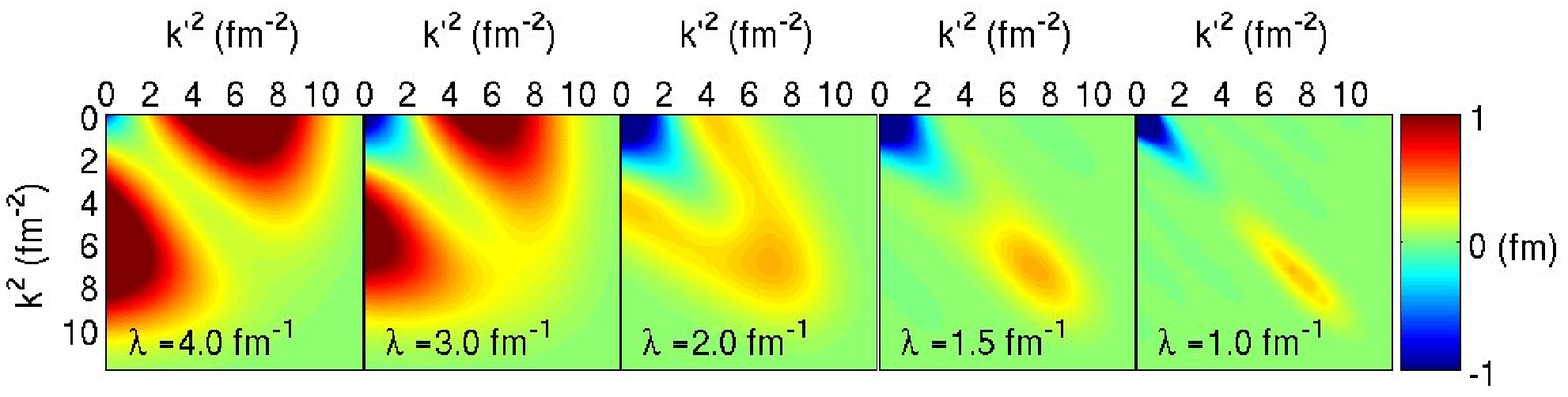}%
 }%
 \\%
 \subfloat[][]{%
  \label{fig:vsrg-c}%
  \includegraphics*[width=6.2in]{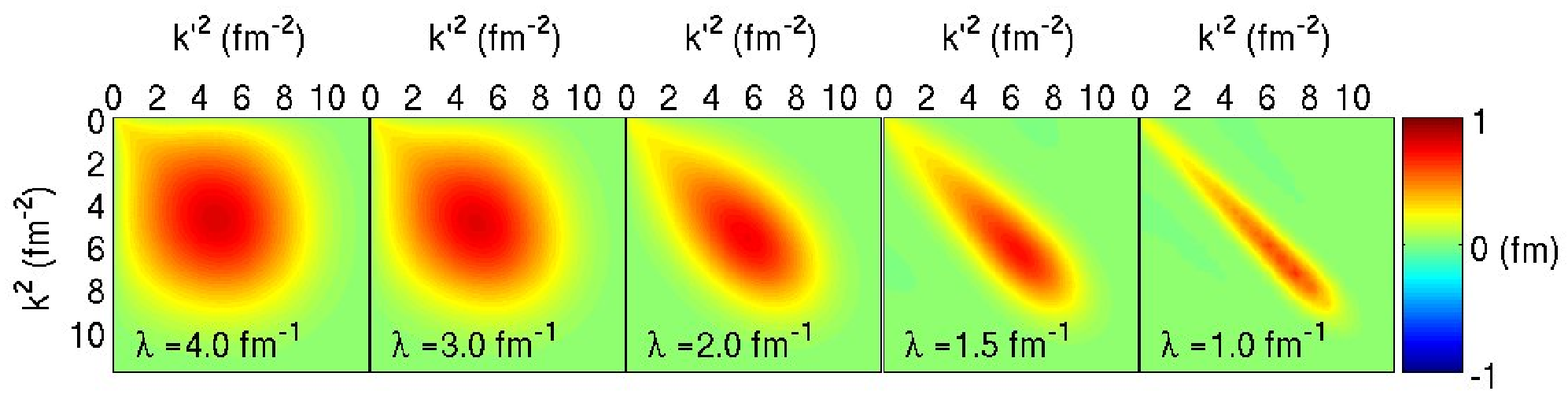}%
 }%
\caption{Contour plots of momentum-space matrix elements for
the SRG evolution with $\lambda$ in the (a) $^1$S$_0$, (b) S-wave
part of the $^3$S$_1$--$^3$D$_1$ and (c) $^1$P$_1$ channels.
The initial potential in (a) is the ($\lm = 600 \mev$) N$^3$LO
potential~\cite{Entem:2003ft} and in (b) and (c) the N$^3$LO potential
with $\lm/\widetilde{\lm} = 500/600 \mev$~\cite{Epelbaum:2004fk}.}
\label{fig:vsrg}
\end{figure}

The evolution of the Hamiltonian according to Eq.~(\ref{eq:diffeq}) as
$\flow$ increases (or $\lambda$ decreases) is illustrated in
Fig.~\ref{fig:vsrg}, using two initial chiral EFT potentials.
On top is $^1$S$_0$ starting from
the harder ($\lm = 600 \mev$) N$^3$LO potential of
Ref.~\cite{Entem:2003ft}, which has significant strength near the
high-momentum diagonal, in the middle is the S-wave part of the
$^3$S$_1$--$^3$D$_1$ channel starting from one of the
potentials of Ref.~\cite{Epelbaum:2004fk}, which has more far
off-diagonal strength initially and comparatively weaker
higher-momentum strength on the diagonal, and on bottom is $^1$P$_1$
with that same potential.  Each of these examples show the
characteristic features of the evolution in $\lambda$, namely the
systematic suppression of off-diagonal strength, as anticipated, with
the width of the diagonal scaling as $\lambda^2$.

\begin{figure}[t]
 \centering
 \subfloat[][]{%
  \label{fig:compare_and_weinberg-a}%
  \includegraphics*[width=3.3in]{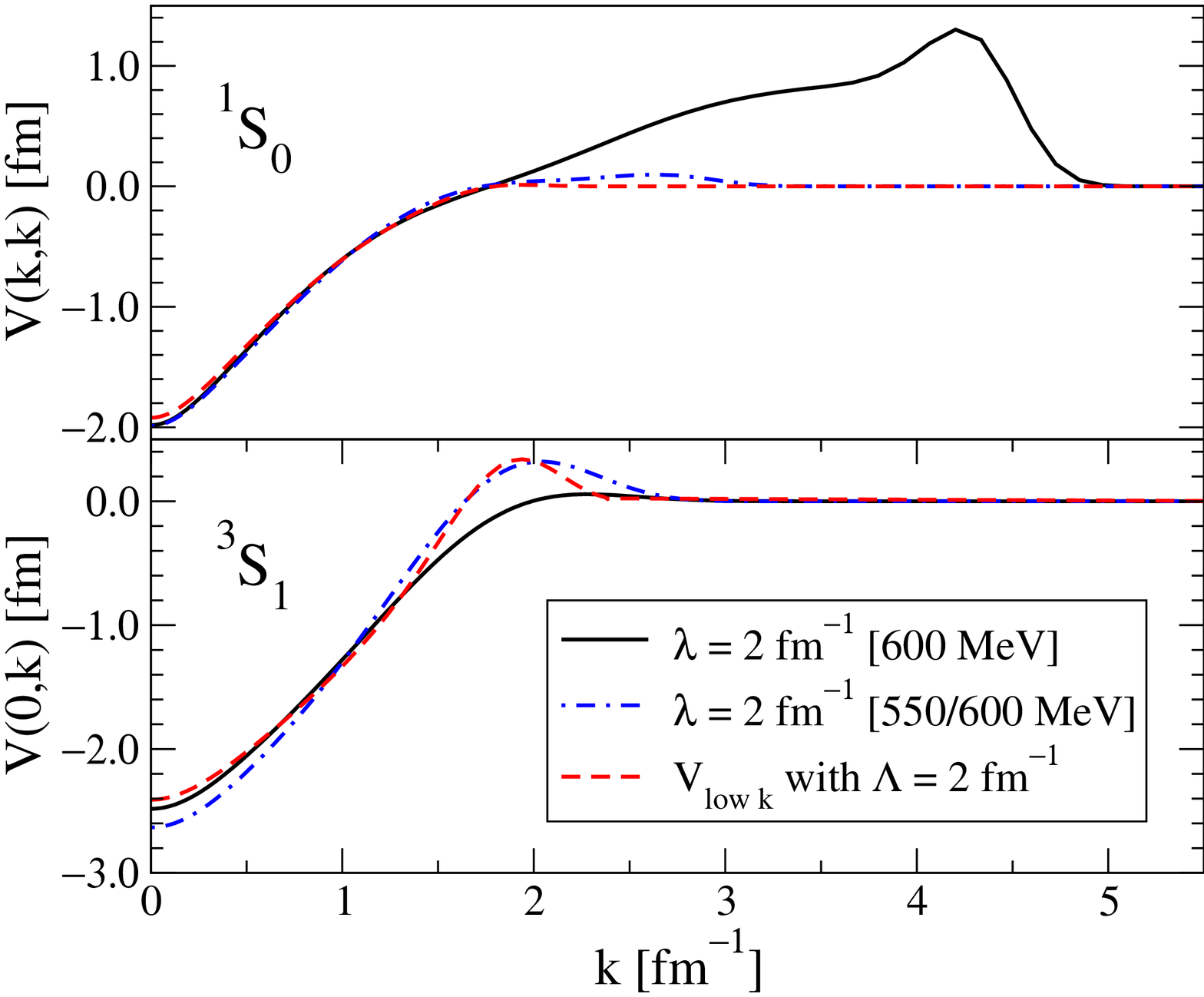}%
 }%
 \hspace*{.4in}%
 \subfloat[][]{%
  \label{fig:compare_and_weinberg-b}%
  \includegraphics*[width=2.8in]{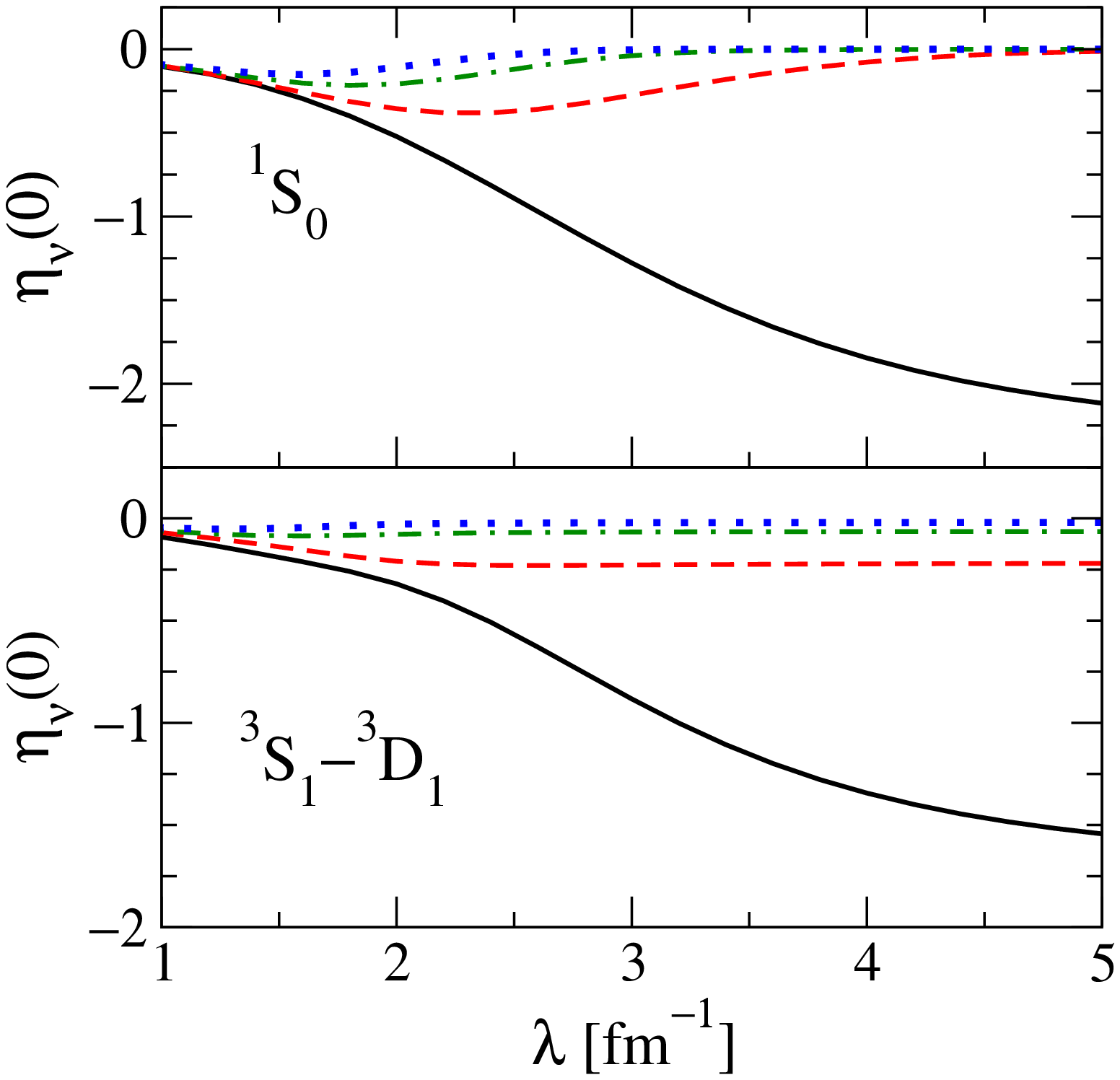}%
 }%
\caption{(a) Low-momentum universality for momentum-space matrix
elements of the evolved SRG potentials at $\lambda = 2 \fmi$ for
$^1$S$_0$ (top, diagonal elements) and $^3$S$_1$ 
(bottom, off-diagonal elements). Also shown is the $\vlowk$ potential
for a smooth regulator with $\Lambda = 2 \fmi$ and $n_{\rm exp}=4$.
(b)~Largest repulsive Weinberg eigenvalues in the $^1$S$_0$
and $^3$S$_1$--$^3$D$_1$ channels as a function of $\lambda$,
with initial potentials as in Fig.~\ref{fig:vsrg}. For details
see Ref.~\cite{Bogner:2006pc}.}
\label{fig:compare_and_weinberg}
\end{figure}

\begin{figure}[t!]
 \centering
 \subfloat[][]{%
  \label{fig:deuteron_convergence-a}%
  \includegraphics*[width=3.0in]{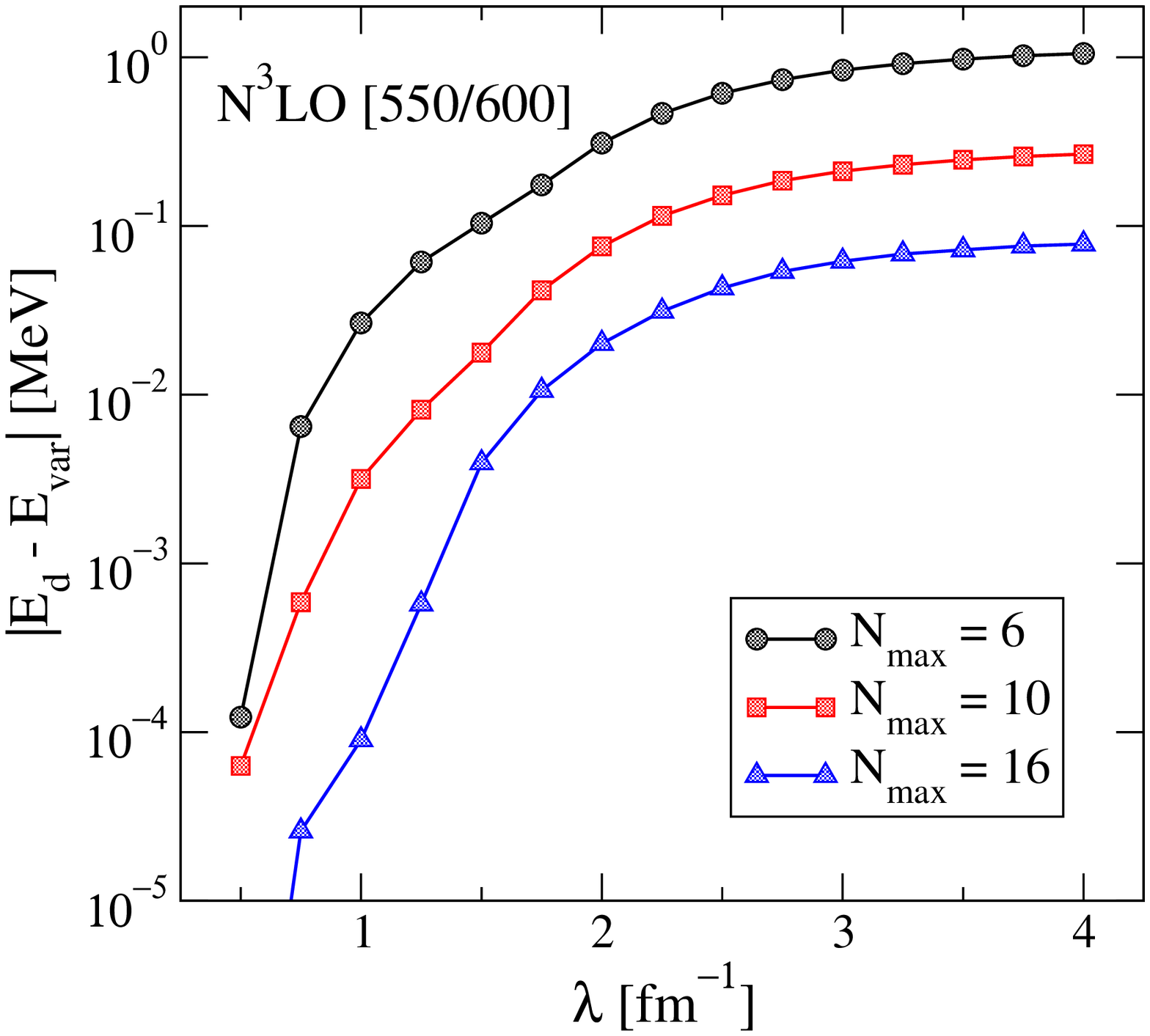}%
 }%
 \hspace*{.4in}%
 \subfloat[][]{%
  \label{fig:deuteron_convergence-b}%
  \includegraphics*[width=3.15in]{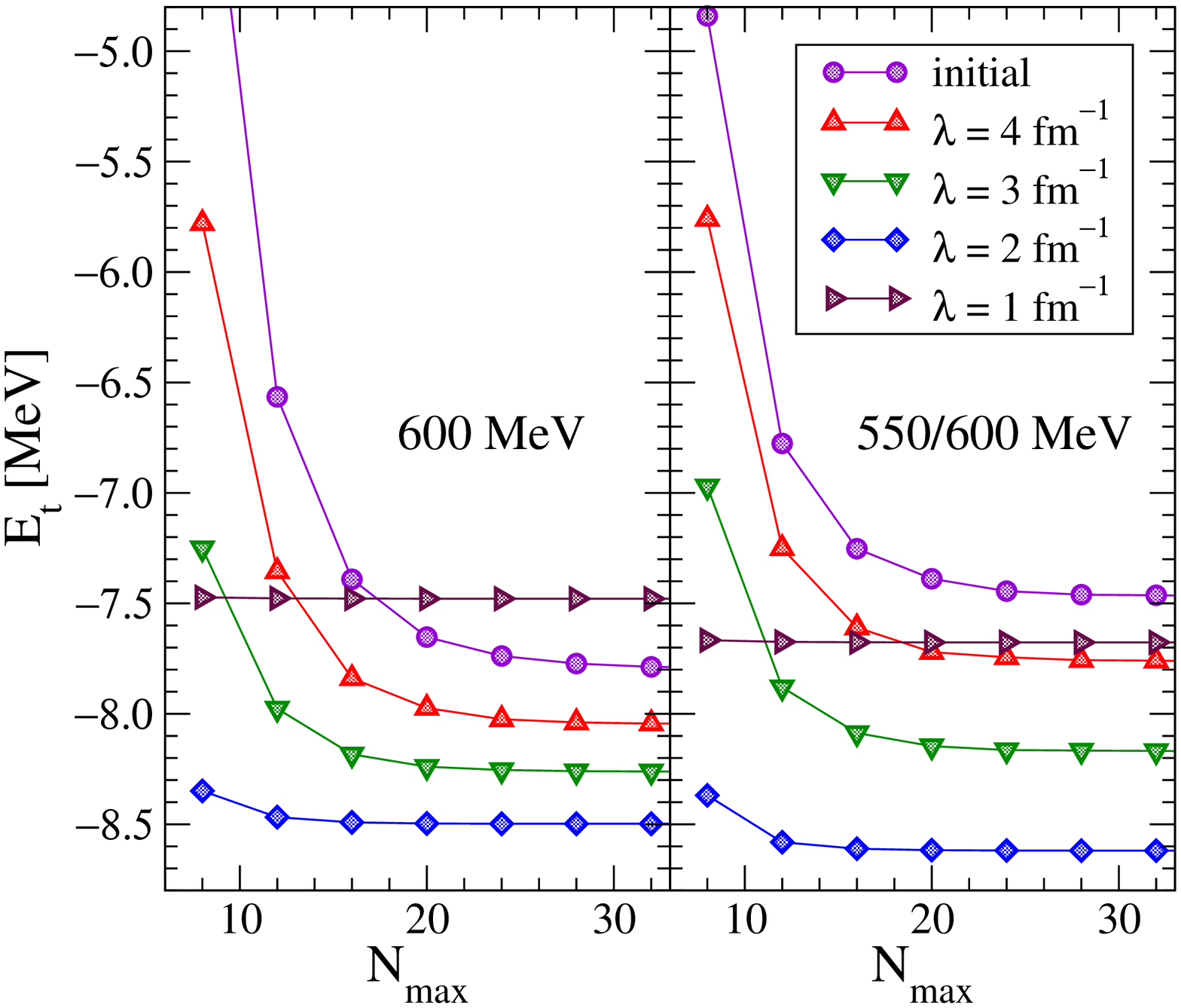}%
 }%
\caption{(a) Absolute error versus $\lambda$ of the deuteron binding
energy from a variational calculation in a fixed-size basis of harmonic
oscillators ($N_{\rm max} \hbar\omega$ excitations). (b)~Variational
triton binding energy for selected $\lambda$ with NN interactions
only, as a function of $N_{\rm max}$. For details see 
Ref.~\cite{Bogner:2006pc}.}
\label{fig:deuteron_convergence}
\end{figure}
 
The SRG-evolved interactions share key similarities (universality,
increased perturbativeness, weaker correlations, etc.) with the
smooth-cutoff $\vlowk$ potentials, even though the decoupling of low
and high momenta is achieved in a somewhat different manner.  As
$\lambda$ is lowered, different initial potentials flow to similar
forms at low momentum (while remaining distinct at higher momentum),
with the universal low-momentum parts numerically similar to the
$\vlowk$ potentials. These observations are illustrated in
Fig.~\subref*{fig:compare_and_weinberg-a} for the diagonal and
off-diagonal matrix elements.

Similarly, the nonperturbative features associated with strong
short-range repulsion and strong short-range tensor forces are
substantially softened as we evolve to lower $\lambda$. The largest
repulsive Weinberg eigenvalues (see Section~\ref{subsec:perturb}) for
$E=0$ are shown as a function of $\lambda$ in
Fig.~\subref*{fig:compare_and_weinberg-b} for the $^1$S$_0$ and
$^3$S$_1$--$^3$D$_1$ channels. In both channels, the large eigenvalues
decrease rapidly as $\lambda$ evolves to $2 \fmi$ and below, as
observed with the $\vlowk$ evolution in Section~\ref{subsec:perturb}.
The more perturbative potentials at lower $\lambda$ induce weaker
short-range correlations in few- and many-body wave functions, which
leads to greatly improved convergence in variational calculations.
This is illustrated in Figs.~\subref*{fig:deuteron_convergence-a}
and~\subref*{fig:deuteron_convergence-b} via calculations of the
deuteron and triton binding energies by diagonalization in a
harmonic-oscillator basis. The improvement in convergence is similar
to that found with smoothly-cutoff $\vlowk$
interactions~\cite{Bogner:2006vp}.

In Fig.~\subref*{fig:deuteron_convergence-b}, the calculations for
different $\lambda$ converge to different values for the triton
binding energy. This reflects the contributions of the omitted (and
evolving) three-body interactions, and follows a similar pattern to
that seen with NN-only $\vlowk$
calculations~\cite{Nogga:2004ab,Bogner:2006vp}. The consistent
evolution of many-body forces is an important issue for low-momentum
interactions. For the SRG, the evolution of 3N forces is readily
practical, as discussed in Section~\ref{subsec:evolution}.

\subsection{$\vlowk$ from SRG flow equations}
\label{subsec:vlowksrg}

\begin{figure}[t!]
 \centering
 \subfloat[][]{%
  \label{fig:vlowk-a}%
  \includegraphics*[width=4.1cm,clip=]{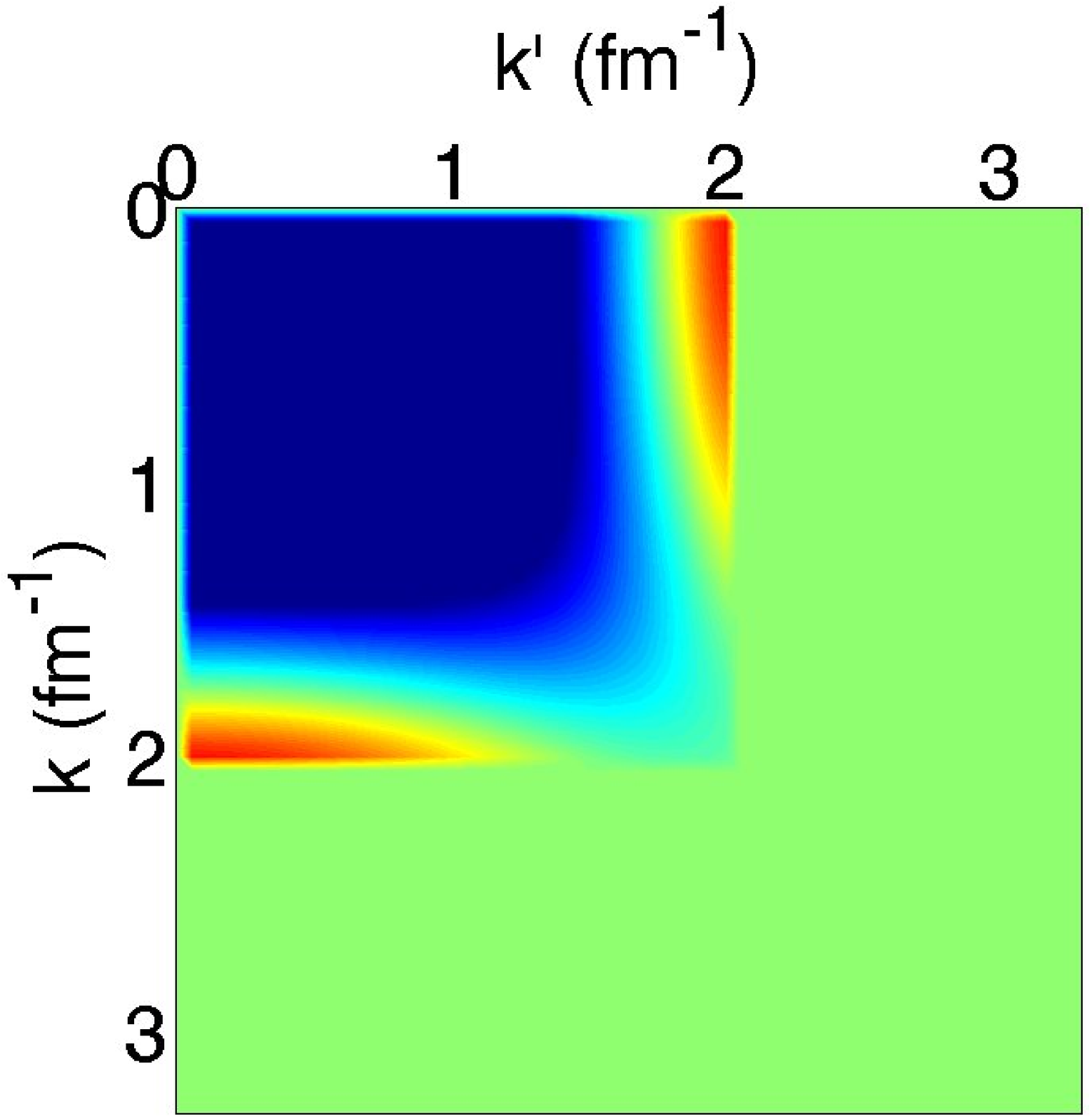}%
  \hspace*{.05in}%
  \raisebox{-2pt}{%
  \includegraphics*[width=5.0cm,clip=]{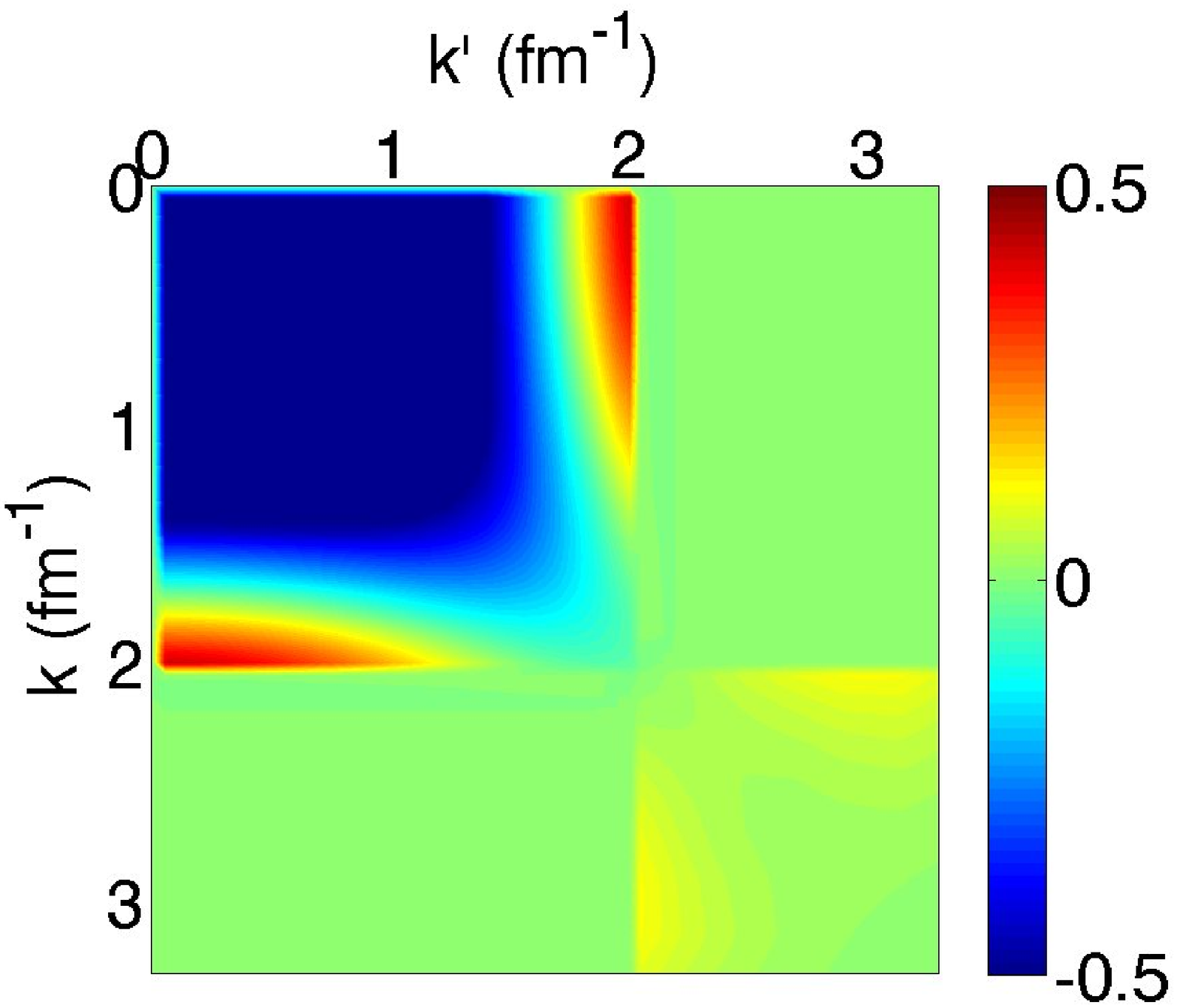}}%
 }%
 \hspace*{.1in}%
 \subfloat[][]{%
  \label{fig:vlowk-b}%
  \includegraphics*[width=3.9cm,clip=]{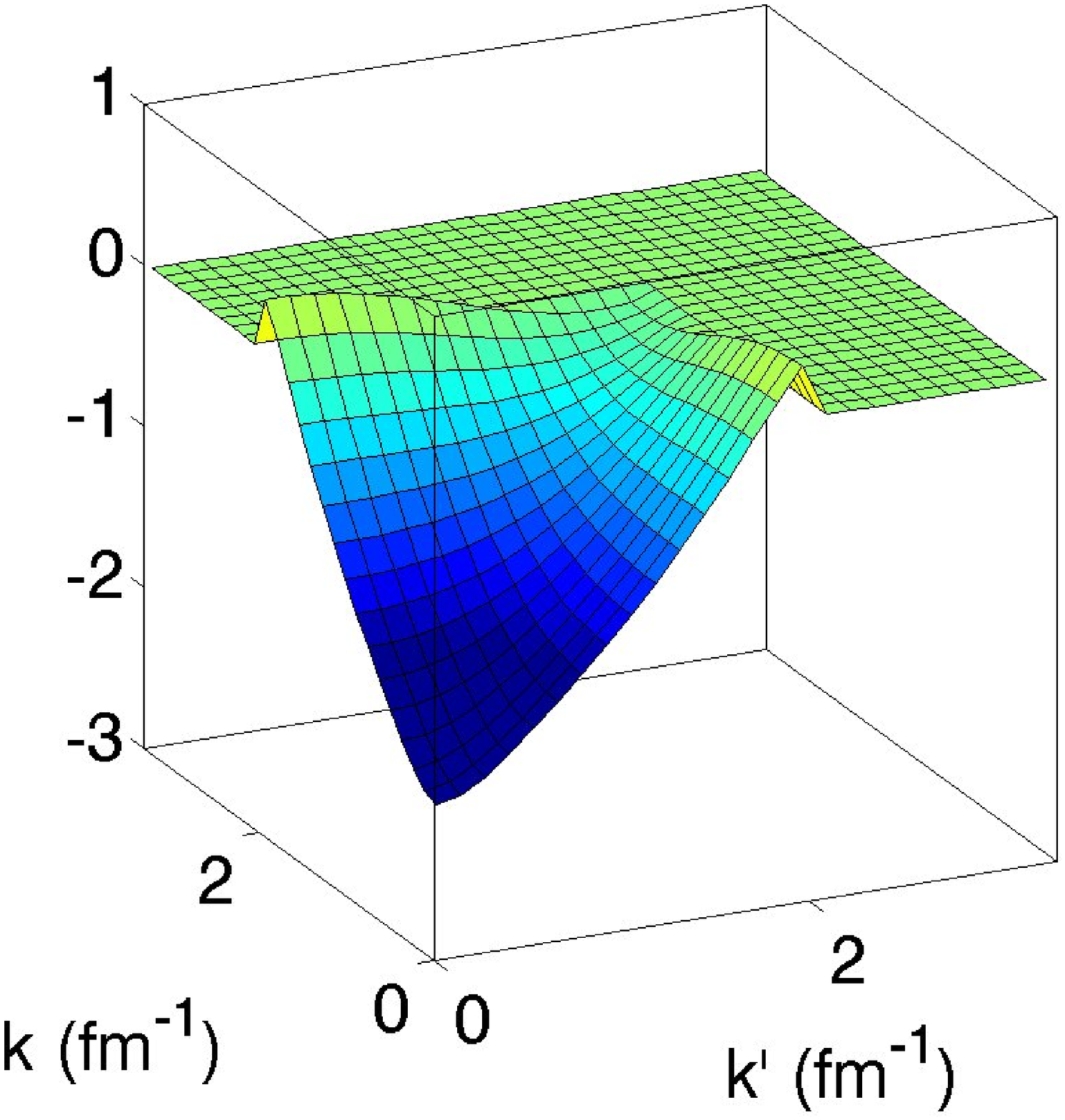}%
  \hspace*{.05in}%
  \raisebox{0pt}{%
  \includegraphics*[width=5.1cm,clip=]{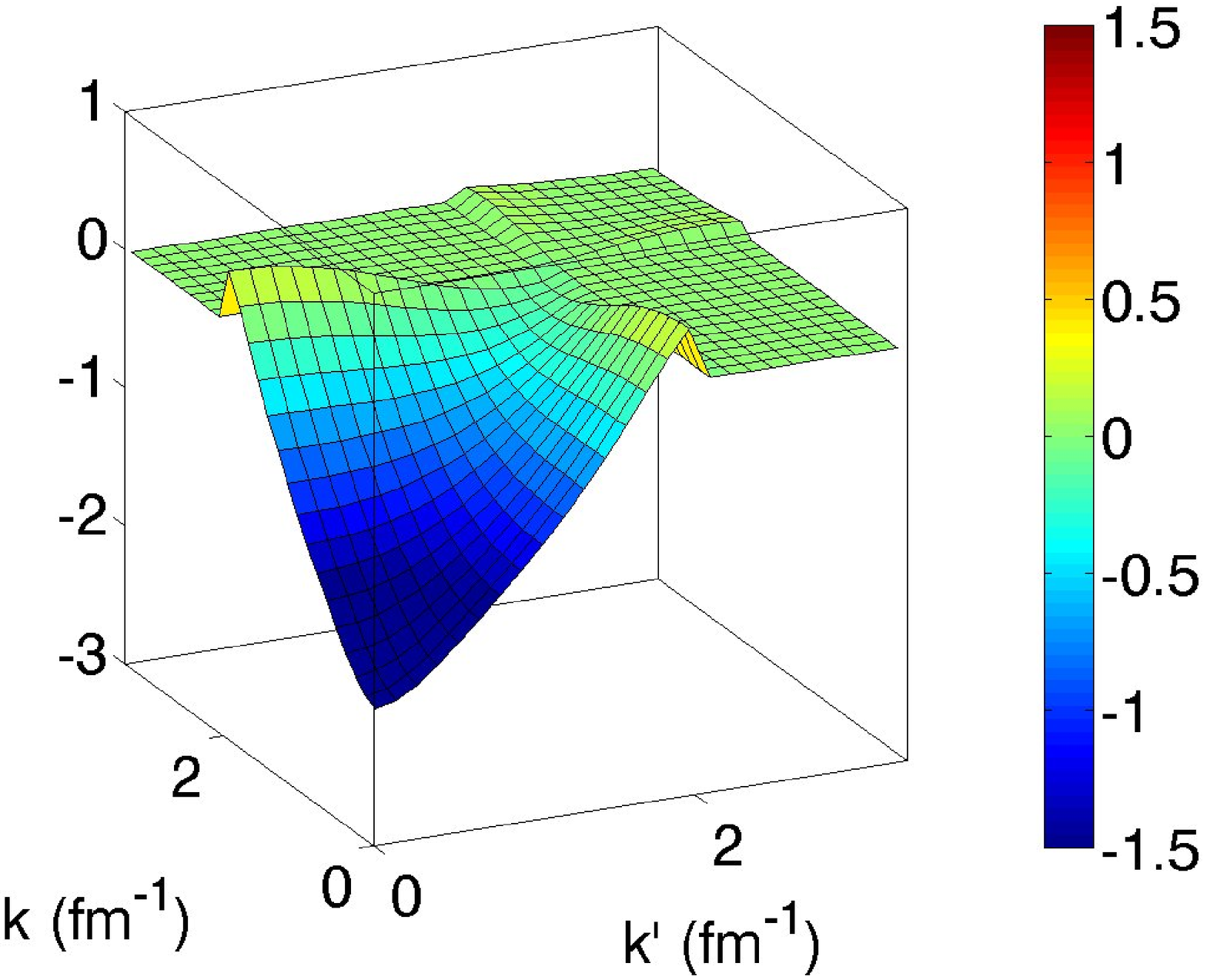}}%
 }%
\caption{(a) and (b):~Comparison of momentum-space $\vlowk$ (left) and
SRG block-diagonal (right) potentials with $\Lambda = 2 \fmi$ evolved
from the N$^3$LO $^3$S$_1$ potential of Ref.~\cite{Entem:2003ft}. The
color and $z$ axes are in fm. For details see Ref.~\cite{Anderson:2008mu}.}
\label{fig:vlowk}
\end{figure}

A powerful feature of the SRG is that the generator $G_s$ can be
tailored to decouple high- and low-momentum physics in different
ways~\cite{Anderson:2008mu}. Block-diagonal decoupling of the sharp
$\vlowk$ form can be generated using SRG flow equations by choosing a
block-diagonal flow
operator~\cite{Gubankova:1998wj,Gubankova:2000cia},
\be
G_s = \biggl(
\begin{array}{cc}
PH_{s}P & 0 \\
0 & QH_{s}Q \end{array} \biggr) \,,
\label{eq:Hbd}   
\ee
with projection operators $P$ and $Q = 1 - P$.  In a partial-wave
momentum representation, $P$ and $Q$ are step functions defined by a
sharp cutoff $\Lambda$ on relative momenta.  This choice for $G_s$,
which means that $\eta_s$ is non-zero only where $G_s$ is zero,
suppresses off-diagonal matrix elements such that the Hamiltonian
approaches a block-diagonal form as $s$ increases. If one considers a
measure of the off-diagonal coupling of the Hamiltonian,
\be
{\rm Tr}[(Q H_s P)^{\dagger} (Q H_s P)]
= {\rm Tr}[P H_s Q H_s P] \geqslant 0 \,,
\label{eq:QHPmeasure}
\ee
then its derivative is easily evaluated by applying the SRG 
equation, Eq.~(\ref{eq:commutator}):
\begin{align}
\frac{d}{ds} \, {\rm Tr}[P H_s Q H_s P] &= 
{\rm Tr}[P\eta_s  Q(Q H_s Q H_s P - Q H_s P H_s P)]
+ {\rm Tr}[(P H_s P H_s Q - P H_s Q H_s Q) Q\eta_s  P] \nonumber \\[1mm]
&= -2 \, {\rm Tr} [(Q \eta_s P)^{\dagger} (Q \eta_s  P)] \leqslant 0
\,.
\end{align}
Thus, the off-diagonal $Q H_s P$ block will decrease in general as $s$
increases~\cite{Gubankova:1998wj,Gubankova:2000cia}.

\begin{figure}[t]
\includegraphics*[width=4.5cm,clip=]{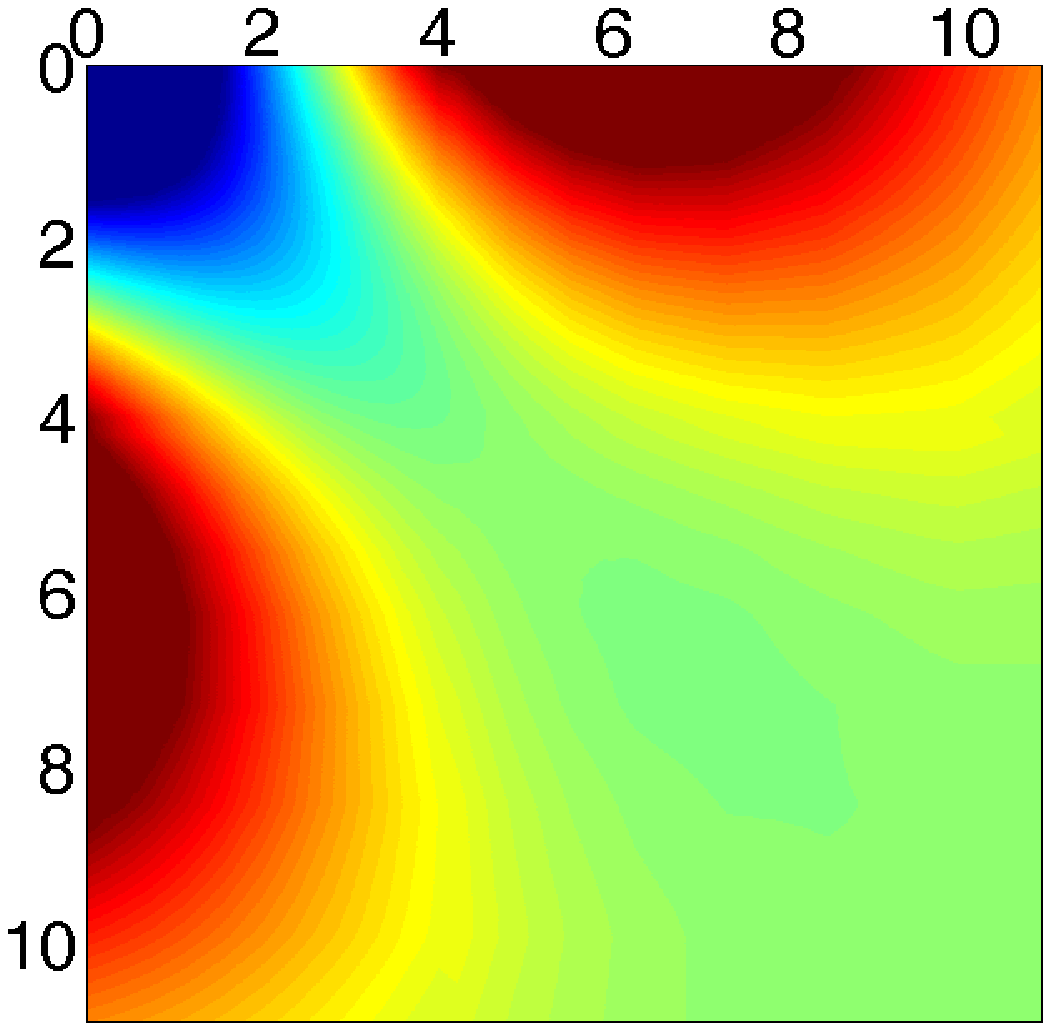}
\includegraphics*[width=4.5cm,clip=]{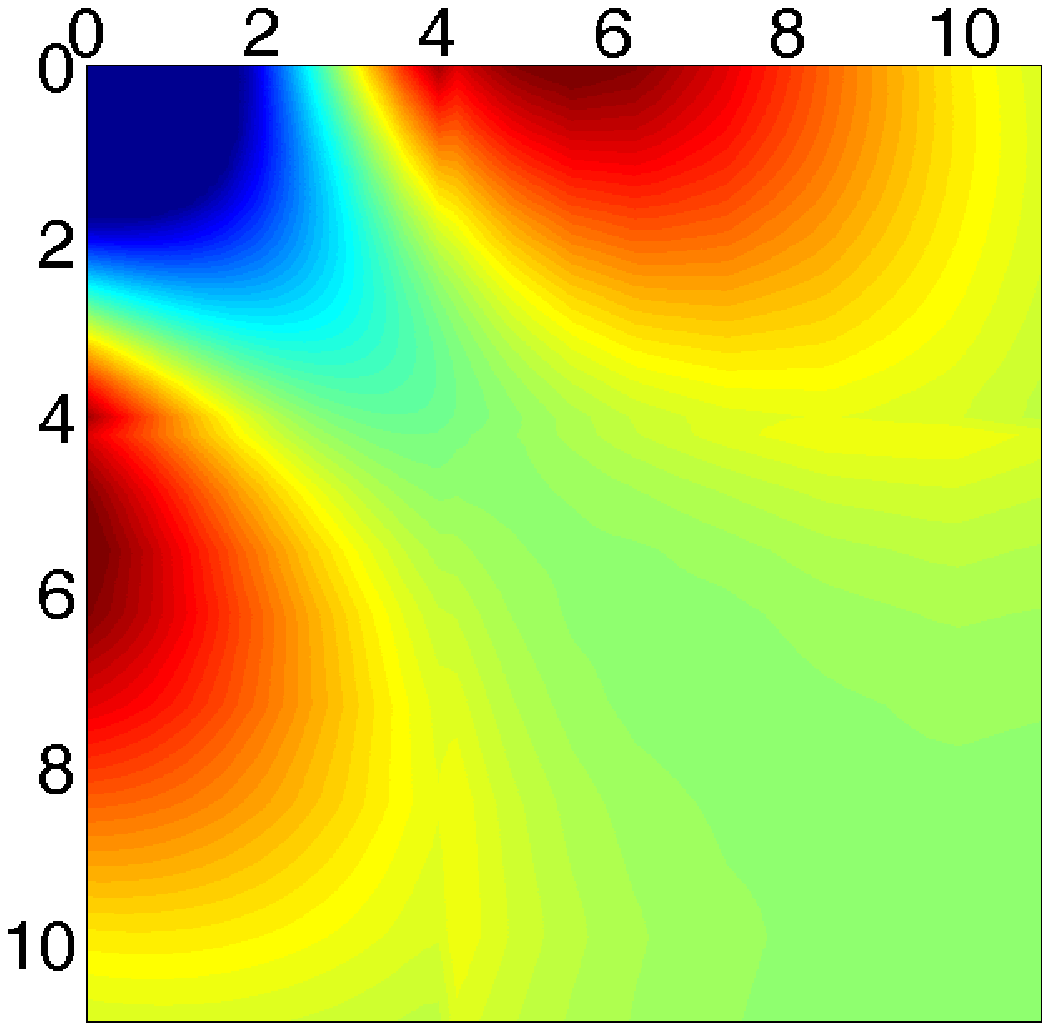}
\includegraphics*[width=4.5cm,clip=]{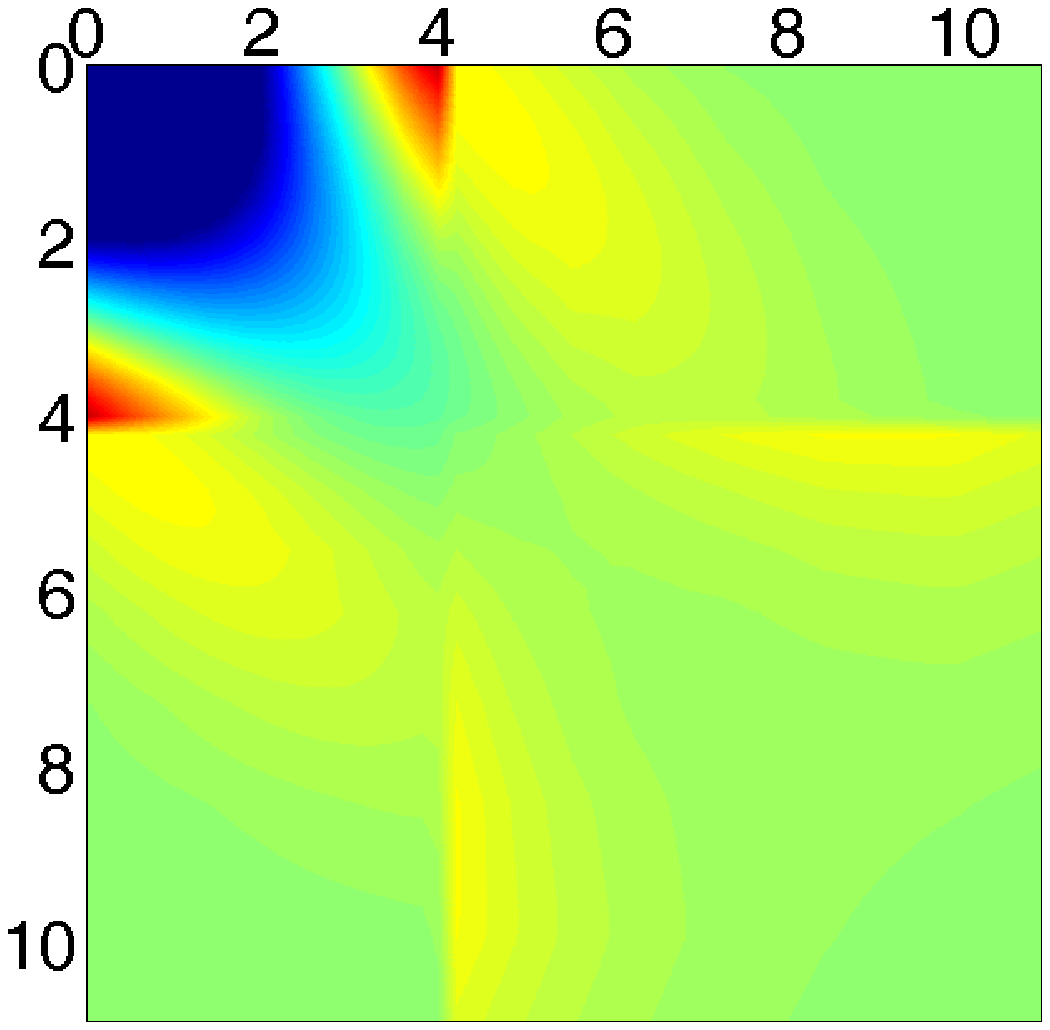}
\includegraphics*[width=4.5cm,clip=]{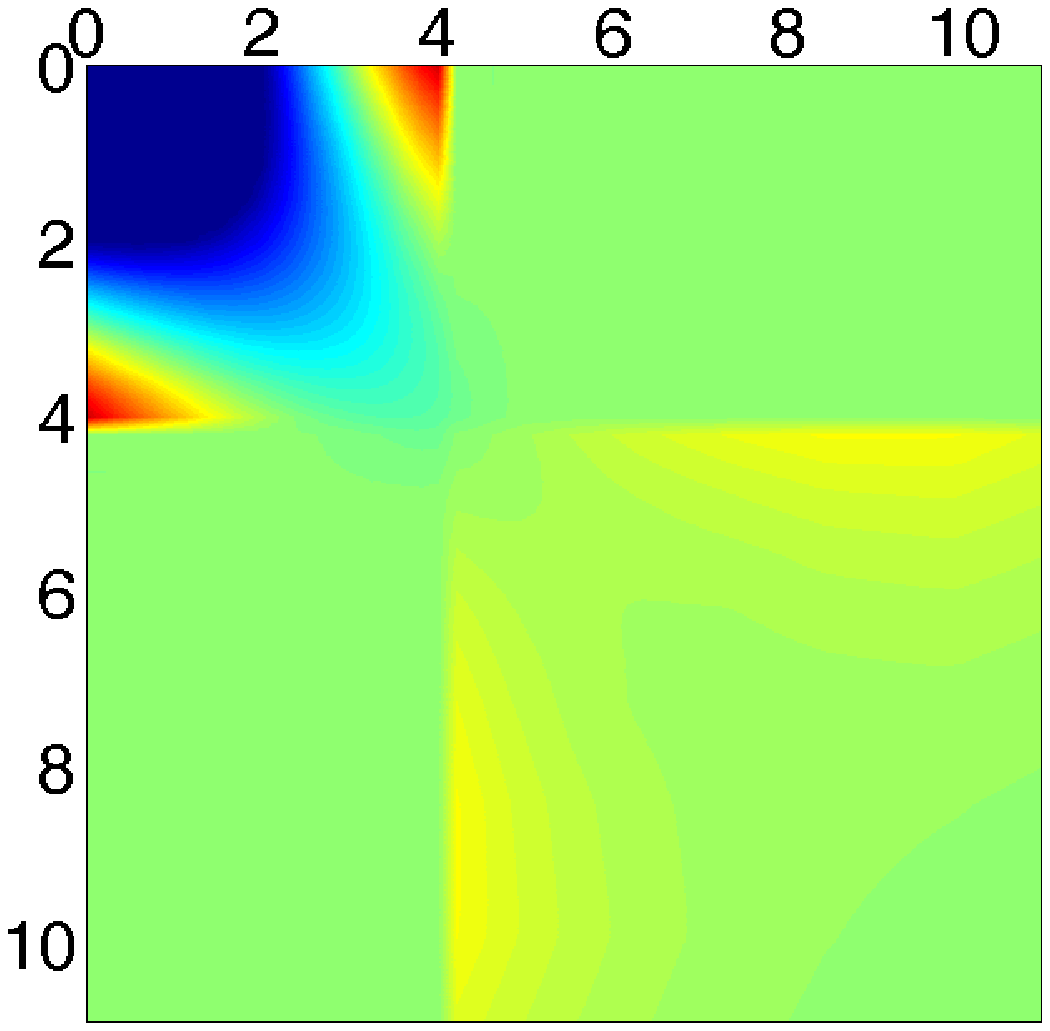}
\caption{Evolution of the $^3$S$_1$ partial wave using the 
SRG block-diagonal flow equation with $\Lambda = 2 \fmi$
at $\lambda = 4, 3, 2$, and $1 \fmi$~\cite{Anderson:2008mu}.
The initial N$^3$LO potential is
from Ref.~\cite{Entem:2003ft}. The axes are in units of $k^2$ from
$0-11 \, {\rm fm}^{-2}$ and the color scale ranges from $-0.5$ to
$+0.5 \fm$ as in Fig.~\protect\subref*{fig:vlowk-a}.}
\label{fig:bd_srg_sharp_3s1}
\end{figure}

The right panels of Figs.~\subref*{fig:vlowk-a}
and~\subref*{fig:vlowk-b} result from evolving the N$^3$LO potential
of Ref.~\cite{Entem:2003ft} using the block-diagonal $G_s$ of
Eq.~(\ref{eq:Hbd}) with $\Lambda = 2\fmi$ to $\lambda = 0.5
\fmi$. The agreement between $\vlowk$ and SRG potentials for momenta
below $\Lambda$ is striking, where a similar degree of universality is
found in other channels.  Deriving an explicit connection
between these approaches is the topic of ongoing research.

The evolution with $\lambda$ of the $^3$S$_1$ channel
is shown in Fig.~\ref{fig:bd_srg_sharp_3s1} (for results in the
$^1$P$_1$ channel see Ref.~\cite{Anderson:2008mu}).   
The evolution of the ``off-diagonal'' 
matrix elements (meaning those outside the $PH_sP$ and $QH_sQ$ blocks)  
can be roughly understood from the
dominance of the kinetic energy on the diagonal.
Let the indices $p$ and $q$ run over indices of the momentum
states in the $P$- and $Q$-spaces, respectively.
To good approximation we can replace $P H_s P$ and
$Q H_s Q$ by their eigenvalues $E_p$ and $E_q$ 
in the SRG equations, yielding~\cite{Gubankova:1998wj,Gubankova:2000cia}
\be
\frac{d}{ds} \, h_{pq} 
\approx \eta_{pq} \, \energy{q} - \energy{p} \, \eta_{pq}
= -(\energy{p}-\energy{q}) \, \eta_{pq} \quad \text{and} \quad
\eta_{pq} \approx \energy{p} \, h_{pq} - h_{pq} \, \energy{q} 
= (\energy{p} - \energy{q})\, h_{pq} \,.
\ee
Combining these two results, we have the evolution of any off-diagonal
matrix element:
\be
\frac{d}{ds} \, h_{pq} \approx - (\energy{p} - \energy{q})^2 \, h_{pq} \,.
\ee
For two-nucleon interactions we can replace the eigenvalues by those
for the relative kinetic energy, which gives the explicit solution
\be
h_{pq}(s) \approx h_{pq}(0) \, e^{-s(p^2 - q^2)} \,.
\label{eq:explicit}
\ee
Thus the off-diagonal matrix elements go
to zero with the energy differences just like with the SRG with
$T_{\rm rel}$; one can see the width of order $1/\sqrt{s} = \lambda^2$
in the $k^2$ plot of the evolving potential in
Fig.~\ref{fig:bd_srg_sharp_3s1} (this is even clearer in other partial
waves, see Ref.~\cite{Anderson:2008mu}).
 
While in principle the evolution to a sharp block-diagonal form means
going to $s = \infty$ ($\lambda = 0$), in practice one must only take
$s$ as large as needed to quantitatively achieve the decoupling
implied by Eq.~(\ref{eq:explicit}).  Furthermore, this should hold for
more general definitions of $P$ and $Q$.  To smooth out the cutoff,
one can introduce a smooth regulator $f_\Lambda$, which can be taken
as an exponential $f_\Lambda(k) = e^{-(k^2/\Lambda^2)^{n_{\rm exp}}}$.
For $\vlowk$ interactions, typical
values used are $n_{\rm exp}=4, 6$, and $8$ (the latter is
considerably sharper but still numerically robust). By replacing
the generator with
\be
G_s = f_\Lambda H_s f_\Lambda + (1-f_\Lambda)H_s(1-f_\Lambda) \,,
\label{eq:Gs}
\ee
one obtains a smooth block-diagonal potential (for examples, see 
Ref.~\cite{Anderson:2008mu}).

\subsection{Related methods}
\label{subsec:other}

There are various other methods used to transform nuclear Hamiltonians
for use in few- and many-body calculations.  Although they are not
cast in the RG language, many of the same principles and features
discussed in Section~\ref{sec:RG} apply.  While space does not permit
detailed discussions of each, here we give a brief synopsis and
comparison, with pointers to the literature for more information.

\begin{figure}[t]
\centering
\includegraphics*[width=3.4in,clip=]{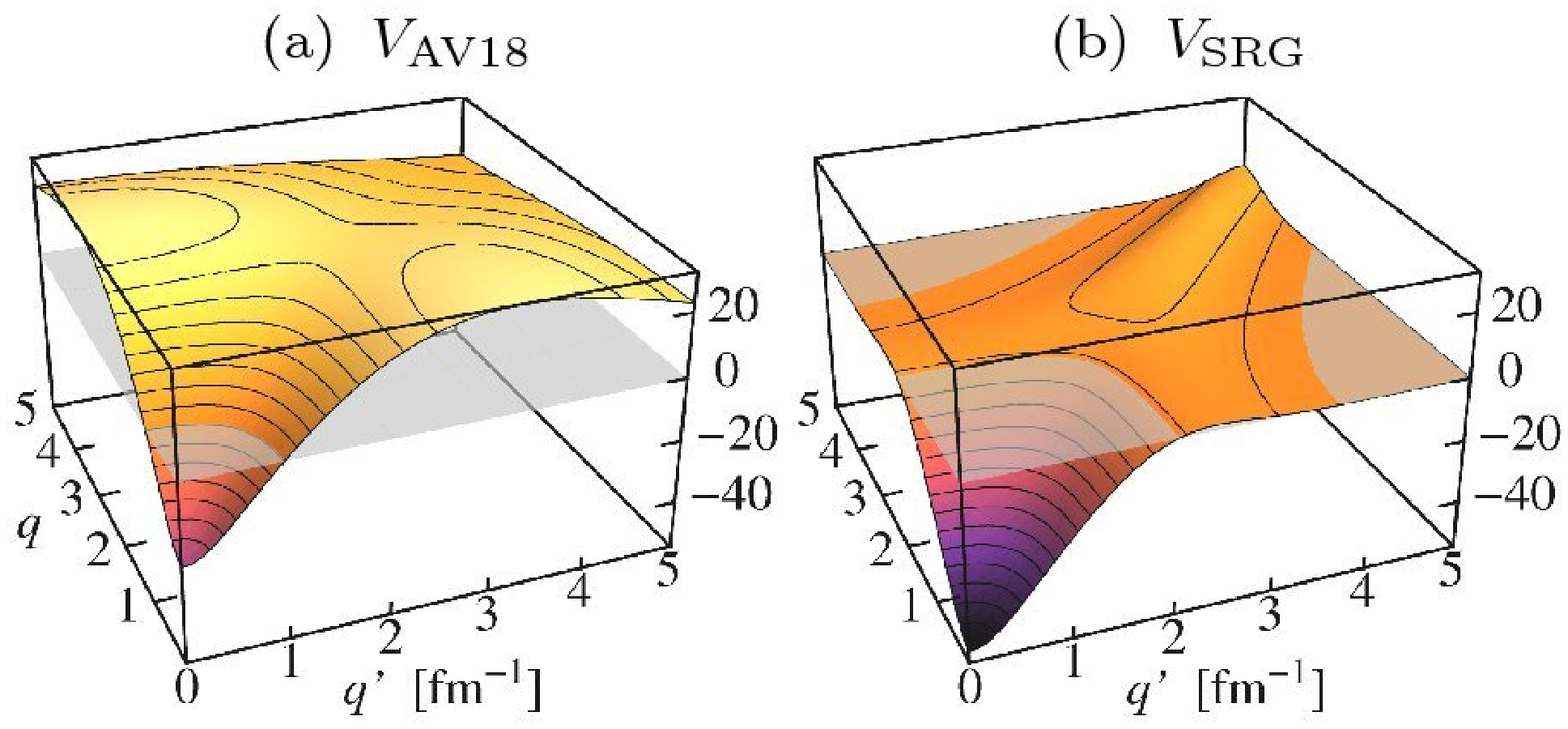}
\includegraphics*[width=3.4in,clip=]{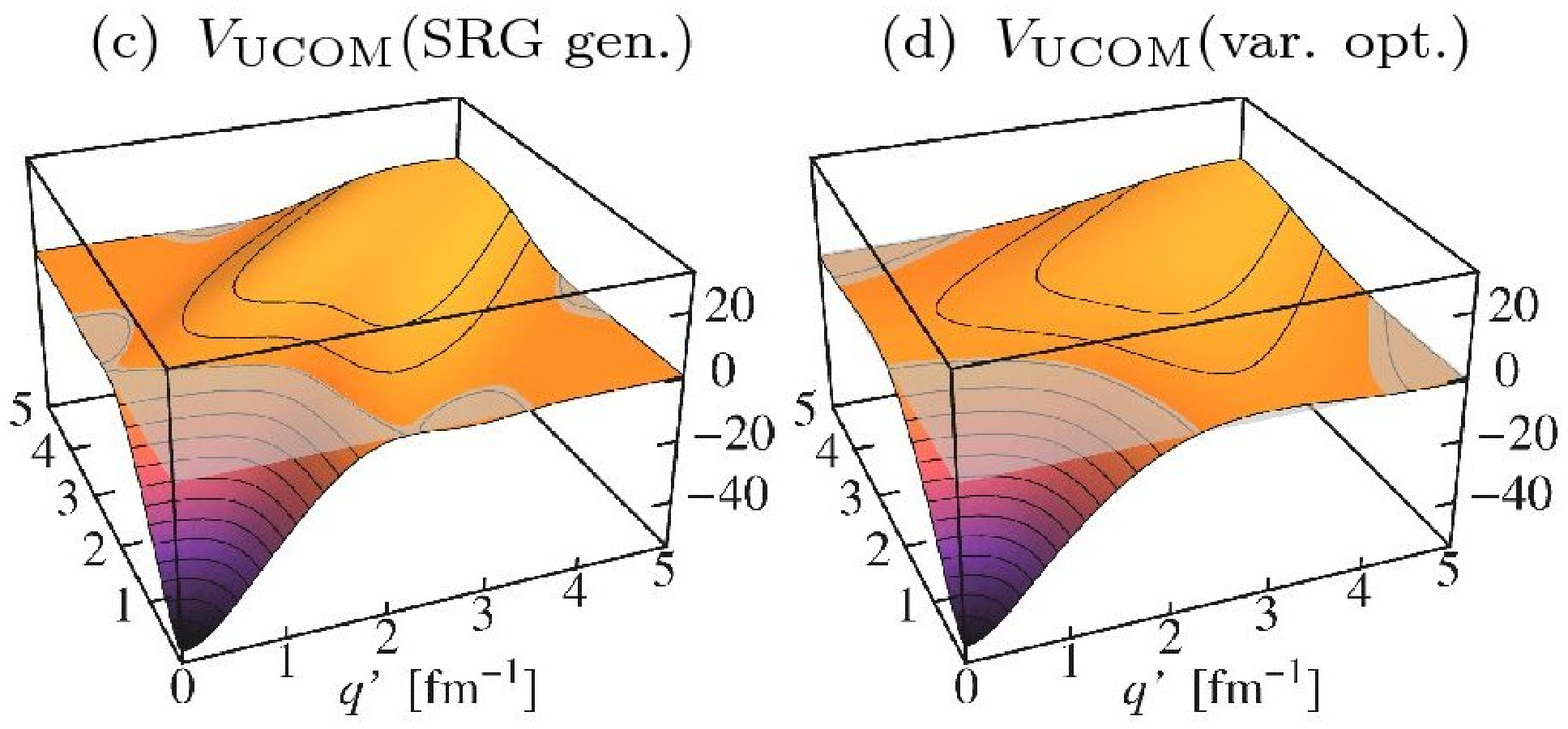}
\caption{Comparison of momentum-space matrix elements of $^1$S$_0$ potentials
(in units of ${\rm MeV} \, {\rm fm}^3$) for (a) Argonne $v_{18}$, (b)
SRG-evolved with $\lambda = 2.4 \fmi$, (c) UCOM-transformed using 
SRG-generated correlators ($\lambda = 2.25 \fmi$), and (d)
UCOM-transformed using variationally optimized
correlators~\cite{Roth:2008km}.
Reprinted with permission from R.~Roth et al.~\cite{Roth:2008km},
copyright (2008) by the American Physical Society.}
\label{fig:ucom}
\end{figure}

\textbf{Bloch-Horowitz.} The energy-dependent Bloch-Horowitz (BH)
equation is used as an intermediate step in deriving
energy-independent sharp and smooth-cutoff $\vlowk$ interactions.  The
BH method has also been used directly in other approaches, where the
energy is determined self-consistently. The harmonic-oscillator-based
effective theory~\cite{Haxton:2002kb,Luu:2004xc,Haxton:2006gw,Haxton:2007hx} is
one such example. While most detailed applications have been limited
to $A=2,3$ because energy-dependent interactions are difficult to
handle in other approaches, perturbative expansions and
clever resummations may overcome this hurdle.

\textbf{UCOM.}  Feldmeier, Neff, Roth and collaborators have developed
a method using unitary transformations designed to remove short-range
``hard core'' and tensor
correlations~\cite{Feldmeier:1997zh,Roth:2004ua,Roth:2005ah,Roth:2005pd}.
The approach is called the Unitary Correlation Operator Method (UCOM).
The result is a soft Hamiltonian that shares favorable features of the
RG-based interactions.  They have also studied parallels (and some
distinctions) of UCOM to the SRG
approach~\cite{Hergert:2007wp,Roth:2008km}, as shown in
Fig.~\ref{fig:ucom}.  Using a UCOM NN potential, which is phase
equivalent to a given initial Hamiltonian, they have investigated
ground-state energies and radii of many nuclei (including heavy nuclei
such as $^{208}$Pb) in Hartree-Fock plus many-body perturbation
theory.  Preliminary results from supplementing the NN potential by a
short-range three-body force with fitted strength are
encouraging~\cite{Roth:2009private}.  As originally formulated, the
UCOM procedure is difficult to carry out beyond the two-body
level. However, the connections to the SRG approach established in
\cite{Hergert:2007wp,Roth:2008km} may help to transform 3N forces.

\textbf{NCSM Lee-Suzuki.}  The Lee-Suzuki transformations used in the
No-Core-Shell-Model (NCSM) are constructed to decouple low from high
energy within a harmonic-oscillator model
space~\cite{Navratil:2000gs}.  A cluster expansion is used to organize
and truncate the many-body components.  This approach has been
successfully used to describe nuclear spectra, especially in $p$-shell
nuclei starting from chiral EFT
interactions~\cite{Nogga:2005hp,Navratil:2007we}.  A detailed overview
of this approach including recent developments is given in
Ref.~\cite{Navratil:2009ut}.  As usually constructed, the Lee-Suzuki
approach is not variational and does not always show smooth
convergence with respect to the model-space size. Unlike the
momentum-space RG evolution, it is not clear if evolving in the
harmonic-oscillator basis is problematic for long-range operators, as
indicated by the sizable effective three- and higher-body
contributions to $B(E2)$ values~\cite{Navratil:2009ut}.

\textbf{UMOA.} In the Unitary Model Operator Approach (UMOA), a
Hermitian, energy-independent effective interaction is derived by
Okubo-type unitary transformations of an initial Hamiltonian (see
Ref.~\cite{Fujii:2005nz} and references therein).  The transformation
is applied twice to make the effective interaction more suitable to
larger nuclei.  In the first transformation, the short-range repulsion
is softened by partitioning the $P$- and $Q$-space with an energy
cutoff on two-body harmonic-oscillator states. In the second
transformation, the softened interaction is transformed so that there
are no matrix elements between the unperturbed ground state and
two-particle--two-hole states for a given closed-shell nucleus.
Calculations have been performed for stable nuclei around
$^{16}$O~\cite{Fujii:2003iy},
$\Lambda$-hypernuclei~\cite{Fujii:2000cg} and for ground-state and
single-particle energies of $^{40}$Ca and
$^{56}$Ni~\cite{Fujii:2009bf}.

\begin{figure}[t]
\centering
\includegraphics*[width=9cm,clip=]{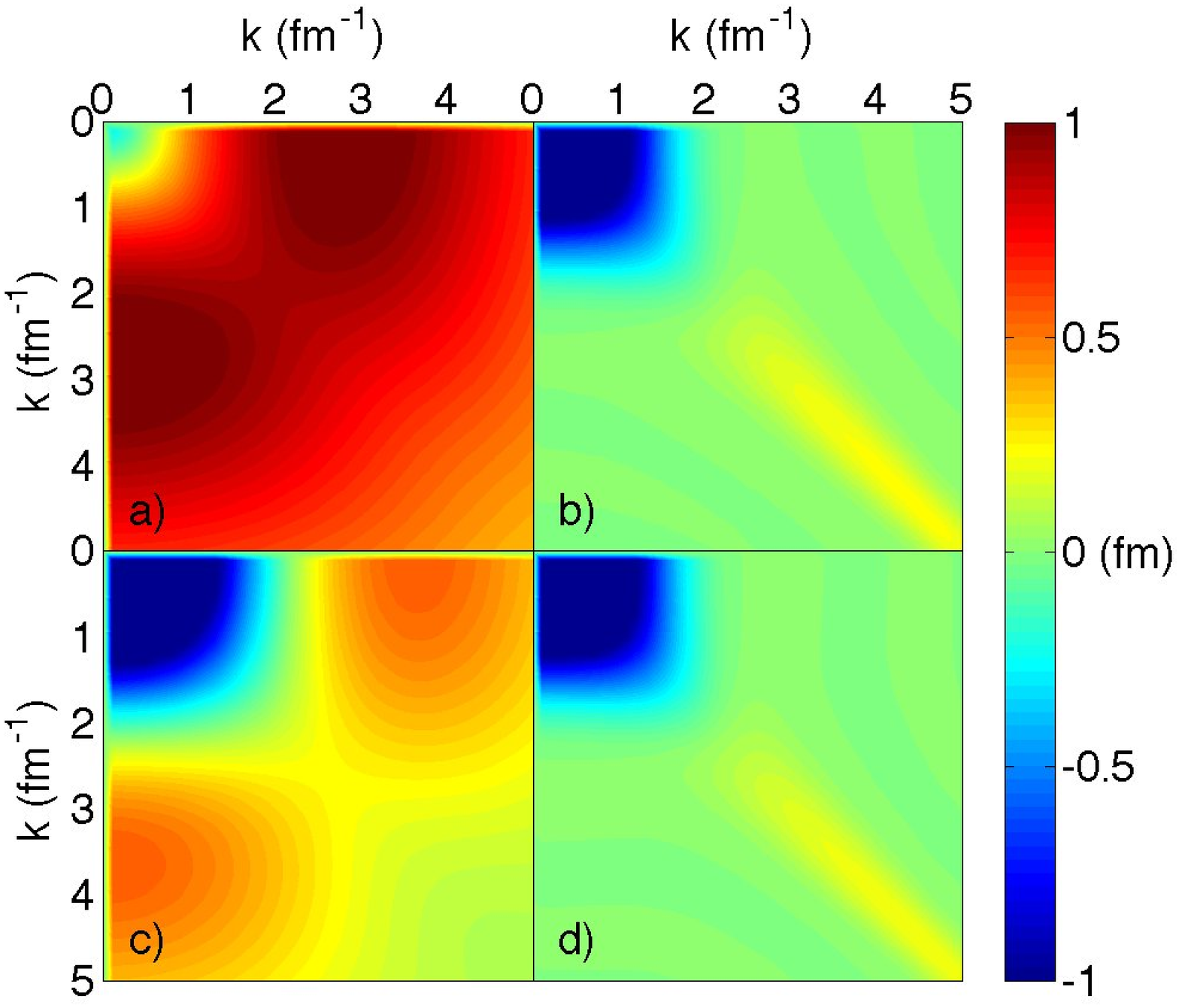}
\hspace*{.05in}
\includegraphics*[width=9cm,clip=]{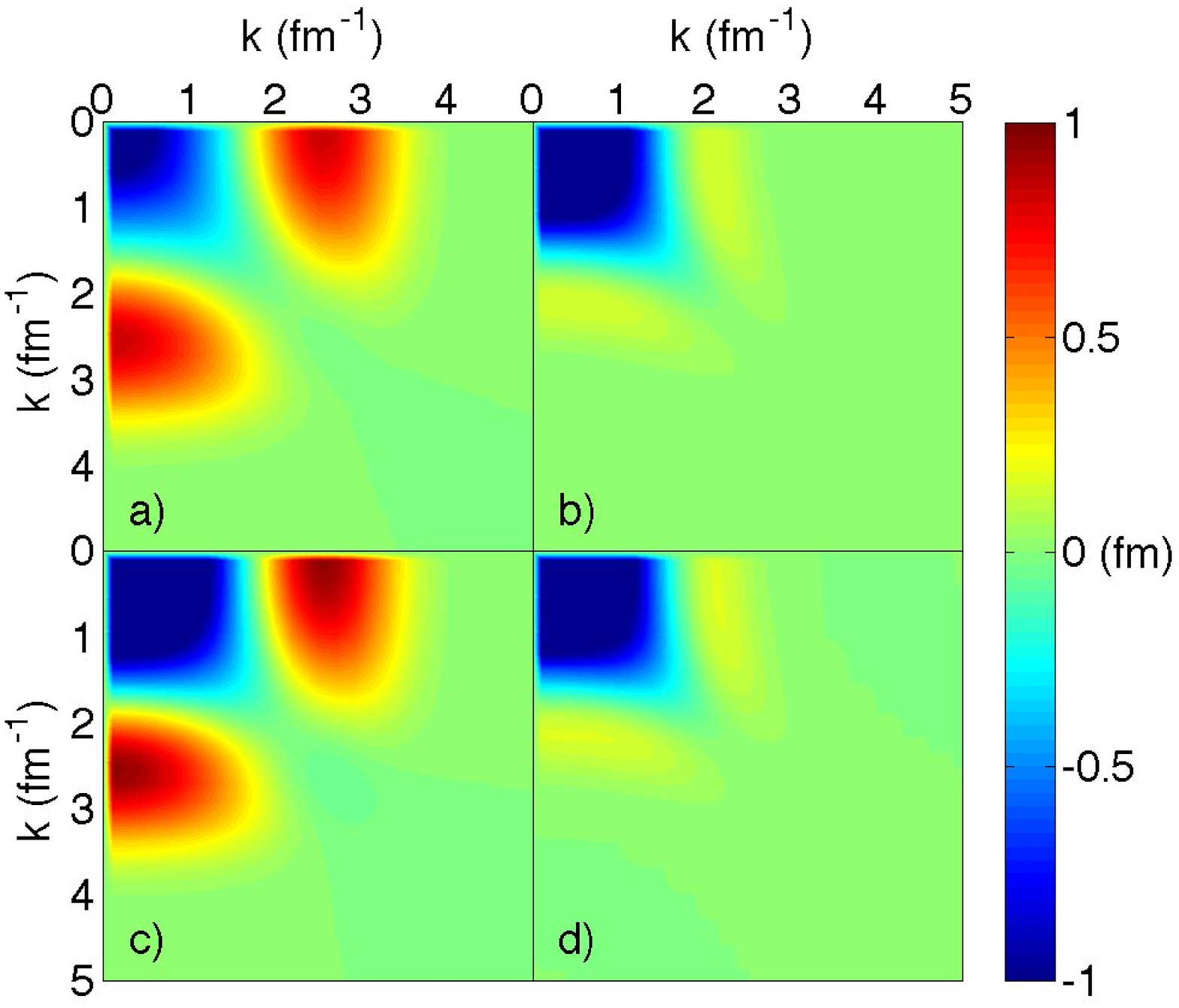}
\caption{$G$ matrix at saturation density for the Argonne $v_{18}$
potential~\cite{Wiringa:1994wb} (left panels) and the N$^3$LO
potential of Ref.~\cite{Entem:2003ft} (right panels) in the $^3$S$_1$
channel. Each set of four panels are a) initial potential, b)
potential evolved by the SRG to $\lambda = 2 \infm$, c) $G$ matrix
based on a), and d) $G$ matrix based on b).}
\label{fig:gmatrix}
\end{figure}

\textbf{$G$ matrix.} The $G$ matrix is the in-medium sum of ladder
diagrams to form a two-body effective interaction from a starting NN
potential.  This formalism has been used in calculations of nuclear
matter~\cite{Baldo99}, for effective interactions in
nuclei~\cite{HjorthJensen:1995ap}, and in self-consistent Green's
function approaches~\cite{Dickhoff:2004xx}.  It has been suggested
that $\vlowk$ NN interactions are essentially the same as a $G$ matrix
but there are substantial differences.  First, $\vlowk$ at a given
cutoff $\Lambda$ is a free-space interaction, equivalent in practice
to any low-energy Hamiltonian (with no ``starting energy''
dependence).  Second, significant coupling between low and high
momenta remains in the $G$ matrix. This is demonstrated in
Fig.~\ref{fig:gmatrix}, which shows the Argonne $v_{18}$ and an
N$^3$LO potential in the $^3$S$_1$ channel compared to low-momentum
interactions (SRG evolved to $\lambda = 2 \infm$, similar results are
found for $\vlowk$ interactions) and to the $G$ matrices obtained from
each. (The qualitative conclusions do not depend on the particular
choice of starting energy.)  We observe that the summation into the
$G$ matrix has a relatively small effect on low-momentum interactions,
in stark contrast to the Argonne $v_{18}$ potential case.  The
resulting low-momentum matrix elements are indeed similar, but it is
also apparent that there is substantial off-diagonal strength in the
$G$ matrix in this case.  This coupling to high momenta prevents the
expansion of nuclear matter properties from being perturbative in the
$G$ matrix, necessitating a non-perturbative scheme such as the
hole-line expansion. One might expect the softer N$^3$LO $^3$S$_1$
potential to avoid this problem, but we see that the off-diagonal
strength, while limited to lower momentum, is actually increased.


\section{Many-body interactions and operators}
\label{sec:manybody}

When evolving nuclear interactions to lower resolution, it is
inevitable that many-body interactions and operators are induced even
if initially absent.  This might be cause for alarm if nuclei could be
accurately calculated with interactions truncated at the two-body
level, as was assumed for part of the history of nuclear structure
calculations.  However, chiral EFT reveals the natural scale and
hierarchy of many-body forces, which dictates their inclusion in
modern calculations of nuclei and nucleonic matter.  Thus the real
concern is whether this hierarchy is maintained as nuclear
interactions are evolved.  In this section, we review the current
status of RG technology to include many-body interactions and
operators and the currently known impact on the hierarchy.

\subsection{Three-nucleon interactions}
\label{subsec:3nf}

Three-nucleon interactions are a frontier. They are crucial for
binding energies and radii, they play a central role for spin-orbit
effects, spin dependencies, for few-body scattering and the evolution
of nuclear structure with isospin, and they drive the density
dependence of nucleonic matter (see Sections~\ref{sec:infinite}
and~\ref{sec:finite})~\cite{Schwenk:2008su}. Three-nucleon
interactions are also required for
renormalization~\cite{Braaten:1997xx,Hammer:2000xg}.  The construction
of 3N forces based on chiral EFT provides a systematic organization of
the physics and an operator basis that can be used to approximate the evolution
of low-momentum 3N interactions.

In chiral EFT without explicit $\Delta$ isobars, 3N forces first enter
at N$^2$LO (see Fig.~\subref*{chiralEFT-a}) and contain a long-range
$2 \pi$-exchange part $V_c$, an intermediate-range $1 \pi$-exchange
part $V_D$ and a short-range contact interaction
$V_E$~\cite{VanKolck:1994yi,Epelbaum:2002vt}:
\be
\parbox[c]{195pt}{%
\includegraphics[scale=0.6,clip=]{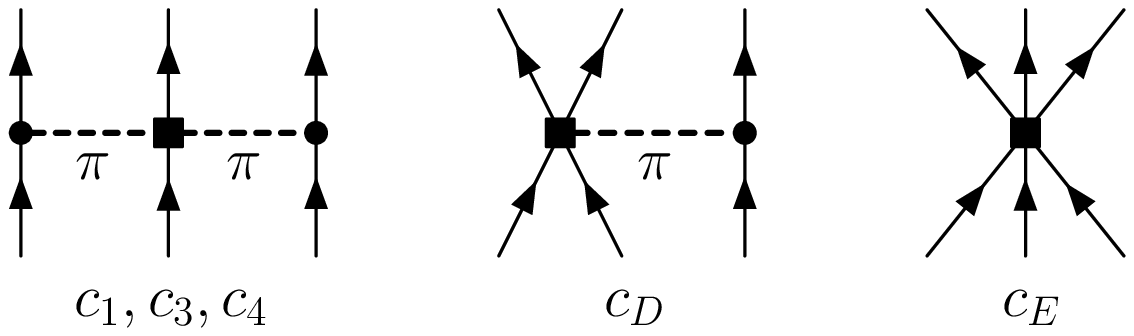}}
\label{3NF}
\ee
The $2 \pi$-exchange interaction is given by
\be
V_c = \frac{1}{2} \, \biggl( \frac{g_A}{2 f_\pi} \biggr)^2 
\sum\limits_{i \neq j \neq k} 
\frac{({\bm \sigma}_i \cdot {\bf q}_i) ({\bm \sigma}_j \cdot 
{\bf q}_j)}{(q_i^2 + m_\pi^2) (q_j^2 + m_\pi^2)} \: F_{ijk}^{\alpha\beta} \,
\tau_i^\alpha \, \tau_j^\beta \,,
\label{Vc}
\ee
where ${\bf q}_i = {\bf k}'_i - {\bf k}_i$ denotes the difference of initial
and final nucleon momenta ($i, j$ and $k=1,2,3$) and
\be
F_{ijk}^{\alpha\beta} = \delta^{\alpha\beta} \biggl[ - \frac{4 c_1 
m_\pi^2}{f_\pi^2} + \frac{2 c_3}{f_\pi^2} \: {\bf q}_i \cdot {\bf q}_j
\biggr]
+ \sum_\gamma \, \frac{c_4}{f_\pi^2} \: \epsilon^{\alpha\beta\gamma}
\: \tau_k^\gamma \: {\bm \sigma}_k \cdot ( {\bf q}_i \times {\bf q}_j)
\,, \label{Vc2}
\ee
while the $1 \pi$-exchange and contact interactions are given respectively by
\begin{align}
V_D &= - \frac{g_A}{8 f_\pi^2} \, \frac{c_D}{f_\pi^2 \lm_\chi}
\: \sum\limits_{i \neq j \neq k} 
\frac{{\bm \sigma}_j \cdot {\bf q}_j}{q_j^2 + m_\pi^2} \: ({\bm \tau}_i
\cdot {\bm \tau}_j) \, ({\bm \sigma}_i \cdot {\bf q}_j) \,, 
\label{VD} \\
V_E &= \frac{c_E}{2 f_\pi^4 \lm_\chi} \: \sum\limits_{j \neq k} ({\bm \tau}_j
\cdot {\bm \tau}_k) \,. \label{VE}
\end{align}
Typical values for applying Eqs.~(\ref{Vc})--(\ref{VE}) are $g_A =
1.29$, $f_\pi = 92.4 \mev$, $m_\pi = 138.04 \mev$ and $\lm_\chi = 700
\mev$.  In the RG calculations based on chiral EFT interactions
discussed here, the 3N force contributions are regulated as in
Ref.~\cite{Bogner:2009un} using
\be
f_{\text{R}}(p,q) = \exp \biggl[ - \frac{(p^2+3 q^2/4)^2}{\lm_{\rm 3NF}^4}
\biggr] \,,
\label{reg}
\ee
with a 3N cutoff $\lm_{\rm 3NF}$ that is allowed to vary
independently of the NN cutoff. Here, $p$ and $q$ are  Jacobi
momenta.
The exchange terms of the 3N force are included by means of the
antisymmetrizer
\beqn
{\mathcal A}_{123} = (1 + P_{12} P_{23} + P_{13} P_{23}) (1 - P_{23}) 
= 1 - P_{12} - P_{13} - P_{23} + P_{12} P_{23} + P_{13} P_{23} \,,
\eeqn
where $P_{ij}$ is the exchange operator for spin, isospin and momenta 
of nucleons $i$ and $j$.  The regulator $f_{\text{R}}(p,q)$ is totally
symmetric when expressed in the nucleon momenta ${\bf k}_i$, and thus
the direct and exchange terms contain the same regulator.

The low-energy $c_i$ couplings for the long-range parts relate $\pi$N,
NN and 3N interactions, which means they can be fitted in different
processes.  The determination of the $c_i$ couplings from $\pi$N
scattering agrees reasonably well with the extraction from NN partial
waves~\cite{Epelbaum:2008ga}.  However, there are large uncertainties
in the values (and some controversy, see Ref.~\cite{Entem:2003cs}).
More accurate determinations are needed for nuclear structure, where
$c_3$ and $c_4$ are particularly important (see
Fig.~\subref*{fig:EOS_compare-b} for an example of the impact of the
$c_i$ uncertainties).  The remaining $c_D$ and $c_E$ couplings are
usually fit to the $^3$H binding energy and another observable in $A
\geqslant 3$.  These leading 3N interactions generally improve the
agreement of theory with experiment in nucleon-deuteron
scattering~\cite{Epelbaum:2005pn,Epelbaum:2008ga} and in the spectra
of light nuclei~\cite{Navratil:2007we} (see Figs.~\subref*{fig:3nfobs}
and~\ref{fig:n3lonuclei}).  The subleading 3N interactions at N$^3$LO
are parameter free and several parts have been calculated to
date~\cite{Ishikawa:2007zz,Bernard:2007sp}.  
The subleading 3N interactions at N$^3$LO
are parameter free and several parts have been calculated to
date~\cite{Ishikawa:2007zz,Bernard:2007sp}, including 
two-pion--one-pion-exchange and pion-ring diagrams,
spin-orbit forces, and contributions that involve NN
contacts~\cite{Epelbaum:2008ga}. 
One consequence is that operator structures of the form in
Eq.~\eqref{Vc} are generated at N$^3$LO, 
effectively weakening the values of the $c_i$ couplings~\cite{Epelbaum:2008ga}.

\begin{table}[t]
\centering
\resizebox{7.25in}{!}{
\begin{tabular}{l|rrrrr|rrrrr|c|c}
& \multicolumn{5}{c|}{$^3$H} & \multicolumn{5}{c|}{$^4$He} & $\max$ & $^4$He \\
\multicolumn{1}{c|}{$\lm$} &
\multicolumn{1}{c}{$T$} & \multicolumn{1}{c}{$\vlowk$} &
\multicolumn{1}{c}{$V_c$} & \multicolumn{1}{c}{$V_D$} &
\multicolumn{1}{c|}{$V_E$} & \multicolumn{1}{c}{$T$} &
\multicolumn{1}{c}{$\vlowk$} & \multicolumn{1}{c}{$V_c$} &
\multicolumn{1}{c}{$V_D$} & \multicolumn{1}{c|}{$V_E$} &
\multicolumn{1}{c|}{$|V_{\rm 3N}/\vlowk|$} & 
\multicolumn{1}{c}{$k_{\rm rms}$} \\ \hline
$1.0$ & $21.06$ & $-28.62$ & $0.02$ & $0.11$ & $-1.06$ & 
$38.11$ & $-62.18$ & $0.10$ & $0.54$ & $-4.87$ & $0.08$ & $0.55$ \\
$1.3$ & $25.71$ & $-34.14$ & $0.01$ & $1.39$ & $-1.46$ & 
$50.14$ & $-78.86$ & $0.19$ & $8.08$ & $-7.83$ & $0.10$ & $0.63$ \\
$1.6$ & $28.45$ & $-37.04$ & $-0.11$ & $0.55$ & $-0.32$ & 
$57.01$ & $-86.82$ & $-0.14$ & $3.61$ & $-1.94$ & $0.04$ & $0.67$ \\
$1.9$ & $30.25$ & $-38.66$ & $-0.48$ & $-0.50$ & $0.90$ &
$60.84$ & $-89.50$ & $-1.83$ & $-3.48$ & $5.68$ & $0.06$ & $0.70$ \\ 
$2.5(a)$ & $33.30$ & $-40.94$ & $-2.22$ & $-0.11$ & $1.49$ &
$67.56$ & $-90.97$ & $-11.06$ & $-0.41$ & $6.62$ & $0.12$ & $0.74$ \\
$2.5(b)$ & $33.51$ & $-41.29$ & $-2.26$ & $-1.42$ & $2.97$ &  
$68.03$ & $-92.86$ & $-11.22$ & $-8.67$ & $16.45$ & $0.18$ & $0.74$ \\
$3.0(*)$ & $36.98$ & $-43.91$ & $-4.49$ & $-0.73$ & $3.67$ & 
$78.77$ & $-99.03$ & $-22.82$ & $-2.63$ & $16.95$ & $0.23$ & $0.80$
\end{tabular}}
\caption{\label{tab:3Nparts} Expectation values of the kinetic energy $T$, 
$\vlowk$ and the different 3N contributions (long-range $2
\pi$-exchange part $V_c$, $1 \pi$-exchange part $V_D$ and contact
interaction $V_E$) for $^3$H and $^4$He. All energies are in MeV and
momenta are in fm$^{-1}$. $(a)$~and $(b)$~denote two possible
solutions for $\lm = 2.5 \, {\rm fm}^{-1}$ and $(*)$ indicates that
the $^4$He fit is approximate, for details see
Ref.~\cite{Nogga:2004ab}. Also listed is the ratio of maximum 3N to
$\vlowk$ contribution and an average relative momentum $k_{\rm
rms}$~\cite{Schwenk:2004hz}.}
\end{table}

\begin{figure}[t]
\centering
\includegraphics[width=4.0in,clip=]{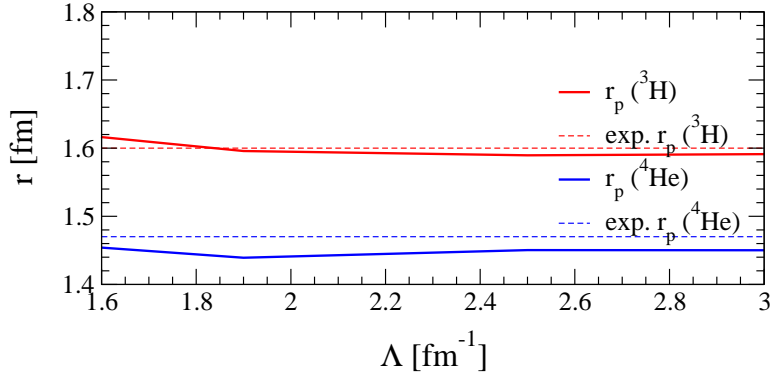}
\caption{The $^3$H and $^4$He radii are approximately cutoff
independent with low-momentum NN and 3N 
interactions~\cite{Andreas,Schwenk:2008su}.\label{radii}}
\end{figure}

Instead of explicitly evolving 3N interactions to low momentum, one
can take the N$^2$LO 3N forces as a truncated basis and assume that
the $c_i$ couplings are not modified by the RG evolution. This is the
strategy adopted in
Refs.~\cite{Nogga:2004ab,Bogner:2005sn,Bogner:2009un}.  In this
scheme, $V_{\rm 3N}(\lm)$ is constructed by fitting the $c_D$ and
$c_E$ couplings to the $^3$H binding energy and the $^4$He binding
energy~\cite{Nogga:2004ab} or matter radius~\cite{Bogner:2009un} for a
range of cutoff values. When the $^4$He radius is fit, the predicted
$^4$He binding energies are very reasonable.  Fit values for $c_D$ and
$c_E$ are found to be natural and are tabulated in
Refs.~\cite{Nogga:2004ab} (with Argonne $v_{18}$~\cite{Wiringa:1994wb}
as initial NN potential) and~\cite{Bogner:2009un} (with the N$^3$LO
potential of Ref.~\cite{Entem:2003ft}).  For cutoffs $\lm \lesssim 2.0
\fmi$, there are linear dependencies in the fitting, which are
consistent with a perturbative $V_D$ and $V_E$ contribution: $E_{\rm
gs} = E(\vlowk + V_c) + c_D \langle D{\rm-term} \rangle +
c_E \langle E{\rm-term} \rangle$ (where $\langle \ldots \rangle$
denotes the matrix elements of the operators). This has been verified
explicitly and also for $V_c$.

Thus, when evolved to lower cutoffs, low-momentum 3N interactions
become perturbative in light nuclei~\cite{Nogga:2004ab,Bogner:2009un}
which means $\langle \Psi^{(3)} | V_{\rm 3N} | \Psi^{(3)} \rangle
\approx \langle \Psi^{(2)} | V_{\rm 3N} | \Psi^{(2)} \rangle$, where
$| \Psi^{(n)} \rangle$ are exact solutions including up to $n$-body
forces.  It is instructive to look at the expectation values of
different contributions to the $^3$H and $^4$He ground-state energies,
as listed in Table~\ref{tab:3Nparts}.  The expectation values are
consistent with chiral EFT power counting estimates, which means that
the hierarchy of many-body forces is maintained. Note also that the
sign of the $c_i$ contribution is cutoff dependent. This contribution
is repulsive in nuclear and neutron matter for low-momentum
interactions (see Sections~\ref{subsec:nuclear}
and~\ref{subsec:neutron}), which is counter to the intuition from
conventional NN and 3N potentials~\cite{Akmal:1998cf,Pieper:2001ap}.

As discussed in Section~\ref{subsec:cutoff} (see Fig.~\ref{Tjon}),
neglecting 3N interactions leads to a universal correlation
(empirically known as Tjon-line) between the $^3$H and $^4$He binding
energies~\cite{Nogga:2004ab,Platter:2004zs}.  This is an example of
how cutoff variation can provide lower bounds for theoretical
uncertainties due to neglected many-body interactions or an incomplete
many-body treatment.  Another example is shown in Fig.~\ref{radii},
where the radii for $^3$H and $^4$He are shown as a function of
$\vlowk$ cutoff $\Lambda$.  With 3N forces included, there
is only a small cutoff dependence that shows the scale of the residual
dependence on higher-order shorter-range many-body interactions.

\subsection{Three-nucleon force evolution}
\label{subsec:evolution}

While the use of a chiral EFT basis to approximate low-momentum 3N
interactions is convenient and computationally efficient, it has not
been verified against explicitly evolved three-body forces. This will
be possible using SRG flow equations, which offer a tractable approach
to evolving 3N forces (that avoids solving for any $T$ matrices).
The SRG flow equation,
\be
\frac{dH_\flow}{d\flow} =  [ [G_s, H_\flow], H_\flow] \,,
\label{eq:commutator2}
\ee
can be applied directly in the three-particle space.  The right side
involves only the Hamiltonian and the generator $G_s$, which can be
evaluated in a basis without solving bound state or scattering
equations.  A potential issue is the role of spectator nucleons; we
consider two solutions: first a decoupling of the 3N part in
momentum-space representation and then a direct solution in a 
harmonic-oscillator basis. In both cases, we take $G_s = \Trel$ and
return at the end to other choices.

To show the basic idea in momentum space, we adopt a notation in which
$\Vtwo12$ means the two-body interaction between particles 1 and 2
while $\Vthree$ is the irreducible three-body potential.  We start
with the Hamiltonian including up to three-body interactions (keeping
in mind that higher-body interactions will be induced as we evolve in
$s$ but will not contribute to three-body systems):
\be
H_s = \Trel + \Vtwo12 + \Vtwo13 + \Vtwo23 + \Vthree \equiv \Trel + V_s
\,.
\label{eq:H}
\ee
(Note: all of the potentials depend implicitly on $s$.)
The relative kinetic energy operator $\Trel$ can be decomposed 
in three ways:
\be
\Trel = \Ttwo12 + T_3 = \Ttwo13 + T_2 = \Ttwo23 + T_1 \,,
\label{eq:Trel}	
\ee
and $T_i$ commutes with $\Vtwo{j}{k}$,  
\be
[T_3, \Vtwo12] = [T_2, \Vtwo13] = [T_1, \Vtwo23] = 0 \,,
\label{eq:Tcomm}
\ee
so the commutators of $\Trel$ with $\Vtwo{j}{k}$ become
$[\Trel,\Vtwo12] = [\Ttwo12,\Vtwo12]$ and similarly for $\Vtwo13$ and
$\Vtwo23$.

Because we define $\Trel$ to be independent of $s$, the SRG flow
equation, (\ref{eq:commutator2}), for the three-body Hamiltonian $H_s$
simplifies to
\be
\frac{dV_s}{ds} = 
\frac{d\Vtwo12}{ds} + \frac{d\Vtwo13}{ds} + \frac{d\Vtwo23}{ds}
+ \frac{d\Vthree}{ds} = [[\Trel, V_s], H_s] \,,
\label{eq:dVs}
\ee
with $V_s$ defined by Eq.~(\ref{eq:H}).  The corresponding equations
for each of the two-body potentials (which are completely determined
by their evolved matrix elements in the two-body systems) are
\be
\frac{d\Vtwo12}{ds} = [[\Ttwo12, \Vtwo12], (\Ttwo12 + \Vtwo12)] 
\,, \label{eq:Vtwoa} 
\ee
and similarly for $\Vtwo13$ and $\Vtwo23$.  After expanding
Eq.~(\ref{eq:dVs}) using Eq.~(\ref{eq:H}) and the different
decompositions of $\Trel$, it is straightforward to show that the
derivatives of two-body potentials on the left side cancel precisely
with terms on the right side, leaving
\begin{align}
\frac{d\Vthree}{ds} &=
[[\Ttwo12,\Vtwo12], (T_3 + \Vtwo13 + \Vtwo23 + \Vthree)]
+ [[\Ttwo13,\Vtwo13], (T_2 + \Vtwo12 + \Vtwo23 + \Vthree)]
\nonumber \\
&+ [[\Ttwo23,\Vtwo23], (T_1 + \Vtwo12 + \Vtwo13 + \Vthree)]
+  [[\Trel,\Vthree],H_s] \,.
\label{eq:diffeqp}
\end{align}
The importance of these cancellations is that they eliminate the
``dangerous'' delta functions, which make setting up the integral
equations for the three-body system problematic \cite{Glockle:1983}.
We emphasize that the $s$-dependence of the two-body potentials on the
right side of Eq.~\eqref{eq:diffeqp} is completely determined by
solving the two-body problem in Eq.~\eqref{eq:diffeq}. This is in
contrast to RG methods that run a cutoff on the total energy of the
basis states (as in the Bloch-Horowitz or Lee-Suzuki approaches). Such
methods generate ``multi-valued'' two-body interactions, in the sense
that the RG evolution of two-body operators in $A>2$ systems depends
on the excitation energies of the unlinked spectator particles
\cite{Navratil:1996vm,Luu:2004xc}.
 
Further simplifications of Eq.~(\ref{eq:diffeqp}) follow from
antisymmetrization and applying the Jacobi identity, but this form is
sufficient to make clear that there are no disconnected pieces.  The
problem is thus reduced to the technical implementation of a
momentum-space decomposition analogous to Eq.~(\ref{eq:diffeq}).  A
diagrammatic approach is introduced in
Refs.~\cite{Bogner:2007qb,Jurgenson:2008jp} to handle this
decomposition.  Work is in progress on evolving 3N forces in momentum
space.  It has been verified that this formalism leaves eigenvalues
invariant for three-particle systems described by simple model
Hamiltonians, such as a two-level system of
bosons~\cite{Bogner:2007qb}.

\begin{figure}[t]
 \centering
 \subfloat[][]{%
  \label{fig:h3_srg-a}%
  \includegraphics*[width=3.0in,clip=]{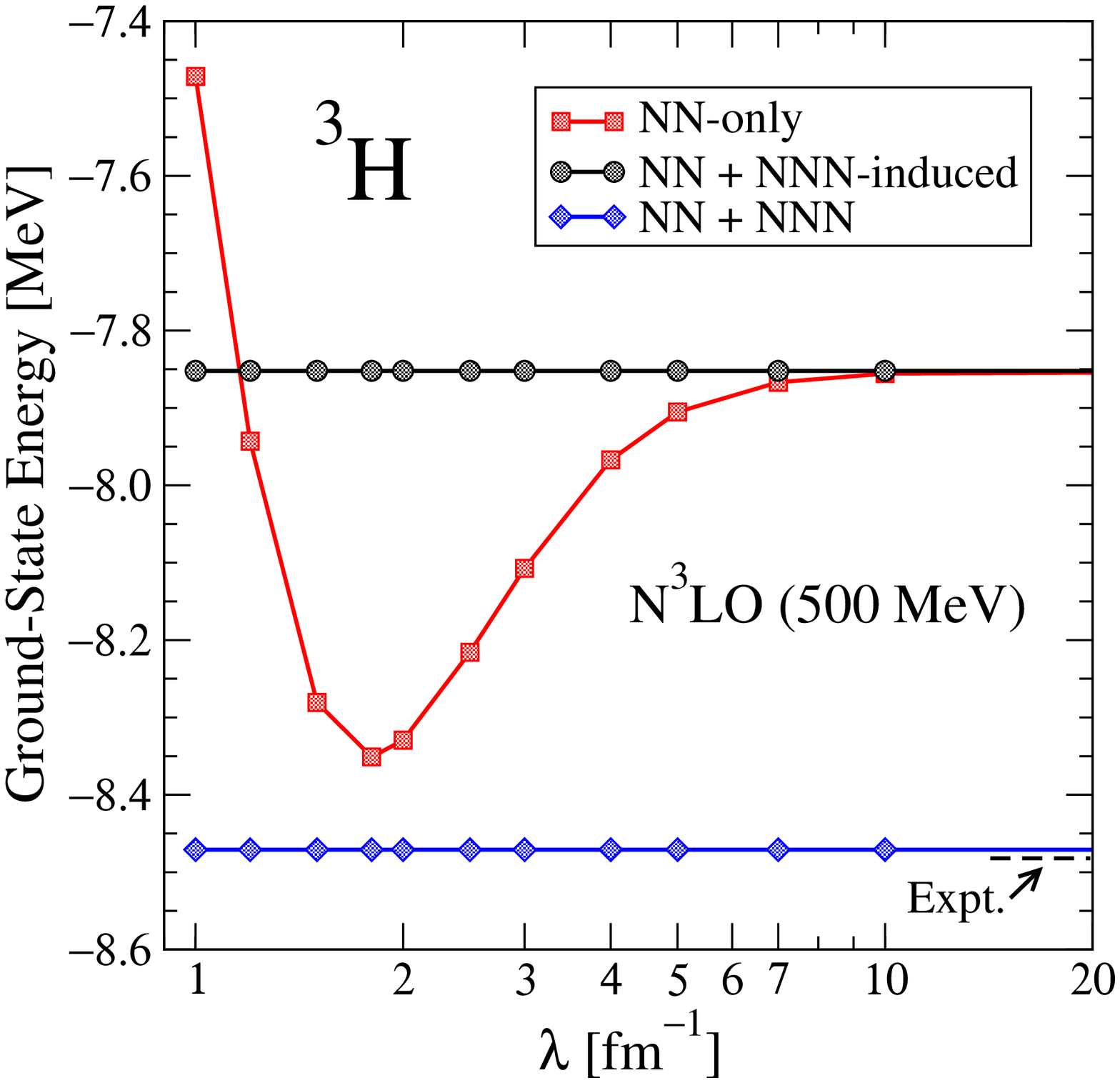}%
 }%
 \hspace*{.4in}%
 \subfloat[][]{%
  \label{fig:h3_srg-b}%
  \includegraphics*[width=3.0in,clip=]{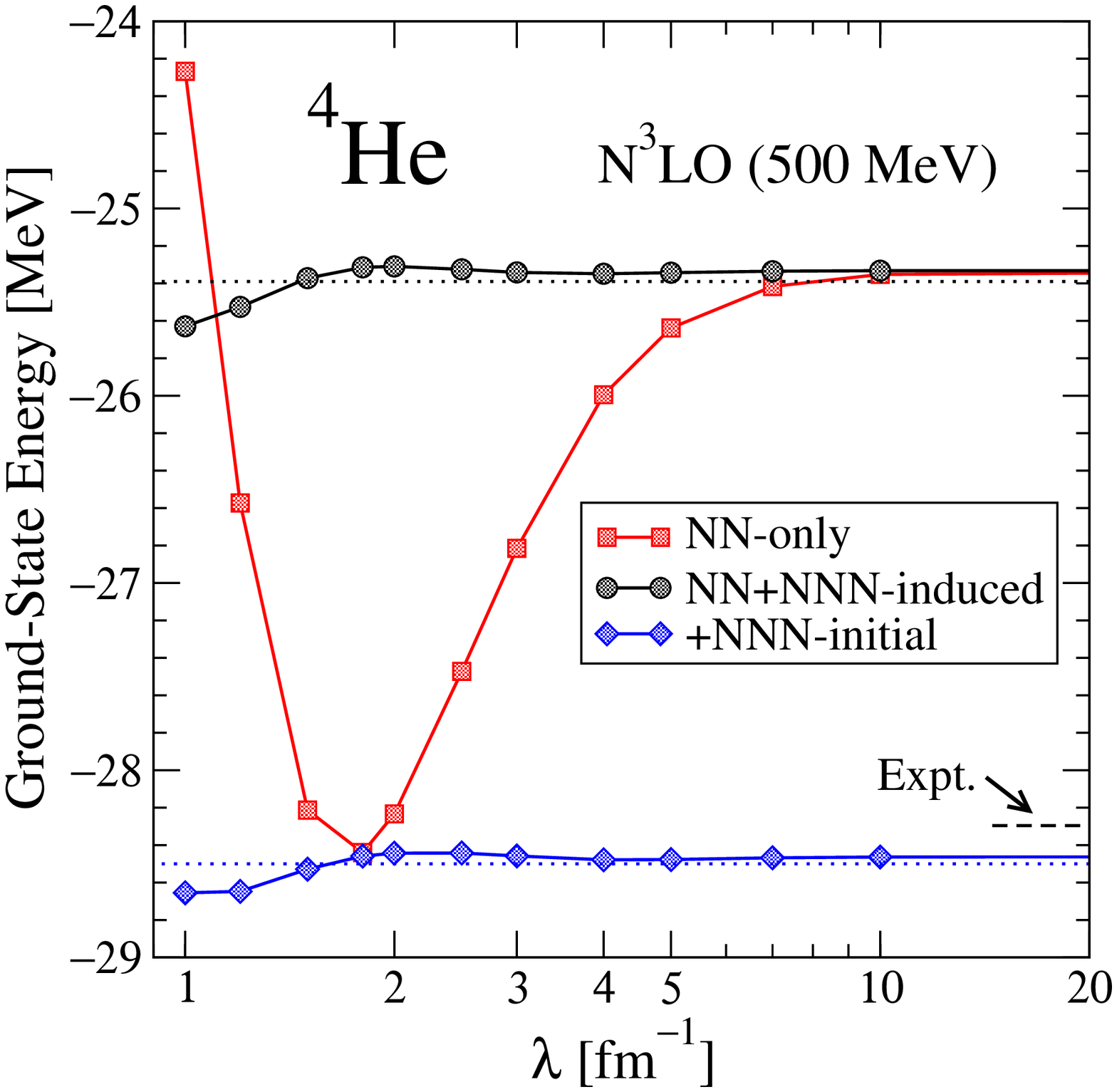}%
 }%
\caption{Ground-state energy of (a)~$^3$H and (b)~$^4$He as a
function of $\lambda$. For details see Ref.~\cite{Jurgenson:2009qs}.}
\label{fig:h3_srg}
\end{figure}

To summarize, because only the Hamiltonian enters the SRG flow
equations, there are no difficulties from having to solve $T$ matrices
(bound state plus scattering wave functions) in all three-body
(including breakup) channels, as required by the analogous three-body
$\vlowk$ evolution equations.  In a momentum basis the presence of
spectator nucleons requires solving separate equations for each set of
$\la V^{(n)}_\lambda \ra$ matrix elements.  But while it is natural to
solve Eq.~\eqref{eq:commutator2} in momentum space, it is an operator
equation so one can use any convenient representation.  A convenient
choice is to evolve in a discrete basis, where spectators are handled
without a decomposition.  This results in coupled first-order
differential equations for the matrix elements of the flowing
Hamiltonian $H_\flow$ where the right side of
Eq.~\eqref{eq:commutator2} is evaluated using simple matrix
multiplications.

Such calculations have been recently performed in the Jacobi
coordinate harmonic-oscillator basis of the No-Core Shell Model
(NCSM)~\cite{Navratil:2009ut}. This is a translationally invariant,
antisymmetric basis for each $A$, with a complete set of states up to
a maximum excitation of \nmax$\hbar\Omega$ above the minimum energy
configuration, where $\Omega$ is the harmonic oscillator
parameter. The 3N evolution used builds directly on
Ref.~\cite{Jurgenson:2008jp}, which presents a one-dimensional
implementation of the approach along with a general analysis of the
evolving many-body hierarchy.

One starts by evolving $H_\lambda$ in the $A=2$ subsystem, which
completely fixes the two-body matrix elements $\la
V_\lambda^{(2)}\ra$.  Next, by evolving $H_\lambda$ in the $A=3$
subsystem, the combined two-plus-three-body matrix elements are
determined.  The three-body matrix elements are isolated by
subtracting the evolved $\la V_\lambda^{(2)}\ra$ elements in the $A=3$
basis~\cite{Jurgenson:2008jp}. Having obtained the separate NN and 3N
matrix elements, they can be applied unchanged to any nucleus.  The
inclusion of 3N forces in the initial Hamiltonian is straightforward.
If applied to $A \geqslant 4$ systems, four-body (and higher) forces
will not be included and so the SRG transformation will be only
approximately unitary.  This allows one to study whether the
decreasing hierarchy of many-body forces is maintained and whether the
induced four-body contribution is unnaturally large.

\begin{figure}[t]
 \centering
 \subfloat[][]{%
  \label{fig:he4_convergence-a}%
  \includegraphics*[width=3.0in,clip=]{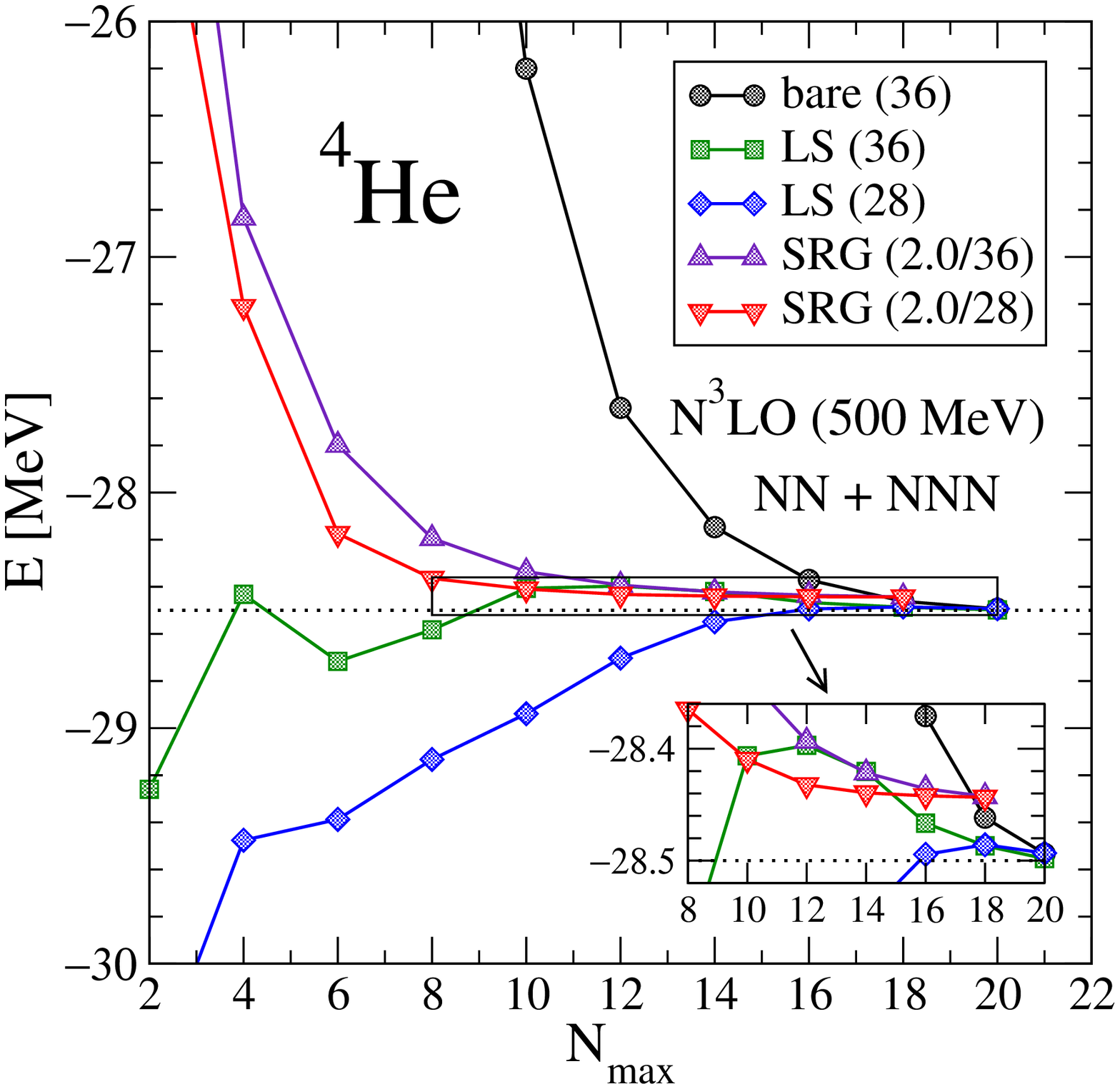}%
 }%
 \hspace*{.4in}%
 \subfloat[][]{%
  \label{fig:he4_convergence-b}%
  \includegraphics*[width=3.0in,clip=]{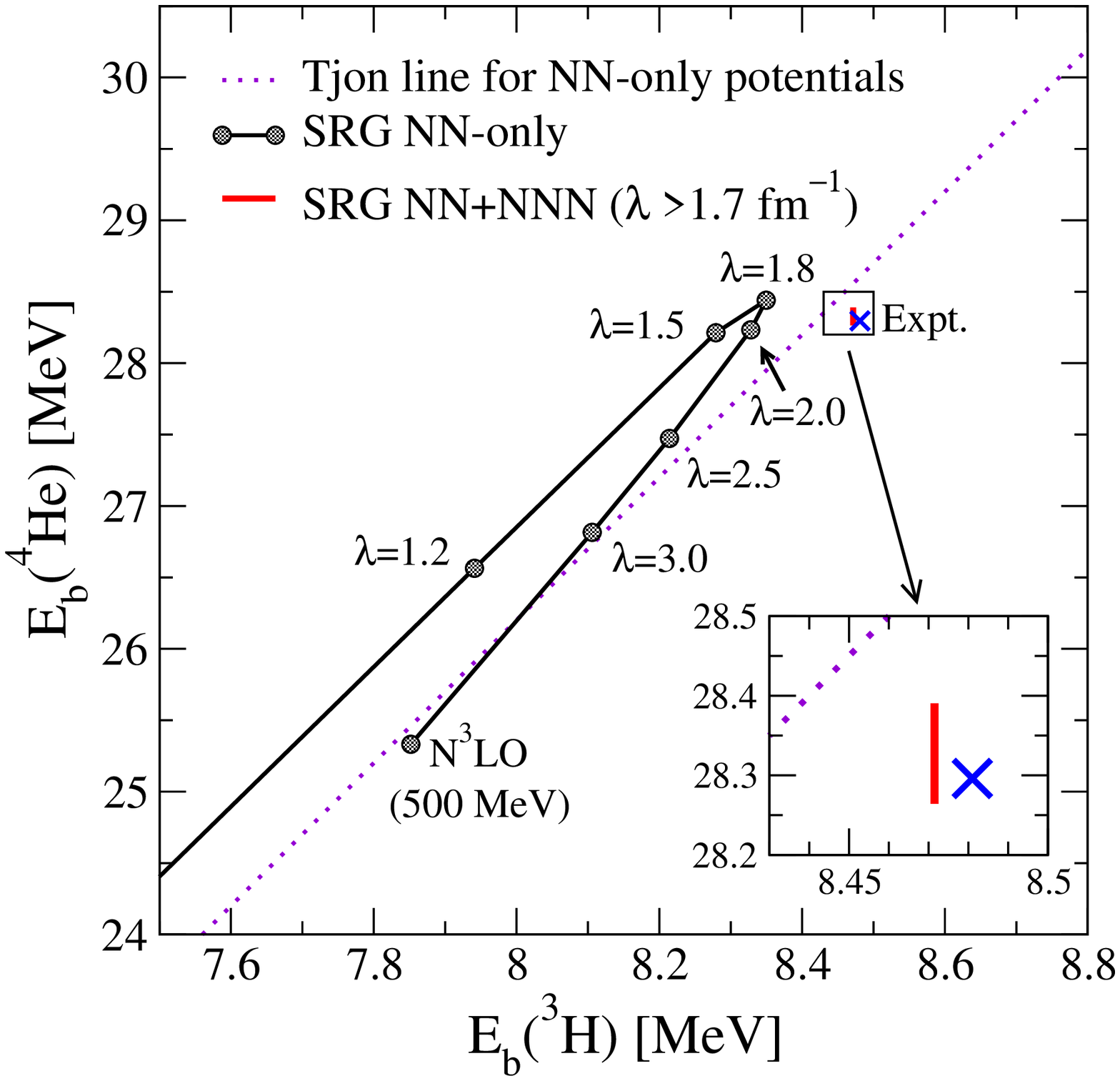}%
 }%
\caption{(a) Ground-state energy of $^4$He as a function of the basis
size \nmax\ based on the N$^3$LO potential of Ref.~\cite{Entem:2003ft}
and N$^2$LO 3N forces from Ref.~\cite{Gazit:2008ma}. Unevolved (bare)
results are compared to those obtained with a Lee-Suzuki (LS) NCSM 
transformation and based on SRG-evolved interactions to $\lambda = 
2.0\,\mbox{fm}^{-1}$ at $\hbar\Omega = 28$ and $36 \mev$. (b)~Binding
energy of $^4$He versus $^3$H. The Tjon line from
phenomenological NN potentials (dotted) is compared with the
trajectory of SRG energies when only NN interactions are kept
(circles). When 3N interactions (initial and induced) are included,
the trajectory lies close to experiment (cross) for $\lambda \geqslant
1.8\,\mbox{fm}^{-1}$ (see inset). For details see
Ref.~\cite{Jurgenson:2009qs}.}
\label{fig:he4_convergence}
\end{figure}

Results from Ref.~\cite{Navratil:2009ut} for the SRG evolution are
shown in Fig.~\ref{fig:h3_srg}. The initial NN interaction is the
$(\lm = 500 \mev$) N$^3$LO potential of Ref.~\cite{Entem:2003ft} while
the initial 3N interaction is the N$^2$LO 3N force (discussed in
Section~\ref{subsec:3nf}) in the local form of
Ref.~\cite{Navratil:2007zn} with $c_D$ and $c_E$ couplings fit to the
average of the $^3$H and $^3$He binding energies and to triton beta
decay according to Ref.~\cite{Gazit:2008ma}. Similar results are
expected starting from other initial potentials because of the
low-momentum universality. NCSM calculations with these initial
interactions and the parameter set in Table~I of
Ref.~\cite{Gazit:2008ma} yield energies of $-8.473(4)\,$MeV for $^3$H
and $-28.50(2)\,$MeV for $^4$He compared with $-8.482\,$MeV and
$-28.296\,$MeV from experiment, respectively. So there is a $20\,$keV
uncertainty in the calculation of $^4$He from incomplete convergence
and a $200\,$keV discrepancy with experiment.  The latter is
consistent with the omission of three- and four-body interactions at
N$^3$LO.  These provide a scale for assessing whether induced
four-body contributions are important compared to other uncertainties.
Figure~\ref{fig:h3_srg} shows the SRG evolution as a function of
$\lambda$ of the $^3$H and $^4$He ground-state energies. When the
induced three-body part is included, the $^3$H ground-state energy is
completely independent of $\lambda$. For $^4$He, the deviation from
$\lambda$ independence due to induced four-body forces is less than
about 50\,keV for $\lambda \geqslant 1.5\,\mbox{fm}^{-1}$.

As seen in Fig.~\subref*{fig:he4_convergence-a}, the NCSM calculations
with SRG-evolved interactions are variational and converge smoothly
and rapidly from above with or without an initial three-body force.
This shows the dramatic improvement in convergence rate compared to
the initial chiral EFT interactions (circles), although the latter are
relatively soft. Thus, once evolved, a much smaller \nmax\ basis is
adequate for a desired accuracy and extrapolating in \nmax\ is also
feasible (in contrast to the Lee-Suzuki NCSM results). The impact of
evolving 3N forces is illustrated in
Fig.~\subref*{fig:he4_convergence-b}, where the binding energy of
$^4$He is plotted against the binding energy of $^3$H. The
experimental values define a point in this plane. The SRG NN-only
results follow a trajectory that is analogous to the Tjon line (see
Fig.~\ref{Tjon}). In contrast, the short SRG trajectory including 3N
interactions highlights the small variations from omitted short-range
three- and four-body forces.\footnote{Note that the trajectory for NN
plus 3N-induced results would be a similarly small line at the
N$^3$LO NN-only point.}

These results demonstrate that the SRG is a practical method to evolve
3N (and higher-body) forces in a harmonic-oscillator basis.
Calculations of $A \leqslant 4$ nuclei including 3N forces show the
same favorable convergence properties observed with low-momentum NN
interactions, with a net induced four-body contribution in $A=4$ that
is smaller than the truncation errors of the initial chiral
interactions.  However, because of the stronger density dependence of
four-nucleon forces, it will be important to monitor the size of the
induced four-body contributions for heavier nuclei and nuclear matter.
The SRG evolution is an alternative to the Lee-Suzuki NCSM
transformation and the SRG-evolved harmonic-oscillator
matrix elements can also be used with other many-body methods. The
success of the block-diagonal SRG (see Section~\ref{subsec:vlowksrg})
in evolving a $\vlowk$ two-body interaction using flow equations
motivates an extension to three- and higher-body forces (and
operators). Work on this is in progress.

\subsection{In-medium SRG}
\label{subsec:normal}

The SRG methods described so far carry out the evolution to low
momentum at zero density in free space. A convenient feature of the
free-space evolution is that it does not have to be repeated for each
different nucleus or nuclear matter density; the evolved NN
interaction is completely determined by solving the two-body SRG
equation in free space, the evolved 3N interaction is completely
determined by solving the three-body SRG equation in free space, and
so on.

An interesting alternative is to perform the SRG evolution directly in
the $A$-body system of interest. In contrast to the free-space SRG,
the in-medium evolution must be repeated for each nucleus or
density. As with all RG methods, evolving to low momentum generates
many-body interactions that must be truncated for practical
calculations. However, unlike the free-space evolution, the in-medium
SRG has the appealing feature that one can approximately evolve
$3,...,A$-body operators using only two-body machinery. The key to
this simplification is the use of normal-ordering with respect to a
finite-density reference state. That is, starting from the
second-quantized Hamiltonian with two- and three-body interactions,
\be
H = \sum_{12} T_{12} \ad_1 a_2 + \frac{1}{(2!)^2} \sum_{1234}
\, \langle12|V|34\rangle \ad_1\ad_2a_4a_3
+ \frac{1}{(3!)^2} \, \sum_{123456} \langle123|V^{(3)}|456\rangle
\ad_1\ad_2\ad_3a_6a_5a_4 \,,
\label{eq:Ham}
\ee
all operators are normal-ordered with respect to a finite-density 
Fermi vacuum $|\Phi\rangle$ (for example, the Hartree-Fock ground
state or the non-interacting Fermi sea in nuclear matter), as 
opposed to the zero-particle vacuum. Wick's theorem can then be
used to {\it exactly} write $H$ as
\be
H = E_0 + \sum_{12} f_{12} \{\ad_1a_2\} 
+ \frac{1}{(2!)^2} \sum_{1234} \, \langle12|\Gamma|34\rangle 
\{\ad_1\ad_2a_4a_3\}
+ \frac{1}{(3!)^2} \sum_{123456} \, \langle123|\Gamma^{(3)}|456\rangle
\{\ad_1\ad_2\ad_3a_6a_5a_4\} \,,
\label{eq:NorderedHam}
\ee
where the zero-, one-, and two-body normal-ordered terms are given by
\begin{align}
E_0 &= \langle \Phi | H | \Phi \rangle = \sum_1 T_{11} n_1 
+ \frac{1}{2} \sum_{12} \, \langle12|V|12\rangle n_1 n_2 
+ \frac{1}{3!} \sum_{123} \, \langle123|V^{(3)}|123\rangle n_1 n_2 n_3
\,, \label{eq:NorderedCoeff1} \\
f_{12} &= T_{12} + \sum_i \langle1i|V|2i\rangle n_i 
+ \frac{1}{2} \, \sum_{ij} \langle1ij|W|2ij\rangle n_i n_j \,,
\label{eq:NorderedCoeff2} \\
\langle12|\Gamma|34\rangle &= \langle12|V|34\rangle  
+ \sum_i \langle12i|V^{(3)}|34i\rangle n_i \,,
\label{eq:NorderedCoeff3}
\end{align}
where $n_i = \theta(\varepsilon_{\rm F} - \varepsilon_i)$ denotes the
sharp occupation numbers in the reference state, with Fermi level or
Fermi energy $\varepsilon_{\rm F}$. By construction, the
normal-ordered strings of creation and annihilation operators obey
$\langle\Phi|\{a^{\dagger}_1\cdots a_n\}|\Phi\rangle = 0$.  It is
evident from Eqs.~\eqref{eq:NorderedCoeff1}--\eqref{eq:NorderedCoeff3}
that the coefficients of the normal-ordered zero-, one-, and two-body
terms include contributions from the three-body interaction $V^{(3)}$
through sums over the occupied single-particle states in the reference
state $|\Phi\rangle$. Therefore, truncating the in-medium SRG
equations to two-body {\it normal-ordered} operators will
(approximately) evolve induced three- and higher-body interactions
through the density-dependent coefficients of the zero-, one-, and
two-body operators in Eq.~(\ref{eq:NorderedHam}).

The in-medium SRG flow equations at the normal-ordered two-body level
are obtained by evaluating $dH/ds = [\eta,H]$ with the normal-ordered
Hamiltonian $H = E_0 + f + \Gamma$ and the SRG generator $\eta =
\eta^{1b} + \eta^{2b}$ (with one- and two-body terms) and neglecting
three- and higher-body normal-ordered terms. For infinite matter,
a natural generator choice is $\eta = [f,\Gamma]$ in analogy
with the free-space SRG. In this case, the explicit form of the SRG
equations simplifies because $\eta^{1b}=0$ and $f_{ij} = f_i \,
\delta_{ij}$. This leads to
\begin{align}
\frac{dE_0}{ds} &= 
\frac{1}{2} \sum_{1234} \, (f_{12} - f_{34})
|\Gamma_{1234}|^2 \, n_1 n_2 \bar{n}_3 \bar{n}_4
\,, \label{eq:NorderedSRG0} \\[1mm]
\frac{df_1}{ds} &= \sum_{234} \, (f_{41} - f_{23})
|\Gamma_{4123}|^2 (\bar{n}_2\bar{n}_3n_4 + n_2n_3\bar{n}_4)
\,, \label{eq:NorderedSRG1} \\[1mm]
\frac{d\Gamma_{1234}}{ds} &= -(f_{12} - f_{34})^2 \,
\Gamma_{1234} + \frac{1}{2} \sum_{ab} \,
(f_{12} + f_{34} - 2f_{ab})
\Gamma_{12ab} \Gamma_{ab34} (1 - n_a - n_b) 
+ \sum_{ab} \, (n_a - n_b) \nonumber \\
&\times \Bigl\{ \Gamma_{a1b3} \Gamma_{b2a4}
\bigl[(f_{a1}-f_{b3}) - (f_{b2} - f_{a4})\bigr]
- \Gamma_{a2b3}\Gamma_{b1a4}
\bigl[(f_{a2}-f_{b3}) - (f_{b1} - f_{a4})\bigr] \Bigr\} \,,
\label{eq:NorderedSRG2}
\end{align}
where the single-particle indices refer to momentum states and include
spin and isospin labels. While the in-medium SRG equations are of second
order in the interactions, the flow equations build up
non-perturbative physics via the successive interference between the
particle-particle and the two particle-hole channels in the SRG
equation for $\Gamma$, Eq.~(\ref{eq:NorderedSRG2}), and between the
two-particle--one-hole and two-hole--one-particle channels for $f$,
Eq.~(\ref{eq:NorderedSRG1}). In terms of diagrams, one can imagine
iterating the SRG equations in increments of $\delta s$. At each
additional increment $\delta s$, the interactions from the previous
step are inserted back into the right side of the SRG equations.
Iterating this procedure, one sees that the SRG accumulates
complicated particle-particle and particle-hole correlations to all orders.

\begin{figure}[t]
 \centering
 \subfloat[][]{%
  \label{fig:SNMinmedium-a}%
  \includegraphics*[width=3.0in,clip=]{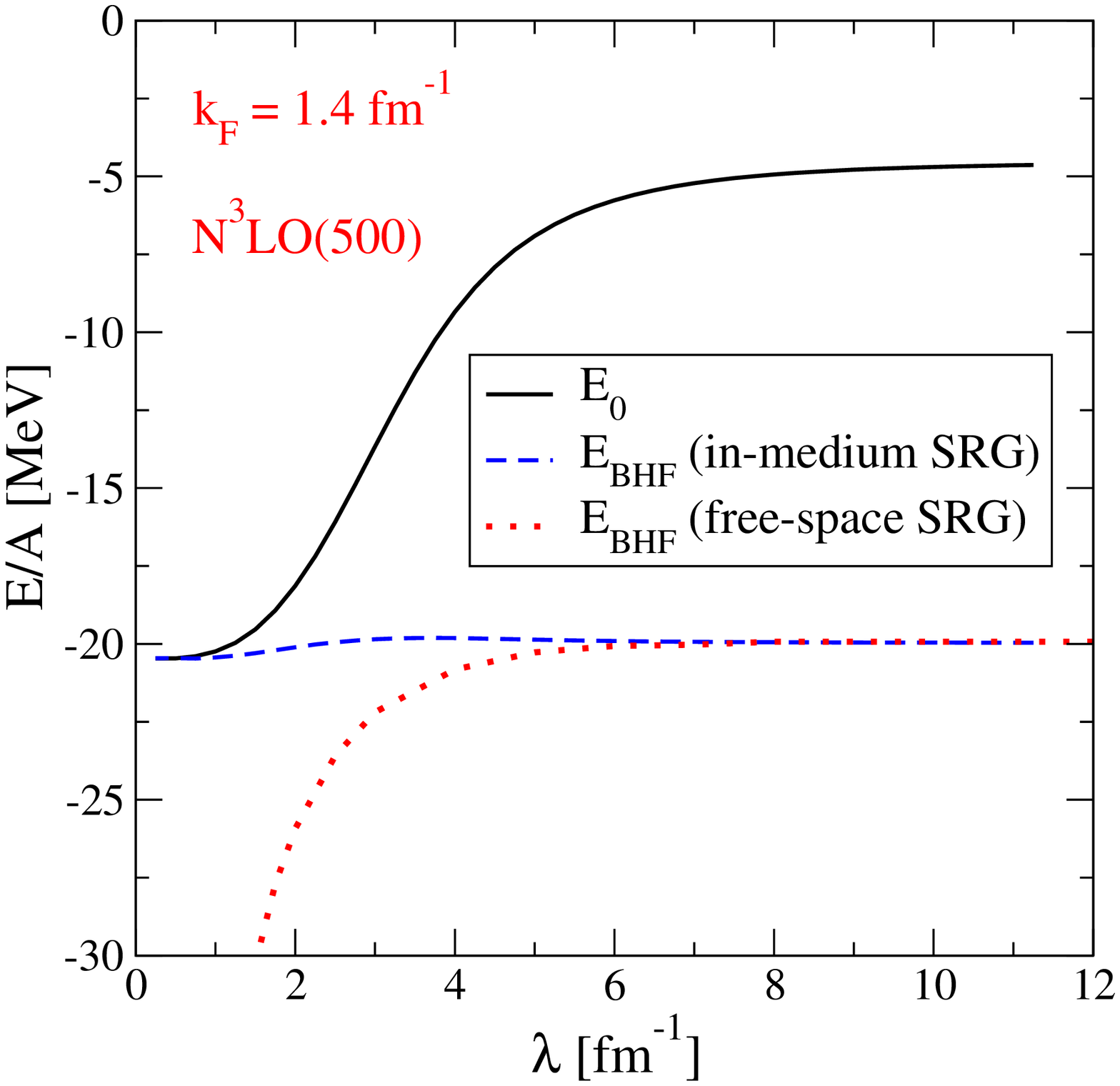}%
 }%
 \hspace*{.4in}%
 \subfloat[][]{%
  \label{fig:SNMinmedium-b}%
  \includegraphics*[width=3.0in,clip=]{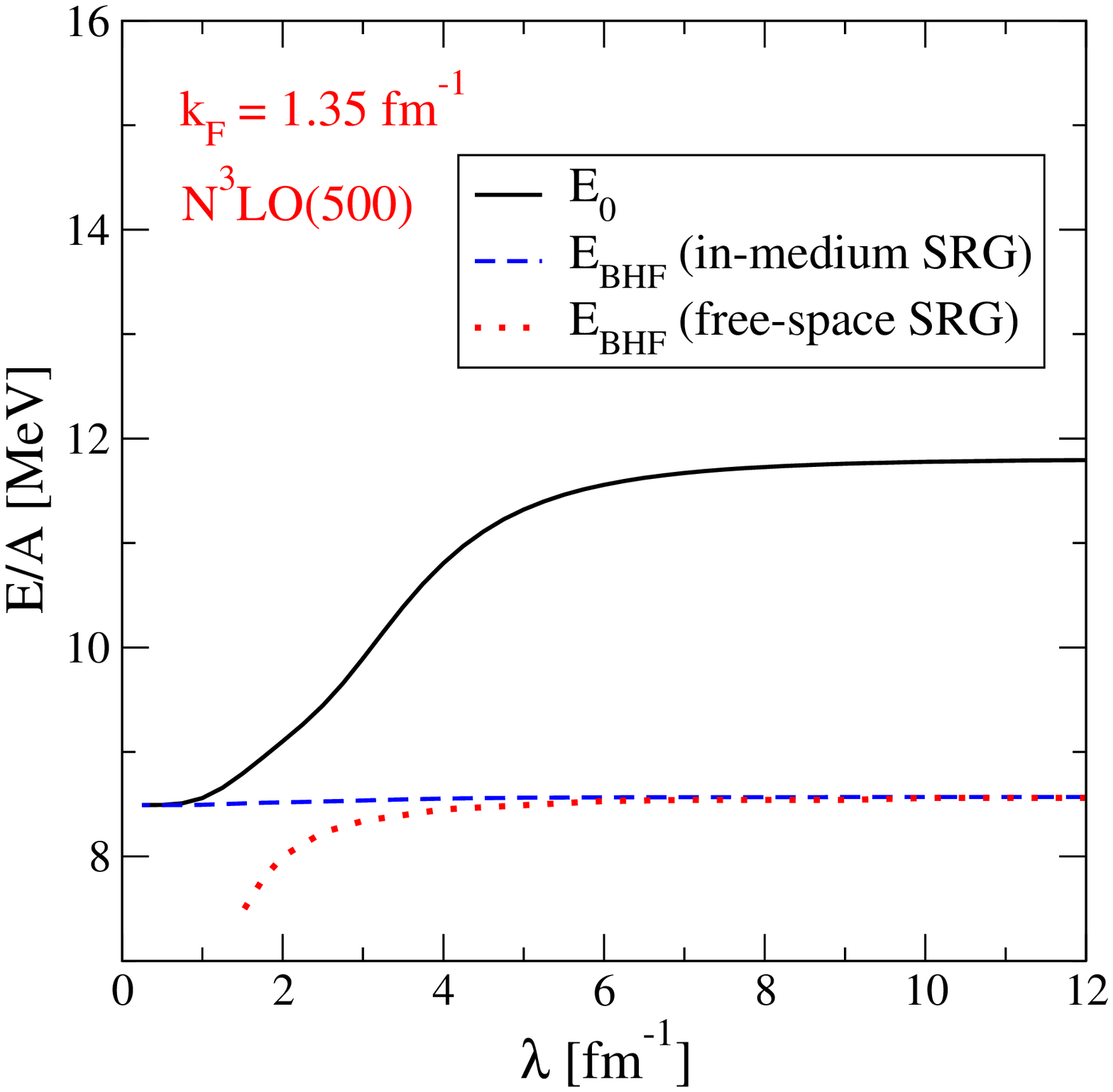}%
 }%
\caption{In-medium SRG evolution of the energy per particle of
(a)~symmetric nuclear matter at a Fermi momentum $k_F=1.4\fmi$ and
(b)~neutron matter at $k_F = 1.35\fmi$~\cite{inmediumNM}. The solid
line denotes the $E_0$ flow of the Hartree-Fock energy. The dashed
line is the energy calculated in the ladder
or Brueckner-Hartree-Fock approximation (BHF) using interactions
evolved with the in-medium SRG at the two-body level, and the
dotted line is the same many-body calculation based on NN-only SRG
interactions evolved in free space. The initial NN interaction is
the N$^3$LO potential of Ref.~\cite{Entem:2003ft}.}
\label{fig:SNMinmedium}
\end{figure}

\begin{figure}[t]
 \centering
 \subfloat[][]{%
  \label{fig:inmediumHe4-a}%
  \includegraphics*[width=3.0in,clip=]{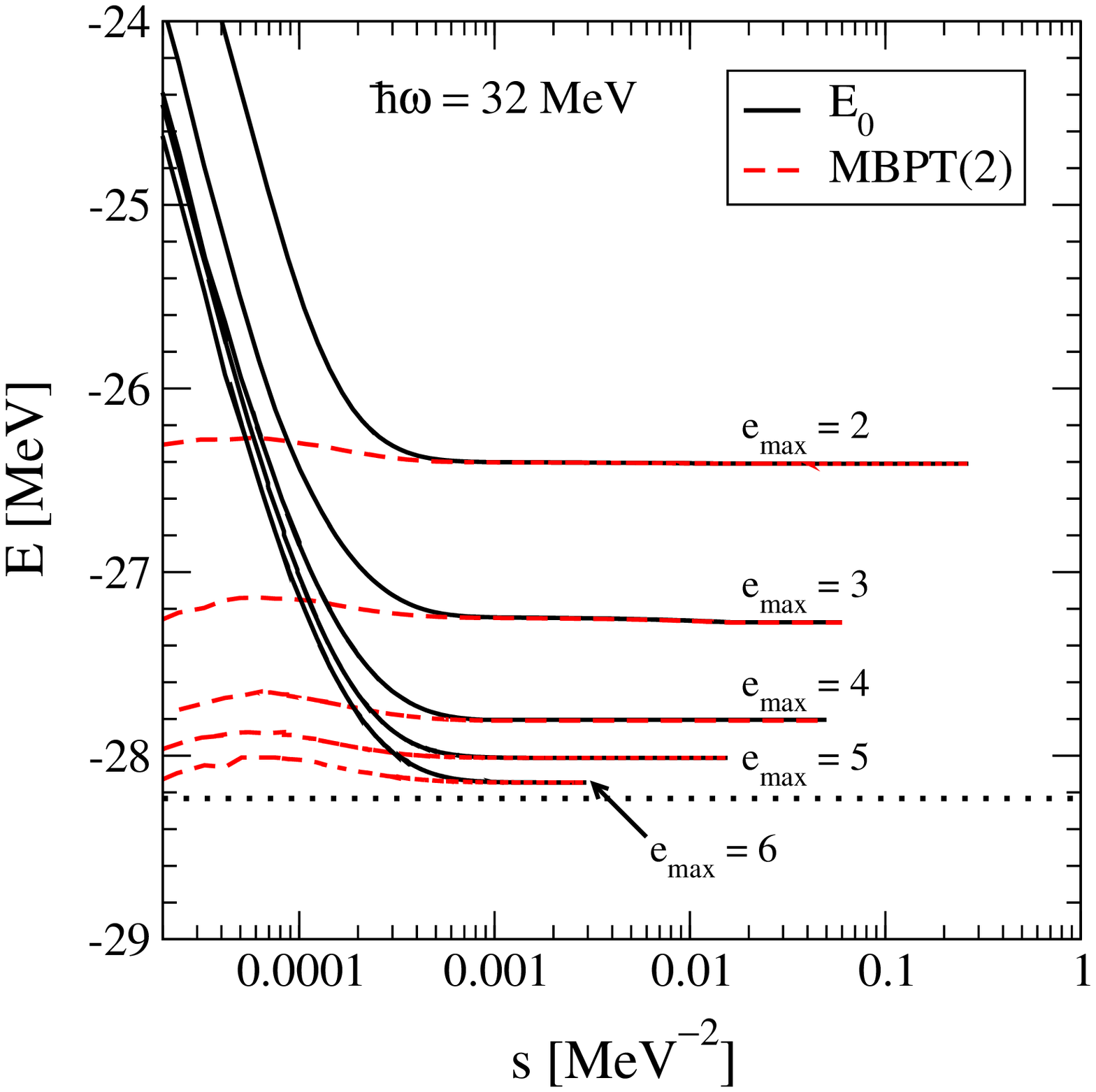}%
 }%
 \hspace*{.4in}%
 \subfloat[][]{%
  \label{fig:inmediumHe4-b}%
  \includegraphics*[width=3.0in,clip=]{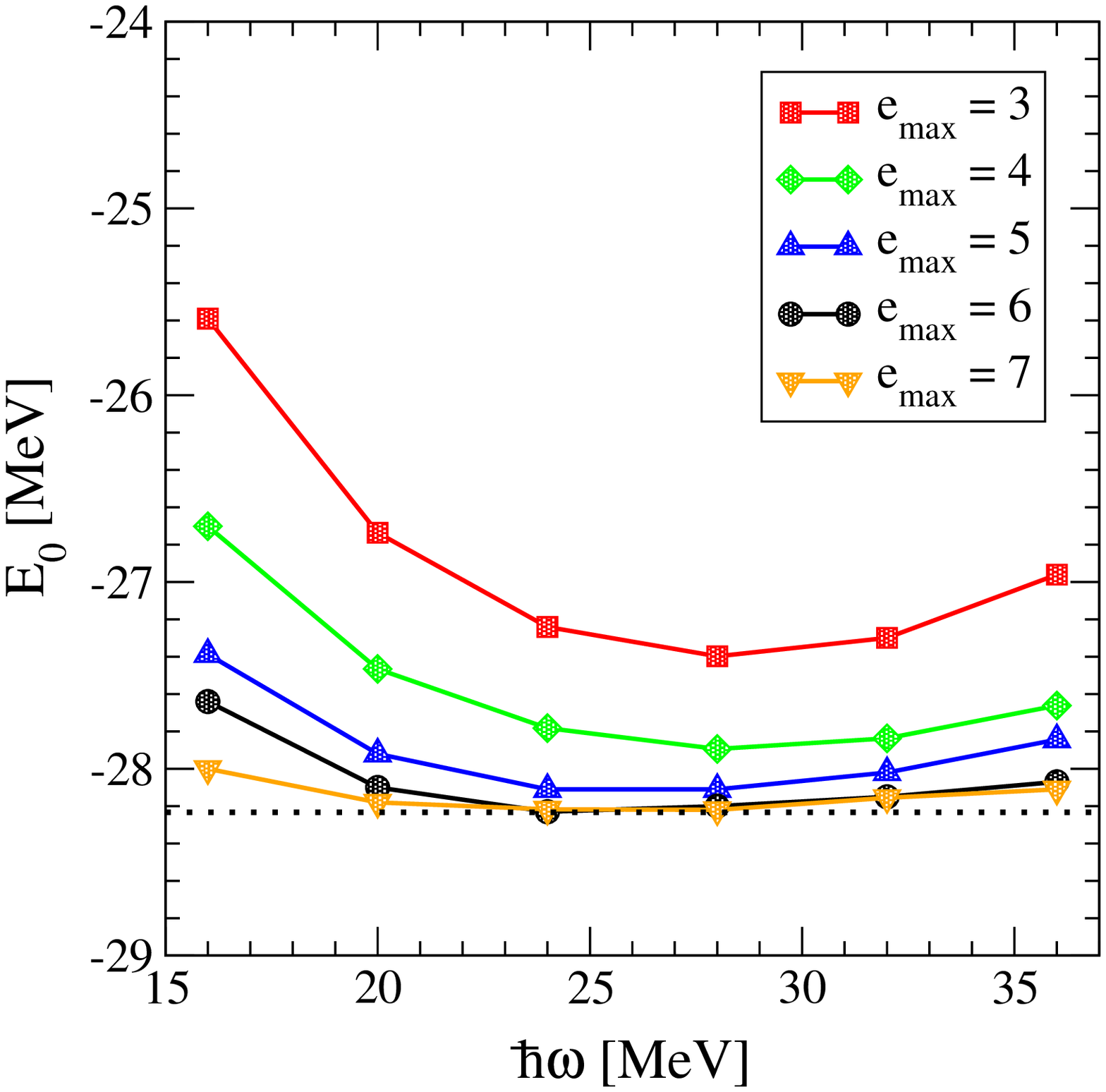}%
 }%
\caption{Ground-state energy of $^4$He based on the in-medium SRG
starting from a free-space SRG potential ($\lambda=2.0\fmi$ from the
N$^3$LO potential of Ref.~\cite{Entem:2003ft})~\cite{inmediumSM}.
(a)~$E_0$ flow (solid line) and including second-order
many-body perturbation theory (MBPT) contributions (dashed line)
as a function of flow parameter $s$. (b)~Convergence behavior as
a function of harmonic oscillator parameter $\hbar \omega$ with
increasing single-particle $e_{\rm max} \equiv
\max(2n+l)$. The horizontal dotted line denotes the ground-state
energy obtained from an exact diagonalization~\cite{Jurgenson:2009qs}.} 
\label{fig:inmediumHe4}
\end{figure}

With this choice of generator $\eta = [f,\Gamma]$, the Hamiltonian is
driven towards the diagonal. This means that Hartree-Fock becomes
increasingly dominant with the off-diagonal $\Gamma$ matrix elements
being driven to zero. As with the free-space SRG, it is convenient for
momentum-space evolution to switch to the flow variable $\lambda
\equiv s^{-1/4}$, which is a measure of the resulting band-diagonal
width of $\Gamma$. In the limit $\lambda \rightarrow 0$,
Hartree-Fock becomes exact for the evolved Hamiltonian; the zero-body
term, $E_0$, approaches the interacting ground-state energy, $f$
approaches fully dressed single-particle energies, and the remaining
diagonal matrix elements of $\Gamma$ approach a generalization of the
quasiparticle interaction in Landau's theory of Fermi
liquids~\cite{Kehrein:2006}. An approximate\footnote{The particle-hole
contributions to the SRG flow equation for $\Gamma$, given by the
last sum and the second line of Eq.~(\ref{eq:NorderedSRG2}), are
neglected in the infinite matter calculations. No such approximation
is made for the $^4$He results presented in
Fig.~\ref{fig:inmediumHe4}.} solution of the $E_0$ flow equation for
symmetric nuclear matter and neutron matter as a function of $\lambda$
is shown in Fig.~\ref{fig:SNMinmedium} for two different Fermi momenta
$k_F$ (corresponding to different densities)~\cite{inmediumNM}. As
expected, the in-medium SRG drives the Hamiltonian to a form where
Hartree-Fock becomes exact in the limit $\lambda \rightarrow 0$. In
contrast to the ladder approximation based on NN-only SRG interactions
evolved in free space, the same many-body calculation using
interactions evolved with the in-medium SRG at the two-body level
gives energies that are approximately independent of $\lambda$. This
indicates that truncations based on normal-ordering efficiently
include the dominant induced many-body interactions via the
density-dependent zero-, one-, and two-body normal-ordered terms.

In a similar manner, the in-medium SRG can be used as an ab initio
method for finite nuclei. Figure~\ref{fig:inmediumHe4} illustrates
this with calculations of the ground-state energy of
$^4$He~\cite{inmediumSM}. As the flow parameter $s$ increases,
Fig.~\subref*{fig:inmediumHe4-a} shows that the $E_0$ flow and
including second-order (in $\Gamma$) many-body perturbation theory
contributions approach each other, as was the case for the infinite
matter results in Fig.~\ref{fig:SNMinmedium}.  In addition, the
convergence behavior with increasing harmonic-oscillator spaces in
Fig.~\subref*{fig:inmediumHe4-b} is very promising. Based on these
preliminary calculations, the in-medium SRG truncated at the
normal-ordered two-body level appears to give accuracies comparable to
coupled-cluster calculations truncated at the singles and doubles
(CCSD) level. Future applications to $^{16}$O and $^{40}$Ca will help
clarify if this similarity to CCSD is systematic, although such a
connection was suggested by White in his work on similar methods in
quantum chemistry~\cite{white:7472}. Finally, we note that the
in-medium SRG is a promising method for non-perturbative calculations
of valence shell-model effective interactions and operators.

\subsection{Effective operators}
\label{subsec:effops}

As discussed in Section~\ref{subsec:intops}, \emph{all} operators in a
low-energy effective theory will evolve under a change of resolution.
Given an initial operator that is consistent with the initial
Hamiltonian (and which will also have many-body components), it is
important to consistently evolve the operators along with the
Hamiltonian. When this evolution is via unitary transformations, the
operator evolution is well defined and in principle straightforward to
implement.

For $\vlowk$ interactions, the use of Lee-Suzuki transformations
provides a convenient formalism to evolve consistent two-body
operators~\cite{Bogner:2007jb}. In the notation of
Ref.~\cite{Stetcu:2004wh,Navratil:1999pw}, the evolution of an
operator~$O$ from a momentum cutoff $\lm_0$ to lower resolution
$\lm$ is given by
\be
O(\Lambda)= \frac{1}{\sqrt{P+\omega^{\dagger}\omega}}
(P+\omega^{\dagger}) \, O \, (\Lambda_0) \,
(P+\omega) \frac{1}{\sqrt{P+\omega^{\dagger}\omega}} \,,
\label{eq:effoperator}
\ee
where the operator $\omega=Q \omega P$ generates the Lee-Suzuki
transformation, the projection operator $P$ projects onto relative
momenta $k < \Lambda$, and $Q = 1-P$ projects onto $\Lambda < k <
\Lambda_0$.  While this approach is straightforward to implement at
the two-body level (and has been tested), the generalization to
higher-body components is not simple.

For the SRG approach, the evolution with $\flow$ of any other operator
$O$ is given by the same unitary transformation that evolves the
Hamiltonian, $O_\flow = U(\flow) O U^\dagger(\flow)$, which means that
$O_\flow$ evolves as
\beqn
\frac{dO_\flow}{d\flow}
= [\eta(\flow),O_\flow] = [ [\Hzero,V_\flow], O_\flow] \,.
\label{eq:opflow}
\eeqn
Just as with the Hamiltonian $H_\flow$, this evolution will
induce many-body operators even if the initial operator is purely
two-body. If we restrict ourselves to two-body operators in relative
momentum space, one has
\beqn
\frac{dO_\flow(k,k')}{d\flow} = \frac{2}{\pi} \int_0^\infty\! q^2 \, dq
\, \Bigl[ (k^2 - q^2) \, V_\flow(k,q) \, O_\flow(q,k')
+ (k'{}^2 - q^2) \, O_\flow(k,q)\, V_\flow(q,k') \Bigr] \,.
\label{eq:Odiffeq}
\eeqn
To evolve a particular $O_\flow$ simultaneously with $V_\flow$, 
we can simply include
the discretized version of Eq.~(\ref{eq:Odiffeq}) as additional
coupled first-order differential equations.

An alternative, more direct, approach
is to construct the unitary transformation at each $s$ explicitly, and then
use it to transform operators.
Let $| \psi_\alpha(s) \rangle$ be an eigenstate of $H_\flow$
with eigenvalue $E_\alpha$ (which is independent of $\flow$).
Then $U(s)$ is given by
\beqn
U(s) = \sum_\alpha | \psi_\alpha(s) \rangle \langle \psi_\alpha(0) | \,.
\label{eq:Ualpha}
\eeqn
In a discretized partial-wave momentum-space basis with momenta
$\{k_i\}$, we can solve for the eigenvectors
of $H = H_{s=0}$ and $H_s$, then construct the matrix elements of
$U(s)$ [which we denote $U_s(k_i,k_j)]$
by summing over the product of momentum-space wave functions:
\beqn
U_s(k_i,k_j) = \sum_\alpha \, \langle k_i | \psi_\alpha(s) \rangle  
\langle \psi_\alpha(0) | k_j \rangle \,.
\eeqn
In practice this is an efficient way to construct the
unitary transformation and subsequently to evolve any operator 
in a few-body space. For example, in the harmonic-oscillator 
basis used for evolving 3N interactions (see 
Section~\ref{subsec:evolution}), one can simply apply 
Eq.~\eqref{eq:Ualpha} in this basis. Proof-of-principle tests 
for operators in a one-dimensional model show that this works
as expected~\cite{Anderson:2010aa}. 

As an example in the two-body system, we consider the momentum
distribution in the deuteron~\cite{Bogner:2007jb}.  The momentum
distribution at relative momentum ${\bf q}$ is the expectation value
of $\adaggera$ (summed over spin substates $M_S$) and is proportional
to the sum of the squares of the normalized S- and D-state parts of
the deuteron wave function, $u(q)^2 + w(q)^2$.  It is not an
observable (see, for example, Ref.~\cite{Furnstahl:2001xq}) but is
useful to illustrate some points.  Because the SRG proceeds via
unitary transformations, by construction no information is lost in the
evolution.  But even more, we can show explicitly that by evolving to
a low-momentum interaction, we decouple the low- and high-momentum
physics in a low-energy state.

\begin{figure}
 \centering
 \subfloat[][]{%
  \label{fig:deutmd_accum-a}%
  \includegraphics*[width=3.5in,clip=]{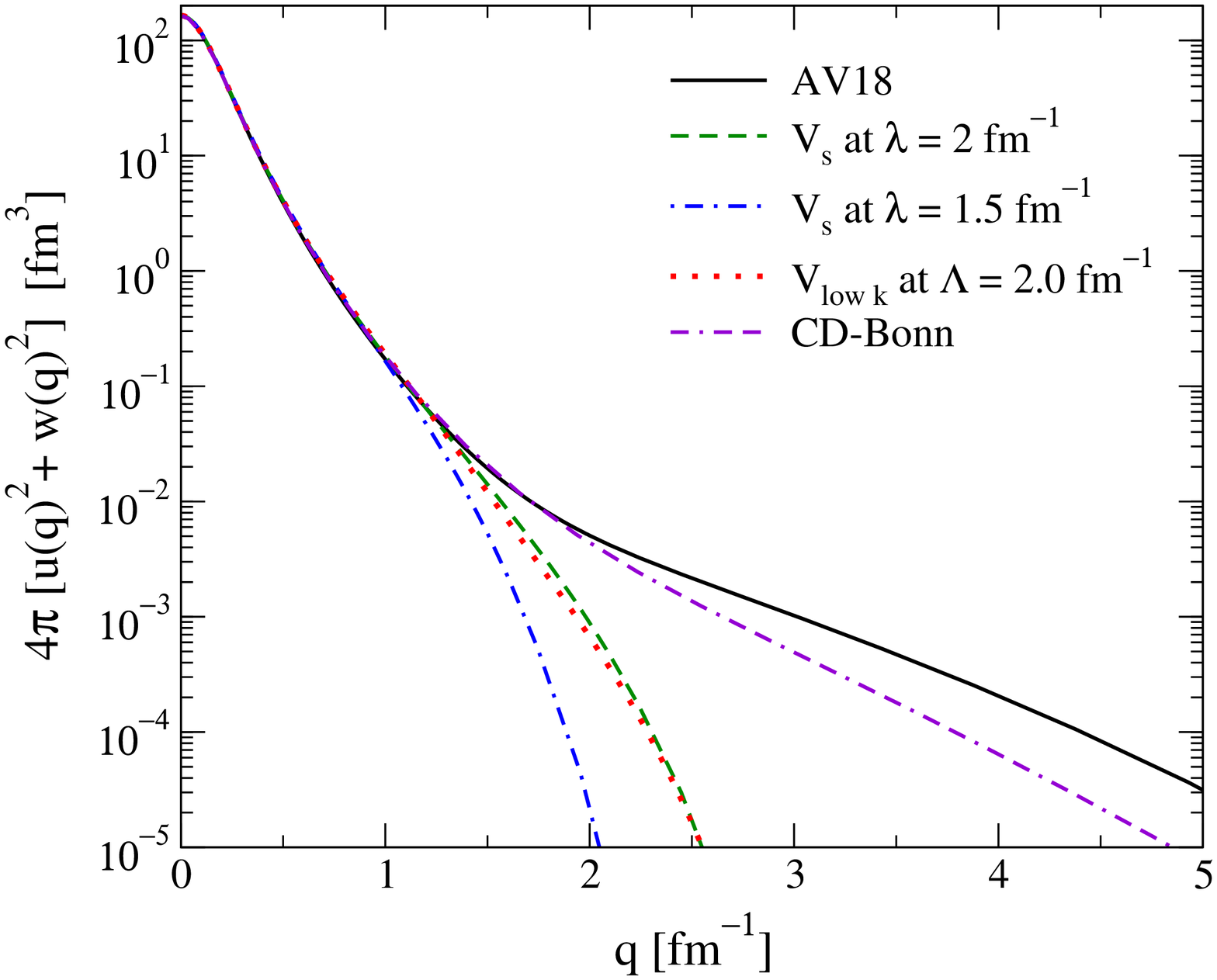}%
 }%
 \hspace*{.2in}%
 \subfloat[][]{%
  \label{fig:deutmd_accum-b}%
  \includegraphics*[width=3.5in,clip]{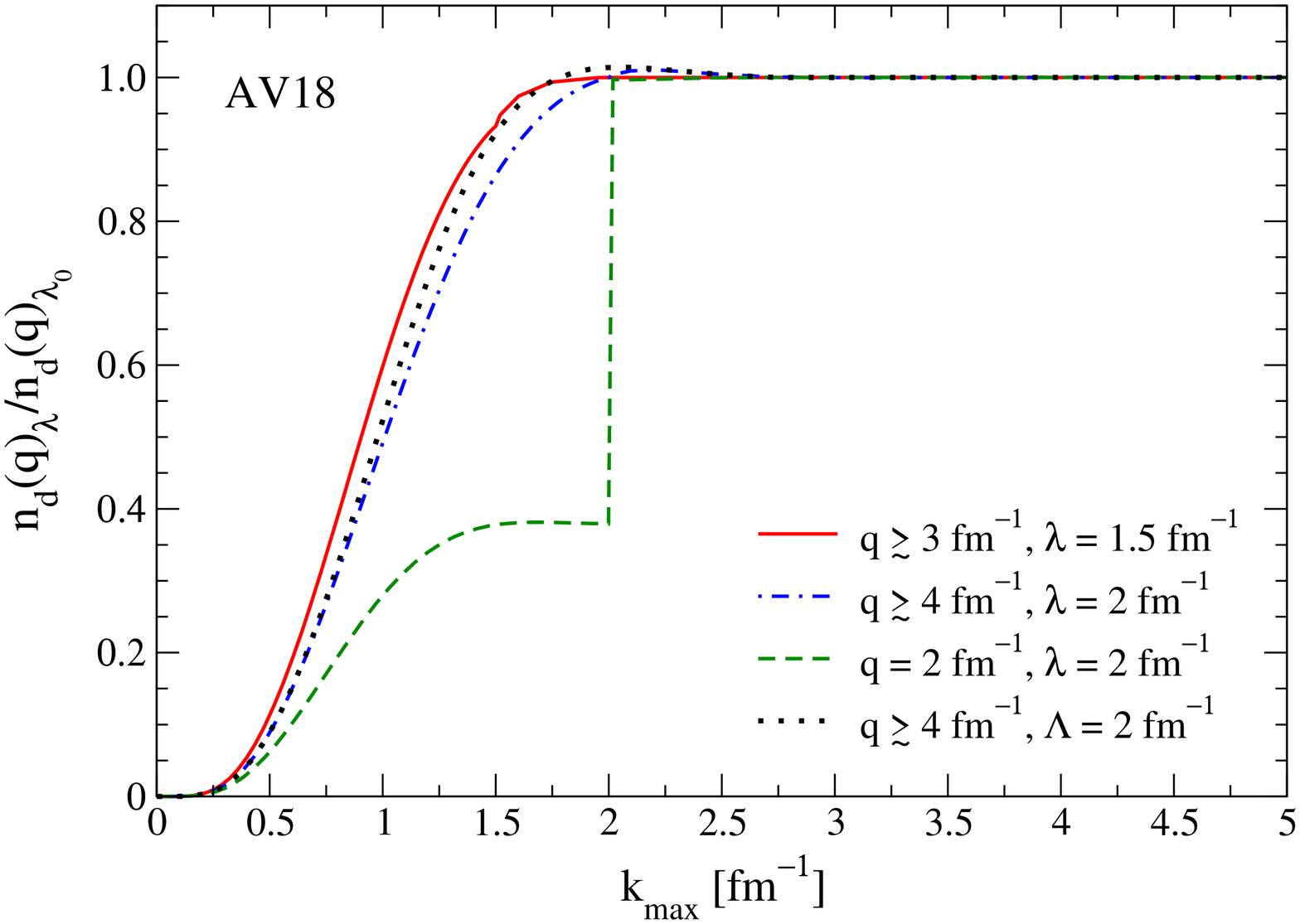}%
 }%
\caption{(a) Deuteron momentum distribution $\langle a^\dagger_{\qvec} 
a^{\protect\phantom\dagger}_{\qvec} \rangle_d = 4\pi [u(q)^2 + w(q)^2]$
using the Argonne $v_{18}$~\cite{Wiringa:1994wb}, 
CD-Bonn~\cite{Machleidt:2000ge} and SRG potentials evolved
from Argonne $v_{18}$ to $\lambda = 1.5\fmi$ and $2\fmi$ (but not
evolving the operator), and a smooth-cutoff $\vlowk$ interaction
with $\Lambda = 2\fmi$ and $n_{\rm exp}=2$.
(b) Ratio of the deuteron momentum distribution at various momenta
$q$ evolved from the Argonne $v_{18}$ potential via the SRG to 
$\lambda=1.5\fmi$ and $2\fmi$, to the corresponding initial momentum 
distribution, as a function of the maximum momentum $\kmax$ in the
deuteron wave functions in the numerator. Note that the unevolved Argonne 
$v_{18}$ result is simply a step function at $q$. For comparison, we also
show the result for a smooth-cutoff $\vlowk$ interaction with $\Lambda 
= 2\fmi$ and $n_{\rm exp}=2$. For details see Ref.~\cite{Bogner:2007jb}.}
\end{figure}

In Fig.~\subref*{fig:deutmd_accum-a}, we plot deuteron matrix elements
of $\adaggera$ for the Argonne $v_{18}$~\cite{Wiringa:1994wb} and
CD-Bonn~\cite{Machleidt:2000ge} potentials, as well as for two SRG and
a smooth-cutoff $\vlowk$ interaction evolved from Argonne $v_{18}$.
We emphasize again that matrix elements of this operator are not
measurable, so one should not ask which of these is the ``correct''
momentum distribution in the deuteron; it is a potential- and
scale-dependent quantity.  It is evident that the $\vsrg$
distributions have substantial momentum components only for $k$ below
$\lambda$, and that the $\vlowk$ distribution is very similar to the
corresponding $\vsrg$ distribution.  
Nevertheless, if we use the SRG- or RG-evolved operator
with these deuteron wave functions, we precisely reproduce the
momentum distribution from the original potential
at \emph{all momenta} and for \emph{all $\lambda$}.

This result by itself is guaranteed by construction.  The more
interesting issue is where the strength of the matrix element comes
from.  For example, for the bare operator and the Argonne $v_{18}$
potential, the momentum distribution at $q = 4\fmi$ comes entirely
from deuteron wave function components at that momentum.  But when
$\lambda = 2\fmi$, it is clear from Fig.~\subref*{fig:deutmd_accum-a}
that the deuteron does not have appreciable momentum components above
$2.5\fmi$ (even though $\vsrg$ has matrix elements near the
diagonal~\cite{Bogner:2006pc}) because of decoupling.  One might
imagine that the operator becomes pathological to compensate.  In
fact, it is also softened.

In Fig.~\subref*{fig:deutmd_accum-b}, we take the ratio of the evolved
operator evaluated with the evolved wave function at $q$ to the
corresponding initial quantity, but include in the numerator only
momenta up to $\kmax$. The numerator is thus: \beqn \int_0^{\kmax}
d{\bf k} \int_0^{\kmax} d{\bf k'} \, \psi_d^\dagger({\bf k};\lambda)
U_\lambda({\bf k,q}) U^\dagger_\lambda({\bf q,k'}) \psi_d({\bf
k'};\lambda) \,.  \eeqn We observe that for all $q$, the ratio
approaches unity for large enough $\kmax$, as dictated by the unitary
transformation.  For $\lambda = 1.5\fmi$ or $2\fmi$, larger values of
$q$ ($q \gtrsim 3 \fmi$ or $q \gtrsim 4 \fmi$, respectively) give
results approximately independent of $q$, with a smooth approach to
unity by $\kmax \approx 1.3 \lambda$. This is consistent with the
operator $U_\lambda({\bf k,q})$ factorizing into $K_\lambda({\bf k})
\, Q_\lambda({\bf q})$ for $k < \lambda$ and $q \gg \lambda$, and thus
the $q$ dependence cancels in the ratio~\cite{Bogner:2007jb}. It
remains to be seen whether this factorization is a general feature
that can be understood using operator product expansion
ideas~\cite{Lepage:1997cs,Anderson:2010aa}.  The immediate point is
that the flow of the operator strength weighted by $\psi_d$ is toward
lower momenta (softened).  Furthermore, there is no sign of
fine-tuning in the evolved operator, because even a rough variational
Ansatz for the wave function is accurate.  More work on operators is
needed, but thus far these observations seem to be
general~\cite{Anderson:2010aa}.


\section{Applications to infinite matter}
\label{sec:infinite}

The physics of nucleonic matter ranges over exciting extremes: from
universal properties at low
densities~\cite{Schwenk:2005ka,Gezerlis:2007fs} that can
be probed in experiments with ultracold atoms~\cite{Giorgini:2008zz};
to understanding nuclear densities in known and new exotic
nuclei~\cite{Bender:2003jk,Drut:2009ce}; to dense matter in
neutron stars and supernovae~\cite{Lattimer:2000nx,Mezzacappa:2005ju}.
At very low densities, the details of pion exchanges are not resolved
and pionless EFT is well suited for capturing large
scattering-length physics with systematic improvements from including effective
range and higher-order terms~\cite{Bedaque:2002mn,Braaten:2004rn}.
For very dilute neutron matter, there are no length scales associated with
the interaction, because the scattering length is large $\kf a_{\rm s}
\gg 1$. As a result, only the density sets the scale and the
properties of neutron matter are
universal~\cite{Furnstahl:2008df}.\footnote{In this regime, nuclear
matter exists as a mixture of nucleons and
nuclei that collapses to clusters at low
temperatures~\cite{Horowitz:2005nd}.} With increasing density the effective
range correction weakens the interactions of nucleons with
higher momenta. This simplifies the many-body problem and has been
used to predict the energy of neutron matter with controlled
uncertainties for densities $\rho \sim
\rho_0/10$~\cite{Schwenk:2005ka} (with saturation density $\rho_0 =
0.16 \fm^{-3}$).

For nuclear densities, Pauli blocking suppresses the nonperturbative
physics due to weakly and nearly bound states and the large scattering
lengths are
dissolved~\cite{Bogner:2005sn,1982NuPhA.379..536R,Friman:1982zz}. The
difficulties in the theory of nuclear matter arise from the strong
short-range repulsion and strong short-range tensor forces in
phenomenological or chiral EFT potentials. The resulting coupling
between low and high momenta is still present in the $G$ matrix, as
discussed in Section~\ref{subsec:other}. This renders the
Bethe-Brueckner-Goldstone expansion
nonperturbative~\cite{Day:1967zz}. As detailed in
Sects.~\ref{subsec:decoupling} and~\ref{subsec:perturb}, the RG
evolution softens these short-range parts, which makes the
many-body calculation more controlled. In this section, we review the
application of low-momentum interactions to infinite matter with
$\rho\sim\rho_0$ and highlight the possibility of perturbative
approaches~\cite{Bogner:2005sn,Bogner:2009un,Hebeler:2009iv}. The
findings can provide a starting point to develop a new power counting
for nuclear matter. We focus on calculations including 3N forces and
so only mention the work of
Refs.~\cite{Kuckei:2002km,Kohler:2005xn,%
Bozek:2006sn,Margueron:2007jc,Gogelein:2008vw,Siu:2009nx,vanDalen:2009vt}.

\subsection{Nuclear matter}
\label{subsec:nuclear}

Over the last fifty years, an accurate prediction of nuclear matter
starting from nuclear forces has been a theoretical milestone on the
way to finite nuclei, but has proved to be an elusive target. Despite
the long-term emphasis on the infinite uniform system, most advances
in microscopic nuclear structure theory over the last decade have been
through expanding the reach of few-body calculations. This has
unambiguously established the quantitative role of 3N forces for light
nuclei ($A \leqslant
12$)~\cite{Nogga:2000uu,Pieper:2001mp,Pieper:2004qh,Pieper:2004qw,%
Navratil:2003ef,Navratil:2007we}. However, until recently few-body
fits have not sufficiently constrained 3N force contributions at
higher density such that nuclear matter calculations are
predictive. Nuclear matter saturation is very delicate, with the
binding energy resulting from cancellations of much larger potential
and kinetic energy contributions.  When a quantitative reproduction of
empirical saturation properties has been obtained, it was imposed by
hand through adjusting phenomenological short-range three-body forces
(see, for example, Refs.~\cite{Akmal:1998cf,Lejeune:2001bg}).

Progress for controlled nuclear matter calculations has long been
hindered by the difficulty of the nuclear many-body problem when
conventional nuclear potentials are used. Recent calculations overcome
these hurdles by combining systematic starting Hamiltonians based on
chiral EFT~\cite{Entem:2003ft,Epelbaum:2004fk} with RG
methods~\cite{Bogner:2003wn,Bogner:2006vp} to soften the short-range
repulsion and short-range tensor components of the initial chiral
interactions~\cite{Bogner:2006tw}. The RG evolution to low momenta
leads to contributions in the particle-particle channel that are well
converged at second order in the potential, suggesting that
perturbative approaches can be used in
place of the Bethe-Brueckner-Goldstone hole-line
expansion~\cite{Bogner:2005sn,Bogner:2009un}. The Weinberg eigenvalue
analysis discussed in Section~\ref{subsec:perturb} provides quantitative
backing to these observations~\cite{Bogner:2006tw,Bogner:2006pc}.

\begin{figure}[t]
\centering
\includegraphics[width=6.5in,clip=]{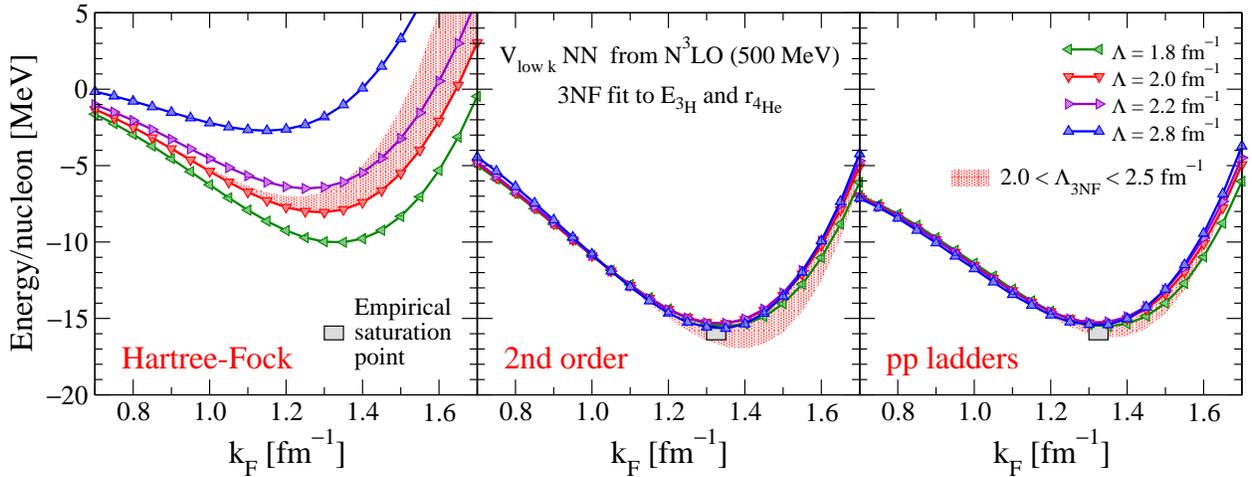}
\caption{Nuclear matter energy per particle as a
function of Fermi momentum $\kf$ at the Hartree-Fock level (left)
and including second-order (middle) and particle-particle-ladder 
contributions (right), based on evolved N$^3$LO NN potentials and
3N forces fit to $E_{\rm^3H}$ and $r_{\rm^4He}$~\cite{Bogner:2009un}. 
Theoretical uncertainties are estimated by the NN (lines) and 3N (band)
cutoff variations.\label{nm_all}}
\end{figure}

The key nuclear matter results are summarized in Fig.~\ref{nm_all},
which shows the energy per particle of symmetric nuclear matter as a
function of Fermi momentum $\kf$, or the density $\rho = 2
\kf^3/(3\pi^2)$. A grey square representing the empirical saturation
point reflects the ranges of nuclear matter saturation properties
predicted by phenomenological Skyrme energy functionals that most
accurately reproduce properties of finite nuclei. Nuclear matter is
calculated in three approximations: Hartree-Fock (left) and including
approximate second-order (middle) and summing particle-particle-ladder
contributions (right). These are the first results for nuclear matter
based on chiral NN and 3N interactions. The technical details are
given in Refs.~\cite{Bogner:2005sn,Bogner:2009un} and work is in
progress to improve the 3N treatment.

The calculations of Fig.~\ref{nm_all} start from the N$^3$LO NN
potential ($\lm=500 \mev$) of Ref.~\cite{Entem:2003ft}. This NN
potential is RG-evolved to smooth-cutoff low-momentum interactions
$\vlowk$ using the techniques of Section~\ref{subsec:smooth}. Based on
the universality of $\vlowk$ discussed in
Section~\ref{subsec:universal}, we do not expect large differences
starting from different N$^3$LO potentials. The N$^2$LO 3N forces of
Section~\ref{subsec:3nf} are taken as a truncated basis for low-momentum
3N interactions,\footnote{This assumes that the $c_i$ coefficients of
the long-range $2 \pi$-exchange part of 3N forces are not modified
by the RG evolution.} where the $c_D$ and $c_E$ couplings have been
fit for various cutoffs to the $^3$H binding energy and the $^4$He
matter radius~\cite{Bogner:2009un}. (In the future, it will be possible
to include consistently evolved three-body forces starting from chiral
EFT using the recent advances in extending the SRG methods beyond the
NN level~\cite{Bogner:2007qb,Jurgenson:2008jp,Jurgenson:2009qs}.) 
The 3N force fits to
the $^4$He radius improve the cutoff independence significantly
compared to fitting to $A=3,4$ energies only (see Fig.~6 in
Ref.~\cite{Bogner:2005sn}). The calculations of Fig.~\ref{nm_all} use
the same 3N regulator of Ref.~\cite{VanKolck:1994yi,Epelbaum:2002vt},
but with a 3N cutoff $\lm_{\rm 3NF}$ that is allowed to vary
independently of the NN cutoff. The shaded regions in
Fig.~\ref{nm_all} show the range of results for $2.0 \fmi <
\Lambda_{\rm 3NF} < 2.5 \fmi$ at fixed $\Lambda = 2.0 \fmi$.  These
predictions are particularly sensitive to uncertainties in the $c_i$
coefficients (also for neutron matter, see
Section~\ref{subsec:neutron}), which raises the possibility of using
nuclear matter to constrain some of the $c_i$ couplings.

\begin{figure}[t]
 \centering
 \subfloat[][]{%
  \label{nmsrg}%
  \includegraphics[width=3.0in,clip=]{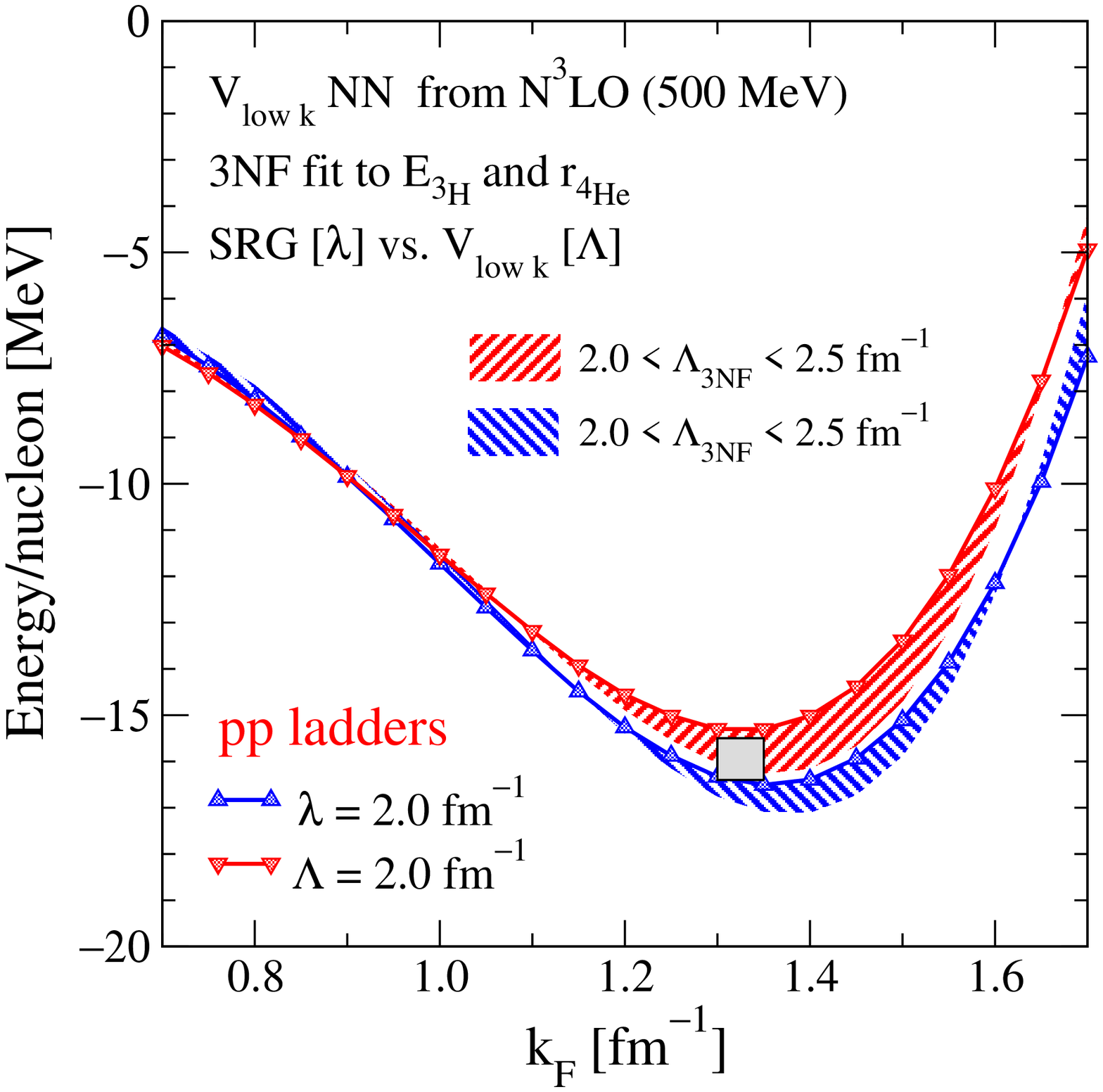}%
 }%
 \hspace*{.4in}%
 \subfloat[][]{%
  \label{NNvs3N}%
  \includegraphics[width=3.0in,clip=]{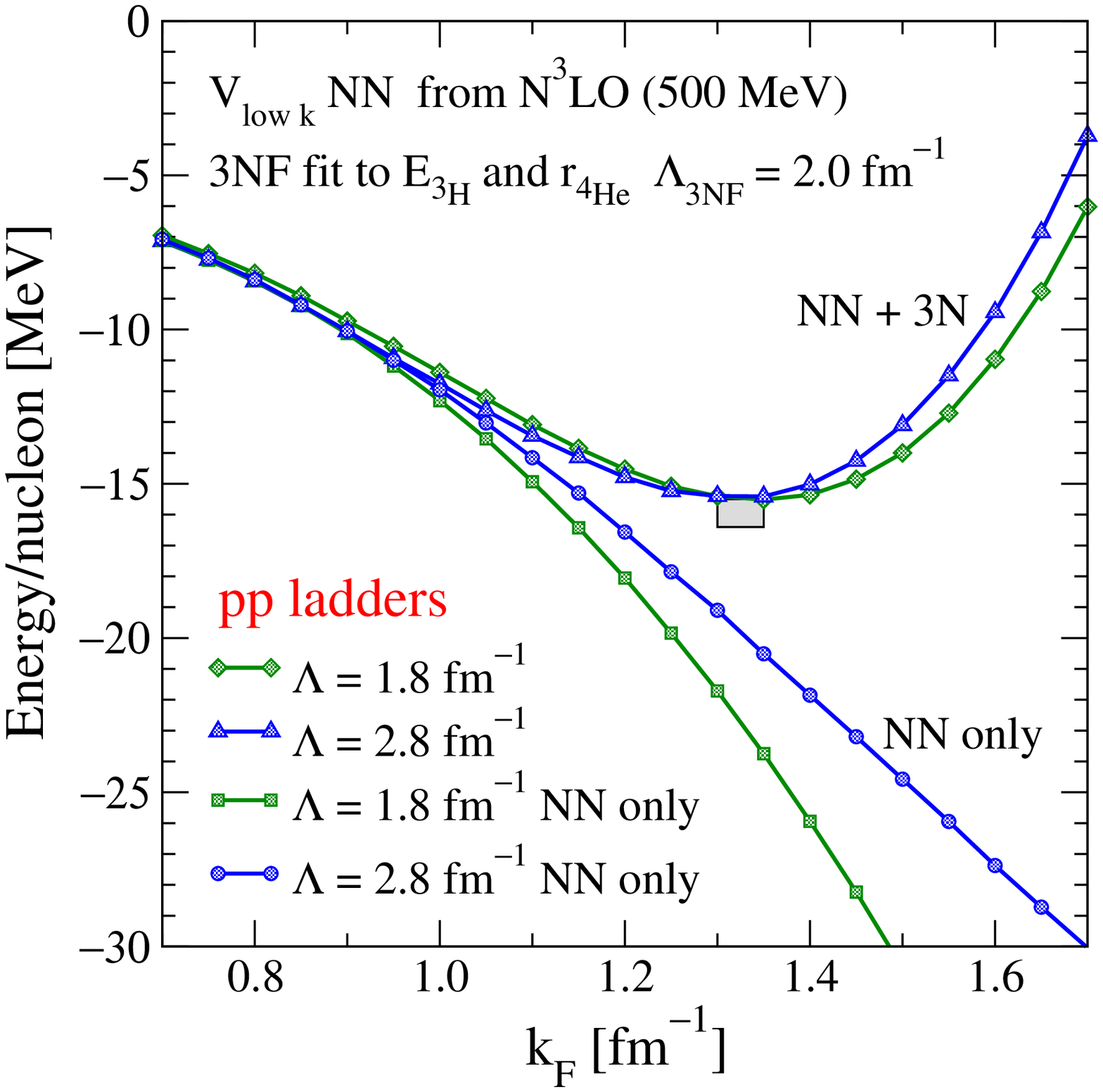}%
 }%
\caption{(a) Nuclear matter energy of Fig.~\ref{nm_all} at the
particle-particle-ladders level compared to
SRG-evolved chiral NN interactions. (b)~Nuclear matter energy of 
Fig.~\ref{nm_all} at the particle-particle-ladders level compared
to NN-only results for two representative NN cutoffs and a fixed 
3N cutoff.}
\end{figure}

The Hartree-Fock results show that nuclear matter is bound even at
this simplest level. A calculation without approximations should be
independent of the cutoffs, so the spread in Fig.~\ref{nm_all} sets
the scale for omitted many-body contributions. The second-order
results show a dramatic narrowing of this spread, with predicted
saturation consistent with the empirical range. The narrowing happens
across the full density range. This is strong evidence that these
encouraging results are not fortuitous, but that cutoff independence
has been reached at the level of $1\mbox{--}2 \mev$ per particle. The
controlled theoretical uncertainties come as a result of the
combination of chiral EFT and RG methods. For all cases, the
compressibility $K = 190\mbox{--}240 \mev$ is in the empirical range. To our
knowledge, these are the first nuclear forces fit only to $A \leqslant
4$ nuclei that predict realistic saturation properties. Similar
nuclear matter energies are found in Fig.~\subref*{nmsrg} when the SRG
is used to evolve the NN potential. This helps support the general
nature of the 3N force fit. On the other hand, the difference of $2
\, \mev$ per particle at saturation and above enlarges the theoretical
uncertainty.

The decreases in cutoff dependence in Fig.~\ref{nm_all} with more
complete approximations is necessary but not sufficient to conclude
that the calculations are under control. Indeed, approximations that
are independent of the cutoff will shift the answer but not widen the
error band from cutoff variation. The theoretical errors arise from
truncations in the initial chiral EFT Hamiltonian, the many-body
approximations, and the approximation of 3N forces. The latter is
particularly uncertain here because it involves long-range
contributions that are independent of the cutoff.  Many-body
corrections to the current approximations include higher-order terms
in the hole-line expansion and particle-hole corrections. An approach
such as coupled cluster theory that can perform a high-level
resummation will be necessary for a robust validation.

\begin{figure}[t]
\centering
\includegraphics[scale=0.45,clip=]{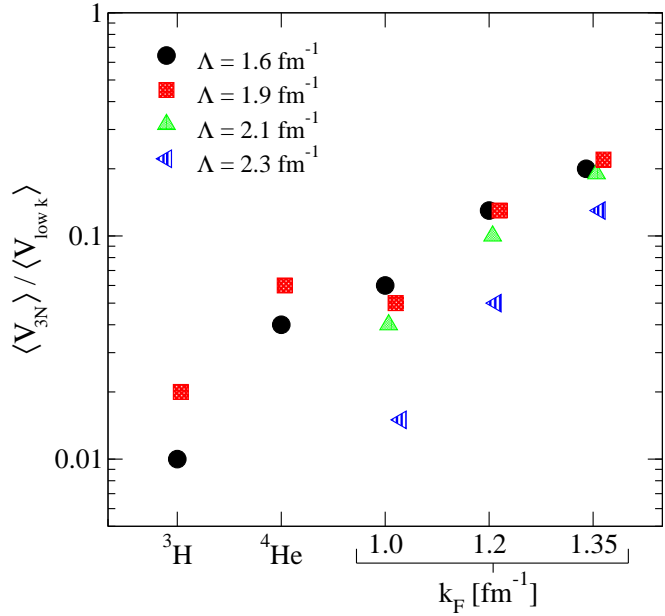}
\caption{Ratio of $\vtn$ to $\vlowk$ expectation values. For nuclear
matter, the expectation values are obtained by the Feynman-Hellman
method and $\langle \vtn \rangle$ corresponds to the total 3N
contribution at second order from Ref.~\cite{Bogner:2005sn}, whereas
the values shown for $^3$H and $^4$He are the largest 3N
contributions from Ref.~\cite{Nogga:2004ab} (because of
cancellations in the light nuclei).\label{expval}}
\end{figure}

The evolution of the cutoff $\Lambda$ to smaller values is accompanied
by a shift of physics. In particular, effects due to iterated tensor
interactions, which peak in the relative momentum range $k \sim 4
\fmi$ (and thus lead to saturation at too high density), are replaced
by three-body contributions. We emphasize that the relative importance
of contributions to observables from the tensor force or from 3N
forces are scale or resolution dependent. Renormalizing the potential
to lower momenta redistributes these contributions and makes the
many-body approach tractable. The role of 3N forces for saturation is
demonstrated in Fig.~\subref*{NNvs3N}. The two pairs of curves show
the differences between the nuclear matter results for NN-only and NN
plus 3N interactions. It is evident that saturation is driven by 3N
forces~\cite{Bogner:2005sn,Bogner:2009un}. Even for $\Lambda = 2.8
\fmi$, which is similar to the lower cutoffs in chiral EFT potentials,
saturation is at too high density without 3N forces. While 3N forces
drive saturation for low-momentum interactions, the 3N contributions
are not unnaturally large. This is demonstrated in Fig.~\ref{expval}
by the ratio of $\langle \vtn \rangle / \langle \vlowk \rangle$ for
the triton, alpha particle, and nuclear matter at various densities.

In Fig.~\ref{nm_all} the particle-particle-ladder sum is little
changed from second order except at the lowest densities shown, where
the presence of two-body bound and nearly bound states necessitate a
nonperturbative summation. The rapid convergence for $\rho\sim\rho_0$
may justify in part the application of chiral EFT to nuclear matter in
perturbation
theory~\cite{Lutz:1999vc,Kaiser:2001jx,Fritsch:2004nx,Kaiser:2006tu},
but a detailed study is needed, in
particular regarding the cutoff dependence and the contributions from
contact interactions and 3N forces beyond $\Delta$ excitations.

While nuclear matter has lost its status to light nuclei as the first
step to nuclear structure, it is still key as a step to heavier
nuclei.  The low-momentum results, while not conclusive, open the door
to ab initio density functional theory based on expanding about
nuclear matter~\cite{Bogner:2008kj,Drut:2009ce}.  This is discussed
further in Section~\ref{subsec:dft}.

\subsection{Neutron matter}
\label{subsec:neutron}

\begin{figure}[t]
\centering
\includegraphics[width=6.5in,clip=]{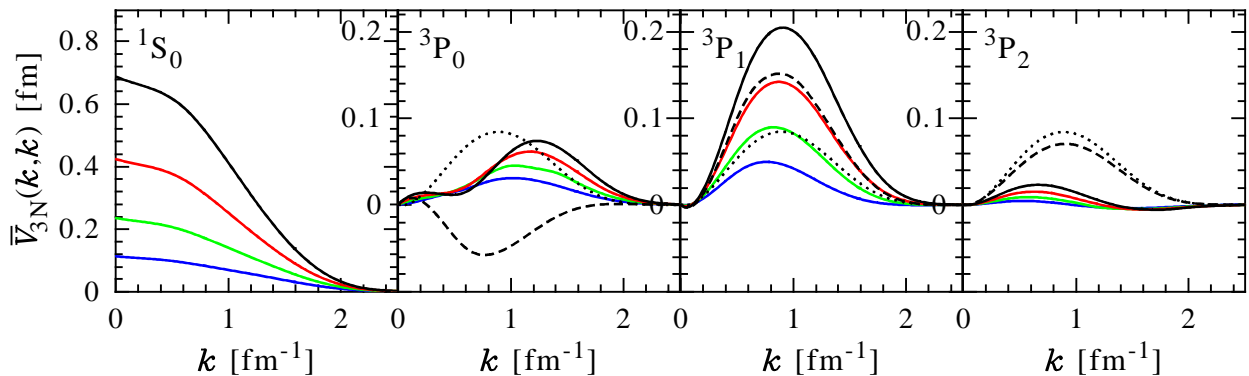}
\caption{Diagonal momentum-space matrix elements of 
the density-dependent two-body interaction $\vbar$ for $P=0$ in the
$^1$S$_0$ and the spin-triplet $P$-wave channels. Results with
$\lm_{\rm 3NF} = 2.0 \fmi$ are shown versus relative momentum $k$
for different Fermi momenta $\kf = 1.0, 1.2, 1.4$ and $1.6 \fmi$
(increasing in strength). For the $P$-wave channels and
$\kf = 1.6 \fmi$, the dotted lines represent the central parts
(degenerate in $J$) of $\vbar$, whereas the dashed lines
include the central plus tensor interactions. For details see
Ref.~\cite{Hebeler:2009iv}.\label{fig:vbar}}
\end{figure}

Neutron matter provides a different perspective from symmetric nuclear
matter, because only the long-range $2\pi$-exchange $c_1$ and $c_3$
terms of N$^2$LO 3N forces
contribute~\cite{Tolos:2007bh,Hebeler:2009iv}.  Density-dependent
two-body interactions $\vbar$ based on these terms for general
momentum and spin configurations and including all exchange diagrams
were derived in Ref.~\cite{Hebeler:2009iv} by summing the third
particle over occupied states in the Fermi sea. This corresponds to
the normal-ordered two-body part of 3N forces. Effective interactions
of this type have been studied in the past using 3N potential models
and approximate treatments (see, for example,
Refs.~\cite{Blatt:1974vv,Coon:1978gr}) and recently in the framework
of in-medium chiral perturbation theory~\cite{Holt:2009ty}. The
partial-wave matrix elements in Fig.~\ref{fig:vbar} show that the
resulting $\vbar$ is dominated in neutron matter by a repulsive
central part. The density dependence of the two-body matrix elements
depends on the partial wave. In the $^1$S$_0$ channel this dependence
can be approximately parameterized by a power of the Fermi momentum.
In the spin-triplet channels this is more complex due to the different
momentum and density dependences of the various operator structures in
$\vbar$. In addition to the density dependence, in general $\vbar$
depends on the two-body center-of-mass momentum, but this dependence
is weak and the $P=0$ approximation for $\vbar$ leads to energies that
agree very well with the exact result at the Hartree-Fock
level~\cite{Hebeler:2009iv}.

\begin{figure}[t]
 \centering
 \subfloat[][]{%
  \label{fig:EOS_compare-a}%
  \includegraphics[width=3.0in,clip=]{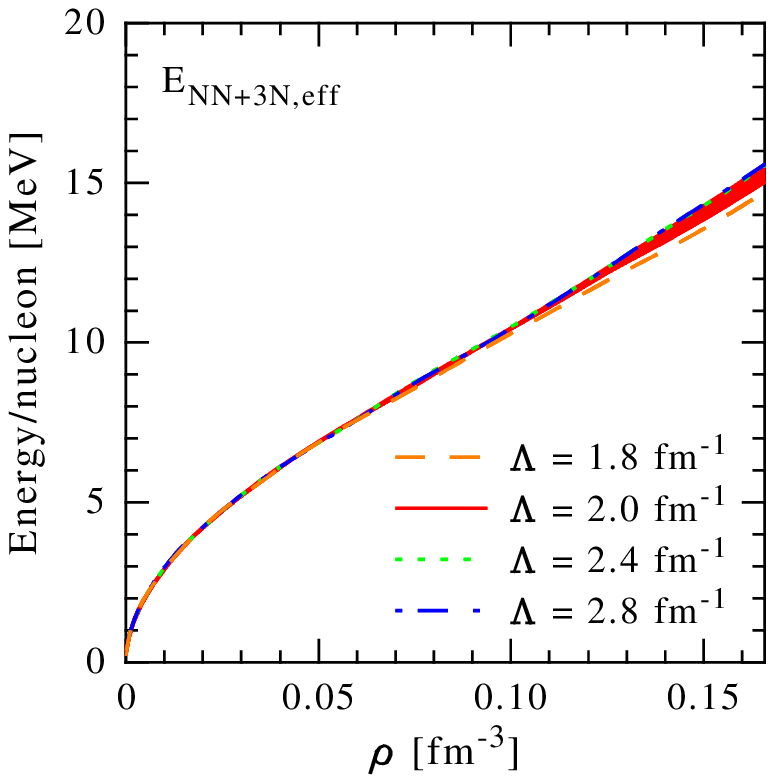}%
 }%
 \hspace*{.4in}%
 \subfloat[][]{%
  \label{fig:EOS_compare-b}%
  \includegraphics[width=3.0in,clip=]{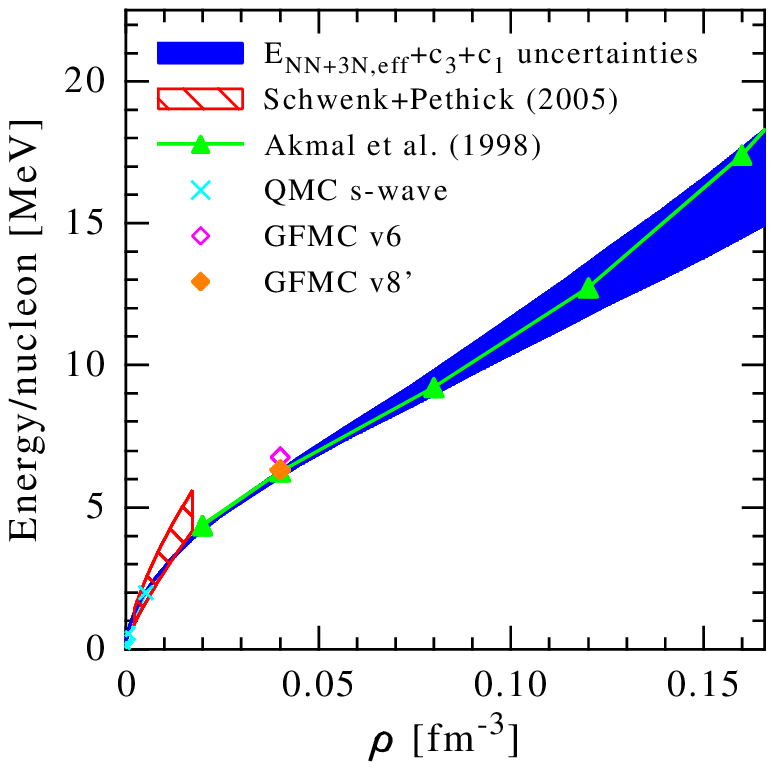}%
 }%
\caption{(a) Energy per particle of neutron matter as a function of
density $\rho$ at second order. The results are based on evolved
N$^3$LO NN potentials and N$^2$LO 3N forces. Theoretical
uncertainties are estimated by varying the NN cutoff (lines) and
the 3N cutoff (band for fixed $\lm = 2.0 \fmi$). (b)~Theoretical
uncertainties of the second-order energy due to the uncertainties
in the $c_1$ and $c_3$ coefficients of 3N forces. The resulting
band is compared to other neutron matter results, for details see
Ref.~\cite{Hebeler:2009iv}.\label{fig:EOS_compare}}
\end{figure}

Using $\vbar$ simplifies the neutron matter calculation based on a
loop expansion around the Hartree-Fock
energy~\cite{Hebeler:2009iv}. Figure~\ref{fig:EOS_compare} shows the
energy of neutron matter as a function of the density $\rho = \kf^3/(3
\pi^2)$ at second-order. These are the first results for neutron
matter based on chiral EFT interactions and including N$^2$LO 3N
forces. As discussed for nuclear matter, the RG evolution of N$^3$LO
NN potentials to low momenta renders the many-body calculation more
controlled. The results of Ref.~\cite{Hebeler:2009iv} suggest that
neutron matter is perturbative at nuclear densities. This is based on
small second-order contributions, self-energy corrections being
negligible, and a generally weak cutoff dependence in
Fig.~\subref*{fig:EOS_compare-a}. For low-momentum interactions,
N$^2$LO 3N forces provide a repulsive contribution to the energy due
to the repulsive central part in $\vbar$.

Figure~\subref*{fig:EOS_compare-b} shows that the theoretical
uncertainties of the neutron matter energy are dominated by the
uncertainties in the $c_i$ coefficients, in particular the $c_3$ part.
The $c_1$ and $c_3$ variation leads to an energy uncertainty of $\pm
1.5 \mev$ per particle at saturation density.  This is larger than the
uncertainties probed by cutoff variations, which are due to the
many-body approximations or from neglected short-range many-body
interactions.  We note that other recent neutron matter calculations
lie within the $c_i$ uncertainty band of
Fig.~\subref*{fig:EOS_compare-b}.  The spread in the energy band
implies microscopic constraints for the symmetry energy and its
density dependence~\cite{Hebeler:2009iv}.  The resulting range for the
symmetry energy $a_4 = (31.6\mbox{--}34.7) \mev$ (where the error is
dominated by the uncertainty of $c_3$) is very useful given the scale
of the empirical range $a_4 = (25\mbox{--}35) \mev$~\cite{Steiner:2004fi}.
Other results for neutron matter with low-momentum interactions
include studies of the equation of state at finite
temperature~\cite{Tolos:2007bh} and of quasiparticle interactions and
Fermi liquid parameters using a particle-hole RG
approach~\cite{Schwenk:2001hg,Schwenk:2002fq,Schwenk:2003bc,Schwenk:2004ut}.

\subsection{Pairing}
\label{subsec:pairing}

Superfluidity plays a central role in strongly-interacting many-body
systems. Pairing in infinite matter impacts the cooling of isolated
neutron
stars~\cite{Yakovlev:2004iq,Blaschke:2004vq,2004ApJS..155..623P} and
of the neutron star crust following X-ray bursts in accreting neutron
stars~\cite{2006MNRAS.372..479C,2008ApJ...687L..87C,Brown:2009kw}. In
addition, studies of superfluidity are used to constrain and improve
calculations of pairing in nuclei (see, for example,
Ref.~\cite{Hebeler:2009dy}). This section provides a brief review of
pairing in infinite matter based on low-momentum interactions.

\begin{figure}[t]
 \centering
 \subfloat[][]{%
  \label{gaps-a}%
  \includegraphics[clip=,width=3.9in]{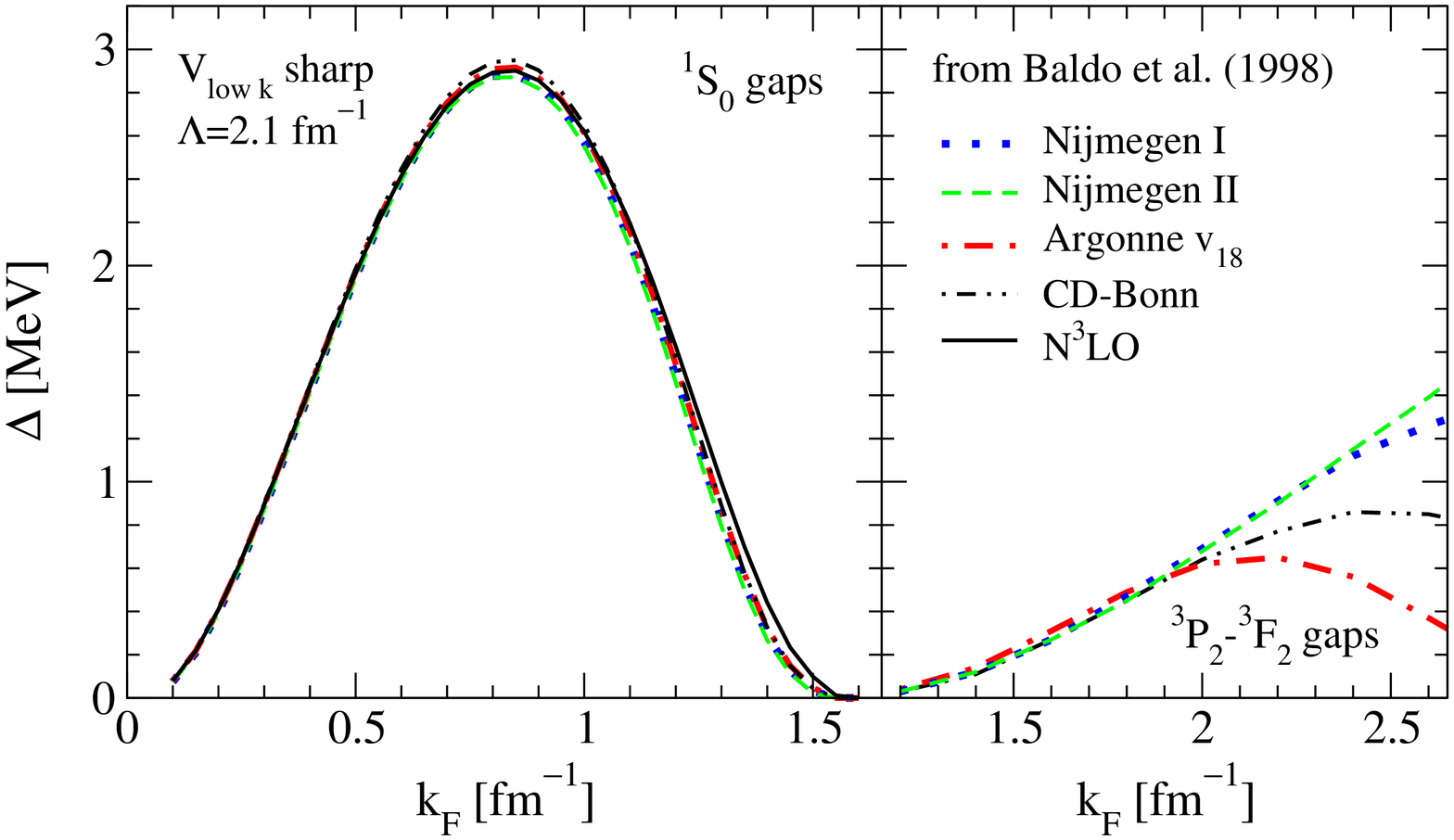}%
 }%
 \hspace*{0.1in}%
 \subfloat[][]{%
  \label{gaps-b}%
  \includegraphics[clip=,width=3.2in]{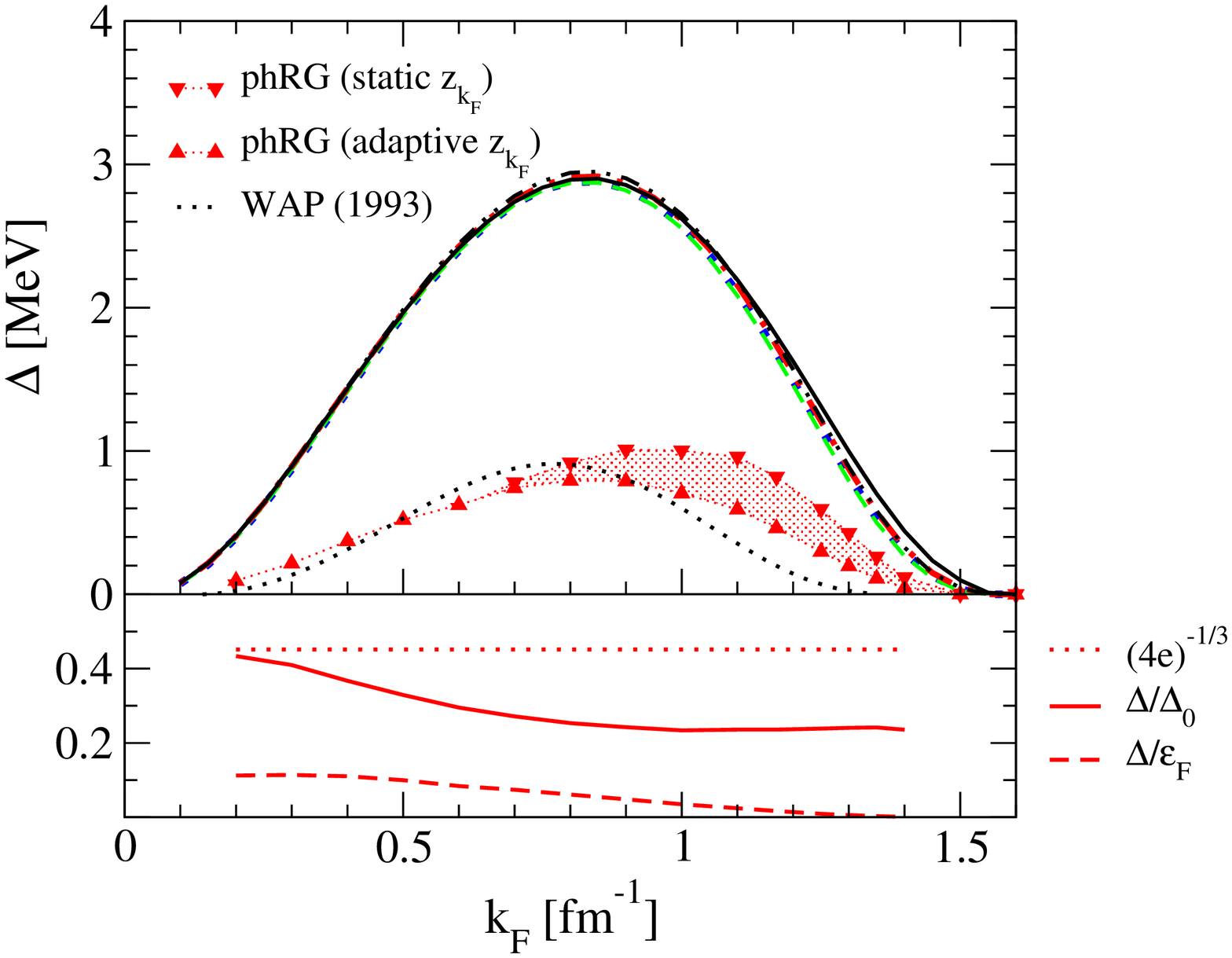}%
 }%
\caption{(a) The $^1$S$_0$ (left panel) and $^3$P$_2-^3$F$_2$
(right panel) superfluid pairing gaps $\Delta$ at the Fermi surface
as a function of Fermi momentum $\kf$ in neutron matter. The gaps
are obtained from charge-dependent NN interactions at the BCS level.
For details see Refs.~\cite{Hebeler:2006kz,Baldo:1998ca}.
(b)~Top panel: Comparison of the $^1$S$_0$ BCS gap to the
results including polarization effects through the phRG, for 
details see Ref.~\cite{Schwenk:2002fq}, and to the results of Wambach 
{\it et al.}~\cite{Wambach:1992ik}. Lower panel: Comparison of the full 
superfluid gap $\Delta$ to the BCS gap $\Delta_0$ and to the Fermi
energy $\ef$.}%
\end{figure}

Figure~\subref*{gaps-a} shows superfluid pairing gaps in neutron
matter, obtained by solving the BCS gap equation with a free
spectrum. At low densities (in the crust of neutron stars), neutrons
form a $^1$S$_0$ superfluid. At higher densities, the S-wave
interaction is repulsive and neutrons pair in the $^3$P$_2$ channel
(with a small coupling to $^3$F$_2$ due to the tensor
force). Figure~\subref*{gaps-a} demonstrates that the $^1$S$_0$ BCS
gap is practically independent of nuclear interactions, and therefore
strongly constrained by NN phase shifts~\cite{Hebeler:2006kz}. This
includes a very weak cutoff dependence for low-momentum interactions
$\vlowk$ with sharp or sufficiently narrow smooth regulators with $\lm
> 1.6 \fmi$. The inclusion of N$^2$LO 3N forces leads to a reduction
of the $^1$S$_0$ BCS gap for Fermi momenta $\kf > 0.6
\fmi$~\cite{Hebeler:2009iv}. This reduction becomes significant for
densities where the gap is decreasing and agrees qualitatively with
results based on 3N potential models (see, for example,
Ref.~\cite{Zuo:2004mc}). Two-nucleon interactions are well constrained
by scattering data for relative momenta $k \lesssim 2
\fmi$~\cite{Bogner:2003wn}. The model dependencies at higher momenta
show up prominently in Fig.~\subref*{gaps-a} in the $^3$P$_2-^3$F$_2$
gaps for Fermi momenta $\kf > 2 \fmi$~\cite{Baldo:1998ca}.

For pairing in nuclei, calculations with $\vlowk$ as pairing
interaction provide a good starting description of neutron-neutron and
proton-proton
gaps~\cite{Duguet:2007be,Lesinski:2008cd,Duguet:2009gc,Hebeler:2009dy},
as discussed in Section~\ref{subsec:dft}. This is very promising and
also consistent with comparisons of the $^1$S$_0$ BCS gaps in neutron
matter with results based on the empirical Gogny
interaction~\cite{Sedrakian:2003cc,Kaiser:2004uj}. In nuclei, however,
partial waves beyond the standard $^1$S$_0$ channel can contribute to
the pairing interaction and play an interesting role for the $T=1,
J=0$ pair formation (because paired nucleons are not in
back-to-back-momentum configurations)~\cite{Baroni:2009eh}.

Understanding many-body effects beyond the BCS level constitutes an
important open problem. For recent progress and a survey of results,
see for instance Ref.~\cite{Gezerlis:2009iw}. Here we review approaches
to this problem using the RG. At low densities, induced interactions
due to particle-hole screening and vertex corrections are significant
even in the perturbative $\kf a$
limit~\cite{Gorkov:1961,Heiselberg:2000ya} and lead to a reduction of
the S-wave gap by a factor $(4e)^{-1/3} \approx 0.45$,
\be
\frac{\Delta}{\ef} = \frac{8}{e^2} \, \exp \biggr\{ \biggr(
\, \begin{minipage}{4.4cm}
\includegraphics[scale=0.5,clip=]{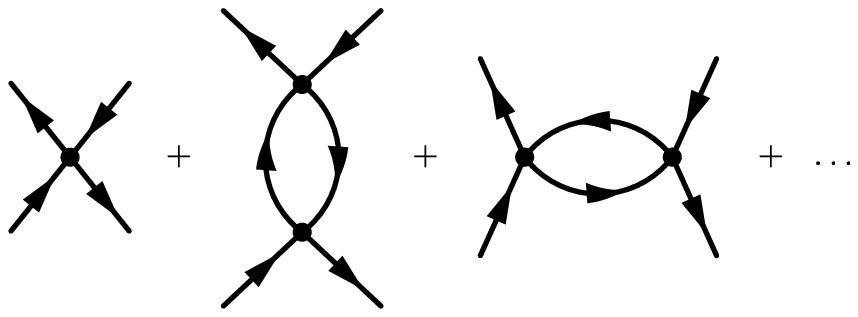}
\end{minipage}\biggr)^{-1} \biggr\}
= (4e)^{-1/3} \, \frac{8}{e^2} \, \exp \biggl\{ \frac{\pi}{2 \kf a} + 
{\mathcal O}(\kf a)\biggr\} \,.
\label{Gorkov}
\ee
This reduction is due to spin fluctuations, which are repulsive for
spin-singlet pairing and overwhelm attractive density
fluctuations.\footnote{In finite systems, the spin and density
response differs. In nuclei with cores, the low-lying response is
due to surface vibrations. Consequently, induced interactions may
be attractive, because the spin response is
weaker~\cite{Barranco:2003kc,Pastore:2008zi}.}

Following Shankar~\cite{Shankar:1993pf}, Ref.~\cite{Schwenk:2002fq}
developed a nonperturbative RG approach for neutron matter, where
induced interactions are generated by integrating out modes away from
the Fermi surface. Starting from $\vlowk$, the solution to the RG
equations in the particle-hole channels (``phRG'') includes
contributions from successive particle-hole momentum shells. The phRG
builds up many-body effects similar to the two-body parquet equations,
and efficiently includes induced interactions to low-lying states in
the vicinity of the Fermi surface beyond a perturbative calculation.
The phRG results~\cite{Schwenk:2002fq} for the $^1$S$_0$ gap are shown
in Fig.~\subref*{gaps-b}, where induced interactions lead to a factor
3--4 reduction to a maximal gap $\Delta \approx 0.8 \mev$. For the
lower densities, the phRG is consistent with the dilute
result\footnote{For $\kf \approx 0. 4 \fmi$, neutron matter is close
to the universal regime, but theoretically simpler due to an
appreciable effective range $\kf r_{\rm e} \approx
1$~\cite{Schwenk:2005ka}.} $\Delta/\Delta_0 = (4e)^{-1/3}$, and at
the larger densities the dotted band indicates the uncertainty due to
an approximate self-energy treatment.

Non-central spin-orbit and tensor interactions are crucial for
$^3$P$_2-^3$F$_2$ superfluidity. Without a spin-orbit interaction,
neutrons would form a $^3$P$_0$ superfluid instead.  The first
perturbative calculation of non-central induced interactions shows that
$^3$P$_2$ gaps below $10 \, {\rm keV}$ are possible (while
second-order contributions to the pairing interaction are not
substantial $| V_{\rm
ind} / \vlowk | < 0.5$)~\cite{Schwenk:2003bc}. This arises from a
repulsive induced spin-orbit interaction due to the mixing with the
stronger spin-spin interaction. As a result, neutron P-wave
superfluidity (in the interior of neutron stars) may be reduced
considerably below earlier estimates. This implies that low-mass
neutron stars cool
slowly~\cite{Yakovlev:2004iq,Blaschke:2004vq,2004ApJS..155..623P}.
Smaller values for the $^3$P$_2$ gap compared to Fig.~\subref*{gaps-b}
are also required for consistency with observations in a minimal
cooling scenario~\cite{Page:2009fu}.


\section{Applications to finite nuclei}
\label{sec:finite}

The advantages of low-momentum interactions discussed in
Section~\ref{sec:infinite} for infinite matter calculations also apply
to finite nuclei. At lower resolution, perturbative methods generally
become more effective, finite-basis expansions converge more rapidly,
and short-range correlations are substantially weakened.  Furthermore,
the freedom to vary the cutoff with the RG provides a useful tool to
diagnose truncation errors and assess the quality of many-body
approximations. In this section, we review applications of
low-momentum interactions to finite nuclei, including ab initio
calculations of light and medium-mass nuclei, shell model
calculations, scattering and reactions, and energy-density functional
approaches.

\subsection{Ab initio calculations}
\label{subsec:abinitio}

\begin{figure}[t]
\centering
\includegraphics*[width=4.0in,clip=]{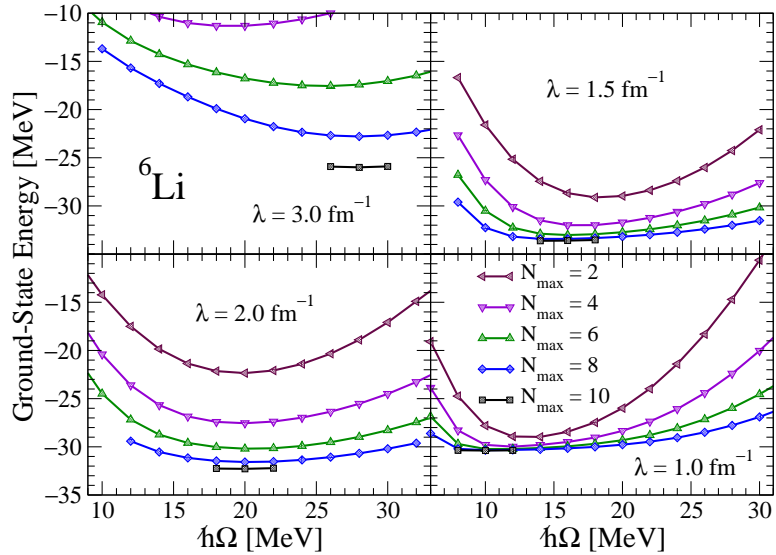}
\caption{ Ground-state energy of $^6$Li versus oscillator parameter
$\hbar\Omega$ for different SRG-evolved interactions with $\lambda =
3.0, 2.0, 1.5$ and $1.0 \fmi$. The initial interaction is the 
\NthreeLO\ NN-only potential of
Ref.~\cite{Entem:2003ft}. The NCSM results clearly show improved
convergence with the maximum number of oscillator quanta $N_{\rm
max}$ for lower cutoffs. Since 3N interactions are neglected, the
different NN calculations converge to different ground-state
energies. For details see Ref.~\cite{Bogner:2007rx}.}
\label{fig:Li6_srg_composite}
\end{figure}

The no-core shell model (NCSM) is a versatile method to calculate
ground-state energies and spectra of light
nuclei~\cite{Navratil:2009ut}. In Ref.~\cite{Bogner:2007rx}, the NCSM
was used to study the convergence of nuclei up to $^7$Li using
RG-evolved chiral NN interactions. Because no additional Lee-Suzuki
transformations are required, the energy calculations satisfy the
variational principle for a given low-momentum Hamiltonian. As shown
in Fig.~\ref{fig:Li6_srg_composite}, the convergence of ground-state
energies improves rapidly for SRG potentials with decreasing
$\lambda$, with similar convergence improvements for $\vlowk$
interactions.

\begin{figure}[t]
 \centering
 \subfloat[][]{%
  \label{fig:3bodyHe4-a}%
  \includegraphics*[width=2.35in,clip=]{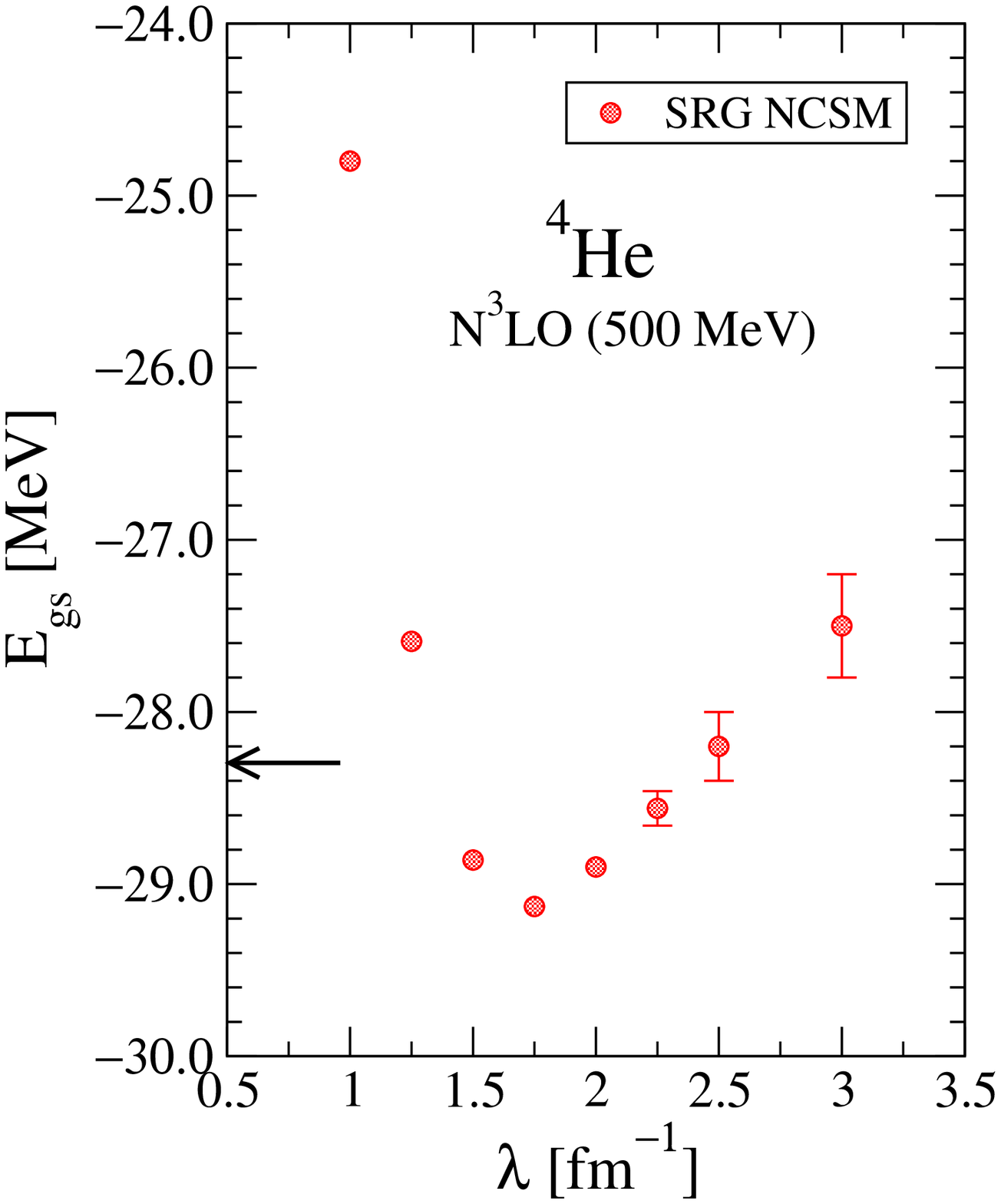}%
 }%
 \hspace*{.1in}%
 \subfloat[][]{%
  \label{fig:3bodyHe4-b}%
  \includegraphics*[width=2.35in,clip=]{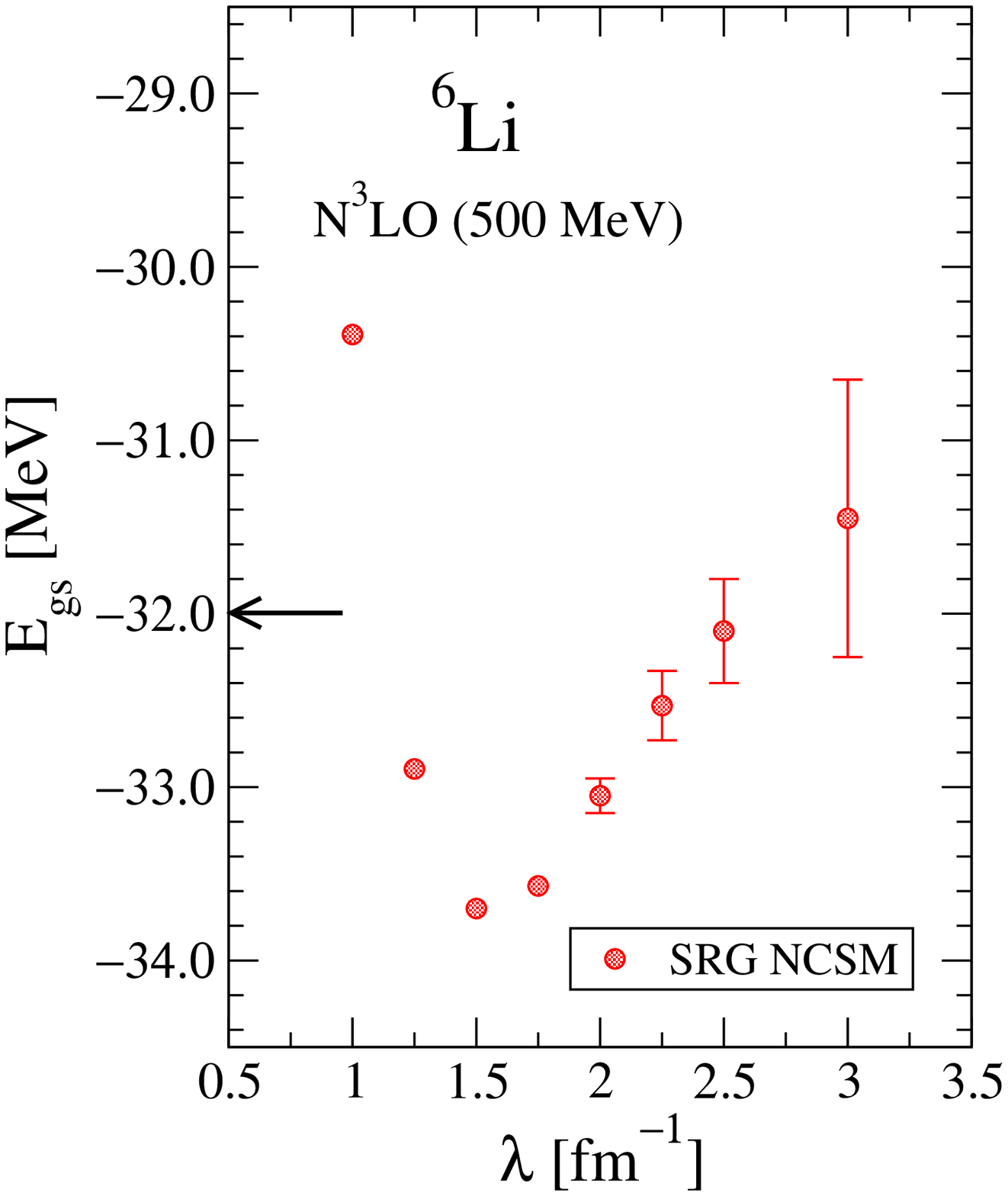}%
 }%
 \hspace*{.1in}%
 \subfloat[][]{%
  \label{fig:3bodyHe4-c}%
  \includegraphics*[width=2.35in,clip=]{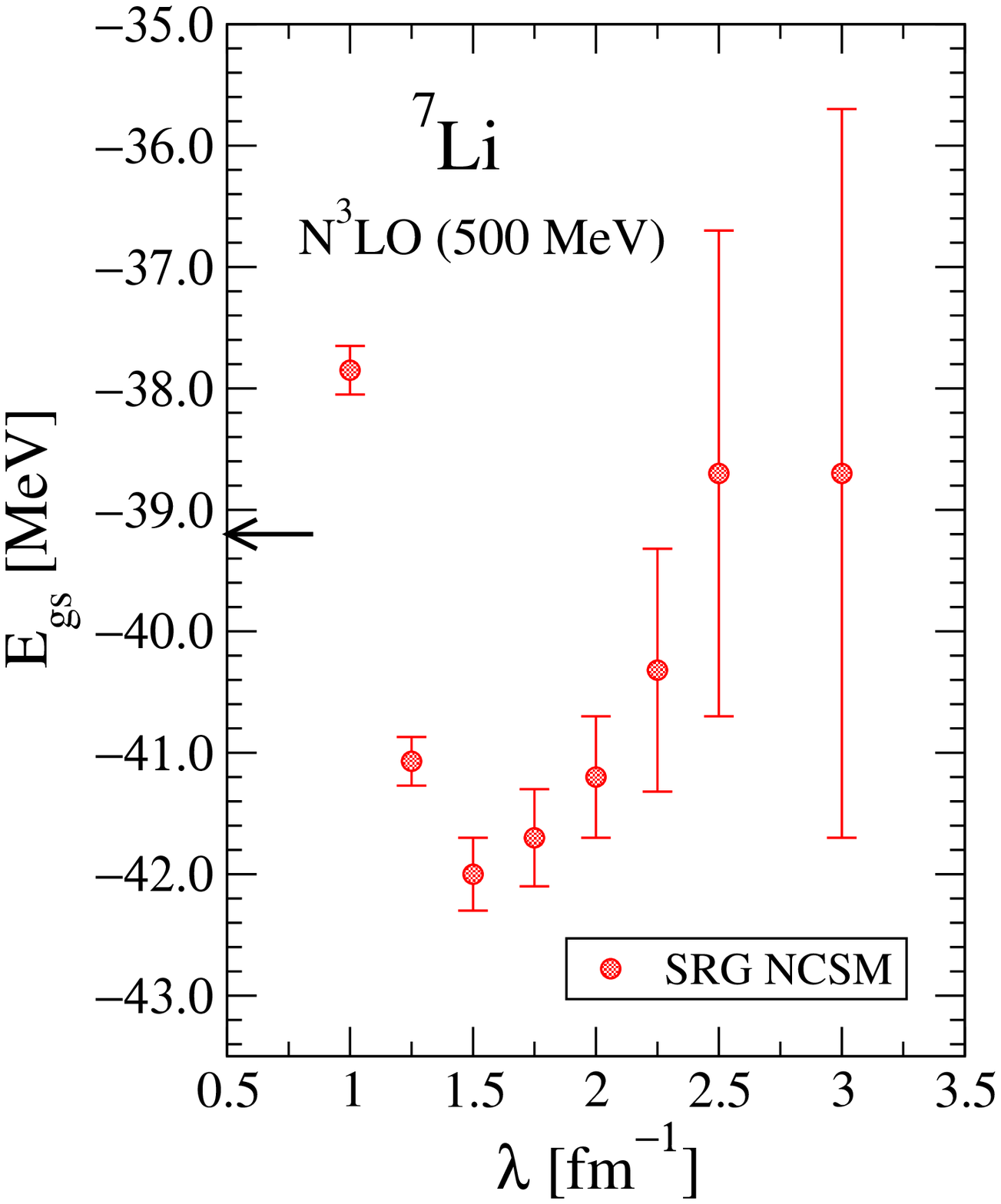}%
 }%
\caption{Ground-state energies of (a) $^4$He, (b) $^6$Li, and (c) 
$^7$Li as a function of $\lambda$ for SRG interactions evolved 
from the N$^3$LO NN-only potential of
Ref.~\cite{Entem:2003ft}. Error bars for larger $\lambda$ values 
are from extrapolations in $N_{\rm max}$. The arrow marks
the experimental energy. The characteristic increase in $E_{\rm gs}$
at small $\lambda$ signals the modification of the long-range 
attractive interaction~\cite{Jurgenson:2008jp}. For details see 
Ref.~\cite{Bogner:2007rx}.}
\label{fig:3bodyHe4}   
\end{figure} 

\begin{figure}[t!]
 \centering
 \subfloat[][]{%
  \label{CCfig-a}%
  \includegraphics[width=3.5in,clip=]{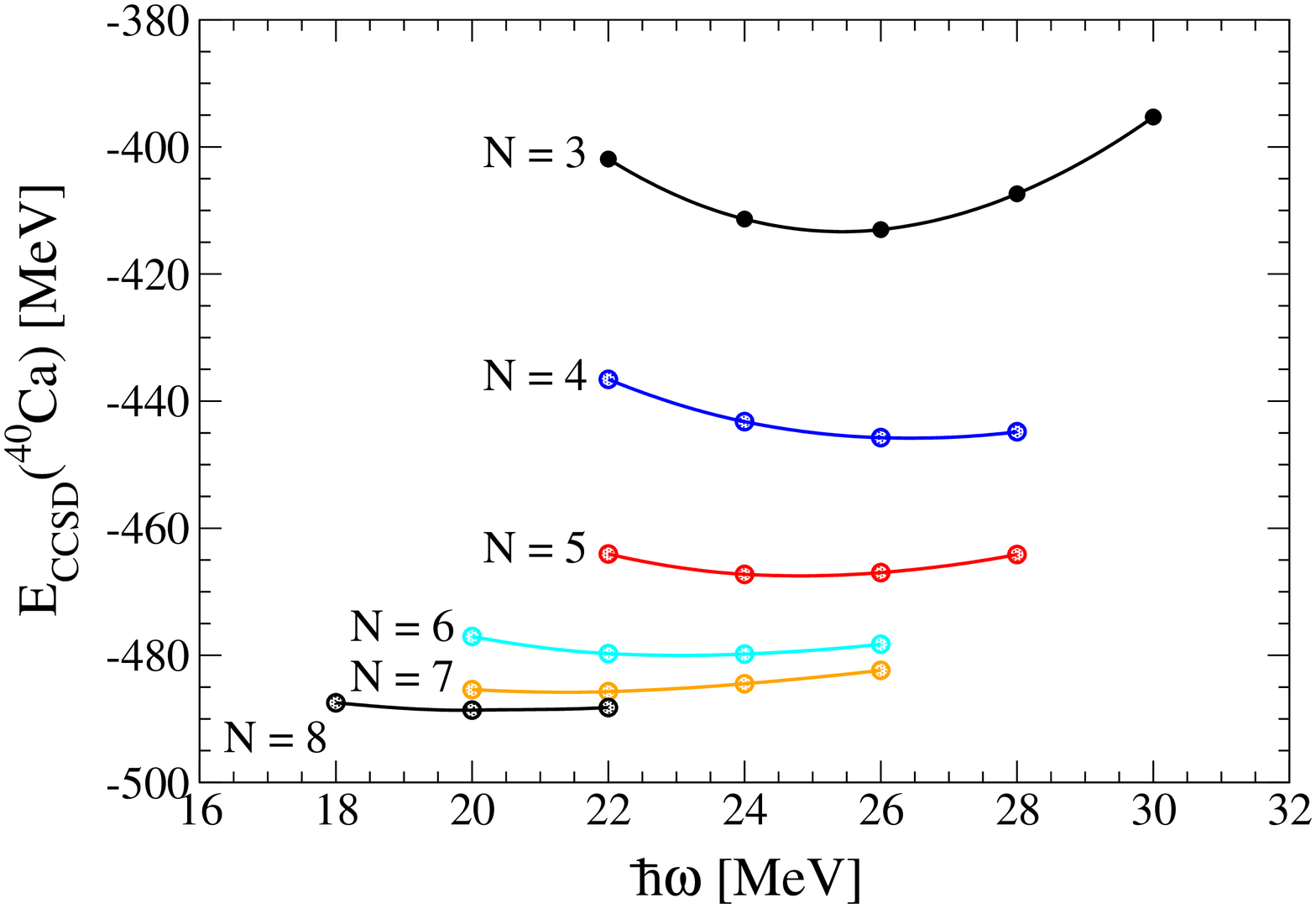}%
 }%
 \hspace*{.2in}%
 \subfloat[][]{%
  \label{CCfig-b}%
  \includegraphics[width=3.1in,clip=]{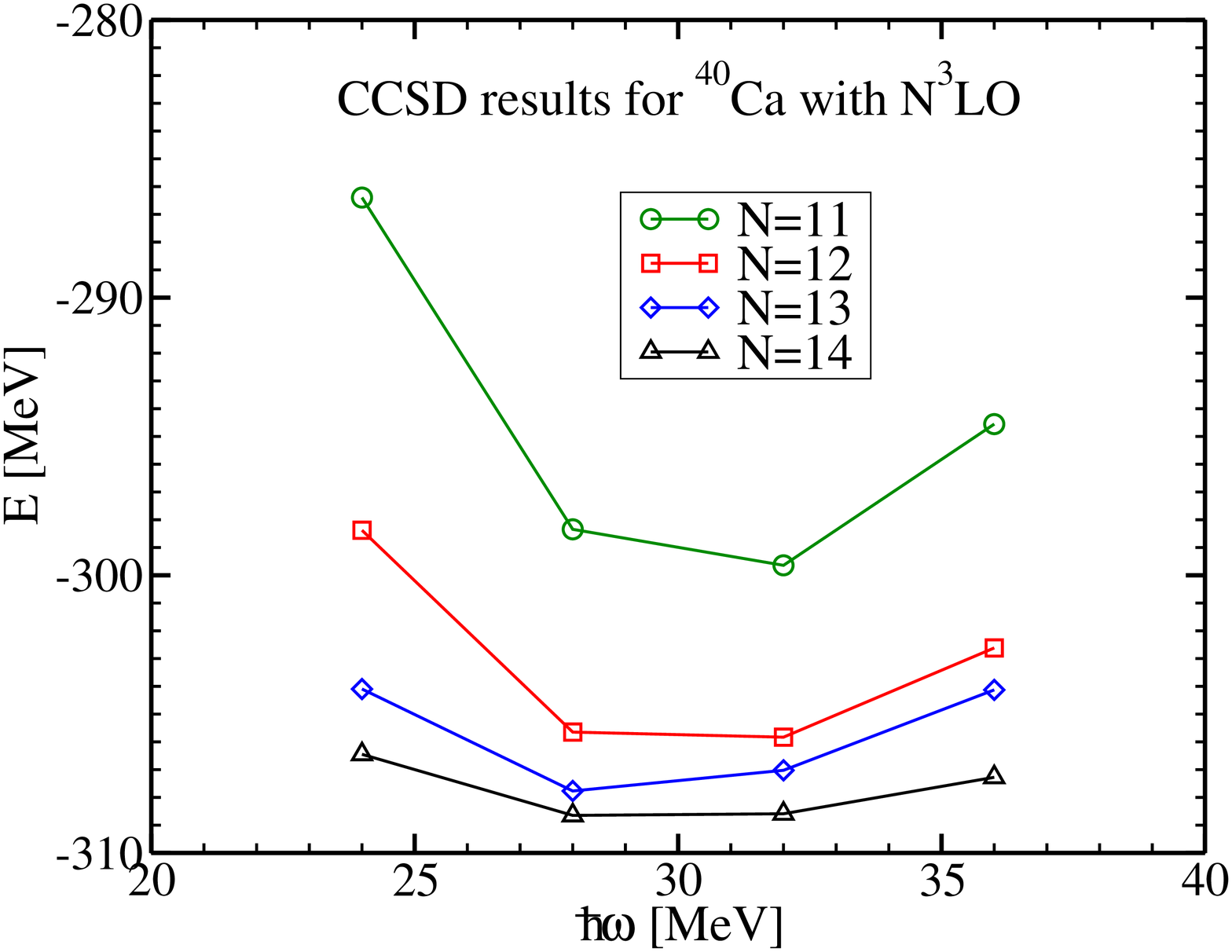}%
 }%
\caption{CCSD ground-state energy of $^{40}$Ca as function of the
oscillator frequency with increasing basis size~$N$,
based on (a)~an RG-evolved low-momentum NN interaction~\cite{Hagen:2007hi} 
and (b)~a N$^3$LO NN potential~\cite{Hagen:2008iw}. 
The difference highlights the importance
of 3N interactions for ground-state energies.\label{CCfig}}
\end{figure}

When the RG equations are truncated at the two-body level, the
evolution is only approximately unitary and converged energies for $A
\geqslant 3$ vary with $\lambda$ or $\lm$. This approximation is systematic,
however, and for useful cutoff ranges, the energy variation shown in
Fig.~\ref{fig:3bodyHe4} is comparable to natural-size truncation errors
in chiral EFT, with no unnaturally large contributions from omitted
three-body forces for  these light nuclei. Further evidence that the
evolution preserves the hierarchy of the underlying chiral EFT is shown
in Figs.~\ref{fig:h3_srg} and \ref{fig:he4_convergence}.

Extensions of ab initio methods to heavier and neutron-rich nuclei are
a frontier of nuclear theory. Coupled-cluster (CC) theory is the prime
method for systems with up to 100 electrons in quantum
chemistry~\cite{Bartlett:2007aa} and a powerful method for nuclei for
which a closed-shell reference state provides a good starting
point~\cite{Gour:2005dm,Hagen:2006pq}. For $^3$H and $^4$He, CC
results agree with the corresponding Faddeev and Faddeev-Yakubovsky
energies~\cite{Hagen:2007hi}. Combined with rapid convergence for
low-momentum interactions, CC theory has pushed the limits of accurate
calculations to medium-mass nuclei and set new benchmarks for $^{16}$O
and $^{40}$Ca~\cite{Hagen:2007hi}. Using an angular-momentum-%
coupled scheme, it is possible to extend CC theory to very large
spaces (15 major shells on a single processor) and to obtain
near-converged ground-state energies for spherical nuclei, $^{40}$Ca,
$^{48}$Ca, and $^{48}$Ni, based on a N$^3$LO NN
potential~\cite{Hagen:2008iw}. The CC developments for medium-mass
nuclei are shown in Fig.~\ref{CCfig}, where the critical importance of
3N forces for ground-state energies is evident.

\begin{figure}[t]
\centering
\includegraphics[width=6.2in,clip=]{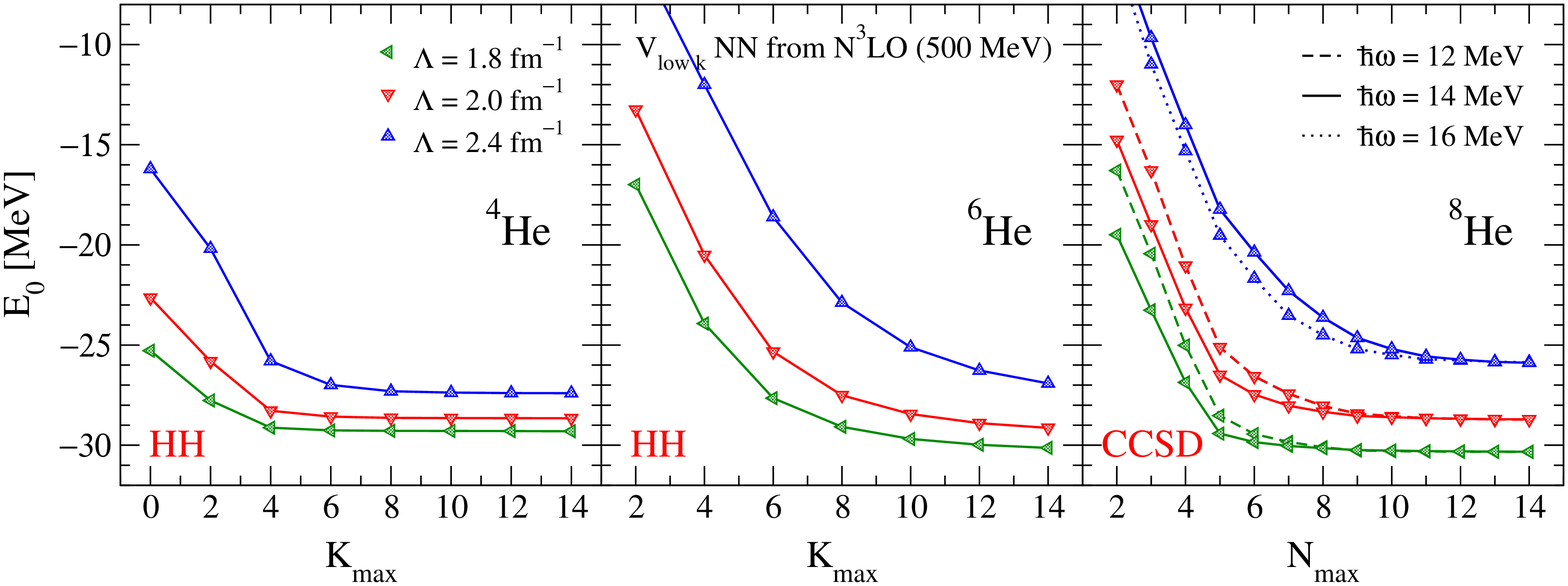}
\caption{Convergence as a function of the basis size for the
HH $^{4}$He and $^{6}$He ground-state
energies and for the CCSD ground-state energy of $^8$He based
on chiral low-momentum NN interactions $\vlowk$ for a range of
cutoffs $\Lambda = 1.8$, $2.0$ and $2.4 \fmi$. For details see
Ref.~\cite{Bacca:2009yk}.\label{4-8He_v2}}
\end{figure}

\begin{figure}[t!]
\centering
\includegraphics[width=3.0in,clip=]{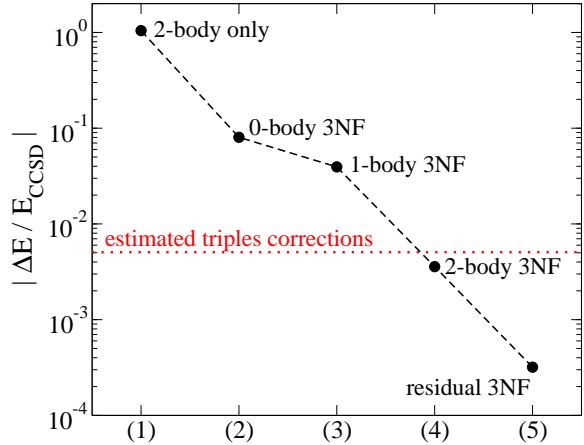}
\vspace*{.1in}
\caption{Relative contributions $|\Delta E / E|$ to the binding energy 
of $^4$He at the CCSD level from $\vlowk$ as well as normal-ordered
zero-, one-, two-body and residual three-body parts of 3N 
forces~\cite{Hagen:2007ew}.\label{3Nparts}}
\end{figure}

Recently, a combination of nuclear and atomic physics techniques led
to the first precision measurements of masses and charge radii of the
helium halo nuclei, $^6$He~\cite{Wang:2004ze,Dilling:talk} and
$^8$He~\cite{Mueller:2008bj,Ryjkov:2008zz}, with two or four
weakly-bound neutrons forming an extended halo around the $^4$He
core. In Fig.~\ref{4-8He_v2}, results are shown for the ground-state
energies of helium nuclei based on chiral low-momentum NN
interactions~\cite{Bacca:2009yk}. This combines the RG evolution with
the exact hyperspherical-harmonics (HH) expansions for $^6$He and CC
theory for $^8$He (see also Ref.~\cite{Hagen:2006pq}), which have the
correct asymptotic behavior of the wave function.  The cutoff
variation in Fig.~\ref{4-8He_v2} highlights the importance of 3N
interactions. For all studied cutoffs, the NN-only results underbind
$^8$He~\cite{Hagen:2006pq,Bacca:2009yk}. Therefore, the helium
isotopes probe 3N effects beyond the overall repulsion 
observed in infinite
nuclear and neutron matter.

Finally, in Fig.~\ref{3Nparts} the first CC results with 3N forces show that
low-momentum 3N interactions are accurately treated as normal-ordered 
zero-, one-, and two-body terms, and that residual 3N forces can be
neglected~\cite{Hagen:2007ew}. This is very promising for developing
tractable approximations to handle many-body interactions (see
Section~\ref{subsec:normal}), and supports the idea that
phenomenological adjustments in shell model interactions
(``monopole shifts'') are due to 3N
contributions~\cite{Zuker:2002pe}. These monopole shifts are enhanced
in neutron-rich nuclei and have a pivotal impact on shell closures and
the location of the neutron drip line, as discussed in
Section~\ref{subsec:shell}.

\subsection{Scattering and reactions}
\label{subsec:scattering}

\begin{figure}[t]
 \centering
 \subfloat[][]{%
  \label{fig:nday-a}%
  \includegraphics[width=3.6in,clip=]{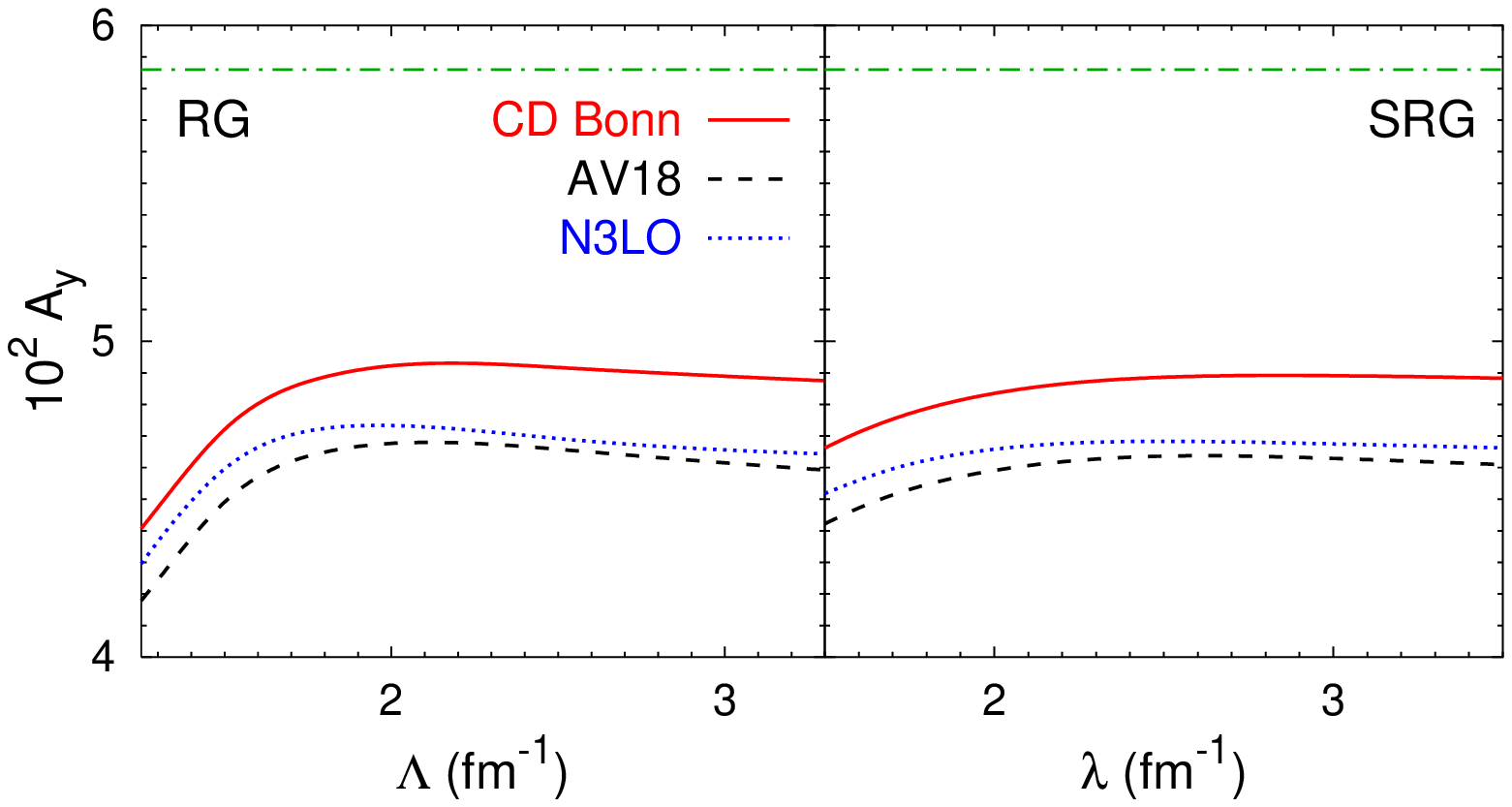}%
 }%
 \hspace*{.1in}%
 \subfloat[][]{%
  \label{fig:nday-b}%
  \includegraphics[width=3.6in,clip=]{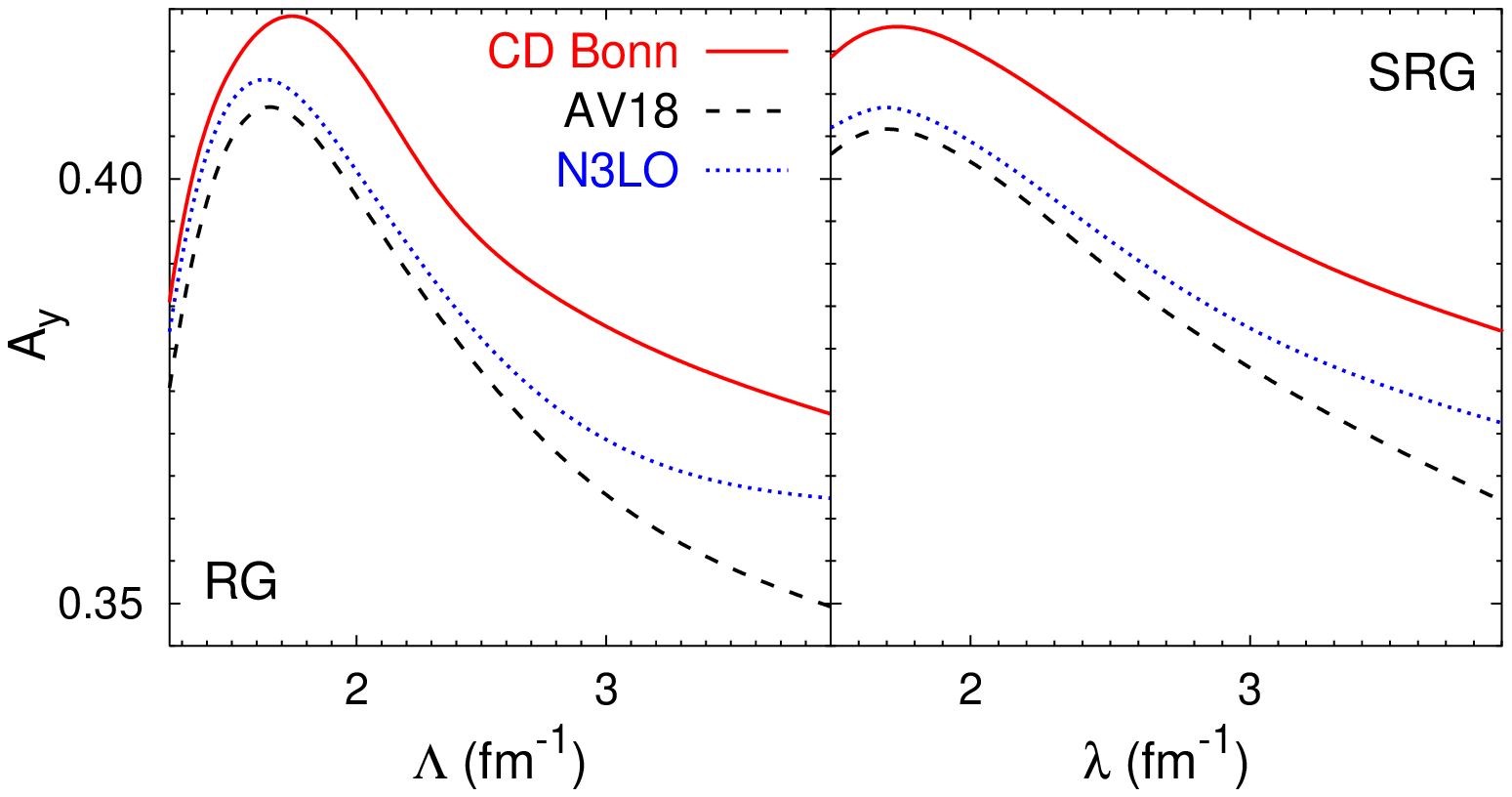}%
 }%
\caption{(a) Neutron analyzing power $A_y$ for neutron-deuteron
scattering at $E_n = 3$ MeV and $\theta_{\rm cm} = 104^\circ$ as
a function of RG cutoff $\Lambda$ (left panel) and SRG parameter
$\lambda$ (right panel). The horizontal line at $10^2 \, A_y = 
5.86$ is the experimental value from \Ref~\cite{mcaninch:93}.
(b)~Maximum of the neutron analyzing power $A_y$ for neutron-$^3$H
scattering at $E_n = 3.5$ MeV as function of RG cutoff $\Lambda$
(left panel) and SRG parameter $\lambda$ (right panel). For details
see Ref.~\cite{Deltuva:2008mv}.\label{fig:nday}}
\end{figure}

In order to probe the sensitivity of 3N and 4N scattering observables
to short-range physics, Ref.~\cite{Deltuva:2008mv} used low-momentum
NN interactions evolved with the RG or SRG to study the variation with
the cutoff $\Lambda$ or $\lambda$. The cutoff dependence due to
truncating the RG equations at the two-body level provides a measure
of the sensitivity to neglected short-range 3N (and higher-body)
forces. Comparing the results shown in Fig.~\ref{fig:nday} one notices
that the cutoff dependence of 3N scattering observables is much weaker
than the one observed for 4N observables. This weak cutoff dependence
in Fig.~\subref*{fig:nday-a} implies that short-range 3N forces are
not likely to solve the long-standing discrepancies of the
nucleon-deuteron $A_y$ in elastic scattering with data. This is indeed
what has been found when short-range or the leading long-range 3N
interactions are included. Nucleon-deuteron $A_y$ in elastic scattering
and the space star differential cross section for breakup barely
change by adding a 2$\pi$-exchange 3N
force~\cite{witala:01,Kievsky:2001fq,kuros:02b}, an effective 3N force
due to the explicit $\Delta$-isobar
excitation~\cite{Deltuva:2003wm,Deltuva:2005zz,Deltuva:2005cc,Deltuva:2005wx},
or N$^2$LO 3N forces~\cite{Epelbaum:2002vt}.  In contrast, the maximum
of the neutron analyzing power $A_y$ for neutron-$^3$H scattering
seems to be more sensitive to short-range many-body forces as
demonstrated by the more pronounced dependence on the cutoff in
Fig.~\subref*{fig:nday-b}. This is consistent with previous
findings~\cite{Deltuva:2006sz} obtained with various NN potentials.

Neutrinos play a crucial role for the physics of stellar collapse,
supernova explosions and neutron
stars~\cite{Raffelt:1996wa,Prakash:2001rx}. The first
calculations~\cite{Lykasov:2008yz,Bacca:2008yr} of neutrino processes
based on chiral EFT and RG-evolved interactions have focused on
neutrino reactions involving two nucleons: neutrino-pair
bremsstrahlung and absorption, $N N \leftrightarrow N N \nu
\overline{\nu}$, which are key for muon and tau neutrino production in
supernovae, and neutrino inelastic scattering, $\nu N N \leftarrow \nu
N N$. These processes are determined by the spin relaxation rate
$1/\tau_\sigma$ using a unified approach to neutrino interactions in
nucleon matter~\cite{Lykasov:2008yz}.  In supernova simulations, the
standard rates for bremsstrahlung are based on the one-pion exchange
approximation to NN interactions~\cite{Hannestad:1997gc}.
Calculations based on chiral EFT and RG-evolved interactions show that
shorter-range non-central forces reduce the spin relaxation rate and
therefore the rate for bremsstrahlung
significantly~\cite{Lykasov:2008yz,Bacca:2008yr}. In addition, the
spin relaxation rate sets the scale for energy transfer in inelastic
scattering, which is presently not included in simulations. The spin
relaxation rates of Ref.~\cite{Bacca:2008yr} lead to a mean-square
energy transfer that is comparable to the incoming neutrino energy,
indicating that NN collisions may be important for energy transfer in
supernovae.

\begin{figure}[t]
 \centering
 \subfloat[][]{%
  \label{fig:ncsmrgm-a}%
  \raisebox{.5in}{\includegraphics[width=3.6in,clip=]{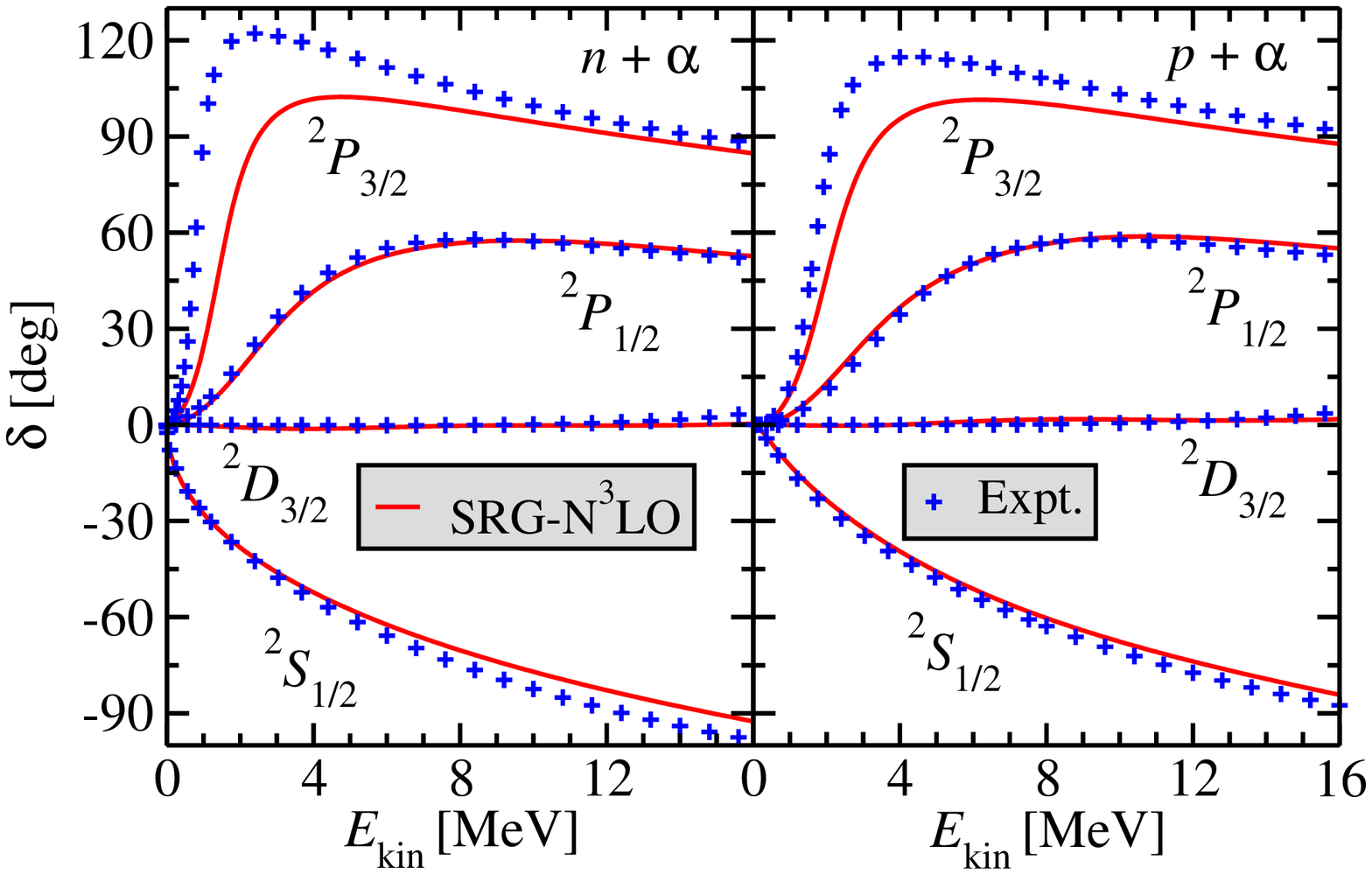}}%
 }%
 \hspace*{.4in}%
 \subfloat[][]{%
  \label{fig:ncsmrgm-b}%
  \includegraphics[width=2.2in,clip=]{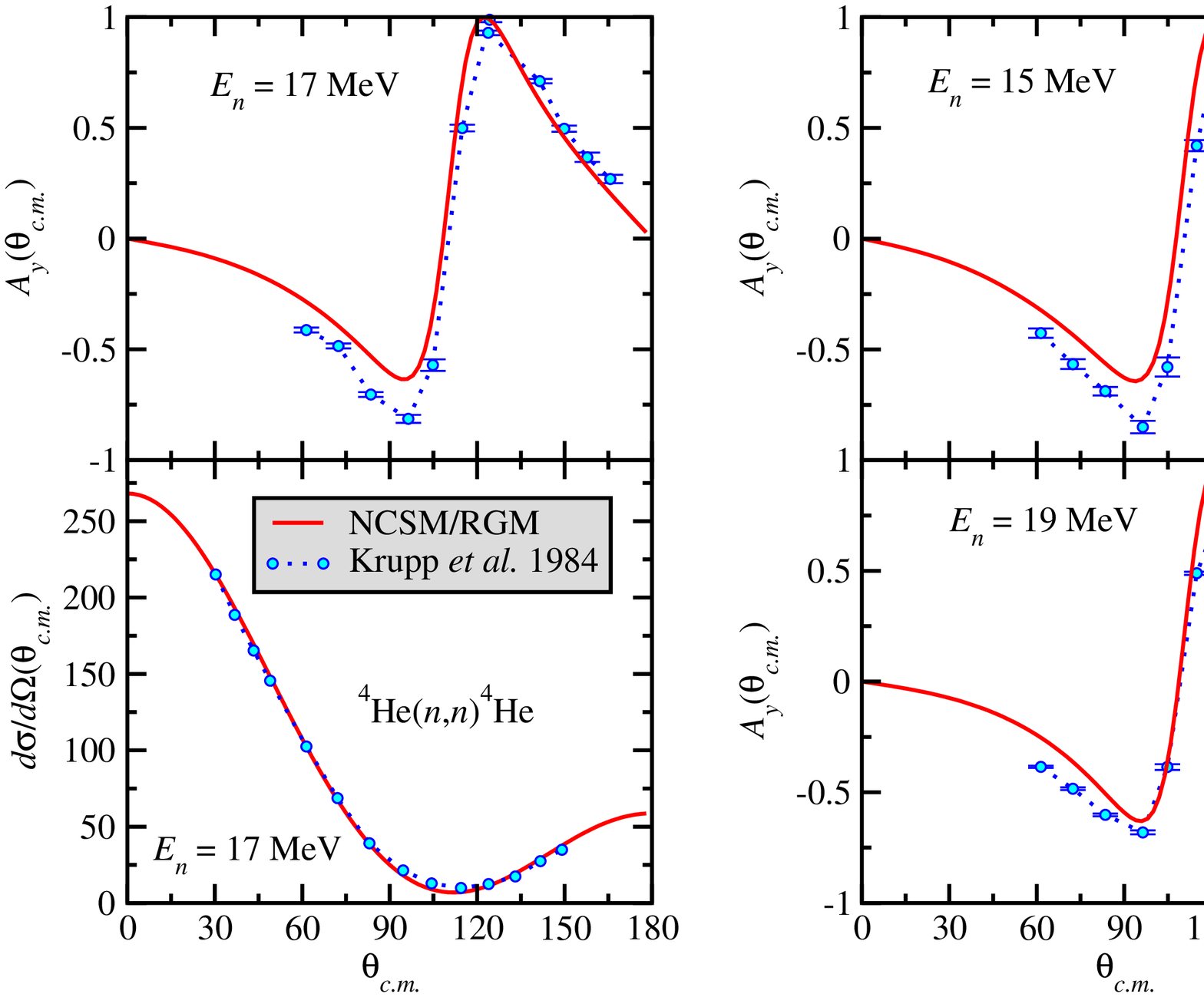}%
 }%
\caption{(a) No-Core Shell Model (NCSM) and Resonating Group Method
(RGM) applied with an SRG-evolved N$^3$LO potential (NN-only, $\lambda
= 2.02 \infm$) to calculate nucleon-alpha elastic scattering phase
shifts at $E_n = 17 \mev$~\cite{Navratil:2009preparation}.
(b)~Comparisons are made to the differential cross section (bottom 
panel) and the analyzing power (top panel) from Karlsruhe 
polarized-neutron experiments~\cite{Navratil:2009preparation}.}
\label{fig:ncsmrgm}
\end{figure}

Quaglioni and Navratil have recently developed an ab initio approach
to nuclear reactions based on merging the No-Core Shell Model (NCSM)
with the Resonating Group Method (RGM), where the many-body Hilbert
space is spanned by cluster wave functions describing a system of two
or more clusters in relative
motion\cite{Quaglioni:2008sm,Quaglioni:2009mn}.
Figure~\ref{fig:ncsmrgm} shows nucleon-alpha elastic scattering at
$E_n = 17 \mev$ using an SRG-evolved N$^3$LO
potential~\cite{Navratil:2009preparation}. As with the ground-state
NCSM calculations, the use of low-momentum interactions in the
NCSM/RGM description of nucleon-nucleus scattering results in faster
convergence with respect to the basis size.  This is a very promising
start.

\subsection{Shell-model approaches}
\label{subsec:shell}

\begin{figure}[t]
\centering
\includegraphics[scale=0.4,clip=]{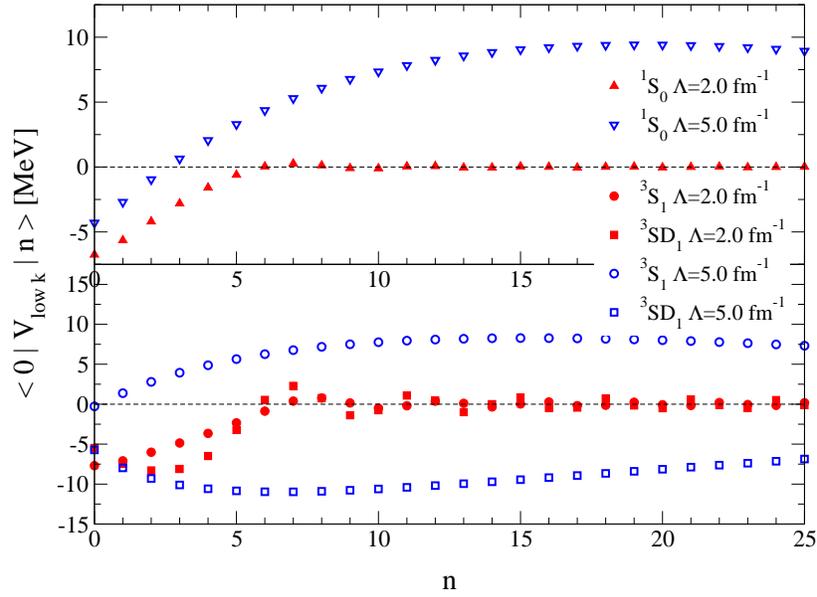}
\caption{Relative harmonic-oscillator matrix elements $\langle n'=0 \,
l' | \vlowk^{S J} | n \, l \rangle$ versus radial quantum number
$n$ for a low-momentum cutoff $\lm = 2.0 \, {\rm fm}^{-1}$ and a
$1.0 \, {\rm GeV}$ cutoff $\lm = 5.0 \, {\rm fm}^{-1}$. Results are
shown for S-wave matrix elements. In both cases, $\vlowk$ is evolved
from the Argonne $v_{18}$ potential and the harmonic oscillator 
parameter is $\hbar \omega = 14 \, {\rm MeV}$~\cite{Schwenk:2004hz}.
\label{fig:relho}}
\end{figure}

Shell-model applications were one of the main motivations for developing
low-momentum interactions~\cite{Bogner:2001yi,Bogner:2002yw}.
The practical benefit of low-momentum interactions in shell model
applications is shown in Fig~\ref{fig:relho}, where we compare S-wave
relative harmonic-oscillator matrix elements for a low-momentum cutoff
$\lm = 2.0 \fmi$ and a $1.0 \, {\rm GeV}$ cutoff $\lm = 5.0 \fmi$. We
find that for lower cutoffs the matrix elements decrease quite rapidly
and become small for $|n - n'| \sim 5-10$. This is not the case for
interactions with high-momentum components, which require basis states
up to $\sim 50$ shells for convergence.

\begin{figure}[t!]
 \centering
 \subfloat[][]{%
  \label{spectra-a}%
  \includegraphics[width=3.0in,clip=]{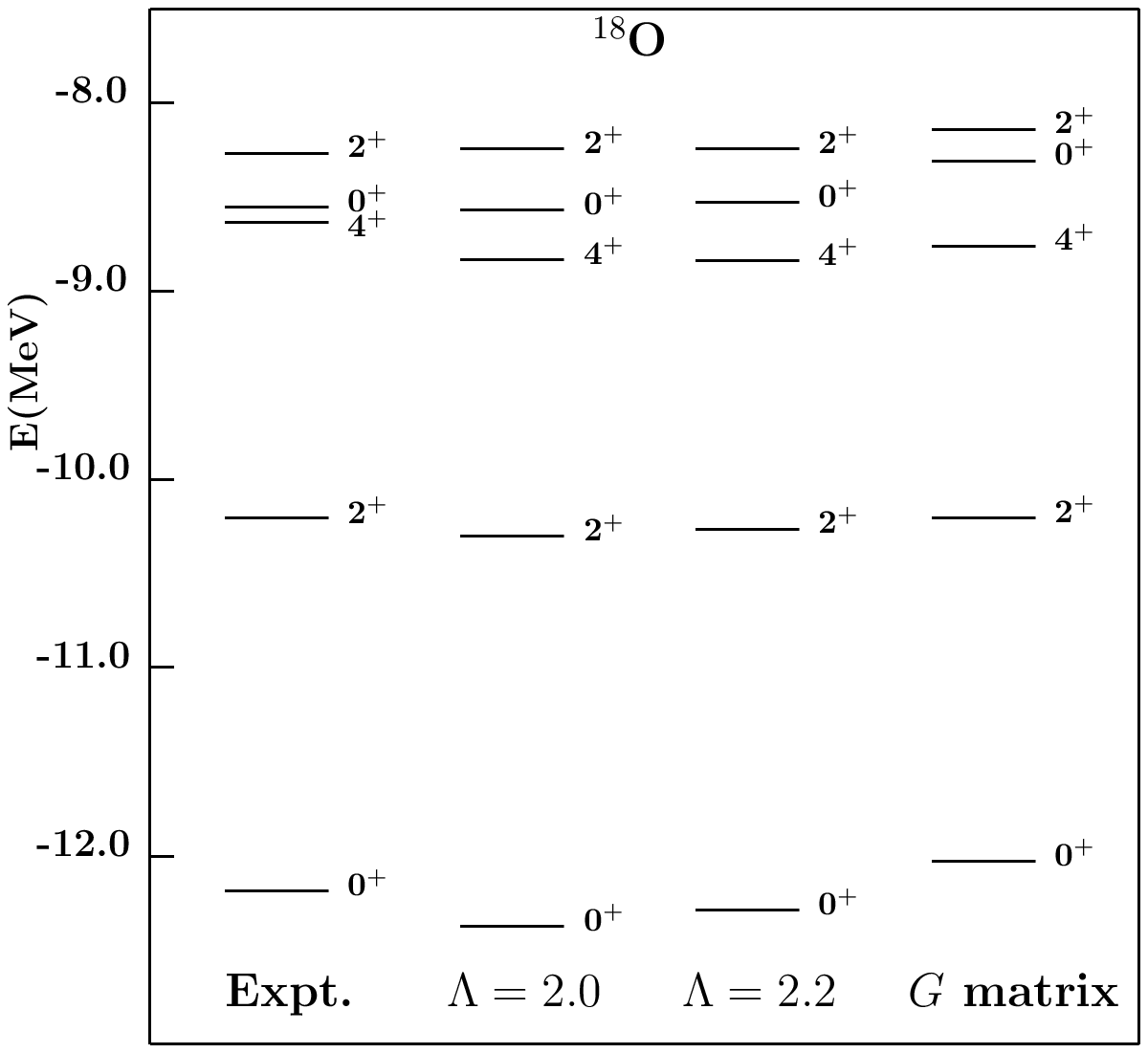}%
 }%
 \hspace*{.4in}%
 \subfloat[][]{%
  \label{spectra-b}%
  \includegraphics[width=3.0in,clip=]{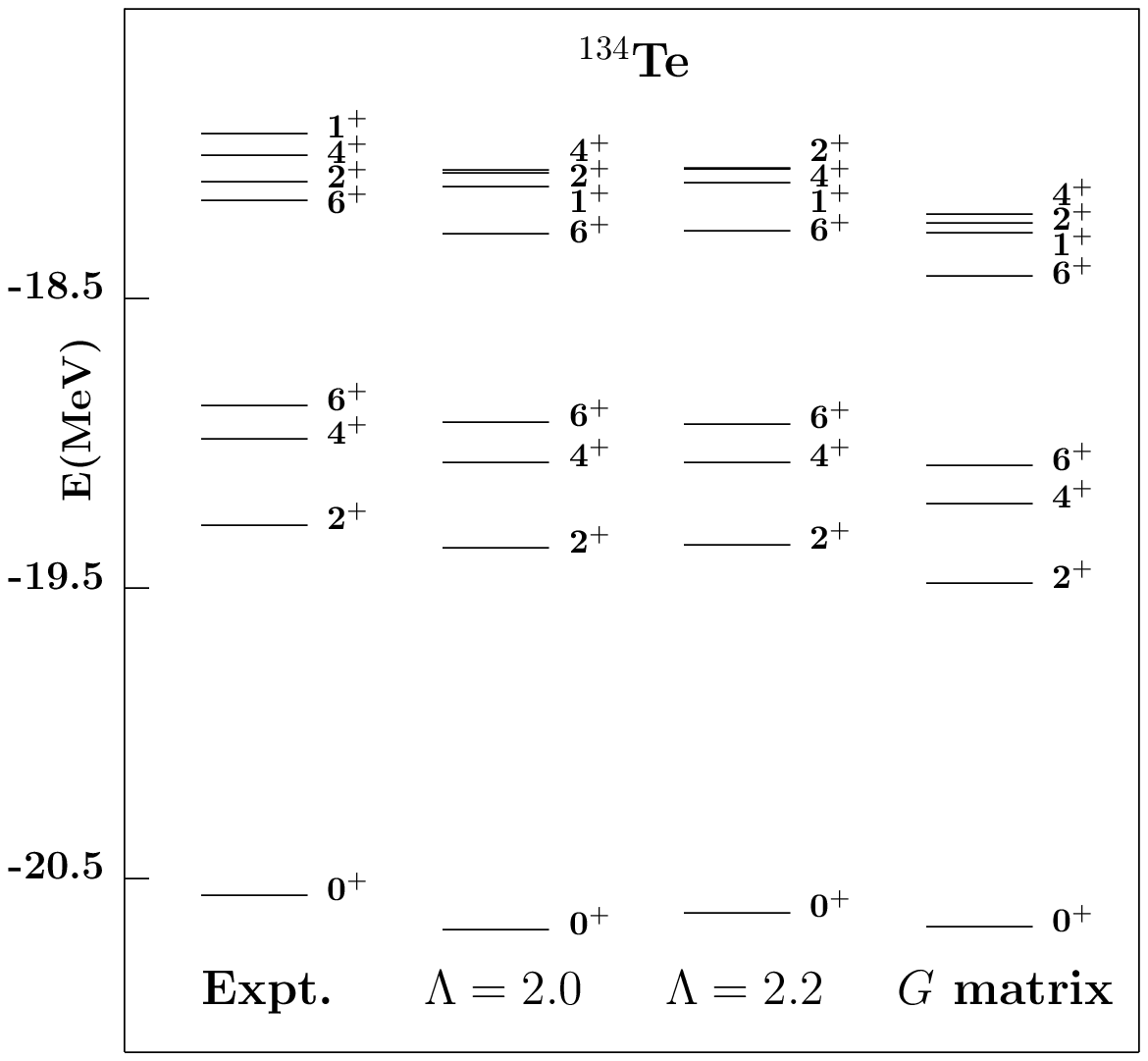}%
 }%
\caption{Low-lying spectra using shell-model effective Hamiltonians
derived from $\vlowk$ and $G$ matrix interactions starting from the
CD-Bonn potential~\cite{Machleidt:2000ge} for (a)~$^{18}$O and
(b)~$^{134}$Te~\cite{Coraggio:2008in}.\label{fig:spectra}}
\end{figure}

In conventional approaches to shell-model effective interactions, the
strong short-range repulsion is tamed by performing a ladder
resummation of the NN potential to obtain a $G$
matrix~\cite{HjorthJensen:1995ap}. However, the $G$ matrix resummation
introduces an inconvenient starting-energy dependence and requires
further approximations in practice. Furthermore, as discussed in
Section~\ref{subsec:other}, $G$ matrices do not decouple low- and
high-energy states. As shown in Fig.~\ref{fig:spectra}, low-lying
spectra of $^{18}$O and $^{134}$Te based on low-momentum NN
interactions are of comparable quality to using $G$ matrix
resummations that are difficult to generalize to 3N forces and involve
starting-energy dependences. In addition, as a consequence of
decoupling, perturbative valence shell-model calculations starting
from low-momentum interactions are under better control than the
corresponding $G$ matrix calculations.
We refer to Ref.~\cite{Coraggio:2008in} for a review on low-momentum
interactions in perturbative calculations of valence shell-model
effective interactions. Moreover, there are promising applications of
low-momentum interactions in the Gamow shell model to handle continuum
states~\cite{Hagen:2006in,Michel:2008pt,Tsukiyama:2009hy}.

\begin{figure}[t]
\centering
\includegraphics[width=6.2in,clip=]{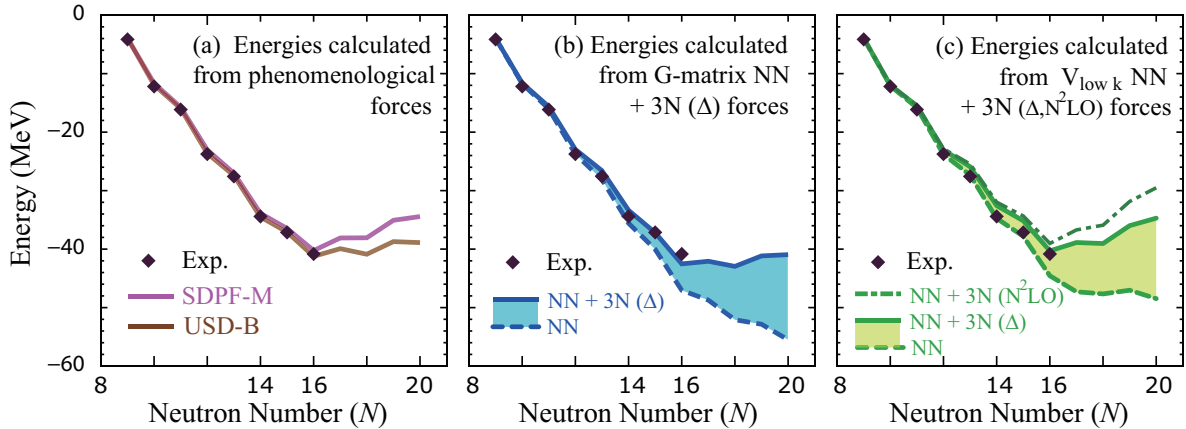}
\caption{Ground-state energies of neutron-rich oxygen isotopes
measured from the energy of $^{16}$O. The experimental energies of
the bound oxygen isotopes $^{16-24}$O are included for
comparison. The left panel (a) shows the energies obtained from the
phenomenological forces SDPF-M~\cite{Utsuno:1999st,Utsuno:2004vr} and
USD-B~\cite{Brown:2006gx}. The middle panel (b) gives the energies obtained
from a $G$ matrix and including Fujita-Miyazawa 3N forces due to $\Delta$
excitations~\cite{Otsuka:2009cs}. The right panel (c) presents the 
energies calculated from $\vlowk$ and 
including chiral EFT 3N interactions at N$^2$LO as well as only
due to $\Delta$ excitations~\cite{Otsuka:2009cs}. The changes due to 
3N forces based on $\Delta$ excitations are highlighted by the shaded
areas.\label{gs}}
\end{figure}

The neutron drip-line, which is the limit of neutron-rich nuclei,
evolves regularly from light to medium-mass nuclei except for a
striking anomaly in the oxygen isotopes.  This anomaly is not
reproduced in theories derived from two-nucleon forces. In
Ref.~\cite{Otsuka:2009cs}, the first microscopic explanation of the
oxygen anomaly based on low-momentum 3N forces was presented. As shown
in Fig.~\ref{gs}, the inclusion of 3N interactions at N$^2$LO or due
to $\Delta$ excitations leads to repulsive contributions to the
interactions among valence neutrons that change the location of the
neutron drip-line from $^{28}$O to the experimentally observed
$^{24}$O.  This 3N mechanism is robust and general, and therefore
expected to impact predictions of the most neutron-rich nuclei and
the synthesis of heavy elements in neutron-rich environments.

\subsection{Density functional theory}
\label{subsec:dft}

While most advances in microscopic nuclear structure theory over the
last decade have been through expanding the reach of few-body
calculations, infinite nuclear matter is still a key step to heavier
nuclei.  In particular, the promising results using low-momentum
interactions open the door to ab initio density functional theory
(DFT) both directly (through orbital-based methods) and based on
expanding about nuclear matter~\cite{Bogner:2008kj}. This is analogous
to the application of DFT in quantum chemistry and condensed matter
starting with the uniform electron gas in local-density approximations
and adding constrained derivative corrections. Phenomenological energy
functionals (such as Skyrme) for nuclei have impressive successes but
lack a (quantitative) microscopic foundation based on nuclear forces
and seem to have reached the limits of improvement with the current
form of functionals~\cite{Bertsch:2004us,Kortelainen:2008rp}.
Furthermore, without theoretical understanding of errors,
extrapolations to the limits of nuclear binding are uncontrolled.

A recent review~\cite{Drut:2009ce} describes various theoretical paths
to ab initio (meaning microscopic) DFT.  One formal constructive
framework for Kohn-Sham DFT is based on effective actions of composite
operators using the inversion method
\cite{FUKUDA94,VALIEV97,Puglia:2002vk,%
Bhattacharyya:2004qm,Bhattacharyya:2004aw,Furnstahl:2004xn}.  This
is an organization of the many-body problem that is based on
calculating the response of a finite system to external, static
sources rather than seeking the many-body wave function.  Other paths
include many-body perturbation theory or RG methods applied to
effective actions~\cite{Drut:2009ce,Polonyi2001,Schwenk:2004hm}.
Either way necessitates a tractable expansion that is controllable in
the presence of inhomogeneous sources or single-particle potentials.
This is problematic for conventional nuclear forces, for which the
single-particle potential needs to be tuned to enhance the convergence
of the hole-line expansion \cite{Day:1967zz,Baldo99}, but is ideally
suited for low-momentum interactions.

The nuclear matter results of Fig.~\ref{nm_all} and
Ref.~\cite{Bogner:2005sn,Bogner:2009un} imply that exchange
correlations are tractable, and this motivates the derivation of an ab
initio density functional based on low-momentum interactions, for
example, using a density matrix expansion~\cite{Bogner:2008kj} (see
also Ref.~\cite{Furnstahl:2008df,Kaiser:2002jz}).  This is one of the
goals of the SciDAC universal nuclear energy density functional
project (UNEDF)~\cite{unedf:2007}.  Future work can use EFT and RG to
identify new terms in the functional, to quantify theoretical errors
for extrapolations, and to benchmark with ab initio methods (such as
with coupled-cluster theory for medium-mass nuclei).

The theoretical errors of nuclear matter calculations based on
low-momentum interactions, while impressively small on the scale of
the potential energy per particle, are far too large to be
quantitatively competitive with existing energy functionals. However,
there is the possibility of fine tuning to heavy nuclei, of using
EFT/RG to guide next-generation functional forms, and of benchmarking
with ab initio methods for low-momentum interactions. Work in these
directions is in progress (see for example Ref.~\cite{Drut:2009ce}).
Overall, these results are very promising for a unified description of
all nuclei and nuclear matter but much work is left to be done.

\begin{figure}[t]
\centering
\includegraphics[width=6.0in,clip=]{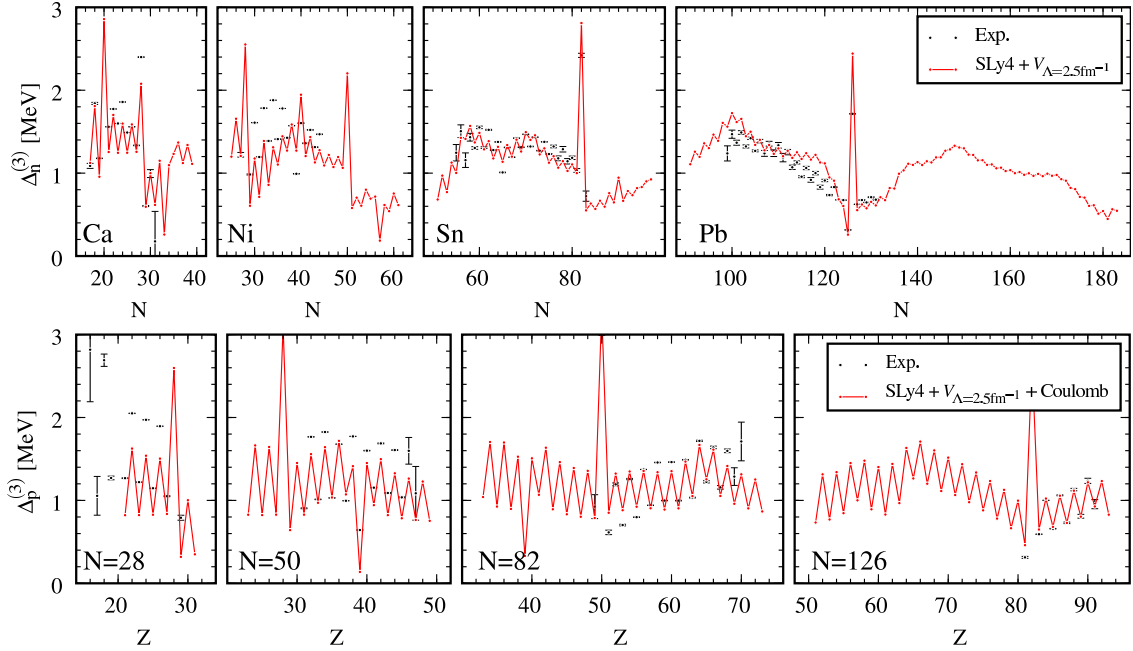}
\caption{Neutron/proton pairing gaps (upper/lower panel) obtained from
calculated and experimental odd-even mass differences along
isotopic/isotonic chains. The calculated gaps use the $^1$S$_0$ part
of $\vlowk$ as pairing interaction in density functional 
calculations (plus Coulomb for proton 
gaps)~\cite{Duguet:2009gc,Lesinski:2009private}.}
\label{nonemppair}
\end{figure}

Recent results using existing energy density functionals in the
particle-hole channel (to build a reasonable self-consistent
single-particle basis) combined with low-momentum interactions
$\vlowk$ in the pairing channel suggest that an ab initio DFT
treatment of pairing is feasible~\cite{Duguet:2007be,%
Lesinski:2008cd,Hebeler:2009dy,Duguet:2009gc,Baroni:2009eh}.
Figure~\ref{nonemppair} shows that the resulting neutron-neutron and
proton-proton pairing gaps in semi-magic nuclei are remarkably
consistent experiment~\cite{Duguet:2009gc,Lesinski:2009private}.
Systematic investigations of theoretical corrections, including 3N
forces, are underway.


\section{Summary and outlook}
\label{sec:summary}

As illustrated in the preceding sections, the use of low-momentum
interactions has positively impacted few- and many-body calculations 
for nuclear structure and reactions in all regions of the nuclear
chart. The advantages of softened potentials are unequivocal and
recent work on evolving 3N forces addresses some of the most pressing
concerns about the RG approach. There remain, however, some
misunderstandings and misconceptions about the application and
implications for nuclear problems of the RG and related methods and
how they connect to EFT. In addition, there is much work in progress
with associated open questions and as yet unrealized opportunities.
Here we reiterate many of our main points by clarifying some
of the misconceptions and surveying the on-going research.

\subsection{Misconceptions and clarifications}
\label{subsec:misconceptions}

In the following we address some of the misleading statements or
common misconceptions associated with low-momentum interactions that we have
encountered in the literature as well as informally.

\begin{itemize}

\item \emph{The nuclear potential is an observable.}  Even if not
  stated explicitly, this is possibly an implicit belief 
  inherited from experience (in both classical and quantum contexts) with
  the Coulomb potential, and the idea that the potential measures the
  energy of two particles at fixed separation.  This intuition breaks
  down, of course, for finite-mass composite particles unless at large
  separations.  Because the short-range part of the potential can be
  modified with a unitary transformation without altering experimental
  predictions, it is not an observable by any meaningful definition.

\item \emph{Both two-nucleon phase shifts and lattice QCD calculations
  show that the $NN$ interaction has a strongly repulsive core.}  We
  reiterate that a short-range quantum mechanical potential is not a
  measurable quantity.  Fitting the change in sign of S-wave phase
  shifts from attractive to repulsive with a local potential leads to
  a repulsive core, but there are an infinity of phase-equivalent
  potentials without this feature.  Similarly, the non-local
  interaction obtained from lattice simulations and converted to
  phase-equivalent local form for comparison to
  phenomenology~\cite{Aoki:2009ji} is based on a non-unique choice of
  nucleon interpolating field~\cite{Beane:2008ia,Hatsuda:2009kq}.
   
\begin{figure}[t]
 \centering
 \subfloat[][]{%
  \label{fig:properties-a}%
  \includegraphics[width=3.2in,clip=]{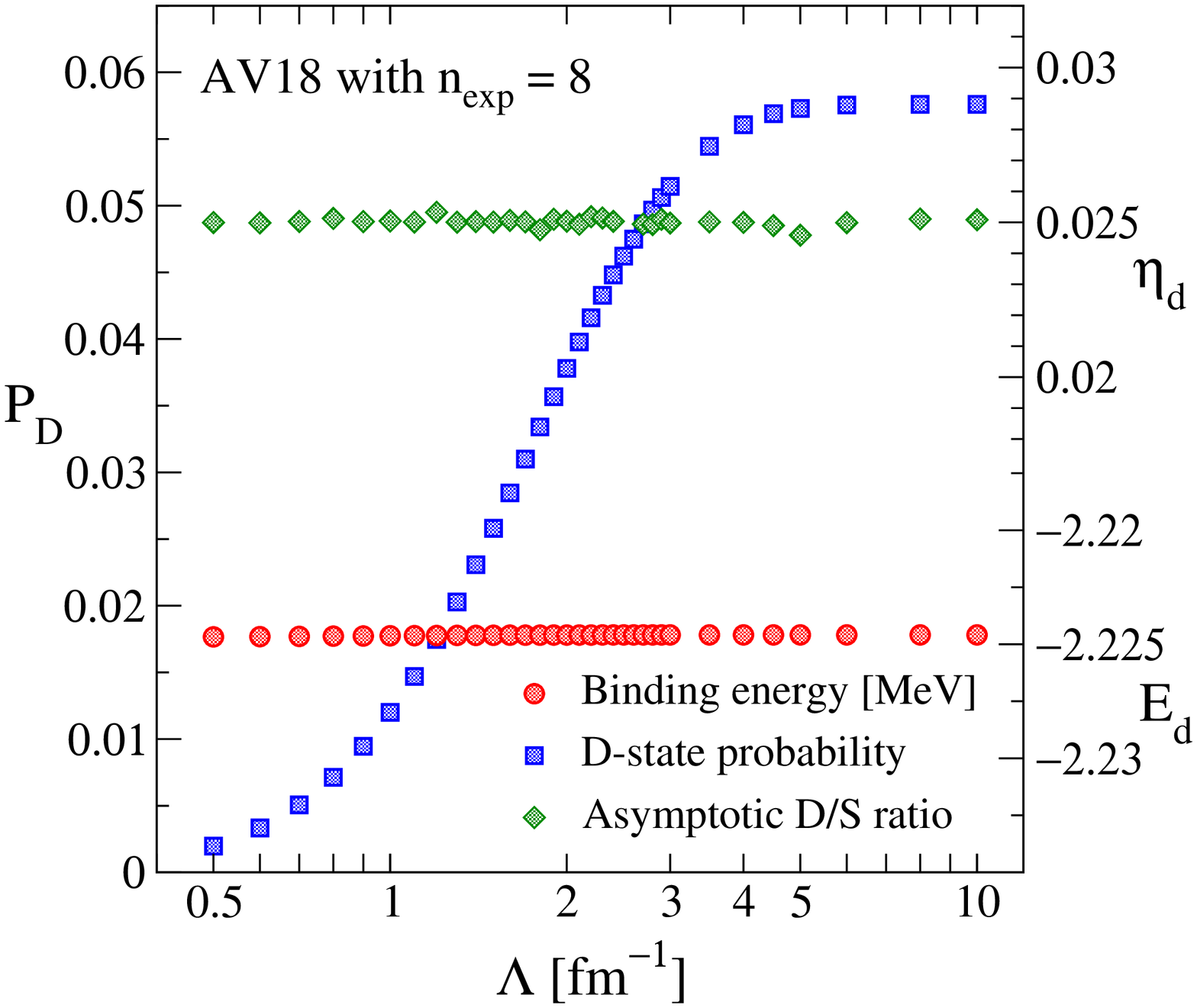}%
 }%
 \hspace*{.4in}%
 \subfloat[][]{%
  \label{fig:properties-b}%
  \includegraphics[width=3.2in,clip=]{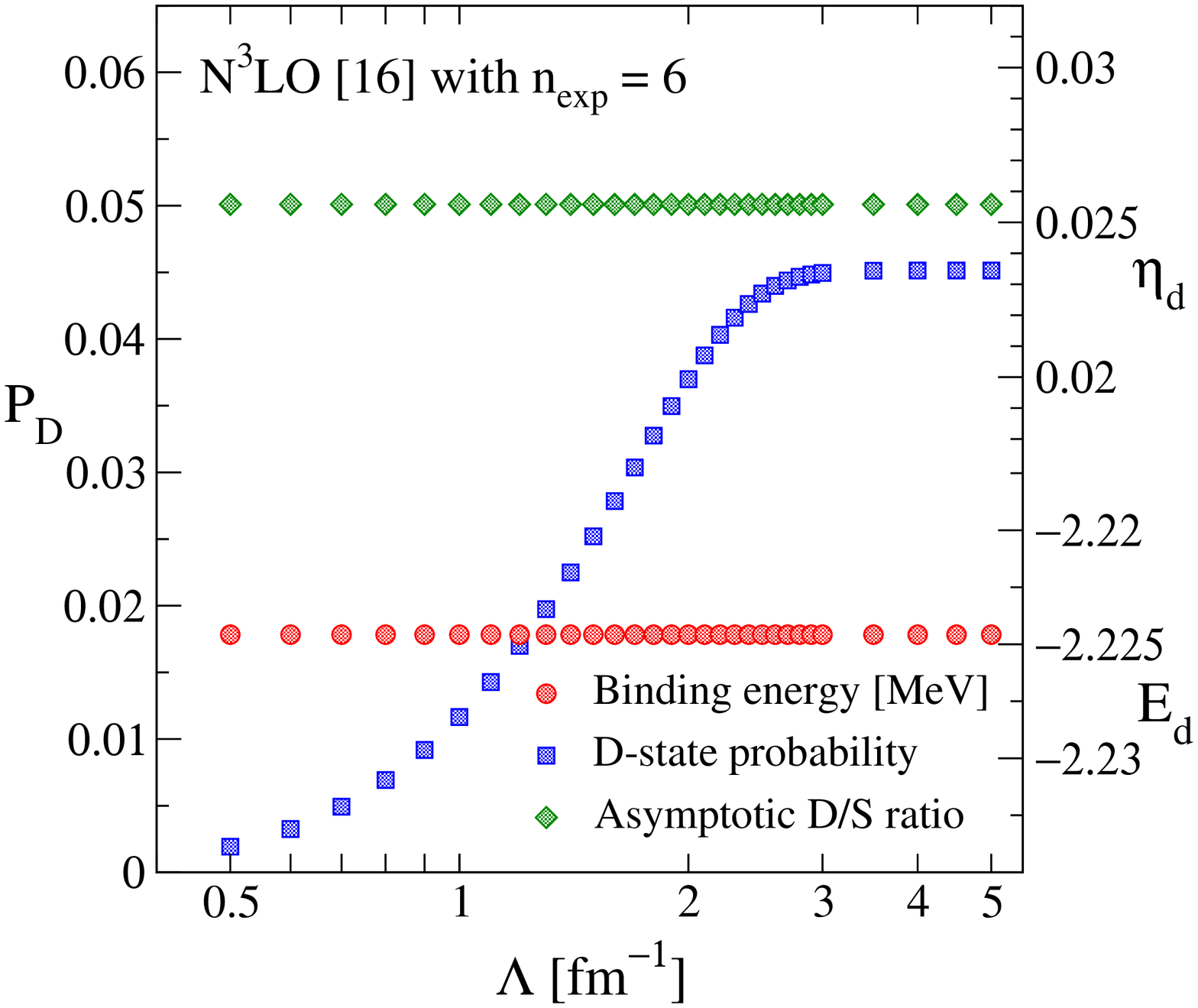}%
 }%
\caption{D-state probability $P_{\rm D}$ (left axis), binding energy 
$E_d$ (lower right axis), and asymptotic D/S-state ratio $\eta_{d}$
(upper right axis) of the deuteron as a function of the 
cutoff~\cite{Bogner:2006vp}, starting 
from (a) the Argonne $v_{18}$~\cite{Wiringa:1994wb} and (b) the
N$^3$LO NN potential of Ref.~\cite{Entem:2003ft} using different
smooth $\vlowk$ regulators. Similar results are found with SRG evolution.}
\label{fig:properties}
\end{figure}

\item \emph{Low-momentum interactions change observable properties
  such as the D-state probability.}  As long as the corresponding
  operators are evolved together with the Hamiltonian, observables
  will remain unchanged.  While it is true that the D-state
  probability of the deuteron is reduced as interactions are evolved
  toward lower cutoffs, this is not an
  observable~\cite{Amado:1979zz,Friar:1979zz}.
  Figure~\ref{fig:properties} illustrates some observable and
  non-observable properties of the deuteron.
  Up to numerical errors, the measurable binding energy
  and the asymptotic D/S-state ratio are independent of the
  low-momentum resolution, while the unobservable D-state probability
  changes dramatically as the short-range part of the tensor force is
  integrated out.

\item \emph{Details of strong-interaction dynamics at high energy are
  important for low-energy nuclear structure and reactions.} More
  precisely, it has been claimed that a good (or at least reasonable)
  description of phase shifts above some energy (such as $E_{\rm lab}
  = 350 \mev$) is required for a Hamiltonian to be useful for nuclear
  structure. While it would seem natural that low- and high-energy
  physics are decoupled, this does not happen automatically and the
  misconception is often driven by experience with conventional NN
  potentials with strong short-range repulsion.  The repulsion causes
  even bound states with very low energies (such as the deuteron) to
  have important contributions to the binding energy and other
  observables from high-momentum components (well above $2\fmi$,
  see Fig.~\ref{fig:deutbinding}) of
  the deuteron wave function.\footnote{We emphasize that the need for
  high-momentum components in these particular calculations
  \emph{does not} imply that the high-energy description is
  correct.}  For example, Ref.~\cite{Benhar:1993ja} presents cross
  sections for electron scattering from the deuteron calculated using
  as input a spectral function that is the momentum distribution times
  a delta function in energy. After noting the effect of excluding
  high momenta on the cross section, the conclusion is that ``the data
  confirm the existence of high-momentum components in the deuteron
  wave function''. Beyond the fact that the short-range parts of wave
  functions are not observables, these conclusions reinforce the
  intuition that there is information in quantitatively reproducing
  high-energy phase shifts that is lost when evolving to low-momentum
  interactions. However, due to decoupling for the RG-evolved 
  interactions, it is completely irrelevant if phase shifts are
  zero above some energy.

\item \emph{Short-range correlations are important for low-energy
  properties of nuclei.} Various electron scattering and proton
  scattering experiments purport to show the existence of short-range
  correlations in nuclei. These results are then interpreted as
  implying that such correlations are an important feature of the
  structure of medium to heavy nuclei (see, for example,
  Refs.~\cite{Frankfurt:2008zv,Frankfurt:2009vv}). As shown for the
  deuteron in Figs.~\subref*{fig:srcorr-a}
  and~\subref*{fig:deuteronwfsp2-a}, the short-range part of a wave
  function is not an observable and is radically changed by
  short-range unitary transformations that preserve all
  observables. Therefore any statement about short-range correlations
  necessarily relies on the (model-dependent) short-distance
  details of the Hamiltonian, which in turn depend on the
  resolution. As has been illustrated in this review, eliminating
  these correlations by evolving to low resolution greatly enhances
  the convergence of few- and many-body calculations.

\item \emph{The low-momentum RG technology fails for observables that
  are sensitive to high-momentum components.} In principle,
  calculated observables~\cite{Schiavilla:2006xx,Alvioli:2007zz} from
  experiments such as $(e,e'pN)$ are unchanged after RG evolution
  whether or not they involve high momentum, because the associated
  operators (which are generally not simple at high momentum) will
  transform to ensure this invariance. In practice there are many
  open questions, such as whether consistent operator evolution is
  difficult to approximate accurately or if factorization
  approximations might be more valid with low-momentum interactions.

\item \emph{Low-momentum interactions are not systematic.} If one
  starts from a chiral EFT interaction determined to a given order in
  the power counting, the truncation error of the initial chiral EFT
  is exactly preserved when many-body interactions are evolved as
  well. If many-body interactions are truncated, the
  error remains of the same natural size for cutoffs used in practical
  calculations. Therefore, low-momentum interactions preserve the
  systematic nature of the initial EFT and EFT improvements are
  immediately folded into the RG evolution. In addition, the control
  of the EFT truncation errors is not lost when two-body forces are
  evolved exactly but the three-body forces are fit using
  chiral 3N interactions as a truncated basis. On the other hand, if
  one starts from phenomenological potentials, then it is correct
  that the low-momentum Hamiltonian is not systematically improvable
  because the starting point is not either.
 
\item \emph{Low-momentum interactions are only useful as long as
  calculations with ``bare'' chiral EFT interactions are not
  possible for all nuclei.} We reiterate that there is nothing
  superior about a ``bare'' interaction as long as the RG-evolved
  interactions reproduce observables with a comparable truncation
  error. Further, converged calculations with different cutoffs are
  very desirable to quantify theoretical uncertainties. Finally, the
  certainty that three-body (and possibly four-body) forces will be
  needed for precision calculations implies that softer interactions
  will continue to be necessary to push to larger nuclei and to
  reactions even with the anticipated gains in computing power. (Note
  that normal-ordering simplifications for many-body forces are more 
  effective for low-momentum Hamiltonians.)

\item \emph{There is only one $\vlowk$ or SRG potential.} In
  fact there are infinitely many different low-momentum interactions,
  corresponding to different initial potentials (often called ``bare''
  interactions) and different choices for the evolution (cutoff value
  and renormalization scheme, such as sharp, smooth and SRG). The same
  holds for chiral EFT interactions, which depend on the regulator and
  cutoff values, as well as on the level of truncation. There
  is an analogous misconception that there is only one N$^3$LO NN
  potential, based perhaps on the dominant use of the $\lm = 500 \mev$
  potential of Ref.~\cite{Entem:2003ft}.  Note that the full advantages
  of low-momentum potentials are only realized if more than one cutoff
  is used.

\item \emph{The RG evolution of nuclear forces to lower momentum
  improves the physics.} To our knowledge, this is not a statement
  that has actually been made by a $\vlowk$ or SRG proponent, but one
  which is often thought to be implied. However, if the initial
  Hamiltonian provides a poor representation of nature at low
  energies, it will not be improved by the RG evolution described
  here. The correct statement is that an exact RG evolution to lower
  momentum \emph{preserves} the observable physics, while making it
  (generally) easier to calculate. If we could easily calculate all
  many-body systems of interest exactly with an initial Hamiltonian,
  then there would be no need for the RG evolution. (Unless changing
  the resolution was useful for other reasons, such as enhancing
  physical insight.)

\item \emph{Evolved low-momentum interactions are strongly non-local.}
  This is misleading because it omits that the non-locality generated
  by RG evolution to lower resolution increases with the momentum
  transfer $q$.\footnote{This contrasts with a Perey-Buck-type
  factorized non-local potential~\cite{Perey:1962aa} such as
  $V({\bf r},{\bf r'}) = V_{\rm local}[({\bf r}+{\bf r'}/2)]
  (\pi^{1/2}\beta)^{-3} \exp[({\bf r}-{\bf r'})^2/\beta^2]$, for
  which the non-locality is independent of the momentum transfer.}
  The local long-range (low $q$) part of low-momentum interactions
  (given by one-pion exchange) is therefore \emph{not} modified by the
  evolution. Given the composite nature of nucleons,
  non-locality at short range (high $q$) is natural.

\item \emph{Low-momentum interactions should reproduce the initial
  NN-only results for $A > 2$.} The RG evolution to lower resolution
  shifts interaction strength from two-body to few-body contributions,
  so it is \emph{guaranteed} that NN-only calculations will disagree
  for $A>2$. A related misconception is: \emph{Low-momentum
  interactions are approximations to ``bare'' potentials.} Because
  induced few-body contributions will be truncated at some level,
  there will be differences from the ``bare'' result. But any ``bare''
  interaction is also a truncated low-energy effective theory, so
  there should be no preference over low-momentum interactions with
  different but still natural few-body contributions.

\item \emph{Soft interactions are not realistic Hamiltonians because
  they overbind nuclei compared to experiment.} 
  This statement tends to follow observations
  that nuclei calculated with NN-only low-momentum interactions are
  increasingly overbound with increasing
  $A$. But this is only true if 3N forces (or the uncertainties from
  neglecting 3N forces) are omitted, which is not justified. The
  test in the end is whether soft Hamiltonians including many-body
  forces reproduce observables.

\item \emph{Many-body forces will explode after integrating out strong
  short-range repulsion.} The calculations reviewed in the preceding
  sections demonstrate explicitly that this is not the case (for
  example, low-momentum 3N interactions are perturbative in light
  nuclei). Many-body contributions are natural in size and the decreasing
  hierarchy is preserved.
  A related
  misconception is: \emph{To include 3N forces with low-momentum
  interactions, it is important to distinguish between ``real'' (or
  ``genuine'') and ``induced'' three-body forces.} The initial and
  evolved 3N forces are both ``effective'' interactions. Because the
  EFT truncation error is preserved in the evolution, there is no
  physics reason to prefer one over the other.  Each Hamiltonian has
  different associated many-body interactions and operators, although
  the long-range parts are the same as they are not modified by the RG
  evolution.

\item \emph{Weren't soft potentials ruled out for nuclear physics long
  ago?}  Soft potentials were explored in the mid-sixties and early
  seventies to improve convergence in nuclear matter and nuclei. In
  particular, separable potentials such as the Tabakin potential were
  considered, as well as unitary transformations of local,
  repulsive-core potentials~\cite{Coester:1970ai,Haftel:1971er}. These
  NN-only soft potentials were abandoned primarily because they did
  not saturate nuclear matter in the empirical range, but at too-high
  densities and with overbinding. In contrast, with low-momentum
  interactions saturation is driven by the (natural-sized) 3N forces
  fit to three- and four-body properties.
  
\item \emph{Nuclear matter saturation with low-momentum interactions
  is largely due to the (long-range) two-pion-exchange part of 3N
  forces. But for conventional NN potentials, two-pion-exchange 3N
  forces are attractive for all nuclei and nuclear
  matter~\cite{Akmal:1998cf,Pieper:2001ap}.} The distinction is because
  decoupling in low-momentum interactions means that 3N
  contributions are from the low-momentum (long-range) parts of 3N
  forces, whereas in calculations with conventional potentials,
  two-pion-exchange 3N forces are also sensitive to the
  (model-dependent) short-range parts of wave functions. This leads to
  a net attraction for hard potentials~\cite{Coon:1978gr}. Due to
  the importance of 3N forces for nuclei, a quantitative understanding
  of this change is an important open problem.
  
\item \emph{$\vlowk$ is just like a $G$ matrix.} The $G$ matrix
  approach leads to a softened version of an initial NN potential and
  one can see strong similarities between a $G$ matrix and $\vlowk$
  \emph{for matrix elements connecting low-energy states} (explicit
  comparisons are shown in Figs.~\ref{fig:gmatrix} and
  \ref{fig:relho}). But the $G$ matrix is an in-medium interaction
  that depends on the nucleus or the density and on a ``starting
  energy'', while low-momentum interactions are (generally) evolved
  in free space. Most important is that the decoupling between low and
  high momenta that is the hallmark of low-momentum interactions is
  \emph{not} achieved by the $G$ matrix construction. The off-diagonal
  contributions, evident in Fig.~\ref{fig:gmatrix}, make $\vlowk$
  qualitatively different from a $G$ matrix. A consequence is that
  while the $G$ matrix is a softened effective interaction, the
  off-diagonal parts will be significant in higher-order
  diagrams. This is why a perturbative expansion in the $G$ matrix
  fails and something like the hole-line expansion is needed (with a
  similar $G$ matrix resummation at the three-body level).
  
\end{itemize}

\subsection{Open problems and opportunities}
\label{subsec:open}

We finish with a survey of work in progress to extend the results
described in this review together with open problems that may lead to
new opportunities.  This is, of course, not a comprehensive list, as
there will be other efforts of which we are unaware and because new
ideas constantly arise.  But we hope this will give a reasonable
vision of what to expect in the next few years.

Chiral EFT power counting of the hierarchy of many-body forces implies
that 3N forces must be included in almost any precision calculation of
nuclear properties; this is evident already from microscopic
calculations of light nuclei.\footnote{Calculations with low-momentum
NN interactions are useful to establish convergence properties and
to use the cutoff dependence to learn about the omitted forces, or
in certain applications where 3N contributions are small, such as
low-density neutron matter.} Ongoing work includes the effects of 3N
forces in one of three ways: matching to a truncated chiral EFT basis
(Section~\ref{subsec:3nf}), explicit evolution with RG flow equations
(Section~\ref{subsec:evolution}), or through normal-ordering and the
in-medium SRG (Section~\ref{subsec:normal}).

The recent evolution of 3N forces with SRG flow equations in a
harmonic-oscillator basis will allow important tests of the matching
approach and whether the low-momentum universality observed at the NN
level (the collapse of different initial potentials) extends to
three-body forces. These studies can also provide useful insights to
more general EFT questions regarding the consistency of fits to NN and
3N interactions and the power counting of three-body forces. In
addition, the SRG can be extended to evolve 3N forces in momentum
space (with the kinetic energy generator) and to generate consistent
3N interactions for $\vlowk$ using the SRG block diagonalization.  We note 
that the formal equivalence between the block-diagonal SRG and
$\vlowk$ is not yet established. Likewise, the very similar results
in few- and many-body calculations using smooth cutoff $\vlowk$ and 
band-diagonal SRG (NN-only) interactions is not yet understood.

The availability of low-momentum 3N interactions opens the door to
many new results in ab initio calculations of finite nuclei using
NCSM, hyperspherical harmonics, and coupled-cluster methods. An
important open problem is to determine the range of resolutions for
which the RG evolution can be used to soften nuclear Hamiltonians
while preserving the hierarchy of many-body interactions.  The study
of residual cutoff dependencies in calculations of finite nuclei and
infinite matter will help answer this question and provide critical
tests of the RG approach.  Work on the consistent evolution of
operators will lead to new opportunities for exploring electroweak
processes in nuclei and for assessing theoretical uncertainties of key
nuclear matrix elements.

In addition, an important direction for future work is to develop and
test tractable approximations to handle many-body forces. The first
applications of low-momentum 3N interactions to the shell model are
very exciting and open many opportunities for exotic nuclei. The
development of the in-medium SRG is promising to include many-body
interactions through normal-ordering and to provide a non-perturbative
approach to valence shell-model effective interactions and operators.
The applications to reactions are another frontier. The NCSM/RGM
combination has shown promise with low-momentum NN interactions (both
$\vlowk$ and SRG) and the extension to include 3N forces is an
additional challenge. Revisiting the nucleon-nucleus optical potential
with low-momentum interactions is also inviting, given the improved
status of Hartree-Fock as a starting point.

Calculations with evolved cutoff-dependent 3N forces open the door to
study 4N interactions. Precision spectroscopy and low-energy reaction
theory will require some control and assessment of the size of 4N
forces. We note that the density dependence of 4N forces is not known
for \emph{any} low-energy nuclear Hamiltonian. Various approximations
for evolving 4N interactions are suggested by analyzing the
diagrammatic form of the SRG~\cite{Bogner:2007qb}.

While the SRG is currently formulated in terms of coupled differential
equations, experience with RG evolution for $\vlowk$ suggest that an
integral form may be more robust numerically. A possible starting
point is the formalism developed in Ref.~\cite{Walhout:1998ig}. There
are also many open questions concerning the relationship of the RG
techniques discussed here to the wide range of RG technology applied
to condensed matter and high-energy problems. For example, what are
transformations that convert energy to momentum dependencies? And is
there a path integral formulation of the SRG?

The low-momentum RG technology that has been developed can also be
usefully applied directly to EFT. Even though the widely available
chiral EFT potentials have cutoffs as low as $500 \mev$, these are not
sharp cutoffs and significant high momentum strength is still
present. As shown in Ref.~\cite{Bogner:2006vp}, an exponential
regulator with $n_{\rm exp} \approx 8$ leads to weaker distortions
than the conventional value of $n_{\rm exp} = 3$ used in the N$^3$LO
NN potentials~\cite{Entem:2003ft,Epelbaum:2004fk}. At still lower
cutoffs, an SRG-type regulator may have additional
advantages. Experience with lower momentum regulators could be merged
with expertise on fitting EFT low-energy constants to optimize the EFT
at lower (softer) cutoffs for testing power counting as well as for
applications to few- and many-body methods. This will help resolve
questions about the utility of $\vlowk$ interactions compared to
unevolved chiral EFT with lower cutoffs and possibly facilitate the
use of error plots for the EFT~\cite{Lepage:1997cs}.

For infinite matter, a power counting is needed for low-momentum
interactions that specifies controlled many-body truncations at finite
density. The Bethe-Brueckner-Goldstone expansion, which involves a
summation of particle-particle ladders into the $G$ matrix, shows (for
potentials with strong short-range repulsion) that a perturbative
expansion in terms of the $G$ matrix does not work and so further
summations are required. (The summation in this case is organized by
the hole-line expansion.) Indications are that both summations are
unnecessary with low-momentum interactions, but a quantitative
power counting (for example, using the momentum dependence of the
interactions together with phase-space constraints) needs to be
demonstrated. This will involve understanding the evolution of
particle-hole physics and what are the non-pertubative aspects of
low-momentum interactions for nuclear structure.

Finally, there are also opportunities for applications beyond ordinary
nuclei.  These include low-momentum interactions between baryons that
carry strangeness for descriptions of hypernuclei (see, for example,
Refs.~\cite{Fujii:2000cg,Schaefer:2005fi,Kohno:2007ng,Dapo:2008qv}).
Renormalization group methods are already applied widely in
condensed matter physics, but the special experience from nuclei may
be of advantage. The application to microscopic inter-atomic
potentials is an example that may prove fruitful.

In closing, we reiterate that methods to derive low-momentum
interactions not based (at least explicitly) on RG evolution are being
pursued with promising results and prospects. The future interplay of
the different approaches will be fruitful. When low-momentum
interactions are coupled with the advances of many-body methods, the
steady improvements in computational algorithms and power, and the
on-going developments in chiral EFT and lattice QCD, the dream of
microscopic calculations connecting QCD to the full range of nuclei
and their reactions is within reach.

\section*{Acknowledgments}

We thank E.~Anderson, S.~Bacca, S.~Baroni, J.~Drut, T.~Duguet,
K.~Hebeler, J.~Holt, E.~Jurgenson, R.~Perry, L.~Platter and K.~Wendt
for useful comments on this review.  This work was supported in part
by the National Science Foundation under Grant Nos.~PHY--0653312,
PHY--0758125, and PHY--0456903, the UNEDF SciDAC Collaboration under
DOE Grant DE-FC02-07ER41457, the Natural Sciences and Engineering
Research Council of Canada (NSERC), and by the Helmholtz Alliance
Program of the Helmholtz Association, contract HA216/EMMI ``Extremes
of Density and Temperature: Cosmic Matter in the Laboratory''.  TRIUMF
receives federal funding via a contribution agreement through the
National Research Council of Canada.

\bibliographystyle{h-elsevier_new} 

\bibliography{vlowk_refs} 

\end{document}